%% file: main.tex
\documentclass[prb,aps,nofootinbib,amssymb,twocolumn,superscriptaddress,10pt,longbibliography]{revtex4-1}

\usepackage{times}
\usepackage{graphicx}
\usepackage{float}
\usepackage{latexsym,amsmath,amssymb,bm,euscript}
\usepackage{color}
\usepackage{subfigure}
\usepackage{epstopdf}
\usepackage[colorlinks=true,linkcolor=blue,citecolor=blue,urlcolor=blue]{hyperref}
\usepackage{hyperref}
\usepackage{comment}
\usepackage{cleveref}
\usepackage{siunitx}
\usepackage{multirow}
\usepackage{svg}
\usepackage{physics}
\svgpath{{./figure/}}
\usepackage{multirow}
\usepackage{bbm}
\usepackage{tabularx}
\usepackage{ltablex}
\usepackage{placeins}
\usepackage{tikz}
\usetikzlibrary{decorations.pathreplacing}
\usepackage{ytableau}
\usepackage{dcolumn}% Align table columns on decimal point
\usepackage{bm}% bold math
\usepackage{array,booktabs,tabularx}
\usepackage[USenglish]{babel}
\usepackage{xcolor}
\usepackage{amsfonts}

\usepackage{placeins}
\usepackage{chemformula}

\usepackage[normalem]{ulem} % For strikethrough (sout)
\usepackage{xcolor} % For colored text

\RequirePackage[normalem]{ulem} %DIF PREAMBLE
\RequirePackage{color}\definecolor{RED}{rgb}{1,0,0}\definecolor{BLUE}{rgb}{0,0,1} %DIF PREAMBLE
\RequirePackage{listings} %DIF PREAMBLE
\RequirePackage{color} %DIF PREAMBLE
\lstdefinelanguage{DIFcode}{ %DIF PREAMBLE
	moredelim=[il][\color{red}\sout]{\%DIF\ <\ }, %DIF PREAMBLE
	moredelim=[il][\color{blue}\uwave]{\%DIF\ >\ } %DIF PREAMBLE
} %DIF PREAMBLE
\lstdefinestyle{DIFverbatimstyle}{ %DIF PREAMBLE
	language=DIFcode, %DIF PREAMBLE
	basicstyle=\ttfamily, %DIF PREAMBLE
	columns=fullflexible, %DIF PREAMBLE
	keepspaces=true %DIF PREAMBLE
} %DIF PREAMBLE
\lstnewenvironment{DIFverbatim}{\lstset{style=DIFverbatimstyle}}{} %DIF PREAMBLE
\lstnewenvironment{DIFverbatim*}{\lstset{style=DIFverbatimstyle,showspaces=true}}{} %DIF PREAMBLE
\makeatletter
\renewcommand\onecolumngrid{% <<<<<<
	\do@columngrid{one}{\@ne}%
	\def\set@footnotewidth{\onecolumngrid}% <<<<<<<<<<<<<<<<
	\def\footnoterule{\kern-6pt\hrule width 1.5in\kern6pt}%
}

\def\ba#1\ea{\begin{equation}\begin{aligned}#1\end{aligned}\end{equation}}

\newcommand{\bmat}[0]{\begin{bmatrix}}
	\newcommand{\emat}[0]{\end{bmatrix}}

\begingroup
\edef\@tempa{\meaning\middle}
\edef\@tempb{\string\middle}
\expandafter \endgroup \ifx\@tempa\@tempb
\def\mid@vertical{\middle|}
\def\mid@dblvertical{\middle\SavedDoubleVert}
\else
\def\mid@vertical{\mskip1mu\vrule\mskip1mu}
\def\mid@dblvertical{\mskip1mu\vrule\mskip2.5mu\vrule\mskip1mu}
\fi

\allowdisplaybreaks

\crefname{appendix}{Appendix}{Appendices}
\crefname{equation}{Eq.}{Eqs.}
\crefname{figure}{Fig.}{Figs.}
\crefname{table}{Table}{Tables}
\crefname{section}{Section}{Sections}
\crefname{enumi}{Case}{Cases}
\creflabelformat{appendix}{[#2#1#3]}
\crefrangelabelformat{figure}{#3#1#4--(#5\crefstripprefix{#1}{#2}#6}
\crefmultiformat{figure}{Figs.~#2#1\xdef\mycreffirstarg{#1}#3}%
{ and~#2#1#3}{, #2#1#3}{ and~#2#1#3}
\makeatletter     
\renewcommand\onecolumngrid{% <<<<<<
	\do@columngrid{one}{\@ne}%
	\def\set@footnotewidth{\onecolumngrid}% <<<<<<<<<<<<<<<<
	\def\footnoterule{\kern-6pt\hrule width 1.5in\kern6pt}%
}

\makeatletter
\AtBeginDocument{\let\LS@rot\@undefined}
\makeatother

\newcommand{\beq}{\begin{equation}}
	\newcommand{\eneq}{\end{equation}}

\renewcommand{\qq}{\mathbf{q}}
\newcommand{\kk}{\mathbf{k}}

\newcommand{\RR}{\mathbf{R}}

\newcommand{\QQ}{\mathbf{Q}}

\newcommand{\hH}{{ \hat{H} }}

\newcommand{\titlePaper}{FeGe as a building block for the kagome 1:1, 1:6:6, and 1:3:5 families: hidden d-orbital decoupling of flat band sectors, effective models and interaction Hamiltonians}

\newcommand{\paperAuthors}{%
	\author{Yi Jiang}
	\thanks{These authors contributed equally to this work.}
	\affiliation{Beijing National Laboratory for Condensed Matter Physics and Institute of Physics, Chinese Academy of Sciences, Beijing 100190, China}
	\affiliation{University of Chinese Academy of Sciences, Beijing 100049, China}
	\affiliation{Donostia International Physics Center (DIPC), Paseo Manuel de Lardizábal. 20018, San Sebastián, Spain}
	
	\author{Haoyu Hu}
	\thanks{These authors contributed equally to this work.}
	\affiliation{Donostia International Physics Center (DIPC), Paseo Manuel de Lardizábal. 20018, San Sebastián, Spain}
	\affiliation{Department of Physics, Princeton University, Princeton, New Jersey 08544, USA}

	\author{Dumitru C\u{a}lug\u{a}ru}
	\thanks{These authors contributed equally to this work.}
	\affiliation{Department of Physics, Princeton University, Princeton, New Jersey 08544, USA}
	
	\author{Claudia Felser}
	\affiliation{Max Planck Institute for Chemical Physics of Solids, 01187 Dresden, Germany}
	
	\author{Santiago Blanco-Canosa}
	\affiliation{Donostia International Physics Center (DIPC), Paseo Manuel de Lardizábal. 20018, San Sebastián, Spain}
	\affiliation{IKERBASQUE, Basque Foundation for Science, 48013 Bilbao, Spain}

	\author{Hongming Weng}
	\affiliation{Beijing National Laboratory for Condensed Matter Physics, and Institute of Physics, Chinese Academy of Sciences, Beijing 100190, China}
	\affiliation{Songshan Lake Materials Laboratory, Dongguan, Guangdong 523808, China}

	\author{Yuanfeng Xu}
	\email{y.xu@zju.edu.cn}
	\affiliation{Center for Correlated Matter and School of Physics, Zhejiang University, Hangzhou 310058, China}
	
	\author{B. Andrei Bernevig}
	\email{bernevig@princeton.edu}
	\affiliation{Department of Physics, Princeton University, Princeton, New Jersey 08544, USA}
	\affiliation{Donostia International Physics Center (DIPC), Paseo Manuel de Lardizábal. 20018, San Sebastián, Spain}
	\affiliation{IKERBASQUE, Basque Foundation for Science, 48013 Bilbao, Spain}
}

\newcommand{\PreserveBackslash}[1]{\let\temp=\\#1\let\\=\temp}

\crefname{appendix}{Appendix}{Appendices}
\crefname{equation}{Eq.}{Eqs.}
\crefname{figure}{Fig.}{Figs.}
\crefname{table}{Table}{Tables}
\crefname{section}{Section}{Sections}
\creflabelformat{appendix}{[#2#1#3]}

\makeatletter
\AtBeginDocument{\let\LS@rot\@undefined}
\makeatother

\makeatletter
\renewcommand\onecolumngrid{% <<<<<<
\do@columngrid{one}{\@ne}%
\def\set@footnotewidth{\onecolumngrid}% <<<<<<<<<<<<<<<<
\def\footnoterule{\kern-6pt\hrule width 1.5in\kern6pt}%
}

\newcommand{\citeSI}[1]{(see \cref{#1})}

\allowdisplaybreaks

\begin{document}

% Note: title and authors are written in macros-private.sty file
\title{\titlePaper}
\paperAuthors

\input{manuscript_bare}

\let\oldaddcontentsline\addcontentsline
\renewcommand{\addcontentsline}[3]{}
%\bibliography{ref}


\let\addcontentsline\oldaddcontentsline

\clearpage

\onecolumngrid
\pagebreak
\thispagestyle{empty}

\newpage
\begin{center}
    \textbf{\large Supplemental Material: \titlePaper}\\[.2cm]
\end{center}

\appendix
\renewcommand{\thesection}{\Roman{section}}
\renewcommand{\thetable}{S\arabic{section}.\arabic{table}}
\renewcommand{\thefigure}{S\arabic{section}.\arabic{figure}}
\renewcommand{\theequation}{S\arabic{section}.\arabic{equation}}

\tableofcontents
\let\oldaddcontentsline\addcontentsline
\newpage
\input{supplement_bare}

\end{document}

%% file: manuscript_bare.tex
\begin{abstract}
The electronic structure and interactions of kagome materials, such as the 1:1 (FeGe) and 1:6:6 (MgFe$_6$Ge$_6$) classes, are complicated and involve many orbitals and bands around the Fermi level. Current theoretical models treat the systems in an $s$-orbital kagome representation, unsuited and incorrect both quantitatively and qualitatively to the material realities. In this work, we lay the basis of a faithful framework of the electronic model for this large class of materials. We show that the complicated ``spaghetti" of electronic bands near the Fermi level can be decomposed into three groups of Fe $d$ orbitals coupled to specific Ge orbitals via symmetry and chemical analysis. Such a decomposition allows for a clear analytical understanding (leading to different results than the simple $s$-orbital kagome models) of the flat bands in the system based on the $S$-matrix formalism of generalized bipartite lattices. Our three minimal Hamiltonians can reproduce the quasiflat bands, van Hove singularities, topology, and Dirac points close to the Fermi level, which we prove by extensive \textit{ab initio} studies. We also obtain the interacting Hamiltonian for the $d$ orbitals in FeGe using the constraint random phase approximation (cRPA) method, which faithfully describes the antiferromagnetic phase. We then use FeGe as a fundamental ``LEGO-like'' building block for a large family of 1:6:6 kagome materials, which can be obtained by doubling and perturbing the FeGe Hamiltonian. We apply the model to its kagome siblings FeSn and CoSn, and also MgFe$_6$Ge$_6$. 
We further extend the formalism developed for the 1:1 family to the 1:3:5 family \ch{AB3Z5} ($A$ = K, Rb, Cs; $B$ = Cr, V, Ti; $Z$ = Sb, Bi), demonstrating the broad applicability of the LEGO-like building block approach. 
Moreover, our method has the potential to be applied to a wider range of materials beyond kagome systems, provided that the relevant LEGO-like building blocks in the crystal and electronic structures can be identified. Our work serves as the first complete framework for the study of the interacting phase diagram of kagome compounds. 
\end{abstract}

\maketitle

\section{Introduction} 
Kagome materials exhibit a rich phase diagram including charge density waves~\cite{CHE22a,DIE21,FER22,KEN21a,LI23,LIA21a,LIU21b,LUO22,MAX13,RAT21,SON21a,SET21,TAN21,TSI22,TSV23,UYK21,UYK22,WAN21b,WAN21d,WAN21g,YU21,ZHU22,ZHA21b,guo2023correlated,LIN21}, superconductivity~\cite{ORT20,CHE21a,CHE21b,DU21,DUA21,FEN21,KAN23a,LI22a,LIU21d,MU21,NAK21,NI21,SHR22,SON21b,WAN21e,WAN21f,WAN23,WU21b,XIA21,XU21b,YIN21,YIN21a,YU21a,ZHA22a}, different magnetic orders\cite{teng2022discovery, mazet2013magnetic, haggstrom1975studies,ISH21,LI21b,LIU23,PAL22, ye2018massive} and topological states\cite{guo2009topological, bolens2019topological, yin2022topological, hu2022topological, liu2019magnetic, liu2018giant, xu2018topological, ghimire2020topology,zhang2022endless,zhou2024chemical}. 
In recent years, kagome superconductors AV$_3$Sb$_5$ (A=K, Cs, and Rb)~\cite{ORT19,CHO21b,KAN21a,ORT21,ORT21a, kautzsch2023structural} 
of the 1:3:5 class (with unusual charge orders~\cite{DEN21,NEU22,JIA21,MIE22,SHU21,WAN21c, guo2022switchable, wagner2023phenomenology} but no soft phonon modes observed~\cite{li2021observation, xie2022electron, liu2022observation, subires2023order}) and \ch{ScV6Sn6}~\cite{ARA22, KOR23, hu2023kagome, CAO23, CHE23, KAN23, TAN23, HU23d, LEE23, TUN23, YI23, HU23e, GU23,GUG23,CHE23,MOZ23} of the 1:6:6 class~\cite{venturini2006filling,fredrickson2008,venturini2008structures, mazet2013magnetic, ghimire2020competing, dally2021chiral} (with soft phonon modes first observed in experiments~\cite{KOR23, CAO23}) have attracted much attention. 
Among the kagome materials, the kagome magnet FeGe~\cite{teng2022discovery, teng2023magnetism, miao2022charge, setty2022electron, yin2022discovery, zhou2023magnetic, chen2023charge, ma2023theory, wang2023enhanced, wu2023novel, chen2023long, chen2023competing, wu2023annealing, wu2023symmetry, zhang2023triple, zhao2023photoemission, shi2023disordered} of the 1:1 class (or, equivalently, the 3:3 class) is particularly attractive: it develops an A-type antiferromagnetic (AFM) order below $T_N=\SI{410}{\kelvin}$~\cite{ohoyama1963new, haggstrom1975mossbauer, forsyth1978low}, and, more interestingly, has a CDW transition at $T_{\text{CDW}}=\SI{100}{\kelvin}$~\cite{teng2022discovery,teng2023magnetism}. 
The kagome 1:1 class materials in space group (SG) 191 have formula TZ, where T are transition metals and Z are main group elements. The 1:6:6 class, however, has the formula MT$_6$Z$_6$ where M are metallic elements, which can be seen as a doubled 1:1 material (which is \ch{T3Z3}) with inserted $M$ atoms. 

A commonly used theoretical model for understanding the non-trivial phase diagram of kagome systems is the $s$-orbital tight-binding (TB) model with nearest-neighbor (NN) hoppings on a kagome lattice\cite{mielke1991ferromagnetism}.
However, such an $s$-orbital TB model is incorrect for FeGe. It is oversimplified and fails to give quantitative descriptions for realistic kagome materials, which have a large number of orbitals near the Fermi level that are entangled together. Moreover, the model also suffers a qualitative fault: the Z element occupies the triangular and honeycomb lattice sites around the kagome T element and electrons from the former can ``hop'' onto the latter. This type of model, in general, should not have flat bands. 

In this work, for the first time, we provide a clear and comprehensive understanding of the complicated ``spaghetti'' of electronic bands for the kagome 1:1 and 1:6:6 materials. Our strategy is to decompose the intricate band structures into several small groups where, within each group, a simple and analytical understanding of the band structures is feasible.  
We first consider the 1:1 class and take FeGe as a representative. We separate the $d$ orbitals of Fe into three groups that are combined with specific orbitals of Ge based on chemical and symmetry principles. 
Three decoupled effective tight-binding models for the three groups of orbitals can then be constructed, where the effective tight-binding models not only quantitatively reproduce the quasi-flat bands, van Hove singularities (vHS), and Dirac points, but also provide an analytical understanding of the origin of flat bands, which are only flat on part of the BZ. 
Moreover, we also provide the full interacting Hamiltonian constructed via the constraint random phase approximation (cRPA) method and identify a hidden $O_h$ symmetry of the interacting term. 
A Hartree-Fock mean-field study of the interacting Hamiltonian accurately reproduces the AFM phase.
We next consider the 1:6:6 family, \ch{MT6Z6}, where we observe that the Hamiltonian for this family can be derived by doubling and perturbing the Hamiltonian of the 1:1 family. By treating FeGe as a ``LEGO-like'' building block, we successfully construct the band structures of the 1:6:6 material \ch{MgFe6Ge6}~\cite{mazet2013magnetic}. Finally, we adapt the formalism to the 1:3:5 family, using modified building blocks, which faithfully reproduce the corresponding band structures. 
Our method can be broadly applied to a wide range of materials beyond kagome systems, provided that the relevant “LEGO-like building blocks” can be identified. The resulting minimal effective models offer a powerful framework for gaining deeper insights into interacting phenomena such as magnetism, transport, and superconductivity, while also providing valuable guidance for experimental research.

\begin{figure}[tbp]
\centering
\includegraphics[width=0.48\textwidth]{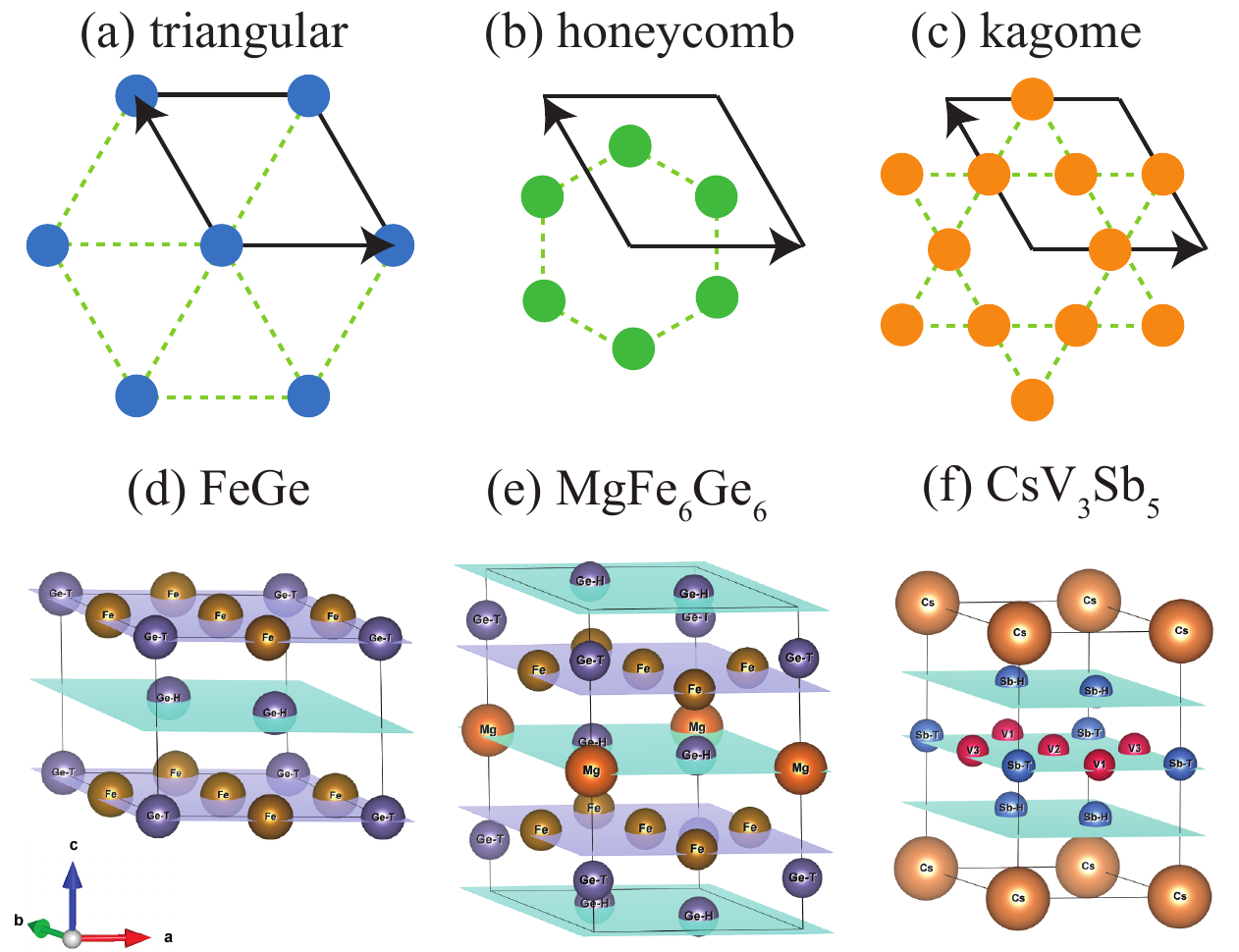}
\caption{\label{Fig:lattice_structure}
Representative 2D lattices and crystal structures of kagome materials in the 1:1, 1:6:6, and 1:3:5 families. The first row illustrates representative 2D hexagonal lattices: (a) triangular, (b) honeycomb, and (c) kagome lattices. (d)-(f) show the crystal structures of prototype kagome materials in the 1:1, 1:6:6, and 1:3:5 families, i.e., FeGe, \ch{MgFe6Ge6}, and \ch{CsV3Sb5}, respectively. In FeGe, the Fe atoms form a kagome lattice and Ge atoms form a triangular (denoted by Ge-T in the plot) and a honeycomb lattice (denoted by Ge-H). \ch{MgFe6Ge6} can be built by ``doubling'' FeGe along $z$-direction and inserting Mg atoms in the middle plane of the honeycomb Ge. The 1:3:5 \ch{CsV3Sb5} has two honeycomb layers of Sb surrounding the kagome V layer. 
}
\end{figure}

\begin{figure}[tbp]
\centering
\includegraphics[width=0.48\textwidth]{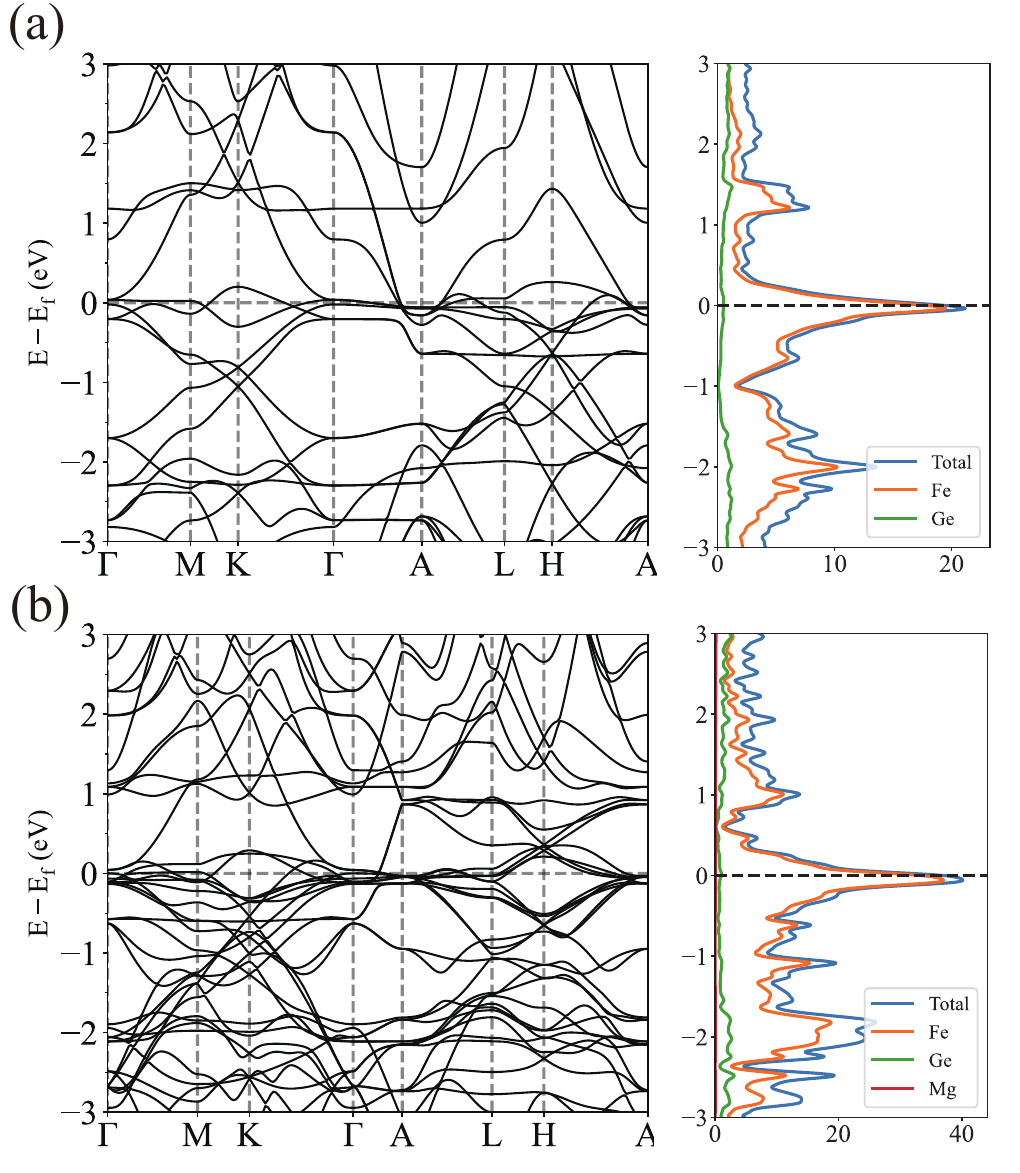}
\caption{\label{Fig:DFT_bands}
The band structure and density of states (DOS) for (a) FeGe and (b) \ch{MgFe6Ge6} in the paramagnetic phase. The band structure of \ch{MgFe6Ge6} can be described as a folding of the FeGe bands along the $k_3$ direction, with small perturbations. The DOS of FeGe and \ch{MgGe6Ge6} are also very close, with a dominant peak at the Fermi level given by the quasi-flat bands.}
\end{figure}

\section{Decomposition of orbitals}
We now introduce the main idea of this work: build minimal effective models that reproduce the complicated band structure of 1:1, 1:6:6, and 1:3:5 classes near the Fermi level $E_f$ by decomposing the orbitals into independent groups. 

We start from the atomic positions in SG 191 $P6/mmm$, which can be divided into three (or, equivalently, maximal Wyckoff positions): triangular, honeycomb, and kagome, depending on their site symmetry groups.
On a two-dimensional (2D) lattice spanned by $\bm{a}_1=(1,0)$, $\bm{a}_2=(-\frac{1}{2}, \frac{\sqrt{3}}{2})$, the triangular lattice has Wyckoff position $1a=(0,0)$ with $C_{6z}$ symmetry, the honeycomb lattice is located at the $2b=(\frac{1}{3}, \frac{2}{3}), (\frac{2}{3}, \frac{1}{3})$ position having $C_{3z}$ symmetry, while the kagome lattice is at the $3c=(\frac{1}{2},0)$, $(\frac{1}{2},\frac{1}{2})$, $(0,\frac{1}{2})$ Wyckoff position with $C_{2z}$ symmetry, as shown in \cref{Fig:lattice_structure}(a)-(c). In 3D systems, atoms can occupy these three sublattices on different $z$ planes. For example, in 1:1 materials like FeGe, three Fe atoms occupy the kagome lattice on the $z=0$ plane, one Ge atom occupies the triangular lattice on the $z=0$ plane (denoted by Ge$^T$) and the other two Ge on the honeycomb lattice on $z=\frac{1}{2}$ plane (denoted by Ge$^H$), as shown in \cref{Fig:lattice_structure}(d). The 1:6:6 materials \ch{MT6Z6} can be seen as two copies of the corresponding 1:1 material \ch{T3Z3} with the M atoms being located at the triangular position of one of the two Ge$^H$ layers. The representative \ch{MgFe6Ge6} is shown as an example in \cref{Fig:lattice_structure}(e). We remark that there also exist more complicated superstructures in the 1:6:6 class that are built from multiple 1:1 materials~\cite{venturini2008structures}, all of which can be understood from the 1:1 family using perturbation theory. The 1:3:5 family, \ch{MT3Z5}, depicted in \cref{Fig:lattice_structure}(f), can be viewed as an extension of the 1:1 (or 3:3) family, with the addition of a honeycomb layer of Z atoms and a triangular layer of M atoms. The two honeycomb Z layers are mirror-symmetric to the central kagome layer.

We then consider the decomposition of orbitals into three groups. In the 1:1 materials, the $d$ orbitals of the kagome lattice contribute most to the bands close to the Fermi energy ($E_f$). As such, we first decompose the five kagome $d$ orbitals into three groups: $(d_{xy}, d_{x^2-y^2})$, $(d_{xz}, d_{yz})$, and $d_{z^2}$, which are then coupled with specific orbitals from the triangular and honeycomb lattices that have dominant hoppings with the former. 
When combined with orbitals from triangular and honeycomb lattices, the kagome $d$ orbitals form generalized bipartite crystalline lattices (BCL) making the $S$-matrix formalism~\cite{cualuguaru2022general, regnault2022catalogue} readily applicable for the identification of various perfectly flat-band limits. 
We emphasize that such a decomposition holds for generic 1:1 materials TZ, since T are transition metals that usually provide $d$ orbitals and Z are main group elements that usually contribute with $s$ and $p$ orbitals.
We apply this strategy to the 1:1 materials including FeGe, FeSn\cite{haggstrom1975studies, sales2019electronic, multer2022imaging, sankar2023observation}, and CoSn\cite{larsson1996single, chen2023visualizing}. With the effective model constructed for the 1:1 materials, the model for 1:6:6 materials can be directly derived by doubling that of the corresponding 1:1 materials and treating the orbitals of the M atoms as a perturbation. \ch{MgFe6Ge6} is used as a representative example in this work. 
The 1:3:5 family will be discussed in \cref{sec:135-family} using modified groups of orbitals due to the different crystal structures.

We emphasize that while the $S$-matrix formalism for constructing flat bands was introduced in Ref.~\cite{cualuguaru2022general}, its application to multi-orbital systems with complex band structures is highly non-trivial. A key contribution of our work is the systematic partitioning of a large set of orbitals into several decoupled (or weakly coupled) groups, guided by symmetry and chemical considerations. Within each group, we identify a bipartite lattice and apply the $S$-matrix formalism to derive the idealized flat band limit. This decomposition not only reduces the complexity of the problem but also enables the construction of minimal, faithful models that effectively capture the essential physics of quasi-flat bands, van Hove singularities, and Dirac points. 
It is important to note that such decomposition is not universally feasible. For systems with low site symmetries on kagome sites or non-symmorphic symmetries, the bands arising from kagome $d$ orbitals are often heavily hybridized. In these cases, constructing minimal decoupled models and bipartite lattice descriptions becomes inherently challenging.

\section{Band structure of $\text{FeGe}$ and $\text{MgFe}_6\text{Ge}_6$}
We first consider FeGe in the 1:1 class and use density functional theory (DFT)\cite{kresse1996efficiency, kresse1993ab1, kresse1993ab2, kresse1994ab, kresse1996efficient} to compute its band structure as a starting point for constructing the effective models. The bands and density of states (DOS) in the PM phases of FeGe are shown in \cref{Fig:DFT_bands}(a). 
This spaghetti-like band structure is very complicated and hosts two quasi-flat bands connected with other bands and multiple vHSs near $E_f$. The highest peak in the DOS is located close to the Fermi level and is mainly contributed by the quasi-flat bands of Fe. For the 1:6:6 class representative \ch{MgFe6Ge6}, the even more complicated band structure and DOS is shown in \cref{Fig:DFT_bands}(b). However, one can observe that the band structure of \ch{MgFe6Ge6} is very close to the spectrum obtained by folding together the bands of FeGe from the $k_3 = 0$ and $k_3 = \pi$ planes onto the $k_3=0$ one. This similarity arises because \ch{MgFe6Ge6} can be viewed as FeGe doubled along the $z$-direction and augmented with an additional Mg atom. This insight inspires us to construct effective Hamiltonians for the 1:6:6 class based on the ones derived for the 1:1 class.

\section{Three minimal effective Hamiltonians}\label{Sec:3models_maintext}
\begin{figure}[tbp]
    \centering
    \includegraphics[width=0.48\textwidth]{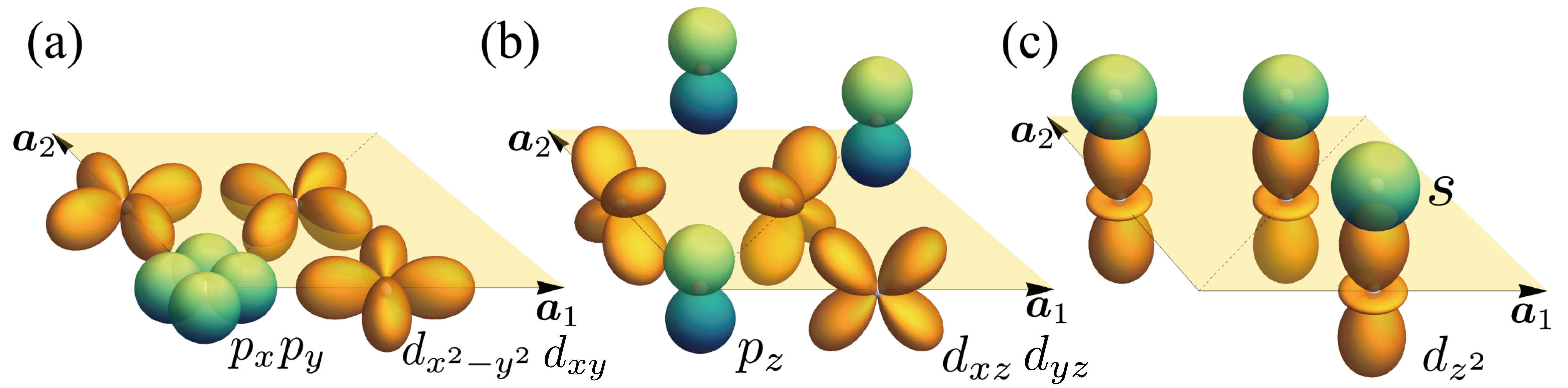}
    \caption{\label{Fig:3orbsets} Decomposition in three groups of orbitals. (a) $d_{x^2-y^2},d_{xy}$ orbitals of kagome Fe and $p_x,p_y$ orbitals of triangular Sn ($d_{x^2-y^2}$ is shown in the plot while $d_{xy}$ is omitted for simplicity). (b) $d_{xz},d_{yz}$ orbitals of kagome Fe and $p_z$ orbital of triangular and honeycomb Sn ($d_{xz}$ is shown and $d_{yz}$ is omitted). (c) $d_{z^2}$ orbital of kagome Fe and the bonding state (equivalent to an $s$ orbital on the $z=\frac{1}{2}$ plane) of honeycomb Sn. The yellow plane denotes the $z=0$ plane in the unit cell. A local coordinate system on the kagome lattice site that rotates with the $C_{6z}$ transformation is used (see \cref{Appendix: local coord} for details) such that the five $d$ orbitals can be separated under the SG 191 symmetries.}
\end{figure}

We first construct maximally localized Wannier functions (MLWFs)\cite{marzari1997maximally, souza2001maximally, marzari2012maximally}, obtain a Wannier TB model, and perform a detailed study of orbital projections, density of states, and orbital fillings using MLWFs for each $d$ orbital. The TB model obtained from the MLWFs, although faithful, still contains more than 20 orbitals and a huge number of hopping parameters. 
It is desirable to build minimal TB Hamiltonians that can not only reproduce the band structure but also help gain insight into the physical properties of the system, including the origin of the quasi-flat bands near $E_f$. 
In general, however, simplifying a system with such a large number of degrees of freedom presents a significant challenge.
Here, based on symmetry and chemical analysis, we divide the orbitals into three groups and construct TB models for them separately, as shown in \cref{Fig:3orbsets}.  

First, the five $d$ orbitals of Fe are split into three groups under $D_{6h}$ symmetry (which is the point group of SG 191), i.e., $(d_{xy}, d_{x^2-y^2})$, $(d_{xz}, d_{yz})$, and $d_{z^2}$. The inplane hopping between the $(d_{xz}, d_{yz})$ and the other three $d$ orbitals is forbidden by $M_z$ symmetry, as the former have opposite $M_z$ eigenvalues to the latter, and $z$-directional hoppings are weak and can be neglected. The hoppings between the $(d_{xy}, d_{x^2-y^2})$ and $d_{z^2}$ orbitals, although not negligible and not forbidden by symmetry, are small compared with other leading hopping terms and will be neglected as an approximation. 
The small hopping between the $(d_{xy}, d_{x^2-y^2})$ and $d_{z^2}$ orbitals can also be explained from a chemistry perspective by the spatial distribution of their probability densities. The $(d_{xy}, d_{x^2-y^2})$ orbitals predominantly occupy the $xy$ planes, whereas the $d_{z^2}$ orbital's density is primarily aligned along the $z$-direction. This spatial distinction leads to a minimal overlap between their wavefunctions.
We then combine the three groups of $d$ orbitals with specific orbitals of Ge based on both chemical and symmetry analysis, as shown in \cref{Fig:3orbsets}. The three resulting groups are
(i) the $d_{xy}, d_{x^2-y^2}$ orbitals of Fe and the $p_x,p_y$ orbitals of Ge$^T$. These orbitals lie on the $z=0$ plane and have large overlaps, forming $\sigma$-like bonds, which can be verified from the Wannier hoppings. They form a BCL with eight bands, and the $S$-matrix formalism\cite{cualuguaru2022general, regnault2022catalogue} can be applied to identify a perfect flat-band limit (attained when the hoppings take specific forms -- to be discussed in the next subsection). 
(ii) the $d_{xz}, d_{yz}$ orbitals of Fe and $p_z$ orbitals of both Ge$^T$ and Ge$^H$. These orbitals all lie along the $z$-direction and have large overlaps, forming $\pi$-like bonds, verified from the Wannier hoppings. 
(iii) the $d_{z^2}$ orbitals of Fe and the $sp^2$ bonding state formed by the $s$, $p_x$ and $p_y$ orbitals of Ge$^H$ (equivalent to an $s$ orbital located on a  kagome lattice in the $z=\frac{1}{2}$ plane).

With the three groups of orbitals at hand, we construct three TB models $H_{i=1,2,3}(\bm{k})$ corresponding to each orbital group, respectively, with the total Hamiltonian being a direct sum:
\begin{equation}
    H(\bm{k})= H_1(\bm{k})\oplus H_2(\bm{k}) \oplus H_3(\bm{k}).
\end{equation}
In what follows, we will discuss each Hamiltonian block individually.

\subsection{$H_1(k)$: Fe $d_{xy}, d_{x^2-y^2}$ and Ge$^T$ $p_x,p_y$ orbitals}
In this group, the $d_{xy}, d_{x^2-y^2}$ orbitals of Fe (denoted as $d_1$, and $d_2$) form the $A_g@3f$, and $B_{1g}@3f$ elementary band representations (EBRs)\cite{bradlyn2017topological}, while the $p_x,p_y$ orbitals of the triangular Ge (denoted as $p_{xy}^t$) form the $E_{1u}@1a$  EBR. 
We construct a TB Hamiltonian of the form (see \cref{Appendix: model_details} for the explicit form of each block): 
\begin{equation}
H_1(\bm{k})=
\left(
\begin{matrix}
H_{p_{xy}^t}(\bm{k}) & S_{p_{xy}^t, d_1}(\bm{k}) & S_{p_{xy}^t, d_2}(\bm{k}) \\
& H_{d_1}(\bm{k})  & S_{d_1,d_2}(\bm{k}) \\
H.c.  &  & H_{d_2}(\bm{k}) \\
\end{matrix}
\right).
\label{Eq:H1_ham}
\end{equation}

The model in \cref{Eq:H1_ham} admits two different flat-band limits:
\begin{itemize}

\item When $H_{d_2}=\mu_{d_2}\mathbf{1}_3, S_{d_1,d_2}=0$, there will be one perfectly flat band stemming from the $d_2$ orbitals, as shown in \cref{Fig:TBbands_flatlimit}(a). This flat band can be understood from the $S$-matrix formalism (see \cref{Appendix: AP_SI_Smatrix}): the $d_2$ and $p_{xy}^t$ orbitals form a BCL with the corresponding sublattices containing $N_{d_2}=3$ and $N_p=2$ orbitals per unit cell, resulting in $N_{d_2}-N_p=3-2=1$ flat band. This flat-band limit also agrees with the orbital projections obtained from DFT (see \cref{Appendix: model_details}), where the quasi-flat band near the Fermi level mainly comes from the $d_{x^2-y^2}$ orbital. 

\item When $S_{p_{xy}^t,d_1}=0$ and $H_{d_1}$ has only NN hoppings, in addition to $H_{d_2}=\mu_{d_2}\mathbf{1}_3, S_{d_1,d_2}=0$, an additional perfectly flat band arises compared to the previous limit. This latter flat band stems from $H_{d_1}(\bm{k})$, which is just a NN kagome Hamiltonian decoupled from all the other orbitals. 

\end{itemize}

The TB parameters of $H_1(\bm{k})$ are fitted to the DFT results by considering both the dispersion and the wavefunctions of the bands. As illustrated in \cref{Fig:TBbands_flatlimit}(b), the fitted band structure exhibits quasi-flat bands akin to those observed in the DFT results at $E_f$. Moreover, the fitted hopping parameters are close to the first flat-band limit. 
The small NN hopping in $H_{d_2}(\bm{k})$ can be attributed to the cancellation effects arising from longer range hopping processes involving orbitals not considered in $H_1(\bm{k})$. 
The wavefunction of this quasi-flat band has a high overlap of 97\% with the DFT wavefunction. 
The vHS from around $-0.5$ eV at the $M$ point is also well-fitted in the current model.
Note that the current $H_1(\bm{k})$ Hamiltonian has only inplane hoppings and is $k_3$-independent, leading to a mismatch of IRREPs at the $L$ point, as shown in \cref{Fig:TBbands_flatlimit}(b). This mismatch will be remedied in the combined model by coupling with $p_z$ orbital of honeycomb Ge. The coupling term is perturbed out in the final model which introduces a $k_3$-dependence to $H_1(\bm{k})$.

\begin{figure}[tbp]
    \centering
    \includegraphics[width=0.48\textwidth]{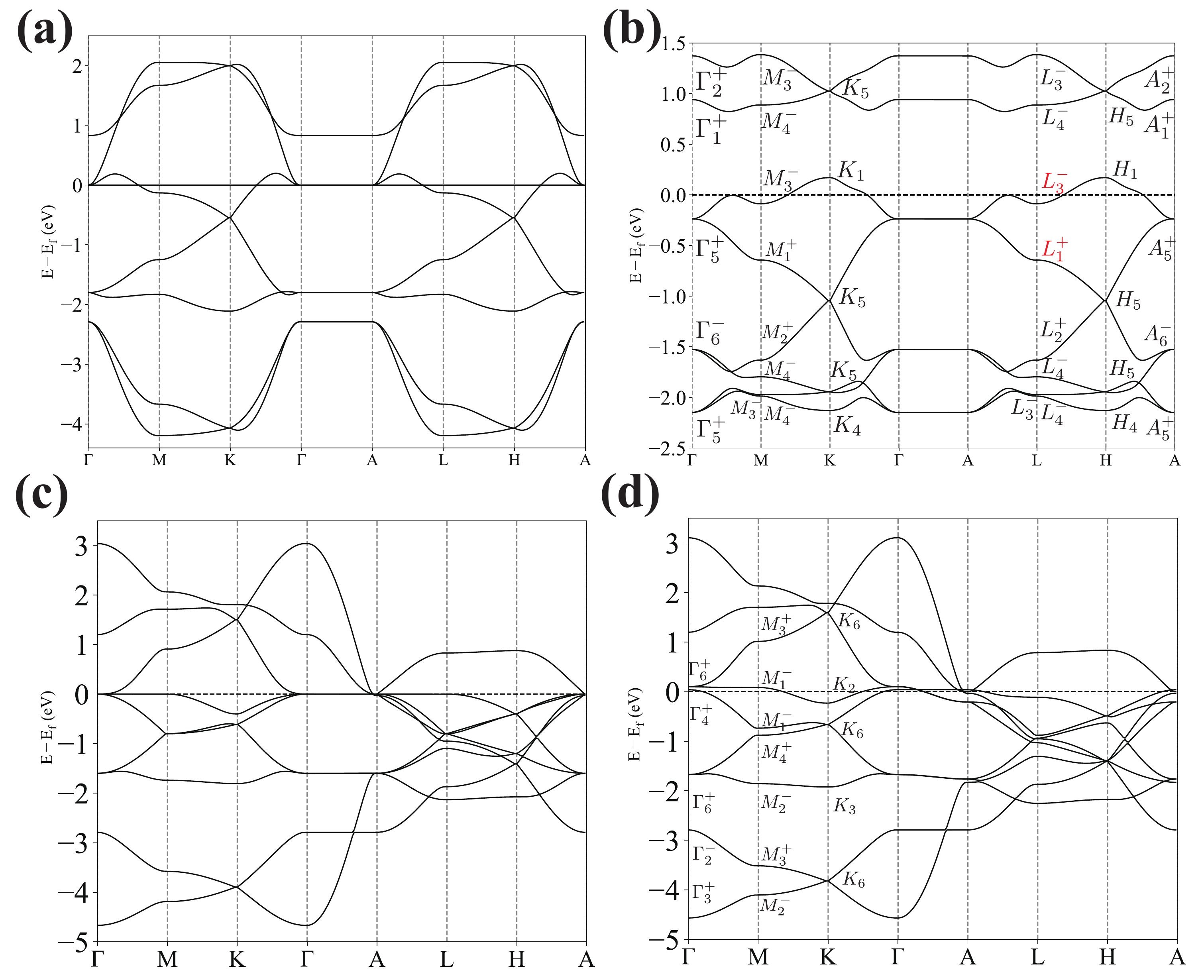}
    \caption{\label{Fig:TBbands_flatlimit} Band structures of the minimal models. 
    (a) Band structure in the first perfectly flat-band limit of $H_1(\kk)$ from \cref{Eq:H1_ham}. The flat band is mainly contributed by the $d_{x^2-y^2}$ orbital. (b) Band structure of $H_1(\kk)$ obtained using realistic parameters extracted by fitting to the DFT bands. The irreducible representations (IRREPs) are marked in the band structure. The IRREPs agree well with the DFT result, except for the two IRREPs marked in red at the $L$ points. This mismatch will be fixed in the final model. 
    (c) Quasi-flat-band limit in $H_2(\kk)$ in \cref{Eq:H2_ham}, where a perfect flat band exists on the $(k_1,0,k_3)$ plane (which includes the $\Gamma$-$M$, $\Gamma$-$A$, and $A$-$L$ lines), which is otherwise in the other regions of the BZ. 
    (d) Band structure of $H_2(\kk)$ obtained using the realistic hopping parameters extracted from DFT. The IRREPs marked on the band structure agree well with the DFT results along the $k_3=0$ plane. }
\end{figure}

\subsection{$H_2(k)$: Fe $d_{xz}, d_{yz}$ and Ge $p_z$ orbitals}
In this group, the $d_{xz}, d_{yz}$ orbitals of Fe  (denoted as $d_3$ and $d_4$) form the $B_{2g}@3f$ and $B_{3g}@3f$ EBRs, respectively, while the $p_z$ orbital of the Ge$^H$ ($p_z^h$) and the $p_z$ orbital of the triangular Ge$^T$ ($p_z^t$) forms the $A_{2}^{\prime\prime}@2d$ and $A_{2u}@1a$ EBRs, respectively. 
The corresponding TB Hamiltonian matrix is given by 
\begin{equation}
    H_2(\bm{k})=
    \left(
    \begin{array}{cccc}
	H_{p_z^h}(\bm{k}) & \bm{0} & \underline{\bm{0}} & S_{p_z^h, d_4}(\bm{k})  \\
        & H_{p_z^t}(\bm{k}) & S_{p_{z}^t, d_3}(\bm{k}) & \underline{\bm{0}} \\
	&  & H_{d_3}(\bm{k})  & S_{d_3,d_4}(\bm{k}) \\
	H.c.& &	& H_{d_4}(\bm{k}) \\
    \end{array}
    \right).
\label{Eq:H2_ham}
\end{equation}
where $\underline{\bm{0}}$ denotes a symmetry-forbidden NN $S$-matrix, while $\bm{0}$ denotes an $S$-matrix that is inessential for fitting the band structure near $E_f$. The explicit form of each TB block is delegated to \cref{Appendix: model_details}. 

We identify two flat-band limits in this model:
\begin{itemize}
\item A perfectly flat-band limit of $H_2(\bm{k})$ can be achieved by setting $S_{d_3,d_4}(\bm{k})=0$ letting $H_{d_4}(\bm{k})$ have only NN hoppings (thus rendering it equivalent to a single-orbital NN kagome Hamiltonian). In this case, the $d_{4}$ orbitals harbor one perfectly flat band near $E_f$. This flat-band limit cannot be explained directly from the $S$-matrix formalism, but is a result of the specific form of the $S_{p_z^h,d_4}(\bm{k})$ matrix: the eigenvector of the flat band of $H_{d_4}(\bm{k})$ is identical to the null vector of $S_{p_z^h, d_4}(\bm{k})$.

\item We also identify a special flat-band limit where one flat band exists on the $(k_1,0,k_3)$ plane (which includes the $\Gamma$-$M$, $\Gamma$-$A$, and $A$-$L$ lines, and other symmetry-related planes), which otherwise disperses in the other regions of the BZ, as shown in \cref{Fig:TBbands_flatlimit}(c). This flat-band limit can be realized by requiring that the $d_3$ and $d_4$ orbitals have the same onsite energy and only inplane NN hoppings satisfying $-t_{d_3}^{NN}=t_{d_4}^{NN}= -t_{d_3,d_4}^{NN}$, where $t_{d_3}^{NN}$ ($t_{d_4}^{NN}$) is the inplane NN hopping between the $d_{3}$ ($d_4$) orbitals, while $t_{d_3,d_4}^{NN}$ is the inplane NN hopping between $d_3$ and $d_4$. 
The resultant flat band is an equal-weight superposition of the $d_3$ and $d_4$ orbitals. 
Such equal-weight decomposition also holds approximately in the DFT band structures (see the orbital projections from \cref{Appendix: model_details}). This quasi-flat-band limit serves as a good approximation for the 1:1 class, including FeGe, FeSn, and CoSn~\cite{SI}. 
\end{itemize}

With these two flat-band limits, we then fit the TB parameters to the DFT results, which are close to the second limit, as shown in \cref{Fig:TBbands_flatlimit}(d). An extremely flat exists along the $\Gamma$-$M$, $\Gamma$-$A$, and $A$-$L$ lines. The overlap of the flat band wavefunctions computed within the fitted $H_2(\bm{k})$ Hamiltonian and within DFT is about $85\%$, showing the faithfulness of the model.

\subsection{$H_3(k)$: Fe $d_{z^2}$ orbitals and Ge$^H$ $sp^2$ bonding state}
\begin{figure*}[htbp]
    \centering
    \includegraphics[width=0.9\textwidth]{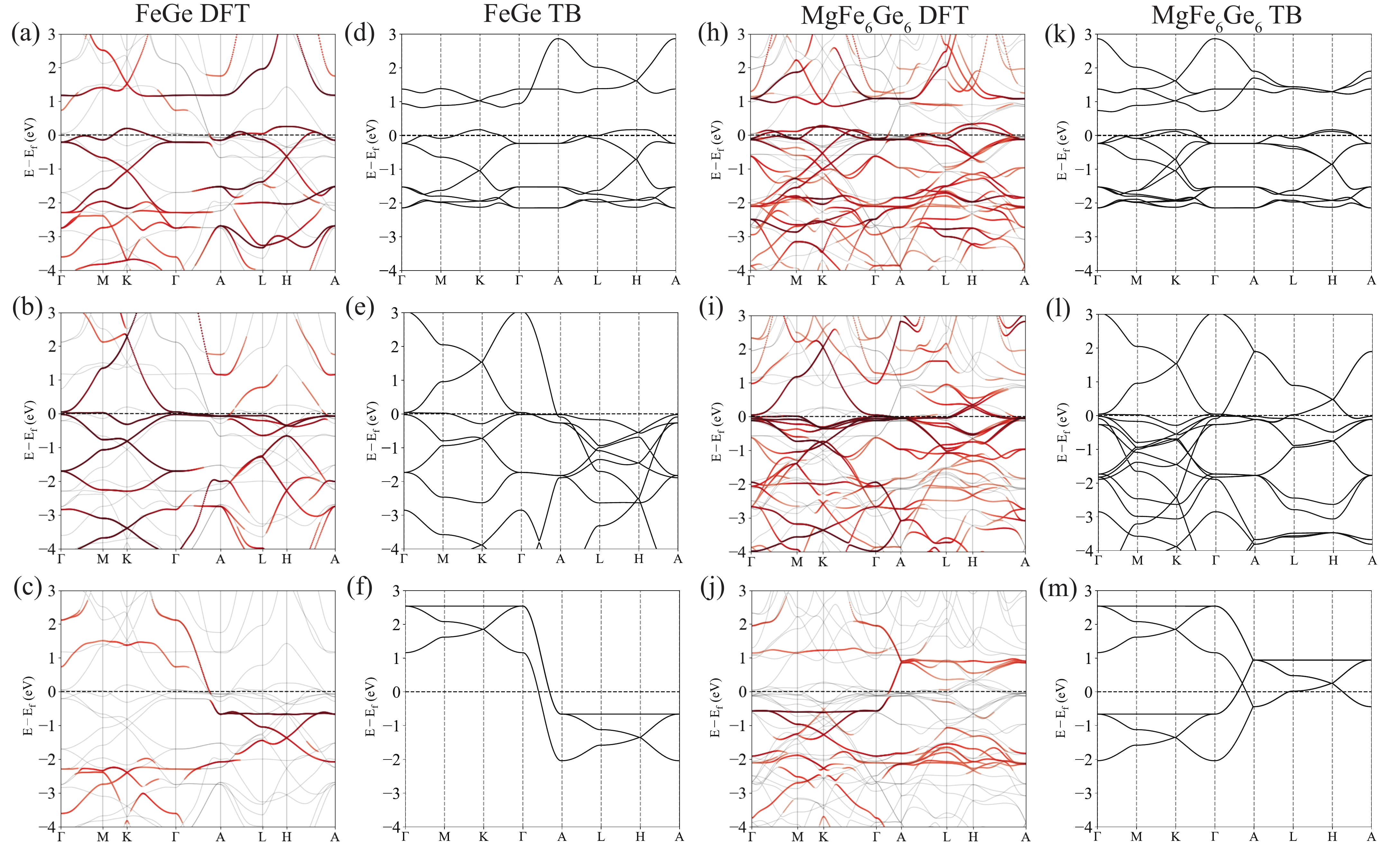}
    \caption{\label{Fig:TBbands} The DFT and TB band structures of FeGe and \ch{MgFe6Ge6}. The first column shows the DFT orbital projections of FeGe from (a) $(d_{xy},d_{x^2-y^2})$ of Fe and $(p_x,p_y)$ of triangular Ge, (b) $(d_{xz},d_{yz})$ of Fe and $p_z$ of honeycomb Ge, and (c) $d_{z^2}$ of Fe. (d)-(f) are the band structures of the three minimal TB Hamiltonian $H_1(\bm{k})$, $H_2(\bm{k})$, and $H_3(\bm{k})$ defined \cref{Eq:TBdecoupled_final} for FeGe. (g)-(i) are the same as (a)-(c), and (j)-(l) are the same as (d)-(f), but for \ch{MgFe6Ge6}. The TB band structures of \ch{MgFe6Ge6} are obtained by folding the corresponding band structure of FeGe along the $k_3$ direction and treating the $s$ orbital of Mg as a perturbation. The TB bands quantitatively match their DFT counterparts near the Fermi level. They are more realistic compared to a single $s$-orbital kagome model, as our models give the correct orbital components for the flat bands and correctly capture the DFT dispersion trends across a wide energy range. }
\end{figure*}

In this group, we couple the $d_{z^2}$ orbital of Fe (denoted as $d_5$) with the $sp^2$ bonding states formed by $s,p_x,p_y$ orbitals of honeycomb Ge. From a symmetry perspective, the Ge bonding states are equivalent to $s$ orbitals located on a kagome lattice in $z=\frac{1}{2}$ plane. These bonding states are introduced to account for the $d_{z^2}$ weight below $-2$ eV and more faithfully reproduce the DFT band structure, as seen from the orbital projections of DFT in \cref{Fig:TBbands}(c). However, for solely capturing the low-energy dispersion near $E_f$, it is more convenient to employ a simpler $d_{z^2}$-only model: the effect of the Ge$^H$ bonding state is to enhance the $d_{z^2}$-orbital weights below $-2$ eV. As a result we here focus on the $d_{z^2}$-only model and let $H_3(\kk)=H_{d_5}(\bm{k})$. 
The explicit form of $H_{d_5}(\kk)$ and the corresponding DFT-fitted dispersion are relegated to \cref{Appendix: model_details}.
We remark that if one uses this simplified model comprising $d_{z^2}$ orbitals only, the Coulomb interaction of $d_{z^2}$ also gets renormalized to smaller values, as the weight of the $d_{z^2}$ orbitals near $E_f$ increases if the bonding states are excluded.

\subsection{Combined model}
Having derived the three minimal models $H_{i=1,2,3}(\bm{k})$, we proceed to combine them into the final model. The resulting final model is further simplified into a direct sum of three decoupled models using the second-order perturbation theory.

An extra coupling term $S_{p_z^h,d_2}(\bm{k})$ between the Fe $d_{x^2-y^2}$ and Ge$^H$ $p_z$ orbitals is introduced to capture the $k_3$-dispersion of $H_1(\kk)$ (see \cref{Appendix: model_details}). This term can be perturbed out as these two orbitals have relatively large onsite energy differences. Similarly, the effects of the $p_z^t$ orbitals within $H_2(\kk)$ can also be treated perturbatively, as $p_z^t$ has a small weight at $E_f$. 
The final Hamiltonian is therefore a direct sum of three models: 
\begin{equation}
    H(\bm{k})= H_1'(\bm{k})\oplus H_2'(\bm{k}) \oplus H_3(\bm{k}), 
\label{Eq:TBdecoupled_final}
\end{equation}
in which $H_1'(\bm{k})$ and $H_2'(\bm{k})$
have the following modified blocks from second-order perturbation: 
$H_{d_2}'(\bm{k}) = H_{d_2}(\bm{k})+H_{d_2, p_z^h}^{(2)}(\bm{k})$, $H_{p_z^h}'(\bm{k}) = H_{p_z^h}(\bm{k})+H_{p_z^h,d_2}^{(2)}(\bm{k})$, and $H_{d_3}'(\bm{k}) = H_{d_3}(\bm{k})+H_{d_3, p_z^t}^{(2)}(\bm{k})$, respectively. $H_{O_1, O_2}^{(2)}$ is used to denote the hopping block of the $O_1$ orbitals arising from the second-order perturbation effects of the $O_2$ orbitals~\cite{SI}. The band structure of $H(\bm{k})$ is shown in \cref{Fig:TBbands}(d)-(f), which quantitatively reproduces the quasi-flat bands, vHS, and Dirac points close to $E_f$ as in the DFT-computed dispersion. 
The Fermi surface (FS) obtained from the TB model shows good agreement with the FS calculated from DFT~\cite{SI}. 
Mismatches between the TB model and the DFT bands mainly appear for energies outside of the $E_f\pm 1$ eV window, which are less relevant to the low-energy physics in both the PM and AFM phases.  
These mismatches arise because, away from $E_f$, other orbitals not considered in our minimal TB model contribute to the bands.
The vHSs in the three sectors (which will move close to $E_f$ in the AFM phase and could be important for the CDW formation) are also well-fitted in the TB model.

\section{Interacting Hamiltonian} 
With the minimal TB Hamiltonian at hand, we compute the Coulomb interaction using the constraint random phase approximation (cRPA) method\cite{aryasetiawan2004frequency, solovyev2005screening, aryasetiawan2006calculations, miyake2009ab, di2023electronic} and construct the interaction Hamiltonian. 
We evaluate the onsite inter-orbital Hubbard interaction $U_{m_i,m_j}$, onsite exchange $J_{m_i,m_j}$, NN and next nearest-neighbor (NNN) Hubbard interaction $U_{m_i,m_j}^{NN/NNN}$, where $m_{i,j}$ are orbital indices. 
In Ref.~\cite{SI}, we tabulate the values of $U_{m_i,m_j}$, $J_{m_i,m_j}$, and $U_{m_i,m_j}^{NN/NNN}$ for the so-called $d-dp$ model\cite{vaugier2012hubbard}, $dp-dp$ model\cite{vaugier2012hubbard}, and $d$-full models, which are constructed using different sets of MLWFs that have different spreads. 
We observe that there is an approximate hidden symmetry of the interacting terms. More precisely, we find that the NN and NNN Coulomb interactions have an approximate spherical symmetry with orbital-independent strengths. The on-site interactions, including the Hubbard interactions and exchange couplings, have an approximate $O_h$ symmetry due to the approximate $O_h$ environment of the Fe atoms given by the surrounding Ge atoms. The corresponding interaction parameters can be further simplified by assuming spherical symmetry and then fitted using Slater integrals\cite{slater1960quantum, sugano2012multiplets, SI}. 
The approximately $O_h$-symmetric environment around each Fe atom also results in close onsite energies for the five $d$-orbitals.

\begin{table}[tbp]
\begin{tabular}{c|ccccc|c|ccccc}
\hline\hline
$U_{ij}$  & $z^2$ & $xz$ & $yz$ & $x^2$ & $xy$ & $J_{ij}$  & $z^2$  & $xz$ & $yz$ & $x^2$ & $xy$ \\\hline
$z^2$     & 4.15 &  3.08 &  3.08 &  2.39  &  2.39 &  $z^2$ & & 0.54& 0.54 & 0.88  &  0.88 \\
$xz$      &       &  4.15 &  2.62 &  2.62  &  2.62 &   $xz$ &   &      & 0.77 & 0.77  &  0.77 \\
$yz$      &       &        &  4.15 &  2.62  &  2.62  &  $yz$ & &      &       & 0.77  &  0.77 \\
$x^2$ &       &        &        &  4.15  &  3.30    &  $x^2$ &   &       &    &    &  0.42 \\
$xy$      &       &        &        &         &  4.15  & $xy$ & &   &       &        &   \\\hline\hline
\end{tabular}
\caption{\label{Tab:d-full-slater-U-F0F2} The onsite Coulomb interaction parameters $U_{ij}$ and $J_{ij}$ of $d$ orbitals of Fe in FeGe, with numbers given in eV. These interactions are computed within the $d$-full model (see Ref.~\cite{SI} for the definition) and then symmetrized with the approximate spherical symmetry.
The averaged NN and NNN density-density interactions have the value $\overline{U}_1=1.41$ eV and $\overline{U}_2=1.22$ eV. In the table, $x^2$ is used to represent the $d_{x^2-y2}$ orbital, and the lower-left half of the table is omitted for simplicity.}
\end{table}

The final interacting Hamiltonian has the form 
\begin{equation}
    \hat{H} = \hat{H}_0 + \hat{H}_{\text{int}},
\label{eq:interacting-model}
\end{equation}
where the second-quantized single-particle Hamiltonian is given by $\hat{H}_0 = \sum_{\bm{k}}\sum_{ij} 
H_{ij}(\bm{k}) c^\dagger_{\bm{k}i}c_{\bm{k}j}$, 
with $H(\bm{k})$ defined in \cref{Eq:TBdecoupled_final}. 
The interacting Hamiltonian contains the onsite Hubbard term, spin-flipping term, and pair-hopping term, defined using orbital-dependent onsite interactions $U_{m_1m_2}$ and $J_{m_1m_2}$ and averaged NN and NNN Hubbard interaction $\overline{U}_1$ and $\overline{U}_2$:
\begin{equation}
\begin{aligned}
    \hat{H}_{\text{int}} &=
    \sum_{im} U_{mm} n_{im\uparrow}n_{im\downarrow}
    + \sum_{i,m\ne m^\prime} 
    U_{mm^\prime} n_{i m\uparrow}n_{i m^\prime\downarrow}\\
    &+ \sum_{i,m<m^\prime \sigma} 
    (U_{mm^\prime} - J_{mm^\prime})
    n_{i m \sigma} n_{i m^\prime \sigma}	\\
    &-\sum_{i, m\ne m^\prime} 
    J_{mm^\prime} c_{im\uparrow}^\dagger c_{i m \downarrow}  c_{i m^\prime \downarrow}^\dagger c_{i m^\prime \uparrow}\\
    &+ \sum_{i, m\ne m^\prime} 
    J_{mm^\prime} c_{im\uparrow}^\dagger c_{im\downarrow}^\dagger c_{i m^\prime \downarrow} c_{i m^\prime \uparrow}\\
    & + \left(\overline{U}_1 \sum_{<ij>}
    + \overline{U}_2 \sum_{\ll ij \gg}\right) \sum_{mm^\prime, \sigma\sigma^\prime} n_{i m \sigma} n_{j m^\prime \sigma^\prime}
\end{aligned}
\end{equation}
where $<ij>$ and $\ll ij \gg$ denote NN and NNN sites. The symmetrized values of interaction parameters are given in \cref{Tab:d-full-slater-U-F0F2}. 
A detailed study of the interaction Hamiltonian is left for future work.

\section{Mean-field study of AFM phase}

In this section, we apply the Hartree-Fock mean-field method to investigate the interacting Hamiltonian in \cref{eq:interacting-model}. We demonstrate that the antiferromagnetic (AFM) phase observed in the experiment and DFT can be accurately reproduced using the minimal model.

The A-type AFM order in FeGe is characterized by the wavevector $\QQ =(0,0,\frac{1}{2})$, which corresponds to a doubled unit cell along the $z$-axis. For $H_3(\kk)$, we use the simplified model $H_{d_5}(\kk)$, which includes only the $d_{z^2}$ orbital. In this case, the exclusion of the $sp^2$ bonding states reduces the effective interaction strength, as the interactions are limited to the $d_{z^2}$ orbitals. To account for this, we normalize the interaction strength for the $d_{z^2}$ orbitals by a factor of about $\frac{1}{2}$, as estimated from the wavefunction in $H_3(\kk)$ (see Ref.~\cite{SI} for more details).

\begin{figure}[htbp]
    \centering
    \includegraphics[width=\linewidth]{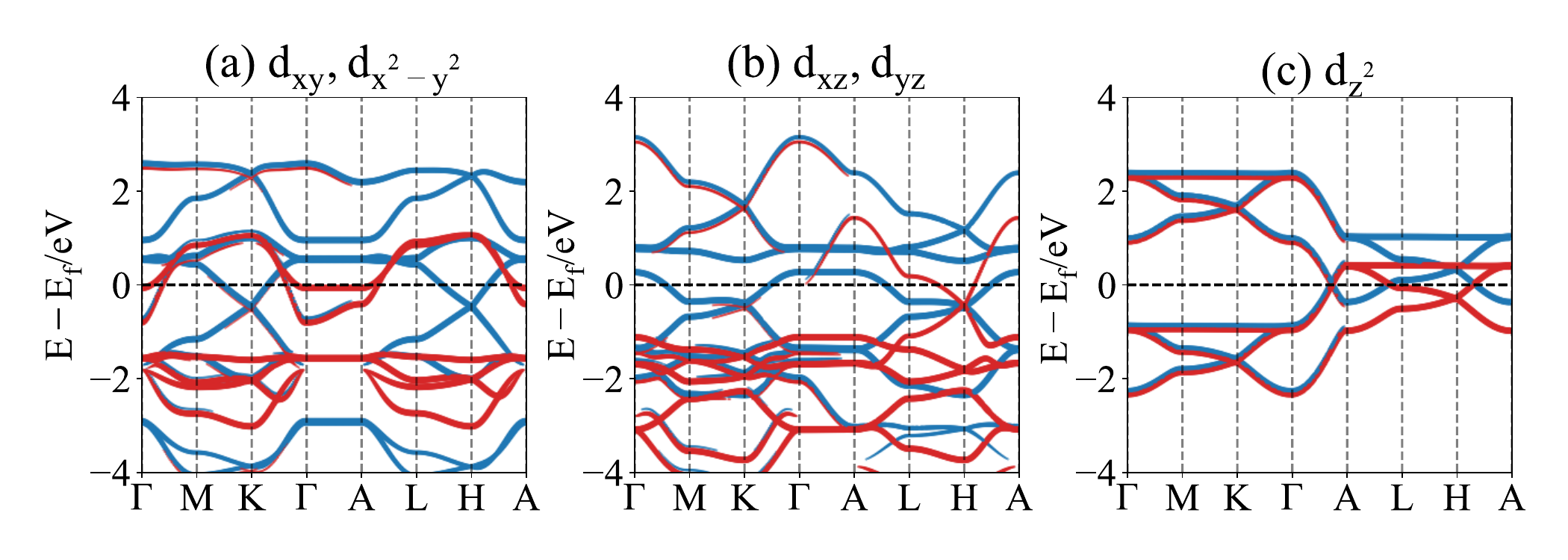}
    \caption{\label{fig:HF-bands} 
    The mean-field band structures of FeGe in the AFM phase, with orbital weights from (a) $(d_{xy}, d_{x^2-y^2})$, (b) $(d_{xz}, d_{yz})$, and (c) $d_{z^2}$. In the figure, the red (blue) colors denote the spin-up (-down) bands from one kagome plane. The other kagome plane is related by $\mathcal{T}\cdot\{E|001\}$, which gives the same dispersion but opposite spin. 
    }
\end{figure}

In \cref{fig:HF-bands}, we present the mean-field band structures in the AFM phase with the projected orbital weights from the $d$-orbitals of Fe. These results show good agreement with the DFT band structure in the AFM phase, as shown in Ref.~\cite{SI}. The calculated magnetic moments and orbital-resolved density of states (DOS) also match well with the DFT results~\cite{SI}.

\section{Application to 1:1 and 1:6:6 class} 

To confirm the generality of our construction for FeGe, we first apply the above formalism to its kagome siblings within the 1:1 class, namely FeSn\cite{haggstrom1975studies, sales2019electronic, inoue2019molecular, kang2020dirac, xie2021spin, han2021evidence, multer2022imaging} and CoSn\cite{sales2019electronic, kang2020topological, liu2020orbital, huang2022flat}. For both compounds, we find a quantitatively good match between DFT and our effective TB models (see \cref{Appendix: compare_11_166}). 
In FeSn and CoSn, a stronger SOC effect is observed, which gaps certain crossing points in the band structure, potentially leading to topological non-trivial states. In Ref.~\cite{SI}, we incorporate an onsite SOC term for the d-orbitals of Fe and Co by fitting the SOC strength to the DFT band structures. The resulting SOC bands in DFT are accurately reproduced using our minimal TB model with the added SOC term.

We then apply the strategy to the larger 1:6:6 class of materials \ch{MT6Z6}, which can be seen as a $z$-direction doubled 1:1 material (\ch{T3Z3}), augmented with an M atom in the middle honeycomb layer. This simple relation motivates us to construct effective models for 1:6:6 materials from the ones of the corresponding 1:1 materials. We use \ch{MgFe6Ge6} as a representative example.

The unit cell of MgFe$_6$Ge$_6$ consists of two FeGe unit cells with an extra Mg atom inducing small displacements for the Fe and triangular Ge atoms, as shown in \cref{Fig:lattice_structure}(e). The DFT band structure and kagome $d$-orbital projections are shown in \cref{Fig:DFT_bands}(b) and \cref{Fig:TBbands}(g)-(i), respectively (see also in \cref{Appendix: compare_11_166} for the unfolded bands). The band structure is close to that of FeGe after a twofold folding along the $ k_3$ direction, resulting from the doubled unit cell in the $z$ direction. 
The minimal TB Hamiltonian of MgFe$_6$Ge$_6$ is constructed starting from the folded Hamiltonian of FeGe and treating the $s$ orbital of Mg as a perturbation, (since the $s$-orbital bands lie high above $E_f$). The resultant band structures shown in \cref{Fig:TBbands}(j)-(l) reproduce remarkably well the main features of the DFT bands near $E_f$. The three-set orbital decomposition and the quasi-flat bands near $E_f$ from FeGe are maintained in \ch{MgFe6Ge6}. 
This strategy is expected to work for all other 1:6:6 materials, and the minimal TB model we build for FeSn and CoSn could also be used to construct their corresponding 1:6:6 materials.

\section{Application to kagome 1:3:5 Family}\label{sec:135-family}

To demonstrate the broad applicability of the LEGO-like building block approach, we extend the formalism to the kagome 1:3:5 family \ch{MT3Z5} (M = K, Rb, Cs; T = Cr, V, Ti; Z = Sb, Bi). 
We focus on three representative materials from this family: \ch{CsCr3Sb5}\cite{liu2024superconductivity, guo2024ubiquitous, li2024correlated, xu2023frustrated, liu2023superconductivity}, \ch{CsV3Sb5}\cite{ORT19,CHO21b,KAN21a,ORT21,ORT21a, kautzsch2023structural}, and \ch{CsTi3Bi5}\cite{yang2023observation, yang2024superconductivity, liu2023tunable, zhou2023physical, yi2023superconducting}. These materials feature kagome layers formed by different transition metals (Cr, V, and Ti), each hosting a distinct number of valence electrons. Consequently, their low-energy physics near the Fermi level varies significantly. In \ch{CsCr3Sb5}, multiple quasi-flat bands appear near $E_f$, while \ch{CsV3Sb5} and \ch{CsTi3Bi5} exhibit several vHSs close to $E_f$. Different electronic structures in these 1:3:5 family materials lead to distinct properties including  superconductivity\cite{liu2024superconductivity} and various types of charge density waves\cite{ORT19}, spin density waves, or nematic transitions\cite{yang2024superconductivity}. 

The 1:3:5 family features two honeycomb layers above and below the kagome layer, which are related by a mirror plane ($M_z$) symmetry. A key adaptation of the LEGO-like building blocks in the 1:3:5 family is the use of $M_z$-even and $M_z$-odd combinations of the $p$ orbitals from the honeycomb Z (Sb or Bi) atoms. The additional honeycomb layer increases the spacing between neighboring kagome layers, significantly reducing interlayer couplings. This reduction accounts for the quasi-2D nature of the band structures observed in the 1:3:5 family. 
By combining with the $d$ orbital decoupling approach introduced in FeGe, we construct three groups of minimal Hamiltonians that effectively decouple the spaghetti-like band structures of the 1:3:5 family. 
In the following, we use \ch{CsCr3Sb5} as a representative example to illustrate the LEGO-like building block method in the 1:3:5 family. The cases of \ch{CsV3Sb5} and \ch{CsTi3Bi5} are discussed in detail in Ref.~\cite{SI}. 
%\cref{app:sec:135-model}.

In \ch{CsCr3Sb5}, the LEGO-like building blocks for the three decoupled Hamiltonians are defined as follows, based on the symmetry analysis:
\begin{itemize}
\item $H_1(\kk)$: Identical to FeGe, which is composed of $(p_x, p_y)$ orbitals from the triangular Sb atoms and $(d_{xy}, d_{x^2-y^2})$ orbitals from kagome Cr atoms, for a total of 8 orbitals that are all inplane and have large overlaps. 

\item $H_2(\kk)$: Composed of $(d_{xz}, d_{yz})$ orbitals from Cr atoms and the $M_z$-odd $(p_x, p_y, p_z)$ orbitals from honeycomb Sb atoms, for a total of 12 orbitals. As the out-of-plane $d_{xz}, d_{yz}$ orbitals of Cr are $M_z$-odd, they couple exclusively to the $M_z$-odd $p$ orbitals of honeycomb Sb, as $z$-directional hoppings are long-range and can be omitted. While the $p_z$ orbitals of triangular Sb also couple to the $d_{xz/yz}$ orbitals, they can be modeled by a simple one-orbital model (see Ref.~\cite{SI})
%Supplementary Material Section VIII 2) 
%\cref{app:sec:135-model-Cr}) 
and are omitted in $H_2(\kk)$. 

\item $H_3(\kk)$: Composed of $d_{z^2}$ orbitals from Cr atoms and the $M_z$-even $(p_x, p_y)$ orbitals from honeycomb Sb atoms, for a total of 7 orbitals. These orbitals form a bipartite lattice and give one perfect flat band when only inter-sublattice coupling is considered. 
\end{itemize}

\begin{figure}[htbp]
    \centering
    \includegraphics[width=\linewidth]{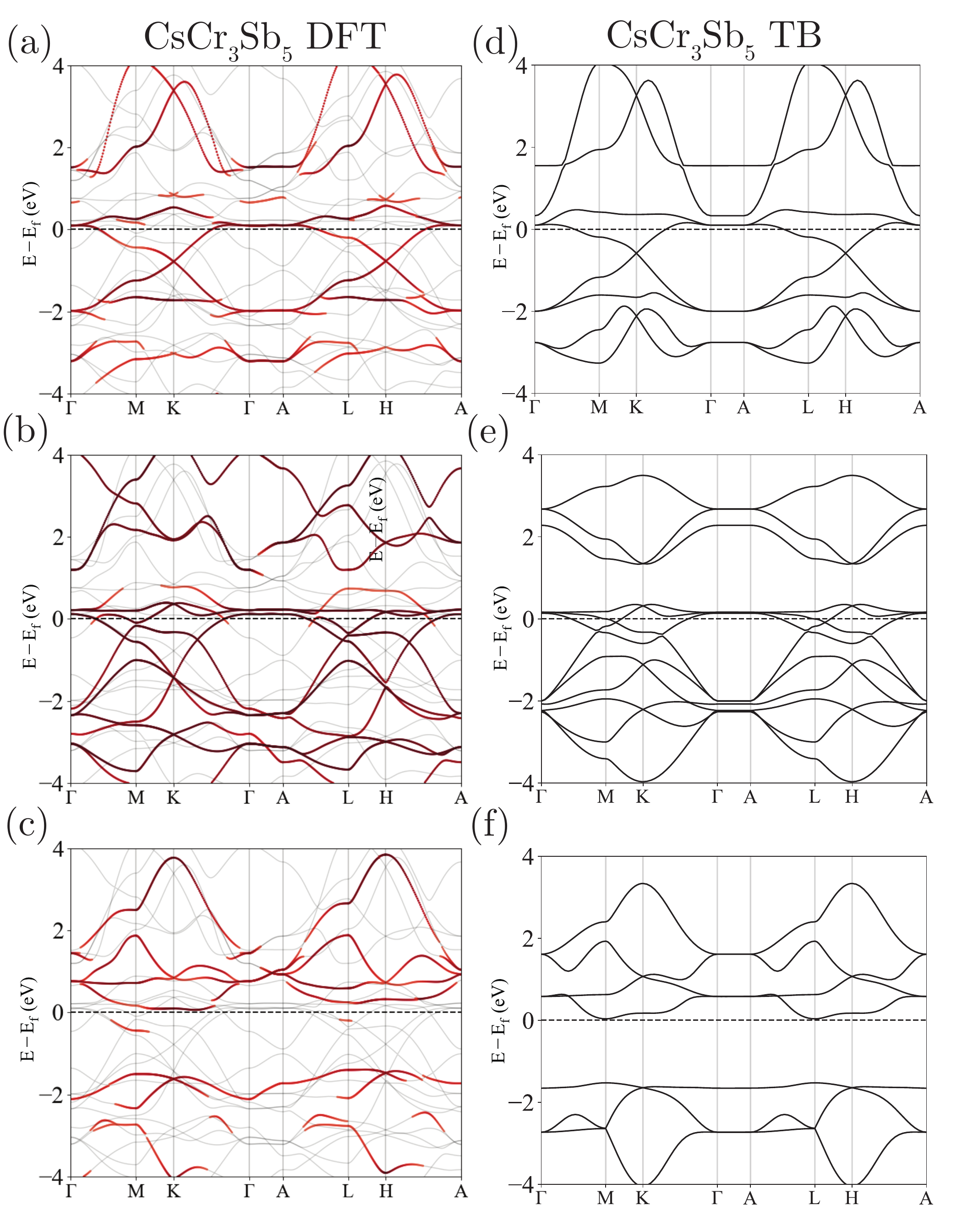}
    \caption{\label{fig:135-bands}  The DFT and fitted minimal TB band structures of the 1:3:5 family \ch{CsCr3Sb5}. (a)-(c) The DFT orbital projections of \ch{CsCr3Sb5} from the three groups of orbitals, where (a) contains $(p_x, p_y)$ orbitals from the triangular Sb atoms and $(d_{xy}, d_{x^2-y^2})$ orbitals from kagome Cr atoms, (b) contains the $(d_{xz}, d_{yz})$ orbitals from Cr and the $M_z$-odd $(p_x, p_y, p_z)$ orbitals from honeycomb Sb, and (c) is composed of $d_{z^2}$ orbitals from Cr atoms and the $M_z$-even $(p_x, p_y)$ orbitals from honeycomb Sb. (d)-(f) shows the band structures of the three fitted minimal TB Hamiltonians $H_{1,2,3}(\kk)$ for \ch{CsCr3Sb5}.
    }
\end{figure}

Using the three minimal TB Hamiltonians defined for \ch{CsCr3Sb5} (see Ref.~\cite{SI}
%Supplementary Material Section VIII
%\cref{app:sec:135-model} 
for details), we fit the parameters to \textit{ab initio} data and obtain the band structures shown in \cref{fig:135-bands}, which exhibit good agreement with the DFT bands. The $H_1(\mathbf{k})$ sector displays a quasi-flat band above $E_f$, similar to the flat band observed in FeGe. However, the $H_2(\mathbf{k})$ and $H_3(\mathbf{k})$ sectors differ significantly from FeGe. In $H_2(\mathbf{k})$, the $d_{xz}, d_{yz}$ orbitals couple strongly with the $M_z$-odd honeycomb $p$-orbitals, resulting in two quasi-flat bands near $E_f$ with a Dirac crossing at the $K$ point. In $H_3(\mathbf{k})$, one quasi-flat band appears above $E_f$, forming part of the four connected honeycomb bands from ($M_z$-even) $p_x,p_y$ orbitals. This quasi-flat band can be understood in terms of the perfect flat band limit, which is identified using the $S$-matrix formalism of the BCL when only inter-sublattice couplings are considered.

For \ch{CsV3Sb5} and \ch{CsTi3Bi5}, we apply similar building blocks and successfully reproduce the \textit{ab initio} band structures using minimal TB models, together with CRPA interactions~\cite{SI}. 
%\cref{app:sec:135-model}). 
This demonstrates that the LEGO-like building block approach developed for FeGe has broad applicability across diverse kagome families and can be tailored to different systems. By decomposing the spaghetti-like band structures of kagome materials into smaller, well-defined building blocks, this method provides deeper insights into their physics, especially the origins of quasi-flat bands and vHSs.

\section{Summary and discussion} 
We have performed a comprehensive first-principle study and have constructed realistic minimal model Hamiltonians for the kagome 1:1 (FeGe, FeSn, and CoSn), 1:6:6 (\ch{MgFe6Ge6}), and 1:3:5 (\ch{CsCr3Sb5}, \ch{CsV3Sb5}, and \ch{CsTi3Bi5}) families. For the first time, we have provided a quantitative understanding of the complicated spaghetti-like band structures of kagome systems. By decomposing the orbitals into three groups and constructing simple but accurate effective models, we have analytically uncovered the origin of the quasi-flat bands near the Fermi level. Our realistic results differ significantly from the simplistic $s$-orbital kagome models usually employed in the literature. 
Moreover, we have shown that the more complicated materials belonging to the 1:6:6 class can be understood using the 1:1 class materials as LEGO-like building blocks and treating the extra atoms as perturbations. 
Our approach can be systematically extended to a broad range of materials beyond kagome systems, as long as the relevant ``LEGO-like building blocks'' are identified through symmetry and chemical analysis.
Our work serves as a complete framework for the theoretical understanding of the band structure in the whole 1:1, 1:6:6, and 1:3:5 classes of kagome materials. The interacting models we have derived can now be solved using various many-body techniques to investigate the magnetic order, CDW, superconductivity, and many other interesting properties of kagome materials, which we leave for future works.

\begin{acknowledgments}
We thank L. Classen, P. M. Bonetti, M. Scherer, C.M. Yue, S.Y. Peng, X.L. Feng, and H.Y. Yang for fruitful discussions. 
Y.J. and H.H. were supported by the European Research Council (ERC) under the European Union’s Horizon 2020 research and innovation program (Grant Agreement No. 101020833), as well as by the IKUR Strategy under the collaboration agreement between Ikerbasque Foundation and DIPC on behalf of the Department of Education of the Basque Government. 
DC acknowledges support from the DOE Grant No. DE-SC0016239.
BAB was supported by the Gordon and Betty Moore Foundation through Grant No. GBMF8685 towards the Princeton theory program, the Gordon and Betty Moore Foundation’s EPiQS Initiative (Grant No. GBMF11070), the Office of Naval Research (ONR Grant No. N00014-20-1-2303), the Global Collaborative Network Grant at Princeton University, the Simons Investigator Grant No. 404513, the BSF Israel US foundation No. 2018226, the NSF-MERSEC (Grant No. MERSEC DMR 2011750), the Simons Collaboration on New Frontiers in Superconductivity, and the Schmidt Foundation at the Princeton University. 
B.A.B. and C.F. are also part of the SuperC collaboration. 
S.B-C. acknowledges financial support from the MINECO of Spain through the project PID2021-122609NB-C21 and by MCIN and by the European Union Next Generation EU/PRTR-C17.I1, as well as by IKUR Strategy under the collaboration agreement between Ikerbasque Foundation and DIPC on behalf of the Department of Education of the Basque Government. 
Y.X. was supported by the National Natural Science Foundation of China (General Program no. 12374163) and the Fundamental Research Funds for the Central Universities (grant no. 226-2024-00200).
H.W. is supported by Chinese Academy of Sciences under grant number XDB33000000, National Key Research and Development Program of China (Grant No.2022YFA1403800), Natural Science Foundation of China (Grant No. 12188101), and the New Cornerstone Science Foundation through the XPLORER PRIZE.
\end{acknowledgments}

\appendix
\counterwithin{figure}{section}
\counterwithin{table}{section}

\section{Local coordinate system}\label{Appendix: local coord}
\begin{figure}[htbp]
	\centering
	\includegraphics[width=0.45\textwidth]{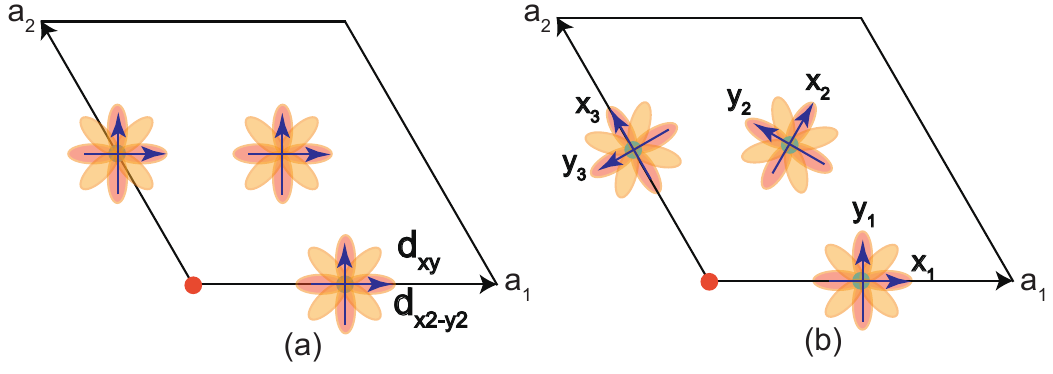}
	\caption{\label{Fig: AP_local_coordinate} Definition of the global (a) and local (b) coordinate systems on the three kagome sites, with $d_{xy}$ and $d_{x^2-y^2}$ placed as representative orbitals.
    In (a), the coordinate systems for the three kagome sites are the same, i.e., along the global $x$ and $y$ directions. In (b), the local coordinate system $(\bm{x}_1, \bm{y}_1)$ at $3f_1$ is the same as the global one, while $(\bm{x}_2, \bm{y}_2)$ at $3f_2$ and $(\bm{x}_3, \bm{y}_3)$ at $3f_3$ are rotated by $\frac{\pi}{3}$ and $\frac{2\pi}{3}$, respectively, relative to the global one.}
\end{figure}

This work adopts a local coordinate system on the $xy$-plane for the kagome sites, under which the five $d$ orbitals of Fe can be decoupled. 
As shown in \cref{Fig: AP_local_coordinate}, the global coordinate systems $(\bm{x}_i,\bm{y}_i)$ on the three kagome sites $3f_1=(\frac{1}{2},0,0), 3f_2=(\frac{1}{2},\frac{1}{2},0), 3f_2=(0,\frac{1}{2},0)$ are simply the Cartesian coordinate, i.e., 
\begin{equation}
    3f_i:\ \bm{x}_i=(1,0,0), \bm{y}_i=(0,1,0),
\end{equation}
while the corresponding local coordinates are defined as:
\begin{equation}
\begin{aligned}
    3f_1:\ &\bm{x}_1=(1,0,0), \bm{y}_1=(0,1,0),\\
    3f_2:\ &\bm{x}_2=(\frac{1}{2},\frac{\sqrt{3}}{2},0),
    \bm{y}_2=(-\frac{\sqrt{3}}{2}, \frac{1}{2},0),\\
    3f_3:\ &\bm{x}_3=(-\frac{1}{2},\frac{\sqrt{3}}{2},0),
    \bm{y}_3=(-\frac{\sqrt{3}}{2}, -\frac{1}{2}, 0).
\end{aligned}
\label{Eq: AP_local_coord}
\end{equation}
The unit vectors are all given in Cartesian coordinates. 

\section{Details of the minimal model}\label{Appendix: model_details}
In this appendix, we give the explicit form of the three minimal TB models $H_{i=1,2,3}(\kk)$ introduced in Sec. \ref{Sec:3models_maintext} in the main text. We employ the local coordinate systems for the kagome sites defined in \cref{Eq: AP_local_coord}. More details can be found in Ref.~\cite{SI}. 

\subsection{$H_1(\kk)$}
For the first group, we use $p_{xy}^t$ to denote the $p_x,p_y$ orbitals of triangular Ge and $d_1, d_2$ to denote the $d_{xy}, d_{x^2-y^2}$ orbitals of Fe in the local coordinates. A tight-binding (TB) model with a few near-neighbor hoppings is then constructed as:
\begin{equation}
H_1(\bm{k})=
\left(
\begin{matrix}
H_{p_{xy}^t}(\bm{k}) & S_{p_{xy}^t, d_1}(\bm{k}) & S_{p_{xy}^t, d_2}(\bm{k}) \\
          & H_{d_1}(\bm{k})  & S_{d_1,d_2}(\bm{k}) \\
H.c.	& 				& H_{d_2}(\bm{k}) \\
\end{matrix}
\right),
\end{equation}
where
\begin{equation}
\begin{aligned}
H_{p_{xy}^t}(\bm{k})&=\mu_{p_{xy}^t} \mathbf{1}_{2},\\
H_{d_1}(\bm{k})&=\mu_{d_1}\mathbf{1}_{3} + 2 t_{d_1}^{NN} H_{\text{Kagome}}^{\text{inplane,NN}}(\bm{k}) \\
& + 2 t_{d_1}^{NNN}
H_{\text{Kagome}}^{\text{inplane,NNN}}(\bm{k}) \\
&+ 2t_{d}^{4N1} \text{Diag}\left[\cos(k_1), \cos(k_1+k_2), \cos(k_2)\right] \\
&+ 2t_{d}^{4N2}
\text{Diag}\left[\cos(k_2)+\cos(k_1+k_2), \right.\\
&\left. \cos(k_1)+\cos(k_2), \cos(k_1)+\cos(k_1+k_2)\right]
\\
H_{d_2}(\bm{k}) &= \mu_{d_2}\mathbf{1}_{3}
+ 2 t_{d_2}^{NN}
H_{\text{Kagome}}^{\text{inplane,NN}}(\bm{k})\\
&+ 2 t_{d_2}^{NNN}
H_{\text{Kagome}}^{\text{inplane,NNN}}(\bm{k})  \\
&+ 2t_{d}^{4N1} \text{Diag}\left[\cos(k_1), \cos(k_1+k_2), \cos(k_2)\right] \\
&+ 2t_{d}^{4N3}
\text{Diag}\left[\cos(k_2)+\cos(k_1+k_2), \right.\\
&\left. \cos(k_1)+\cos(k_2), \cos(k_1)+\cos(k_1+k_2)\right]
\end{aligned}
\end{equation}
\begin{equation}
\begin{aligned}
&S_{d_1,d_2}(\bm{k})=
2 t_{d_1,d_2}^{NN}
\left(
\begin{matrix}
0 &  \cos \frac{k_2}{2}  & -\cos \frac{k_1+k_2}{2} \\
-\cos \frac{k_2}{2} & 0 & \cos \frac{k_1}{2}  \\
\cos \frac{k_1+k_2}{2} & -\cos \frac{k_1}{2} & 0\\
\end{matrix}
\right)\\
&+ 2 t_{d_1,d_2}^{NNN}
\left(
\begin{matrix}
0 &  \cos k_1+\frac{k_2}{2}  & -\cos \frac{k_1-k_2}{2}  \\
-\cos k_1+\frac{k_2}{2}  & 0 & \cos \frac{k_1}{2}+k_2  \\
\cos \frac{k_1-k_2}{2}  & -\cos \frac{k_1}{2}+k_2 & 0\\
\end{matrix}
\right) \\
&+ \sqrt{3}(t_{d}^{4N2}-t_d^{4N3})
\text{Diag}\left[\cos(k_2)-\cos(k_1+k_2), \right.\\
&\left. \cos(k_1)-\cos(k_2), -\cos(k_1)+ \cos(k_1+k_2)\right]
\end{aligned}
\end{equation}
\begin{equation}
\begin{aligned}
&S_{p_{xy}^t, d_1}(\bm{k})= \\
&t_{p_{xy}^t,d_1}^{NN}
\left(
\begin{matrix}
0 &  i\sqrt{3}  \sin(\frac{k_1+k_2}{2}) & i\sqrt{3} \sin(\frac{k_2}{2}) \\
-2i \sin(\frac{k_1}{2}) &  	-i \sin(\frac{k_1+k_2}{2}) & 
i \sin(\frac{k_2}{2}) \\
\end{matrix}
\right) +\\
&t_{p_{xy}^t,d_1}^{NNN}\left(
\begin{matrix}
-2i \sin\frac{k_1+2k_2}{2} & i \sin\frac{k_1-k_2}{2} & -i \sin\frac{2k_1+k_2}{2}\\
0& i \sqrt{3} \sin\frac{k_1-k_2}{2} & i \sqrt{3} \sin\frac{2k_1+k_2}{2}\\
\end{matrix}
\right)\\
&S_{p_{xy}^t, d_2}(\bm{k})=\\
&t_{p_{xy}^t,d_2}^{NN}
\left(
\begin{matrix}
-2i \sin(\frac{k_1}{2}) & -i\sin(\frac{k_1+k_2}{2}) & i \sin(\frac{k_2}{2}) \\
0 &  -i\sqrt{3} \sin(\frac{k_1+k_2}{2}) & -i\sqrt{3} \sin\frac{k_2}{2} \\
\end{matrix}
\right)\\
&H_{\text{Kagome}}^{\text{inplane,NN}}(\bm{k})
=\left(
\begin{matrix}
    0 &  \cos(\frac{k_2}{2})  &\cos(\frac{k_1+k_2}{2}) \\
    & 0 & \cos(\frac{k_1}{2})  \\
    c.c. &  & 0\\
\end{matrix}
\right) \\
&H_{\text{Kagome}}^{\text{inplane,NNN}}(\bm{k}) =
\left(
\begin{matrix}
    0 &  \cos(k_1+\frac{k_2}{2})  &\cos(\frac{k_1-k_2}{2}) \\
    & 0 & \cos(\frac{k_1}{2}+k_2)  \\
    c.c. &  & 0\\
\end{matrix}
\right)
\end{aligned}
\end{equation}
where $c.c.$ denotes complex conjugation. In this model, 
$\mu_{p_{xy}^t}$ and $\mu_{d_i} (i=1,2)$ are the onsite energies of the $p_{x, y}^t$ and $d_{i=1,2}$ orbitals, respectively, while
$t_{p_{xy}^t, d_{i}}^{NN} (i=1,2)$ are the NN inter-sublattice hoppings from the $p_x, p_y$ to the $d_{i}$ orbitals.
The intra-orbital NN and NNN hoppings of the $d_{i}$ orbitals are given by $t_{d_i}^{NN}$ and $t_{d_i}^{NNN} (i=1,2)$, respectively, while
$t_{d_1,d_2}^{NN}, t_{d_1,d_2}^{NNN}$ denote, respectively, the inter-orbital NN and NNN hoppings from the $d_{1}$ to the $d_{2}$ orbitals. Finally,  
$t_d^{4Ni}$ ($i=1,2,3$) are longer-range hoppings for the $d_1$ and $d_2$ orbtials that stretch across the unit cell. 

A few long-range hoppings are considered in the model to give a more faithful reproduction of the DFT bands. The $k_3$-dependence is introduced in the final model by considering the coupling with the $p_z$ orbitals of honeycomb Ge ($p_z^h$):
\begin{equation}
\begin{aligned}
S_{p_z^h, d_2}(\bm{k})&=
t_{p_z^h,d_2}^{NN}\cdot 2i \sin(\frac{k_3}{2})\\
&\times\left(
\begin{matrix}
    e^{\frac{i}{6}(k_1+2k_2)}  & e^{\frac{i}{6}(k_1-k_2)}  & e^{-\frac{i}{6}(2k_1+k_2)} \\
    e^{-\frac{i}{6}(k_1+2k_2)} & e^{-\frac{i}{6}(k_1-k_2)}  & e^{\frac{i}{6}(2k_1+k_2)} \\
\end{matrix}
\right).
\end{aligned}
\end{equation}

\subsection{$H_2(\kk)$}

In this Hamiltonian block, we let $d_3$ and $d_4$ denote the kagome $d_{xz}$ and $d_{yz}$ orbitals of Fe, respectively, as well as $p_z^h$ and $p_z^t$ denote the $p_z$ orbitals of honeycomb and triangular Ge, respectively. 
Considering the NN and possible NNN hoppings, we construct the following 9-band TB model:
\begin{equation}
H_2(\bm{k})=
\left(
\begin{array}{cccc}
H_{p_z^h}(\bm{k}) & \bm{0} & \underline{\bm{0}} & S_{p_z^h, d_4}(\bm{k})  \\
& H_{p_z^t}(\bm{k}) & S_{p_{z}^t, d_3} & \underline{\bm{0}}  \\
&  & H_{d_3}(\bm{k})  & S_{d_3,d_4}(\bm{k}) \\
H.c.& &	& H_{d_4}(\bm{k}) \\
\end{array}
\right),
\end{equation}
where
\begin{equation}
\begin{aligned}
H_{p_z^h}(\bm{k})&=\mu_{p_z^h} \mathbf{1}_{2}\\
& +t_{p_z^h}^{NN} 		
\left(
\begin{matrix}
    0 &  e^{-\frac{i}{3}(2k_1+k_2)} (1+e^{ik_1}+e^{i(k_1+k_2)})\\
    c.c. & 0\\
\end{matrix}
\right)\\
H_{d_i}(\bm{k})&=
    \left(\mu_{d_i} + 2 t_{d_i}^{zNN}\cos(k_3)\right)\mathbf{1}_{3} \\
& + 2t_{d_i}^{NN}
\left(
\begin{matrix}
0 &  \cos(\frac{k_2}{2})  & -\cos(\frac{k_1+k_2}{2}) \\
& 0 & \cos(\frac{k_1}{2})  \\
H.c. &  & 0\\
\end{matrix}
\right)\\
S_{d_3,d_4}(\bm{k})&=
2 t_{d_3,d_4}^{NN}
\left(
\begin{matrix}
0 &  \cos\frac{k_2}{2}  & \cos\frac{k_1+k_2}{2} \\
-\cos\frac{k_2}{2} & 0 & \cos\frac{k_1}{2}  \\
-\cos\frac{k_1+k_2}{2} & -\cos\frac{k_1}{2} & 0\\
\end{matrix}
\right) \\
S_{p_z^h, d_4}(\bm{k})&=
2 t_{p_z^h, d_4}^{NN} \cos(\frac{k_3}{2}) \\
&\times\left(
\begin{matrix}
    -e^{\frac{i}{6}(k_1+2k_2)}  & e^{\frac{i}{6}(k_1-k_2)}  & -e^{-\frac{i}{6}(2k_1+k_2)} \\
    e^{-\frac{i}{6}(k_1+2k_2)} &  -e^{-\frac{i}{6}(k_1-k_2)}  & e^{\frac{i}{6}(2k_1+k_2)} \\
\end{matrix}
\right)\\
H_{p_{z}^t}(\mathbf{k}) &=
\mu_{p_z^t} + 2 t_{p_z^t}^{NN} \cos(k_3),\\
S_{p_z^t,d_3}(\bm{k})&=
2i\cdot t_{p_z^t, d_3}^{NN}
\left(
\begin{matrix}
      \sin(\frac{k_1}{2})  & \sin(\frac{k_1+k_2}{2})  & \sin(\frac{k_2}{2}) \\
\end{matrix}
\right).\\
\end{aligned}
\end{equation}
The TB parameters appearing in $H_2(\kk)$ have similar meanings to those appearing in $H_1(\kk)$ (for example, $t_{d_3,d_4}^{NN}$ is the inter-orbital NN hopping between $d_3$ and $d_4$). 
In $H_2(\kk)$, we use $\bm{0}$ to represent the $S$-matrix that is less relevant for the bands near $E_f$ and is therefore ignored, while $\underline{\bm{0}}$ denote a matrix block with symmetry-forbidden NN hoppings. 
For example, $S_{p_z^t, d_4}=\bm{0}$ is enforced by the $M_{120}(=M_y)$ symmetry, while 
$S_{p_z^h, d_3}=\bm{0}$ is enforced by the $M_{100}$, $M_{010}$, and $M_{110}$ symmetries, as $d_{xz}$ and $d_{yz}$ have opposite representation matrices under these rotations. A detailed examination shows that these symmetries enforce $t_{p_z^h,d_3}^{NN}=0$.

\subsection{$H_3(\kk)$}
In this group, we first construct a 6-band TB model using the $d_{z^2}$ (denoted as $d_5$) orbital of kagome Fe and the $sp^2$ bonding states (denoted as $b$) of the honeycomb Ge:
\begin{equation}
\begin{aligned}
H_3(\bm{k})&=
\left(
\begin{matrix}
H_{b}(\bm{k}) & S_{b, d_5}(\bm{k}) \\
H.c.	&  H_{d_5}(\bm{k}) \\
\end{matrix}
\right),
\end{aligned}
\end{equation}
where
\begin{equation}
\begin{aligned}
&H_{d_5}(\bm{k})= \mu_{d_5} \bm{1}_3 + t_{d_5}^{NN} H_{\text{kagome}}^{\text{inplane},z^2}(k_1,k_2)
\\
&H_{b}(\bm{k}) = (\mu_b + 2 t_{b}^{zNN}\cos(k_3)) \bm{1}_3 + t_b^{NN} H_{\text{kagome}}^{\text{inplane},z^2}(k_1,k_2)
\\
&S_{b,d_5}(\bm{k}) = 2 \cos(\frac{k_3}{2}) \left(t_{b,d_5}^{NN} \bm{1}_3 + t_{b,d_5}^{NNN}
H_{\text{kagome}}^{\text{inplane},z^2}(k_1,k_2) \right)
\\
&H_{\text{kagome}}^{\text{inplane},z^2}(k_1,k_2) =
2\left(
\begin{matrix}
0 & \cos(\frac{k_2}{2}) & \cos(\frac{k_1+k_2}{2})   \\
& 0 & \cos(\frac{k_1}{2})   \\
c.c. & & 0 \\
\end{matrix}
\right). \\
\end{aligned}
\label{Eq: AP-H3-bonding}
\end{equation}
In the model, $\mu_{i=d_5,b}$ are the onsite energies of the $d_5$ and $b$ orbitals, $t_{i=d_5, b}^{NN}$ are the intra-orbital NN intra-kagome hoppings of the $d_5$ and $b$ orbtials, while $t_{b}^{zNN}$ is the $z$-directional hopping of the bonding states. Finally, $t_{b, d_5}^{NN}, t_{b, d_5}^{NNN}$ are the inter-orbital NN and NNN hoppings between the $d_5$ and $b$ orbitals.  

We introduce the bonding states of the honeycomb Ge in this model in order to fit the $d_{z^2}$ orbital weight over a larger energy scale. However, if one is only interested in the low energy physics near $E_f$, it is more convenient to use only the $d_{z^2}$ orbital to build a simpler model. This is because the $d_{z^2}$ orbital weights below $-2$ eV are far from $E_f$ and can therefore be perturbed out. The resulting $d_{z^2}$-only model can be written as: 
\begin{equation}
\begin{aligned}
H_3(\bm{k})=H_{d_5}(\bm{k})=&\left(\mu_{d_5} + 2 t_{d_5}^{zNN} \cos(k_3)\right)\bm{1}_3 \\
&+ t_{d_5}^{NN} H_{\text{kagome}}^{\text{inplane},z^2}(k_1,k_2)
\end{aligned}
\label{Eq:AP-dz2only}
\end{equation}
where $t_{d_5}^{zNN}$ is an extra $z$-directional hopping between the $d_5$ orbitals.

\begin{figure}[tbp]
    \centering
    \includegraphics[width=0.45\textwidth]{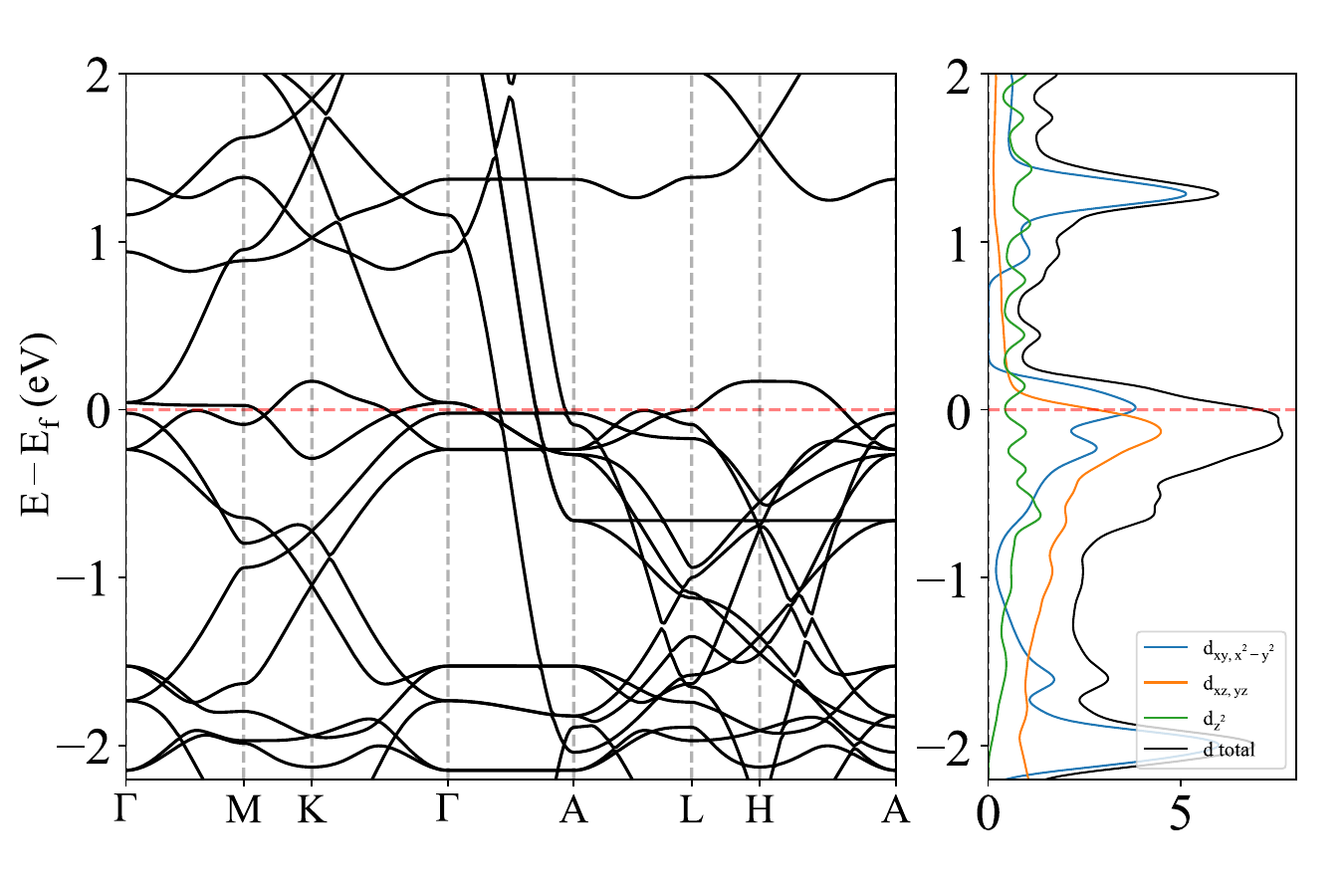}
    \caption{\label{Fig: AP-bands-dos} The band structure and density of states of the minimal TB model \cref{Eq: AP-final-model}.}
\end{figure}

\begin{figure*}[tbp]
    \centering
    \includegraphics[width=1\textwidth]{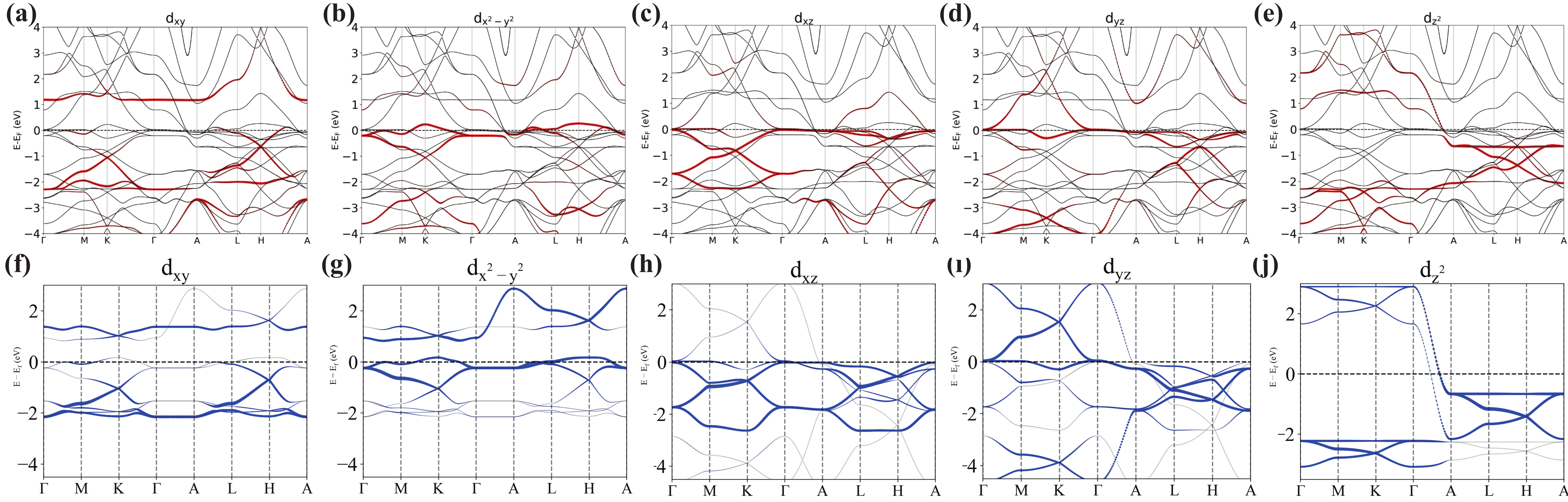}
    \caption{\label{Fig: AP-orbital-bands} The orbital-resolved band structures for the five $d$ orbitals of Fe obtained directly from DFT ((a)-(e)) and computed from the minimal TB model ((f)-(i)). 
    The minimal TB model shows a good agreement with DFT results.}
\end{figure*}

\subsection{Final model}

We obtain the final model for FeGe by combining together the three Hamiltonian blocks derived above. The extra coupling term with the $p_z^h$ orbitals is perturbed out using second-order perturbation theory, as the onsite energy difference between the $p_z^h$ and $d_2$ orbitals is large. 
$p_z^t$ is also perturbed out as it has a relatively small weight at the Fermi level and a large onsite energy difference with the $d_3$ orbitals. We remark that the triangular $p_z^t$ orbital has been reported to be important to the CDW formation in FeGe\cite{miao2022charge, 
wang2023enhanced} and in the 1:6:6 family \ch{ScV6Sn6}\cite{hu2023kagome}. A simple one-orbital model can be built for this triangular $p_z$ orbital separately.

The final model is a direct sum of the three decoupled minimal Hamiltonians
\begin{equation}
H(\bm{k})=H_1'(\bm{k})\oplus H_2'(\bm{k}) \oplus H_3(\bm{k})
\label{Eq: AP-final-model}
\end{equation}
where $H_1'(\bm{k})$ and $H_2'(\bm{k})$ have the following modified TB blocks
\begin{equation}
\begin{aligned}
H_{d_2}(\bm{k}) &\rightarrow H_{d_2}(\bm{k})+H_{d_2, p_z^h}^{(2)}(\bm{k}),\\
H_{p_z^h}(\bm{k}) &\rightarrow  H_{p_z^h}(\bm{k})+H_{p_z^h,d_2}^{(2)}(\bm{k}), \\
H_{d_3}(\bm{k}) &\rightarrow H_{d_3}(\bm{k})+H_{d_3, p_z^t}^{(2)}(\bm{k}).
\end{aligned}
\end{equation}
where
\begin{equation}
\begin{aligned}
H_{d_2, p_z^h}^{(2)}(\bm{k})&=
\frac{1}{\mu_{d_2}-\mu_{p_z^h}}
S^\dagger_{p_z^h, d_2}(\bm{k})S_{p_z^h, d_2}(\bm{k}),\\
H_{p_z^h,d_2}^{(2)}(\bm{k})&=
\frac{1}{\mu_{p_z^h}-\mu_{d_2}}
S_{p_z^h, d_2}(\bm{k})S^\dagger_{p_z^h, d_2}(\bm{k}),\\
H_{d_3, p_z^t}^{(2)}(\bm{k})&=
\frac{1}{\mu_{d_3}-\mu_{p_z^t}}
S^\dagger_{p_z^t, d_3}(\bm{k})S_{p_z^t, d_3}(\bm{k}).
\end{aligned}
\end{equation}
We use $H_{O_1, O_2}^{(2)}$ to denote the second-order perturbation corrections for $O_1$ stemming from the hoppings between $O_1$ and $O_2$. 

The fitted band structure and density of states of the final model are shown in \cref{Fig: AP-bands-dos}. 
The orbital weights for each $d$ orbital of Fe are shown in \cref{Fig: AP-orbital-bands} together with the DFT-computed orbital weights. The TB models reproduced the DFT results remarkably well.  
We use the orbital weights of $d_{z^2}$ from the 6-band model \cref{Eq: AP-H3-bonding} in \cref{Fig: AP-orbital-bands}(j), which reproduces the DFT weights more faithfully. For the simplified 3-band model \cref{Eq:AP-dz2only}, the $d_{z^2}$ weights are simply given by the three upper bands in \cref{Fig: AP-orbital-bands}(j).

\section{Comparison of the band structures in the 1:1 and 1:6:6 families}\label{Appendix: compare_11_166}

\begin{figure*}[tbp]
    \centering
    \includegraphics[width=1\textwidth]{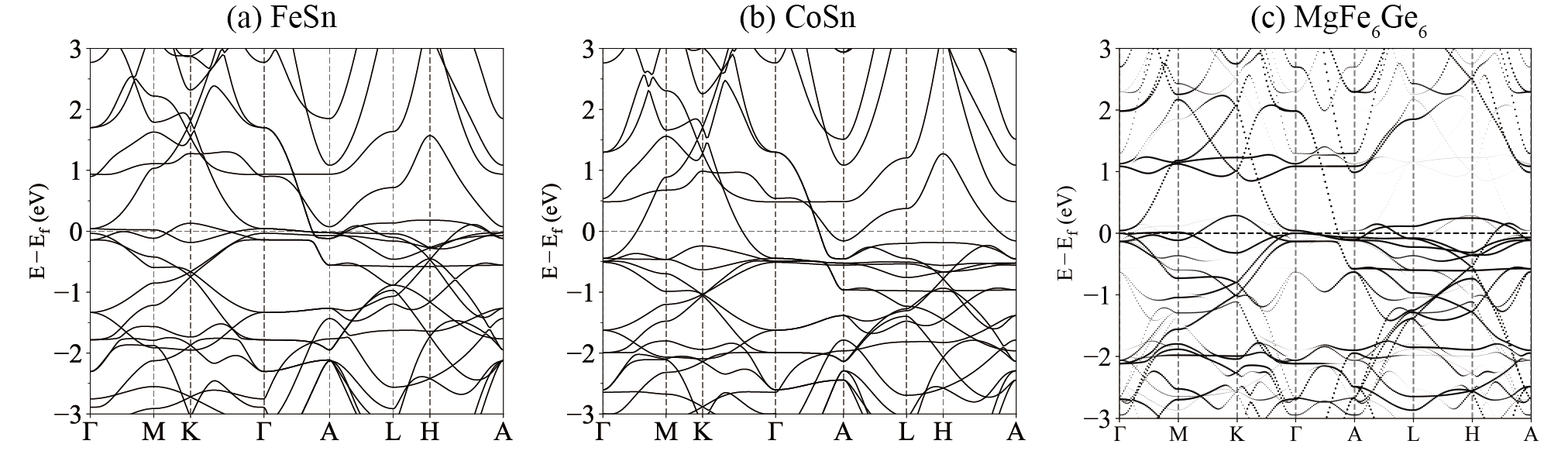}
    \caption{\label{Fig: AP-FeSn-CoSn-bands} DFT band structure of (a) FeSn, (b) CoSn, and (c) \ch{MgFe6Ge6}. The bands are computed in the PM phase without spin-orbital coupling. The bands of \ch{MgFe6Ge6} have been unfolded to the BZ of FeGe for better comparison. These 1:1 and 1:6:6 family kagome materials share very close band structures with varying Fermi levels.}
\end{figure*}

In this appendix, we give a brief comparison of the DFT band structures of the 1:1 family FeSn, CoSn, and the 1:6:6 family \ch{MgFe6Ge6}, as shown in \cref{Fig: AP-FeSn-CoSn-bands}. These 1:1 family kagome materials share very similar band structures with varying Fermi levels. For example, Co has one more valence $d$ electron compared to Fe, so the Fermi level in CoSn is higher and above the two quasi-flat bands. For \ch{MgFe6Ge6}, the bands have been unfolded into the BZ of FeGe for better comparison. The bands of \ch{MgFe6Ge6} are very close to those of FeGe, validating our treatment of \ch{MgFe6Ge6} as a doubled and perturbed version of FeGe.

\section{A brief review of the $S$-matrix formalism and flat band theory}\label{Appendix: AP_SI_Smatrix}

In this section, we give a brief review of the $S$-matrix formalism and the flat-band theory of Ref.\cite{cualuguaru2022general, ma2020spin,regnault2022catalogue}. We only summarize the main results here. Rigorous proofs can be found in Ref.\cite{cualuguaru2022general}.

A bipartite crystalline lattice (BCL) is a periodic lattice with two different sublattices $L$ and $\tilde{L}$. Assume that $N_L$ and $N_{\tilde{L}}$ orbitals per unit cell are placed in the $L$ and $\tilde{L}$ sublattices, respectively. With no loss of generality, we take $N_L\ge N_{\tilde{L}}$. 
The $S$-matrix is the inter-sublattice hopping matrix of the model having dimension $N_{L}\times N_{\tilde{L}}$, denoted by $S(\bm{k})$. The tight-binding (TB) Hamiltonian with only inter-sublattice hoppings has the form
\begin{equation}
    H(\bm{k})=
    \left(
    \begin{matrix}
	\bm{0}_{N_L} &  S(\bm{k}) \\
	S^\dagger(\bm{k}) &  \bm{0}_{N_{\tilde{L}}}  \\
    \end{matrix}
    \right),
\end{equation}
where each entry denotes a matrix block, with $\bm{0}_N$ being the zero matrix of dimension $N\times N$.
This Hamiltonian has chiral symmetry $C$ with the representation matrix
\begin{equation}
    D(C)=\text{Diag}(\bm{1}_{N_L}, -\bm{1}_{N_{\tilde{L}}}),
\end{equation}
i.e., $D(C)H(\bm{k})D^{-1}(C)=-H(\bm{k})$.
The chiral symmetry enforces the dispersion to be chiral-symmetric, which results in at least $N_L-N_{\tilde{L}}$ perfectly flat bands pinned at zero energy. If the rank of the inter-sublattice hopping matrix obeys $\text{rank}(S_{\bm{k}})=r_s\leq N_{\tilde{L}}$, then there will be $2r_s$ chirally-symmetric dispersive bands, leading to $N_L+N_{\tilde{L}}-2r_s$ perfectly flat bands at zero energy.

When intra-sublattice hoppings are added, i.e.,
\begin{equation}
    H(\bm{k})=
    \left(
    \begin{matrix}
	A(\bm{k}) &  S(\bm{k}) \\
	S^\dagger(\bm{k}) &  B(\bm{k})  \\
    \end{matrix}
    \right),
\end{equation}
the Hamiltonian is no longer chiral-symmetric. If $A(\bm{k})$ has a $\bm{k}$-independent eigenvalue of multiplicity $n_a$ ($n_a>N_{\tilde{L}}$), then $H(\bm{k})$ will have at least $n_a-N_{\tilde{L}}$ perfectly flat band, whose energy is not necessarily zero.

%% file: supplement_bare.tex
\section{Crystal structure and band structure of $\text{FeGe}$}\label{Sec:Appendix_bandstructure}
\subsection{Crystal structure}
\begin{figure}[htbp]
	\centering
	\includegraphics[width=1\textwidth]{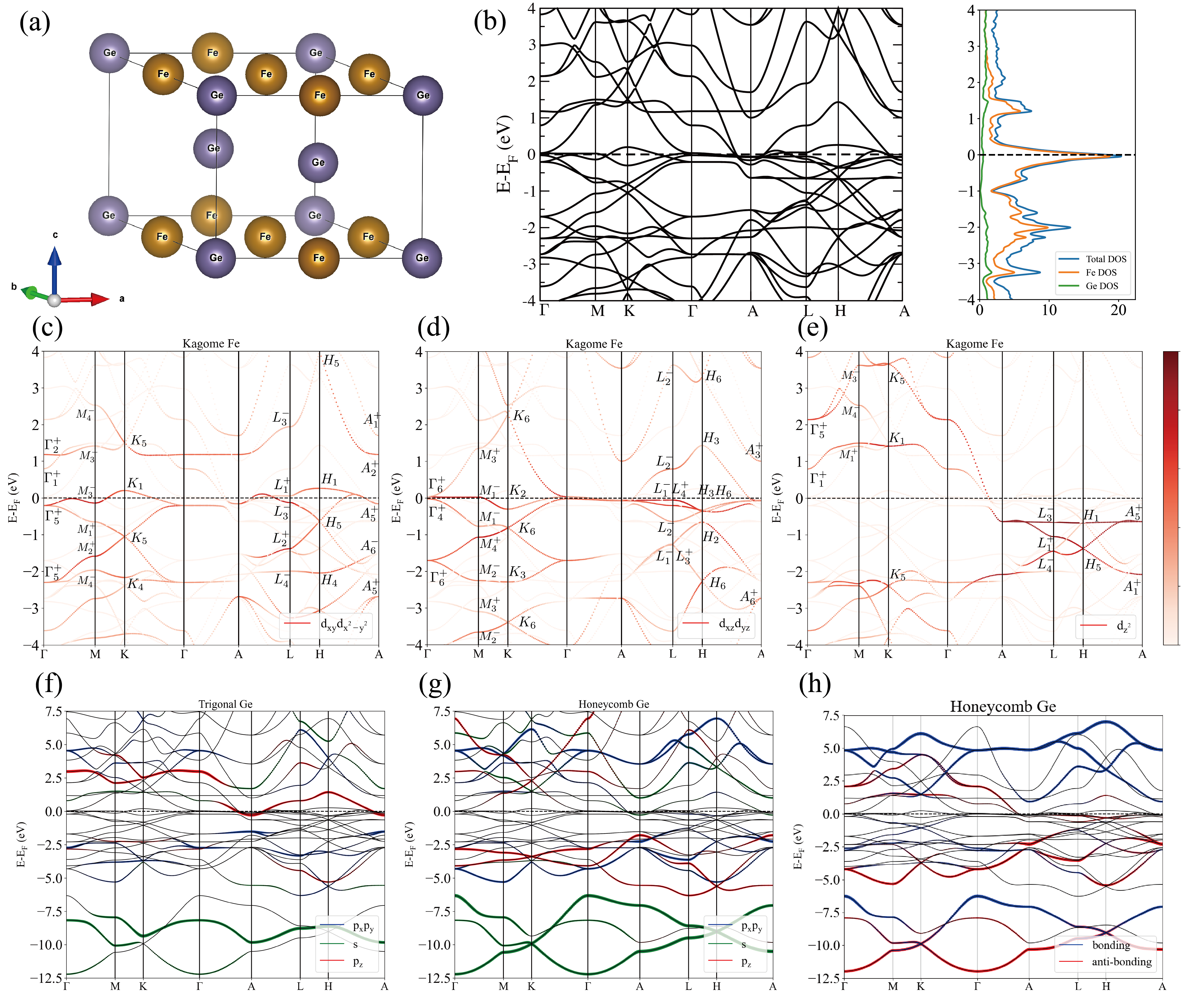}
	\caption{\label{Fig:FeGe_PM_struct_band}(a) Crystal structure and (b) band structure and density of states (DOS) without SOC of FeGe in the paramagnetic phase.
    (c)-(h) are orbital projections of the bands in the PM phase of FeGe, where (c) is the projection of $d_{xy}, d_{x^2-y^2}$ of Fe, (d) $d_{xz}, d_{yz}$ of Fe, (e) $d_{z^2}$ of Fe, (f) $s, p$ orbitals of honeycomb Ge, (g) $s, p$ orbitals of triangular Ge, and (h) the bonding and anti-bonding states formed by $x,p_x,p_y$ orbitals of honeycomb Ge (see \cref{Sec:tb_models_wannier}).  IRREPs are marked in (c)-(e) for three groups of $d$ orbitals. 
 }
\end{figure}

FeGe has space group (SG) 191 $P6/mmm$ symmetry (or more rigorously, Shubnikov space group 191.234 $P6/mmm1^\prime$) in the paramagnetic phase, which can be generated by $C_{6z}, C_{2,110}$, and inversion $P$, together with the time-reversal symmetry (TRS) $\mathcal{T}$. 
The lattice constants given in Ref.\cite{teng2022discovery} are $a=4.985\AA$ and $c=4.049\AA$. The crystal structure is shown in \cref{Fig:FeGe_PM_struct_band}(a), with the three basis vectors of the conventional cell taken as 
\begin{equation}
\begin{aligned}
    &\bm{a}_1=(a,0,0),\quad
    \bm{a}_2=(-\frac{1}{2}a, \frac{\sqrt{3}}{2}a, 0),\quad 
    \bm{a}_3=(0,0,c), \\
&\bm{b}_1 = \frac{4\pi}{\sqrt{3}a}(\frac{\sqrt{3}}{2}, \frac{1}{2}, 0),\quad 
\bm{b}_2 = \frac{4\pi}{\sqrt{3}a}(0, 1, 0),\quad 
\bm{b}_3 =\frac{2\pi}{c} (0, 0, 1),
\end{aligned}
\label{eq_conv_cell}
\end{equation}

where the coordinates are given in the Cartesian coordinate system.
We divide the atoms of FeGe into three lattices:
\begin{itemize}
	\item Fe atoms at Wyckoff position $3f$ form a kagome lattice, with site symmetry group $D_{2h}$.
	\item Ge atom at Wyckoff position $1a$ forms a triangular lattice, with site symmetry group $D_{6h}$.
	\item Ge atoms at Wyckoff position $2d$ form a honeycomb lattice, with site symmetry group $D_{3h}$.
\end{itemize}
In \cref{SG191-wyckoff}, we list the Wyckoff positions and their site symmetry groups, with coordinates written in the basis of $(\bm{a}_1,\bm{a}_2,\bm{a}_3)$. In \cref{SG191-PG-IRREPs}, we tabulate the IRREPs and their basis functions of these site symmetry groups. 

Notice that although five $d$ orbitals form two 2D and one 1D irreducible representations (IRREPs) in $D_{6h}$, they can only form 1D IRREPs in $D_{2h}$ without spin-orbital coupling (SOC) (for each spin), which is the site symmetry group of the kagome lattice $3f$. When a global coordinate system is used to define orbitals at different kagome sites, e.g., the Cartesian coordinates, the $d_{xz}$ and $d_{yz}$ orbitals are coupled (the same holds for $d_{xy}$ and $d_{x^2-y^2}$) and form a 2D reducible representation. For example, $d_{xz}$ at $(\frac{1}{2},0,0)$ will transform into a linear combination of $d_{xz}$ and $d_{yz}$ at $(\frac{1}{2},\frac{1}{2},0)$ under $C_6$. However, if local coordinate systems are defined at each kagome site, for example, the Cartesian coordinates at $(\frac{1}{2},0,0)$, a $C_6$-rotated Cartesian coordinate at $(\frac{1}{2},\frac{1}{2},0)$, and a $C_3$-rotated Cartesian coordinate at $(0,\frac{1}{2},0)$, then the $d_{xz}$ and $d_{yz}$ orbitals defined at each local coordinate are fully decoupled and form 1D IRREPs. A more detailed discussion of the local coordinates is left in Sec.\ref{Sec:tb_models_wannier}.

\begin{table}[htbp]
    \centering
    \begin{tabular}{c|c|c|c}
	\hline\hline
	Lattice  & Wyckoff & Site symmetry & Coordinates \\\hline
	\multirow{3}{*}{triangular} & $1a$ & $D_{6h}(6/mmm)$ & $(0, 0, 0)$	 \\\cline{2-4}
        & $1b$ & $D_{6h}(6/mmm)$ & $(0, 0, \frac{1}{2})$ \\\cline{2-4}
        & $2e$ & $C_{6v}(6/mm)$ & $(0, 0, z), (0, 0, -z)$ \\\hline
	\multirow{2}{*}{Honeycomb} & $2c$ & $D_{3h}(-6m2)$ & $(\frac{1}{3}, \frac{2}{3}, 0), (\frac{2}{3}, \frac{1}{3}, 0)$\\\cline{2-4}
        & $2d$ & $D_{3h}(-6m2)$ &  $(\frac{1}{3}, \frac{2}{3}, \frac{1}{2}), (\frac{2}{3}, \frac{1}{3}, \frac{1}{2})$\\\hline
	\multirow{4}{*}{Kagome} & $3f$ & $D_{2h}(mmm)$ & $(\frac{1}{2},0,0), (\frac{1}{2},\frac{1}{2},0), (0,\frac{1}{2},0)$\\\cline{2-4}
        & $3g$ & $D_{2h}(mmm)$ & $(\frac{1}{2},0,\frac{1}{2}), (\frac{1}{2},\frac{1}{2},\frac{1}{2}), (0,\frac{1}{2},\frac{1}{2})$\\\cline{2-4}
        & \multirow{2}{*}{$6i$} & \multirow{2}{*}{$C_{2v}(2mm)$} & $(\frac{1}{2},0,z), (\frac{1}{2},\frac{1}{2},z), (0,\frac{1}{2},z)$,\\
        & & & $(\frac{1}{2},0,-z), (\frac{1}{2},\frac{1}{2},-z), (0,\frac{1}{2},-z)$\\\hline\hline
    \end{tabular}
    \caption{\label{SG191-wyckoff} Wyckoff positions and site symmetry groups in SG 191. For a complete list of Wyckoff positions, see \textit{Bilbao Crystallographic Server} (BCS)\cite{aroyo2006bilbao1, aroyo2006bilbao2}. }
\end{table}

\begin{table}[htbp]
	\centering
	\begin{tabular}{ccc||ccc||ccc}
		\hline\hline
		$D_{6h}$ & Dim & Basis &
        $D_{3h}$ & Dim & Basis &
        $D_{2h}$ & Dim & Basis \\\hline
		$A_{1g}$ & 1 & $s$, $d_{z^2}$ &
        $A_1^\prime$ & 1 & $s$ &
        $A_g$ & 1 & $d_{z^2}, d_{x^2-y^2}$\\
		$E_{1g}$ & 2 & $(d_{xz}, d_{yz})$ & $A_2^{\prime\prime}$ 
         & 1 & $p_z$ &
        $B_{1g}$ & 1 & $d_{xy}$\\
		$E_{2g}$ & 2 & $(d_{xy}, d_{x^2-y^2})$ & $E^\prime$ & 2 &  $(p_x, p_y)$ &
        $B_{2g}$ & 1 & $d_{xz}$\\
		$A_{2u}$ & 1 & $p_z$ & & & &
        $B_{3g}$ & 1 & $d_{yz}$ \\
		$E_{1u}$ & 2 & $(p_x, p_y)$ & & & & & \\\hline\hline
	\end{tabular}
	\caption{\label{SG191-PG-IRREPs} Irreducible representations with their dimension and basis functions of point group $D_{6h}$, $D_{3h}$, and $D_{2h}$. For a complete list of IRREPs, see BCS\cite{aroyo2006bilbao1, aroyo2006bilbao2}. }
\end{table}

\subsection{Magnetic order}

FeGe is reported to be paramagnetic above the N\'eel temperature $T_N \approx 410$ $K$, and forms an A-type antiferromagnetic (AFM) order below $T_N$\cite{teng2022discovery, teng2023magnetism}, with the magnetic moment on each Fe atom being around $1.7$ $\mu_B$\cite{ohoyama1963new, haggstrom1975mossbauer, forsyth1978low}. In \cref{Fig:FeGe_AFM_struct_band}(a), we show the crystal structure with AFM order.

The structure of the AFM phase has the symmetry of type-IV Shubnikov SG, i.e., magnetic space group (MSG) 192.252 $P_{c}6/mcc$ in BNS setting, or MSG 191.13.1475 $P_{2c}6/mm^\prime m^\prime$ in the OG setting.
Using the convention of OG setting (where operations are written in the original non-magnetic unit cell), FeGe has an anti-unitary translation $\mathcal{T}\cdot \{E|001\}$ and a unitary halving subgroup $P6/mcc$. The anti-unitary translation $\mathcal{T}\cdot \{E|001\}$ connects two kagome layers and reverses the direction of spin, which will lead to spin-degeneracy in the AFM band structure (in the AFM BZ).

We also discuss the spin-space group (SSG)\cite{brinkman1966theory, Litvin1974, yang2021symmetry, jiang2023enumeration,xiao2023spin,ren2023enumeration} of FeGe. SSGs give a complete symmetry description of magnetic materials with weak spin-orbit coupling, in which the operations have (partially) unlocked real space and spin rotations. Denote an SSG operation as $\{R_s||R_l|\bm{\tau}\}$, where $R_s$ is the spin rotation and $\{R_l|\bm{\tau}\}$ is the real-space rotation and translation. We use a simplified notation that the inversion in spin space is TRS, i.e., $\{P||E|\bm{0}\}=\mathcal{T}$. 
The SSG $\mathcal{S}$ of FeGe can be identified as the collinear group 191.2.1.1.L \cite{jiang2023enumeration}, which has a pure-lattice operation group $\mathcal{H}$ being SG 191, a pure-spin operation group owned by all collinear orders $\mathcal{S}_0=\{C_{\infty z}|| E|\bm{0}\} + \{M_x C_{\infty z}|| E|\bm{0}\}$, 
and a quotient group $Q=\mathcal{S}/(\mathcal{H} \times \mathcal{S}_0)$.
The quotient group $\mathcal{Q}$ contains non-trivial SSG operations and is generated by $\{M_z||E|001\}$ (a lattice translation $\bm{\tau}=(0,0,1)$ accompanied by a $M_z$ in spin space that reverses the spin). This operation is equivalent to $\mathcal{T}\cdot\{E|001\}$ when combining $\{M_z||E|001\}$ with $\{C_{2z}||E|\bm{0}\}$. 
The pure-spin operation group $\mathcal{S}_0$ for collinear orders contains arbitrary-angle spin rotation $C_{\infty z}$ along the $z$-axis, together with all mirrors $M_x C_{\infty z}$ with mirror planes passing the $z$-axis. These two types of operations maintain the collinear magnetic order (assumed to be along the $z$-direction). The pure lattice group $\mathcal{H}$ describes a doubled unit cell due to the $A$-type AFM order, with the translation group basis $\bm{A}_3=(0,0,2c)=2\bm{a}_3$ where $\bm{a}_3$ is the original translation basis in the third direction. The quotient group generator $\{M_z||E|001\}$ is the combination of $\bm{a}_3$ with a spin reflection $M_z$.

Compared with MSG, the collinear SSG of FeGe contains extra pure-spin rotations along the $z$-axis. However, as the MSG of FeGe can be obtained by combining certain operations in the original SG 191 with TRS but does not eliminate any spacial symmetry, the SSG here does not show much advantage. In \cref{Sec:FeSn_CoSn} when discussing the SSG of FeSn which has inplane AFM order, the SSG is more advantageous and has much more symmetry than the corresponding MSG. 
In this case, the extra symmetries given by SSG could lead to extra band degeneracy and more symmetry constraints on other physical properties including the anomalous Hall effect (AHE) and non-linear optical responses.

\subsection{Band structures}
\subsubsection{Paramagnetic phase}
We use the Vienna \textit{ab-initio} Simulation Package (VASP)\cite{kresse1996efficiency, kresse1993ab1, kresse1993ab2, kresse1994ab, kresse1996efficient} to perform the \textit{ab-initio} computations for PM and AFM phases of FeGe.

The paramagnetic (PM) band structure without spin-orbit coupling (SOC) is shown in \cref{Fig:FeGe_PM_struct_band}(b), where the bands near the Fermi level are mainly contributed by the $d$ orbitals of Fe according to the orbital projections shown in \cref{Fig:FeGe_PM_struct_band}(c)-(g). We label the band IRREPs for three groups of $d$ orbitals in (c)-(e), where the IRREP labels follow the convention in \textit{Bilbao Crystallographic Server} (BCS)\cite{aroyo2006bilbao1, aroyo2006bilbao2}. 

From the orbital projections, it can be seen that there are two quasi-flat bands near $E_f$, one from $(d_{xy}, d_{x^2-y^2})$ while the other one from $(d_{xz}, d_{yz})$ of Fe. $d_{z^2}$ orbitals also contribute a extremely flat band on $k_3=\pi$ plane at about $-0.5$ $eV$. The bands of $(d_{xy}, d_{x^2-y^2})$ in energy range $[-2, 0.5]$ eV are roughly the same on the $k_3=0,\pi$ planes, with von Hove singularities (vHS) at $M$ and $L$, and Dirac points at $K$ and $H$. However, for $(d_{xy}, d_{x^2-y^2})$ and $d_{z^2}$, their bands on the $k_3=0,\pi$ planes change significantly.

\subsubsection{Antiferromagnetic phase}\label{Sec:SI_AFM_bands}

\begin{figure}[htbp]
	\centering
	\includegraphics[width=1\textwidth]{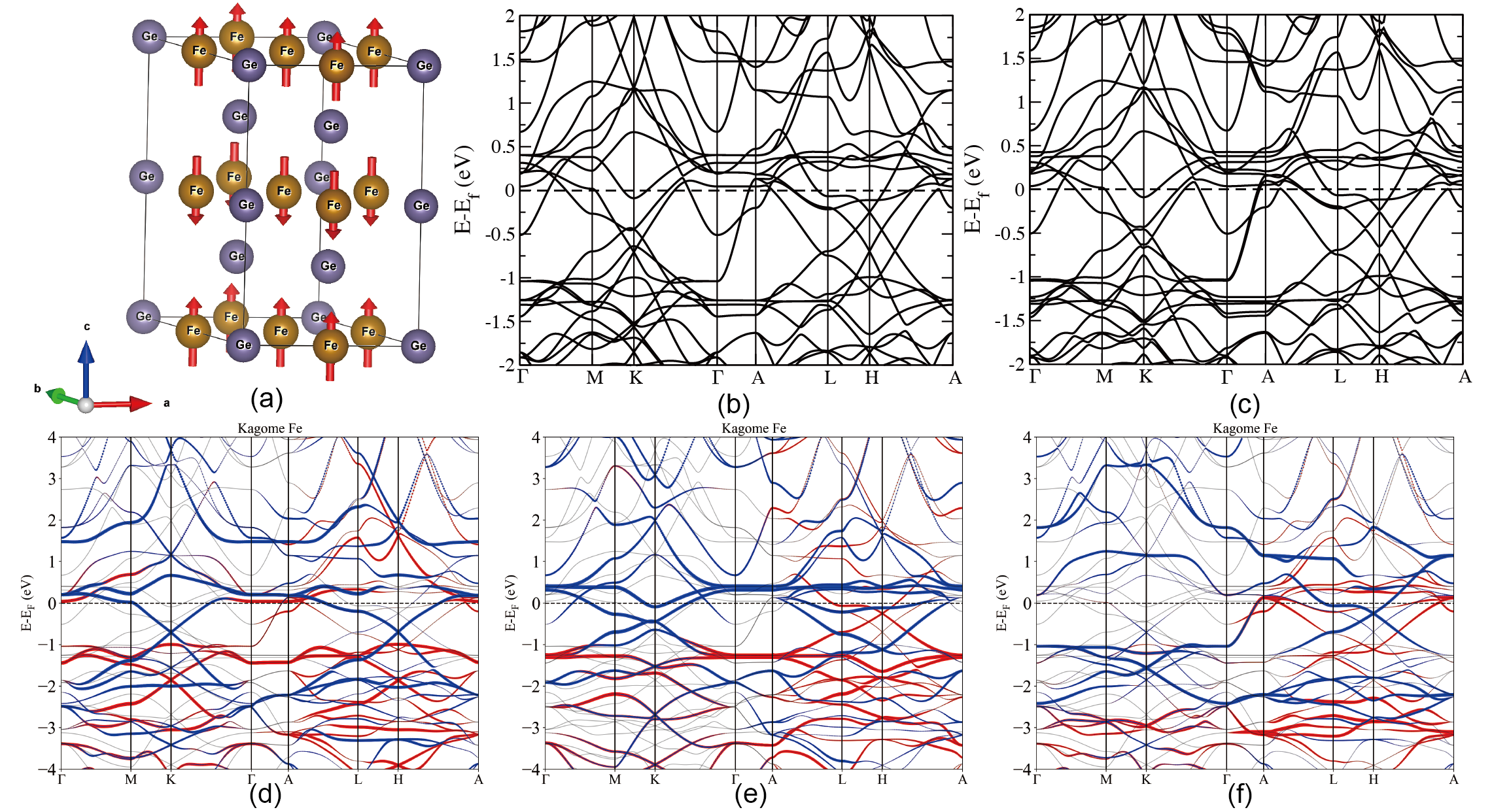}
	\caption{\label{Fig:FeGe_AFM_struct_band}(a) Crystal structure, (b) band structure without SOC, and (c) band structure with SOC of FeGe in the AFM phase.
    (d)-(f) are orbital projections of the bands in the AFM phase of FeGe, where (c) is the projection of $d_{xy}, d_{x^2-y^2}$ of Fe, (d) $d_{xz}, d_{yz}$ of Fe, (e) $d_{z^2}$ of Fe, where blue and red colors denote $d$ orbitals of opposite spins of three Fe on one kagome plane.}
\end{figure}

In \cref{Fig:FeGe_AFM_struct_band}(b), (c) the AFM band structure without and with SOC computed from DFT. SOC has negligible effects on the band structure near $E_f$. The computed magnetic moment on each Fe atom is about $1.53$ $\mu_B$ from DFT, which is close to the experiment value of around $1.7$ $\mu_B$ in literature\cite{ohoyama1963new, haggstrom1975mossbauer, forsyth1978low}. The bands in the AFM phase are two-fold degenerate with the $d$ orbital bands consisting of spin-up orbitals from one kagome layer and spin-down orbitals from the other kagome layer, related by the anti-unitary translation $\mathcal{T}\cdot \{E|001\}$.

In \cref{Fig:FeGe_AFM_struct_band}(d)-(f), we show the $d$ orbital projections of Fe in the AFM bands without SOC. We use blue and red colors to denote $d$ orbitals of opposite spins from three Fe atoms on one kagome layer. The three Fe atoms on the other kagome layer have identical orbital projections, with spin-up and down bands reversed. 
There are two vHSs that are close to the Fermi level, with the one at $M$ mainly coming from the $d_{xy}$ and $d_{x^2-y^2}$ orbitals while the one at $L$ mainly from $d_{z^2}$ but has some mixing with $d_{xz}$ and $d_{yz}$ orbitals of Fe at kagome sites. The vHS at $M$ is 19 meV higher than the Fermi level, while the one at $L$ is 69 meV lower.

\subsubsection{DFT calculation details}
The ab-initio calculations in this work are performed using the Vienna ab-initio Simulation Package (VASP)\cite{kresse1996efficient} with generalized gradient approximation of Perdew-Burke-Ernzerhof (PBE) exchange-correlation potential\cite{PBE1996}. The $8\times8\times8$ $k$-mesh and an energy cutoff of 500 eV are used for both self-consistency and cRPA computations (\citeSI{Sec:CRPA}). 
The MLWFs are obtained using WANNIER90\cite{mostofi2008wannier90}. The local coordinate system defined in \cref{Eq:local_coordinate} is used in constructing MLWFs.

\section{Tight-binding (TB) models of FeGe}\label{Sec:SI_FeGe_TB}

In this section, we construct tight-binding (TB) Hamiltonians for FeGe. We first use maximally localized Wannier functions to construct faithful TBs as references. Then we construct three decoupled groups of minimal TB models that capture the main features of FeGe.

\subsection{TB from Wannier functions}\label{Sec:tb_models_wannier}

We first construct maximally localized Wannier functions (MLWFs) by Wannier90\cite{marzari1997maximally, souza2001maximally, marzari2012maximally, pizzi2020wannier90} to obtain onsite energies and hoppings parameters of orbitals and use the Wannier TB to calculate the filling and magnetic moments of each orbital in both PM and AFM phases.

\begin{figure}[htbp]
	\centering
	\includegraphics[width=0.6\textwidth]{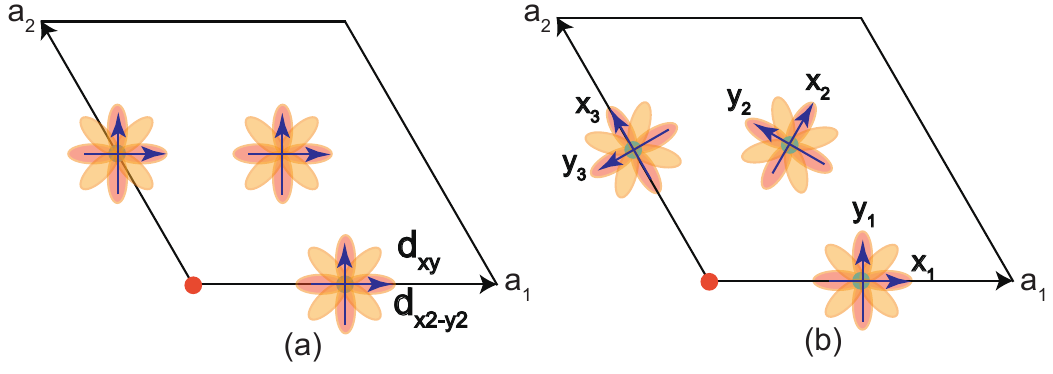}
	\caption{\label{Fig:local_coordinate} Definition of (a) global and (b) local coordinate systems on three kagome sites, with $d_{xy}$ and $d_{x^2-y^2}$ placed as representative orbitals.
    In (a), the coordinate system at three kagome sites is the same, i.e., along the global $x$ and $y$ direction, while in (b), the local coordinate system $(x_1, y_1)$ at $3f_1$ is the same as the global one, while $(x_2, y_2)$ at $3f_2$ and $(x_3, y_3)$ at $3f_3$ is rotated $\frac{\pi}{3}$ and $\frac{2\pi}{3}$ from the global one, respectively.}
\end{figure}

To construct MLWFs, it is more convenient to use local coordinates on kagome sites $3f$ to decouple the five $d$ orbitals.
As shown in \cref{Fig:local_coordinate}, the global coordinate system on each kagome site is the Cartesian coordinate, i.e., 
\begin{equation}
    3f_i:\ \bm{x}_i=(1,0,0), \bm{y}_i=(0,1,0), \bm{z}_i=(0,0,1),
\end{equation}
while the local coordinates at each kagome site $3f_i$ are defined as (same for kagome site $3g$):
\begin{equation}
    \begin{aligned}
        3f_1:\ &\bm{x}_1=(1,0,0), \bm{y}_1=(0,1,0),\bm{z}_1=(0,0,1)\\
        3f_2:\ &\bm{x}_2=(\frac{1}{2},\frac{\sqrt{3}}{2},0),
        \bm{y}_2=(-\frac{\sqrt{3}}{2}, \frac{1}{2},0),
        \bm{z}_2=(0,0,1)\\
        3f_3:\ &\bm{x}_3=(-\frac{1}{2},\frac{\sqrt{3}}{2},0),
        \bm{y}_3=(-\frac{\sqrt{3}}{2}, -\frac{1}{2}, 0),
        \bm{z}_3=(0,0,1).
    \end{aligned}
    \label{Eq:local_coordinate}
\end{equation}
where the coordinates are written in Cartesian coordinates. 

\begin{figure}[htbp]
	\centering
	\includegraphics[width=1\textwidth]{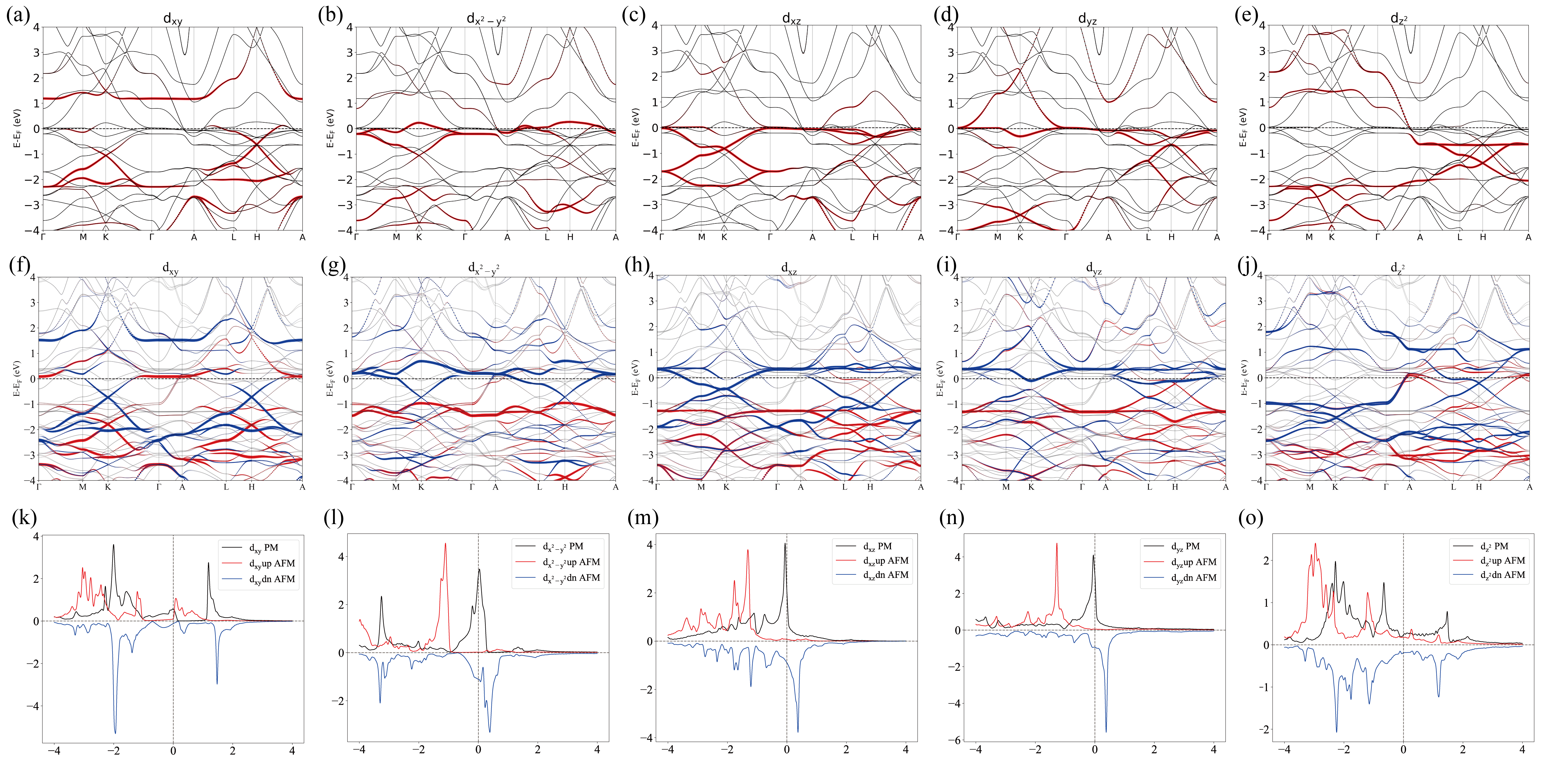}
	\caption{\label{Fig:wannier_d_projections} The band projections of five $d$ orbital of Fe in PM phase (a)-(e) and AFM phase (f)-(j), and density of states (DOS) in (h)-(l), obtained from MLWFs where local coordinates defined in \cref{Eq:local_coordinate} are used in order to separate $d$ orbitals. In (f)-(j), the blue and red colors denote two opposite spins, where only three Fe atoms on the same kagome layer are considered, as the other three Fe atoms have the reversed spin distributions enforced by the anti-unitary translation $\mathcal{T}\cdot\{E|0,0,1\}$. }
\end{figure}

We first construct MLWFs in both the PM and AFM phases using a large orbital group, i.e., $s, p, d$ orbitals of Fe and $s,p$ orbitals of Ge, in order to give a faithful representation of the DFT band structure. In \cref{Tab:filling_magmom_wannier}, we show the filling number of each orbital in the PM and AFM phases and the magnetic moments of $d$ orbitals in the AFM phase. The total fillings in the PM and AFM phases are close to 36, which is the number of valence electrons in FeGe used in DFT. 
We also plot the orbital projections and density of stats of five $d$ orbitals separately in both the PM and AFM phase in \cref{Fig:wannier_d_projections}.

Note that the filling in Wannier TB is not exactly the number of electrons, because FeGe is a metallic system with numerous bands crossing $E_f$. The total filling (per PM unit cell) in the PM and AFM phases calculated from Wannier TB is 36.26 and 36.15, respectively (close to the number of valence electrons used in DFT, i.e., 36 from $s,p,d$ of 3 Fe and $s,p$ of 3 Ge). 
When constructing MLWFs, the dispersion and orbitals weights in the fitted Wannier bands could be slightly different from DFT, and the number of valence electrons is not fixed. In this case, the total filling of the orbitals considered may be slightly different. 
If we enforce the total filling to be the same as DFT in Wannier, $E_f$ would be about 25 $meV$ lower, which is insignificant as the Wannier filling does not deviate much from the DFT value.

\begin{table}[htbp]
\begin{tabular}{c|c|c|c|c|c|c|c|c|c|c|c}
\hline\hline
& Orbitals & $s$  & $p_z$ & $p_x$ & $p_y$ & $d_{z^2}$ & $d_{xz}$ & $d_{yz}$ & $d_{x^2-y^2}$ & $d_{xy}$ & Total \\ \hline
\multirow{3}{*}{PM filling}  & Fe       & 0.35 & 0.22  & 0.25  & 0.26  & 1.54      & 1.72     & 1.56     & 1.42          & 1.57     & 8.88  \\ \cline{2-12} 
& Tri-Ge   & 1.28 & 0.36  & 0.75  & 0.75  & -         & -        & -        & -             & -        & 3.12  \\ \cline{2-12} 
& Hon-Ge   & 1.26 & 0.74  & 0.60  & 0.60  & -         & -        & -        & -             & -        & 3.25  \\ \hline
\multirow{3}{*}{AFM filling} & Fe       & 0.24 & 0.25  & 0.25  & 0.25  & 1.66      & 1.63     & 1.48     & 1.49          & 1.62     & 8.87  \\ \cline{2-12} 
& Tri-Ge   & 1.19 & 0.44  & 0.76  & 0.76  & -         & -        & -        & -             & -        & 3.14  \\ \cline{2-12} 
& Hon-Ge   & 1.13 & 0.83  & 0.65  & 0.65  & -         & -        & -        & -             & -        & 3.20  \\ \hline
AFM magmom                   & Fe       & -    & -     & -     & -     & 0.19      & 0.38     & 0.42     & 0.45          & 0.12     & 1.56  \\ \hline
PM DOS@$E_f$ & Fe & - & - & - & - & 0.02 & 0.21 & 0.31 & 0.33 & 0.06 & 0.93
\\\hline
AFM DOS@$E_f$ & Fe & - & - & - & - & 0.08 & 0.20 & 0.21 & 0.23 & 0.05 & 0.77 \\\hline
PM filling $\in[-2,2]$ & Fe & 0.01 & 0.04 & 0.06 & 0.05 & 0.73 & 1.25 & 0.89 & 0.66 & 0.84 & 4.54
\\\hline\hline
\end{tabular}
\caption{\label{Tab:filling_magmom_wannier}The filling, magnetic moments (magmom), and density of states (DOS) at the Fermi level ($E_f$) of each orbital in the PM and AFM phase. All values are computed using the TB model from MLWFs constructed using $s,p,d$ orbitals of Fe and $s,p$ orbitals of Ge. The filling numbers are averaged over atoms on the same Wyckoff positions. In the PM phase, the fillings have timed 2 to account for two degenerate spins. The DOS at $E_f$ is normalized to 1 for the orbitals considered in the model.  
In the last row, we also list the fillings computed for bands $\in[-2,2]$ eV, which will be used as the reference for constructing minimal TB models in the next section. The total filling (per PM unit cell) in the PM and AFM phases is 36.26 and 36.15, respectively, which is close to the number of valence electrons used in DFT, i.e., 36 from $4s^1 3d^7$ of 3 Fe ($3p^6$ are core states and not included in the Wannier TB, and the $p$ orbitals of Fe in the table are $4p$) and $4s^2 4p^2$ of 3 Ge. 
Notice that the fillings of $4s$ and $4p$ orbitals of Fe are very small, which is a result of charge transfer between $3d$ of Fe and other orbitals of Ge. 
}
\end{table}

We then construct another group of MLWFs using a smaller number of orbitals, i.e., $d$ orbitals of Fe and $s,p$ orbitals of Ge, which have 27 orbitals in total. Other orbitals, e.g., $s,p$ orbitals of Fe, have main distributions far from the Fermi level $E_f$, thus having little effect on the low-energy physics.

To construct this group of MLWFs, we first replace the $s, p_x, p_y$ orbitals of honeycomb Ge by $s$ (bonding) and $p_y$ (anti-bonding) orbitals at kagome sites $3g$. 
This can be understood from the analog with graphene, where the carbon atoms at honeycomb sites $2d$ have their $p_z$ orbitals form Dirac cones near $E_f$ while $s,p_x,p_y$ orbitals form $sp^2$ hybrid bonding and anti-bonding states that distribute mainly below and above $E_f$, respectively. 
From the elementary band representations (EBRs)\cite{bradlyn2017topological, cano2018building, elcoro2021magnetic}, $A_1^\prime(s)@2d+E^\prime(p_x,p_y)@2d$ is (irrrep-)equivalent to $A_g(s)@3g+B_{2u}(p_y)@3g$, which means the $sp^2$ hybrid orbitals on honeycomb sites $2d$ is equivalent to the $s$ and $p_y$ orbitals on kagome site $3g$. The real-space invariant\cite{xu2021three} for $s,p_x,p_y@2d$ and $s,p_y@3g$ are both zero, which verifies that they can transform to each other. 
The $s@3g$ orbitals form the bonding states distributed mainly below $E_f$ while $p_y@3g$ orbitals form the anti-bonding states mainly above $E_f$, which can be verified from the IRREPs and orbital projections, as shown in \cref{Fig:FeGe_PM_struct_band}(h). 
The reason for this basis transformation is that the $p_x,p_y$ orbitals of honeycomb Ge form only one EBR but have distributions both below and above $E_f$ and cannot be decomposed into two EBRs. Thus we include the $s$ orbitals to form two separated $sp^2$ hybrid states, which will facilitate further analysis. We also remark that the inclusion of $s$ orbitals of Ge is helpful to increase the locality of $d$ orbitals and reduces long-range hoppings. A hopping truncation of $0.1$ results in a reasonably good band structure compared with DFT.

The onsite energy and nearest neighbor (NN) hoppings of this 27-band tight-binding model from MLWFs are listed in \cref{Table:wannier_hoppings} for reference. 
It can be seen that the onsite energies of five $d$ orbitals of Fe have close values of about $-1$ eV, which results from a quasi-cubic crystal field formed by six neighboring Ge atoms around a Fe atom. In \cref{Sec:approximated_interaction}, we will use this quasi-cubic crystal field to simplify the Coulomb interaction matrix, and more details can be found there. 
For the onsite energy, after transforming into the quasi-cubic local coordinates, the $T_{2g}$ and $E_g$ splitting also holds approximately, with a splitting of about 200 meV. 
The $p$ orbitals of triangular and honeycomb Ge, however, have very different onsite energies, which results from their different chemical environments, i.e., triangular Ge has Fe atoms on the same plane but honeycomb Ge does not.

\begin{table}[htbp]
\begin{tabular}{c|c|c|c|c|c|c|c|c|c|c|c|c}
\hline\hline
Orbital       & $d_{xy}$ & $d_{x^2-y^2}$ & $d_{xz}$ & $d_{yz}$ & $d_{z^2}$ & Tri $s$ & Tri $p_x$ & Tri $p_y$ & Tri $p_z$ & Hon $p_z$ & $sp^2$-$b$ & $sp^2$-$ab$ \\ \hline
onsite energy & -1.15    & -1.00         & -0.89    & -0.84    & -1.10     & -8.08     & -0.42  & -0.42   & 0.60      & -1.41       & -4.64     & 0.93     \\ \hline
$d_{xy}$    & 0.53     & 0.01          & -        & -        & -0.16     & -  & 0.00 & 0.73 & -            & -     & -       & -         \\ \hline
$d_{x^2-y^2}$ & -0.01    & -0.47   & -  & -  & -0.18    & -0.66   & -0.86 & 0.00   & -  & -0.54      & 0.23      & -    \\ \hline
$d_{xz}$  & - & -   & -0.20    & -0.16    & -  & -  & -  & -   & 0.58  & -     & -      & -         \\ \hline
$d_{yz}$  & -  & -  & 0.16     & 0.29     & -   & -  & -   & -  & -   & 0.77  & -  & -1.17 \\ \hline
$d_{z^2}$     & 0.16     & -0.18   & -   & -   & -0.22     & 0.51      & 0.68     & 0.00   & -        & 0.05      & -1.07      & -      \\ \hline\hline
\end{tabular}
\caption{\label{Table:wannier_hoppings}The onsite energy and NN hoppings of the 27-band TB model from Wannier functions. The second row lists the onsite energy of each orbital, and the other rows are the NN hoppings. `Tri $p_i$' and `Hon $p_i$' denote the $p_i$ orbitals at triangular and honeycomb sites and $sp^2$-$b$ (-$ab$) denote the bonding (anti-bonding) states formed by $s,p_x,p_y$ orbitals of honeycomb Ge. The onsite hopping from $d_{x^2-y^2}$ to $d_{z^2}$ is $-0.17$, which is not listed in the table. The label `-' in the table means the NN hopping is forbidden by symmetries. The hopping parameter $t_{ij}$ is from the $d_i@3f_1$ orbital of each row to $d_j@3f_2, s,p@1a, p_z@2c_2, b,ab@3g_1$ labeled for each column. 
All numbers are in eV.}
\end{table}

\subsection{Separation of orbitals into three groups}

The 27-band TB model obtained from MLWFs can serve as a good starting point for further studies. However, it still contains a large number of orbitals and hopping parameters. The longer-range hoppings beyond NN cannot be discarded directly, as the resultant band structure will deviate significantly from the original TB model. 
It is desirable to construct minimal TB models that contain fewer orbitals and have only NN or a few longer-range hoppings but could reproduce important features in the band structure, including the flat bands and van Hove singularities.

To fulfill this goal, we extract three groups of orbitals from the 27 orbitals of the Wannier TB, with the hoppings between orbitals from different groups being relatively weak. Therefore, we can construct the TB models for each group separately. The three groups of orbitals are

\begin{itemize}
    \item $(d_{xy},d_{x^2-y^2})$ orbitals of Fe on the kagome sites and $(p_x,p_y)$ orbitals of triangular Ge, i.e., 8 orbitals in total.
    \item $(d_{xz},d_{yz})$ orbitals of Fe on kagome sites, $p_z$ orbital of honeycomb Ge and $p_z$ orbital of triangular Ge, i.e., 9 orbitals in total. 
    \item $d_{z^2}$ orbitals of Fe and the $sp^2$ bonding states of honeycomb Ge, i.e., 6 orbitals in total. 
\end{itemize}

\begin{figure}[htbp]
	\centering
	\includegraphics[width=0.8\textwidth]{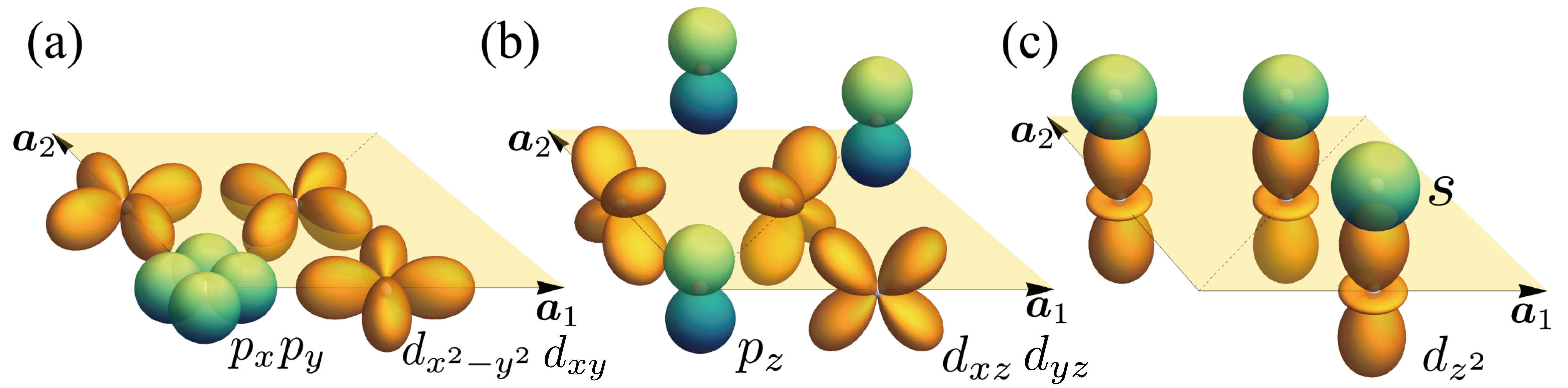}
	\caption{\label{Fig:3setorb_SI} Decomposition of three sets of orbitals. (a) $d_{x^2-y^2},d_{xy}$ of kagome Fe and $p_x,p_y$ of triangular Sn ($d_{xy}$ is omitted in the plot for simplicity). (b) $d_{xz},d_{yz}$ of kagome Fe and $p_z$ of triangular and honeycomb Sn ($d_{yz}$ is omitted for simplicity). (c) $d_{z^2}$ of kagome Fe and the bonding state (equivalent to $s$ orbital on $z=\frac{1}{2}$ plane) of honeycomb Sn.}
\end{figure}

We now justify the reason for this separation. 
We first consider the separation of five $d$ orbitals of Fe. Under symmetries of SG 191, $d$ orbitals on kagome sites naturally split into three groups, i.e., $(d_{xy}, d_{x^2-y^2})$, $(d_{xz}, d_{yz})$, and  $d_{z^2}$. The hoppings between $(d_{xz}, d_{yz})$ and the other two groups of $d$ orbitals are forbidden on the $z=0$ plane, as $M_z$ is an onsite symmetry and these orbitals have opposite $M_z$ eigenvalues, while the inter-kagome layer hoppings are much smaller as the distance is long. 
Moreover, the hoppings between $(d_{xy}, d_{x^2-y^2})$ and $d_{z^2}$ are also relatively small from the 27-band Wannier TB model in \cref{Table:wannier_hoppings}. This comes from the different shapes of the wavefunctions, i.e., the amplitudes of $(d_{xy}, d_{x^2-y^2})$ wavefunctions mostly come from the $xy$ planes,  while the amplitudes of $d_{z^2}$ wavefunctions mostly come from the $z$-axis. This results in a small overlap between the wavefunction of $d_{z^2}$ and $d_{xy},d_{x^2-y^2}$ orbitals. 
Notice that since $d_{x^2-y^2}$ and $d_{z^2}$ belong to the same trivial IRREP of $D_{2h}$, they have a small onsite hopping of -0.17 eV. A proper linear combination of $d_{x^2-y^2}$ and $d_{z^2}$ could remove the onsite hopping (note the simple equal-weight combination fails as they do not have the same onsite energy), but would make the basis more complicated. Thus we omit it for simplicity. 
Therefore, we conclude that three groups of $d$ orbitals can be analyzed separately. 
As for the $s$ and $p$ orbitals of Fe, they are far from the Fermi level $E_f$ and are omitted.

We then consider the orbitals of Ge, which do not have many contributions near $E_f$. They are used as auxiliary orbitals in order to describe the dispersion of $d$ orbitals of Fe near $E_f$ more faithfully. 
For triangular Ge, we consider $(p_x,p_y)$ and $p_z$ orbitals, and omit its $s$ orbital which is far from $E_f$, as shown in Table. \ref{Table:wannier_hoppings} (although tri-$s$ has a large hopping (-0.66$eV$) to $d_{x^2-y^2}$, the large onsite difference makes the effect of tri-$s$ on $d_{x^2-y^2}$ negligible). For honeycomb Ge's, we consider the $p_z$ and the bonding states formed by $s,p_x,p_y$ orbitals (see \cref{Sec:tb_models_wannier} for more details about the bonding states). These orbitals have non-negligible hybridization with $d$ orbitals of Fe. Denote these orbitals as $p_x^t, p_y^t, p_z^t, p_z^h, b$, where the superscript $t$ and $h$ denote triangular and honeycomb, and $b$ denotes the bonding states of honeycomb Ge. 

Finally, we combine the three groups of $d$ orbitals with specific orbitals of Ge.
\begin{enumerate}
    \item $(d_{xy}, d_{x^2-y^2})$ orbitals of Fe and $(p_x^t, p_y^t)$ orbitals of triangular Ge. These orbitals form a bipartite crystalline lattice (BCL) with 8 bands. 
    The reason to combine $(d_{xy}, d_{x^2-y^2})$ with $(p_x^t, p_y^t)$ orbitals is that they all lie on the $z=0$ plane and have large overlaps which result in $\sigma$-like bonds. This can be verified from the large hopping parameters from the Wannier TB in \cref{Table:wannier_hoppings}, as well as the overlap of orbital projections in DFT in \cref{Fig:FeGe_PM_struct_band}.

    \item $(d_{xz}, d_{yz})$ orbitals of Fe and $p_z$ orbitals of honeycomb and triangular Ge. These orbitals are all along the $z$-direction and have large overlaps that form $\pi$-like bonds, verified from Wannier hoppings and orbital projections. The $sp^2$ anti-bonding states of honeycomb Ge, although have large hoppings with $d_{yz}$, lie high above the Fermi level. We do not consider them directly but add $z$-directional hoppings to $d_{yz}$, which can be seen as the indirect hopping given by the anti-bonding states.

    \item $d_{z^2}$ orbitals of Fe and the $sp^2$ bonding states of honeycomb Ge. For this group of orbitals, the bands close to $E_f$ come mainly from $d_{z^2}$. However, the $d_{z^2}$ dispersion has a large splitting about $5$ eV on the $k_3=0$ plane but has an almost standard $s$-orbital kagome dispersion with a flat band on the $k_3=\pi$ plane (\cref{Fig:FeGe_PM_struct_band}). This band structure cannot be faithfully reproduced using $d_{z^2}$ orbitals only (with NN or NNN hoppings the $d_{z^2}$ bands are connected and cannot give the split bands on $k_3=0$ plane) but needs the assistance of other orbitals, i.e., the bonding states, which have the largest hopping with $d_{z^2}$. However, if we are only interested in the low energy physics near $E_f$, we can perturb out the bonding states and use the $d_{z^2}$ alone to build the TB model. In this case, the Coulomb interaction of $d_{z^2}$ also needs to be renormalized to a smaller value as the weight of $d_{z^2}$ near $E_f$ is increased when the bonding states are perturbed out.
    We also remark that the couplings between $d_{z^2}$ and the first group of orbitals, although existing in the Wannier TB, are not considered as an approximation due to their relatively small hopping values and thus minor effects for bands near $E_f$. 
\end{enumerate}

In the following sections, we first use the three aforementioned groups of orbitals to construct three minimal TB models that can reproduce the band structures in the PM phase and use the $S$-matrix formalism\cite{cualuguaru2022general} of generalized bipartite lattices to explain the quasi-flat bands near the Fermi level. We then combine these models together with extra coupling terms between $d_{x^2-y^2}$ and honeycomb $p_z$ and use second-order perturbations to perturb out the coupling terms and obtain three decoupled models. 

In \cref{Sec:TB_H1}, we build $H_1(\kk)$ model for $(d_{xy}, d_{x^2-y^2})$ of Fe and $(p_x^t, p_y^t)$ of triangular Ge, which has only inplane hoppings and is $k_3$-independent. In \cref{Sec:TB_H2}, we build $H_2(\kk)$ for $(d_{xz}, d_{yz})$ of Fe and $p_z$ of both triangular and honeycomb Ge. In \cref{Sec:TB_H3}, we first build $H_3(\kk)$ for $d_{z^2}$ and $sp^2$ bonding states of honeycomb Ge. We then perturb out the bonding states and construct an effective $H_3(\kk)$ that only contains $d_{z^2}$. In \cref{Sec:Full_TB_model}, we build the combined model by first coupling honeycomb $p_z^h$ with $d_{x^2-y^2}$ to introduce the $k_3$-dependence for $d_{x^2-y^2}$ and then perturb out this extra coupling term such that three $H_{i}(\kk)$ are still decoupled. We then further perturb out the triangular $p_z^t$ in $H_2(\kk)$ which has negligible distribution near $E_f$. The final minimal TB model is the direct sum of three $H_{i=1,2,3}(\kk)$ with perturbation terms.

\subsection{Minimal TB models for three groups of orbitals}\label{sec_minimal_TB}

We first give some preliminaries on the TB Hamiltonians. 
We adopt the Fourier transform convention of Bloch function 
\begin{equation}
	|\psi_j(\bm{k})\rangle=\frac{1}{\sqrt{N}}\sum_{\bm{R}} e^{i\bm{k}\cdot (\bm{R}+\bm{\tau}_j)} |\phi_{\bm{R}j}\rangle.
\end{equation}
where $|\phi_{\bm{R}j}\rangle$ denotes the Wannier function basis, and $\bm{\tau}_j$ is the Wyckoff position of these Wannier functions in the unit cell. In this convention, the hopping matrix is
\begin{equation}
	H_{ij}(\bm{k})=\langle \psi_i(k)|\hat{H}_t|\psi_j(k)\rangle=\sum_{\bm{R}} e^{i\bm{k}\cdot (\bm{R}+\bm{\tau}_j-\bm{\tau}_i)}H_{ij}(\bm{R})
\end{equation}
where $H_{ij}(\bm{R})=\langle \phi_{\bm{0},i}|\hat{H}|\phi_{\bm{R}j}\rangle$ is the hopping between Wannier functions. 
Note that the Hamiltonian in this convention is not periodic in the Brillouin zone (BZ), i.e., $H(\bm{k}+\bm{G})=V^\dagger(\bm{G})H(\bm{k})V(\bm{G})$, where $V(\bm{G})_{ij}=e^{i\bm{G}\cdot \bm{\tau}_i}\delta_{ij}$ is the embedding matrix. 

The TB Hamiltonian satisfies the symmetry constraints 
\begin{equation}
\begin{aligned}
    D^{-1}(g)H(\bm{k})D(g)&= H(g^{-1}\bm{k}),\quad g\text{ unitary},\\
    D^{-1}(g)H^*(\bm{k})D(g)&= H(g^{-1}\bm{k}),\quad g\text{ anti-unitary},\\
\end{aligned}
 \label{eq_sym_constraint}
\end{equation}
where $g\in G$ are symmetry operations of (magnetic) space group $G$, and $D(g)$ the representation matrix of $g$ under the TB basis $\psi_i(\bm{r})$ defined by $g\psi_i(\bm{r})=\psi_i(g^{-1}\bm{r})=D_{ij}(g)\psi_j(\bm{r})$, and $g\bm{k}$ denotes the action of $g$ on $\bm{k}$. 
We use the \textit{MagneticTB}\cite{zhang2022magnetictb} package to generate the TB Hamiltonians, where \cref{eq_sym_constraint} is used to solve the independent hopping parameters in the Hamiltonian that satisfy the symmetry.

For a high-symmetry point (HSP) $\bm{k}_0$ with little group $G_{\bm{k}_0}$, assume $g^{-1}\bm{k}=\bm{k}+\bm{G}_g$, then
\begin{equation}
\begin{aligned}
	D^{-1}(g)H(\bm{k})D(g)=H(\bm{k}+\bm{G}_g)&=V^\dagger(\bm{G}_g)H(\bm{k})V(\bm{G}_g)\\
	\Rightarrow 
	[H(\bm{k}), D(g)V^\dagger(\bm{G}_g)]&=0
\end{aligned}
\end{equation}
Thus $D^\prime(g)=D(g)V^\dagger(\bm{G}_g)$ has common eigenstates with $H(\bm{k})$. This relation can be used to compute the IRREPs of bands of TB models. 

When the bases of a TB Hamiltonian are divided into two groups, e.g., $p$ and $d$, which have representation matrices $D_p(g)$ and $D_d(g)$ of operation $g$, the S-matrix $S_{p, d}(\bm{k})$ which describes the coupling between $p$ and $d$ orbitals satisfies
\begin{equation}
    D_p^{-1}(g)S_{p,d}(\bm{k})D_d(g)=S_{p,d}(g^{-1}\bm{k}).
\end{equation}
This equation can be used to check whether the elements of the S-matrix are forbidden by symmetries.

\subsubsection{$H_1(k)$: $d_{xy}, d_{x^2-y^2}$ of Fe and $p_x, p_y$ of triangular Ge}\label{Sec:TB_H1}

The first group of orbitals consists of
$d_{xy}, d_{x^2-y^2}$ orbitals of Fe and $p_x, p_y$ orbitals of Ge at triangular sites, which form a bipartite crystalline lattice (BCL) of 8 bands. The band representations (BRs) are the superposition of three elementary BRs (EBRs), i.e., $E_{1u}@1a, A_g@3f, B_{1g}@3f$, 
where $E_{1u}$, $A_g$, and $B_{1g}$ are the IRREPs formed by $(p_x, p_y)$, $d_{xy}$, and $d_{x^2-y^2}$ orbitals, respectively, as shown in \cref{SG191-PG-IRREPs}. We use $3f_1$, $3f_2$, and $3f_3$ to denote the three kagome lattice sites listed in \cref{SG191-wyckoff}.

We first use a global coordinate system on kagome sites to define orbitals, which is more straightforward and is the default choice in many softwares\cite{zhang2022magnetictb}, and then transform it into the local coordinate system defined in \cref{Eq:local_coordinate}. 
We choose the TB basis under the global coordinate system as
\begin{equation}
   \psi_{\text{glob}}=(p_x@1a, p_y@1a, d_{xy}@3f_1, d_{x^2-y^2}@3f_1, d_{xy}@3f_2, d_{x^2-y^2}@3f_2, d_{xy}@3f_3, d_{x^2-y^2}@3f_3),
   \label{eq_TB1_global_basis}
\end{equation}

The generators of SG 191 have the following $O(3)$ matrices under the conventional basis defined in \cref{eq_conv_cell}:
\begin{equation}
C_{6z}=
\left(
\begin{matrix}
	1 & -1 & 0 \\
	1 & 0 & 0 \\
	0 & 0 & 1 \\
\end{matrix}
\right),\quad
C_{2,110}=
\left(
\begin{matrix}
	0 & 1 & 0 \\
	1 & 0 & 0 \\
	0 & 0 & -1 \\
\end{matrix}
\right),\quad
P=
\left(
\begin{matrix}
	-1 & 0 & 0 \\
	0 & -1 & 0 \\
	0 & 0 & -1 \\
\end{matrix}
\right),\quad
\end{equation}
and representation matrices under the TB basis defined in \cref{eq_TB1_global_basis} using the global coordinate system
\begin{equation}
	D_{\text{glob}}(C_{6z})=
	\left(
	\begin{matrix}
		\frac{1}{2} & -\frac{\sqrt{3}}{2} &0 & 0& 0& 0& 0&0 \\
		\frac{\sqrt{3}}{2} & \frac{1}{2} & 0& 0& 0& 0& 0& 0\\	
		0& 0& 0& 0& 0& 0& -\frac{1}{2} & \frac{\sqrt{3}}{2} \\
		0& 0& 0& 0& 0& 0& -\frac{\sqrt{3}}{2} & -\frac{1}{2} \\
		0& 0& -\frac{1}{2} & \frac{\sqrt{3}}{2}  & 0& 0& 0&0 \\
		0& 0& -\frac{\sqrt{3}}{2} & -\frac{1}{2}  & 0& 0& 0&0 \\
		0& 0& 0& 0& -\frac{1}{2} & \frac{\sqrt{3}}{2}  & 0& 0 \\
		0& 0& 0& 0& -\frac{\sqrt{3}}{2} & -\frac{1}{2}  &0 &0\\		
	\end{matrix}
	\right),\quad
\end{equation}
\begin{equation}
	D_{\text{glob}}(C_{2,110})=
	\left(
	\begin{matrix}
		-\frac{1}{2} & \frac{\sqrt{3}}{2} &0 & 0& 0& 0& 0&0 \\
		\frac{\sqrt{3}}{2} & \frac{1}{2} & 0& 0& 0& 0& 0& 0\\	
		0& 0& 0& 0& 0& 0& \frac{1}{2} & -\frac{\sqrt{3}}{2} \\
		0& 0& 0& 0& 0& 0& -\frac{\sqrt{3}}{2} & -\frac{1}{2} \\
		0& 0& 0& 0& \frac{1}{2} & -\frac{\sqrt{3}}{2}  & 0& 0 \\
		0& 0& 0& 0& -\frac{\sqrt{3}}{2} & -\frac{1}{2}  &0 &0\\		
		0& 0& \frac{1}{2} & -\frac{\sqrt{3}}{2}  & 0& 0& 0&0 \\
		0& 0& -\frac{\sqrt{3}}{2} & -\frac{1}{2}  & 0& 0& 0&0 \\
	\end{matrix}
	\right),\quad
 \label{eq_TB1_rep_global}
\end{equation}
\begin{equation}
	D(P)=\text{Diag}(-1,-1,1,1,1,1,1,1)
\end{equation}
The TRS in the NSOC setting is $D(T)=\bm{1}_8$, where the complex conjugation is absorbed in the definition of symmetry constraints in \cref{eq_sym_constraint}.

Under the global coordinate system, the two $d$ orbitals are entangled under SG 191 operations, as can be seen from the representation matrices. In order to separate them, we transform the basis into local coordinates as shown in \cref{Fig:local_coordinate}(b), with the coordinates at each kagome site defined in \cref{Eq:local_coordinate}. 
Choose the TB basis under the local coordinate system as
\begin{equation}
    \psi_{\text{loc}}=(p_x@1a, p_y@1a, d_{xy}@3f_1, d_{xy}@3f_2, d_{xy}@3f_3,
    d_{x^2-y^2}@3f_1, d_{x^2-y^2}@3f_2, d_{x^2-y^2}@3f_3), 
   \label{eq_TB1_local_basis}
\end{equation}
Here the $d_{xy},d_{x^2-y^2}$ are under local coordinates and are rotated from the original $d$ orbitals, related to $\psi_{\text{glob}}$ by
$\psi_{\text{loc}}^T=C_2 C_1 \psi_{\text{glob}}^T$, where $C_1$ is the global to local coordinate transformation and $C_2$ is a rearrangement of basis, defined by
\begin{equation}
C_1=
\left(
\begin{matrix}
1 & 0 & & & & & & \\
0& 1 &  & & & & & \\	
& & 1& 0& & & &\\
& & 0& 1& & & &\\
& & & & -\frac{1}{2} & -\frac{\sqrt{3}}{2}  & & \\
& & & & \frac{\sqrt{3}}{2} & -\frac{1}{2}  &   &\\		
& & & & & & -\frac{1}{2} & \frac{\sqrt{3}}{2}   \\
& & & & & & -\frac{\sqrt{3}}{2} & -\frac{1}{2}  \\
\end{matrix}
\right),\quad
C_2=
\left(
\begin{matrix}
    1 & 0 & & & & & & \\
    0 & 1 & & & & & &\\	
    & & 1& 0& 0& 0& 0 & 0\\
    & & 0& 0& 1& 0& 0 & 0 \\
    & & 0& 0& 0 & 0 & 1& 0 \\
    & & 0& 1& 0 & 0 & 0 &0\\		
    & & 0& 0& 0 &1 & 0 & 0 \\
    & & 0& 0& 0 &0 & 0 & 1 \\
\end{matrix}
\right).
\label{eq_TB1_coord_transform}
\end{equation}

The representation matrices $D_{\text{loc}}(g)$ under $\psi_{\text{loc}}$ is related to $D_{\text{glob}}(g)$ by
\begin{equation}
    D_{\text{loc}}(g)=C_2 C_1 D_{\text{glob}}(g)C_1^{-1} C_2^{-1},
\end{equation}
which results in
\begin{equation}
D_{\text{loc}}(C_{6z})=
\left(
\begin{matrix}
    \frac{1}{2} & -\frac{\sqrt{3}}{2} & & & & & &  \\
    \frac{\sqrt{3}}{2} & \frac{1}{2} &  & & & & & \\	
    & & 0& 0& 1& & & \\
    & & 1& 0& 0& & &  \\
    & & 0& 1& 0& & &  \\
    & & & &  & 0 & 0 &1\\		
    & & & &  &1 & 0 & 0 \\
    & & & &  &0 & 1 & 0 \\
\end{matrix}
\right),\quad
D_{\text{loc}}(C_{2,110})=
\left(
\begin{matrix}
-\frac{1}{2} & \frac{\sqrt{3}}{2} & & & & & &  \\
\frac{\sqrt{3}}{2} & \frac{1}{2} &  & & & & & \\	
& & 0& 0& -1& & & \\
& & 0& -1& 0& & &  \\
& & -1& 0 & 0& & &  \\
& & & &  & 0 & 0 &1\\		
& & & &  &0 & 1 & 0 \\
& & & &  &1 & 0 & 0
\end{matrix}
\right),
\label{eq_TB1_rep_local}
\end{equation}
and $D_{\text{loc}}(P)=D_{\text{glob}}(P), D_{\text{loc}}(T)=D_{\text{glob}}(T)$.

In the following, we first give the general form of the TB Hamiltonian constrained by symmetries operations defined in \cref{eq_TB1_rep_local}, and then fit the hopping parameters to the DFT results, which matches correctly with the IRREPs and the main feature of the orbital projections in DFT. The hopping parameters obtained are effective hoppings that can be seen as the superposition of hoppings from several near neighbors. Remark that the near-neighbor hoppings obtained in the Wannier TB (\cref{Table:wannier_hoppings}) cannot be used directly, as there exist longer-range hoppings and also hybridizations with other orbitals in the Wannier TB. Moreover, as the onsite energies of Fe $d$ and Ge $p$ in the Wannier TB are close, a direct second-order perturbation does not work well. Thus we only use hoppings in the Wannier TB as a reference and refit the hoppings in the minimal TB model.

Considering a few near neighbor hoppings of the kagome lattice and hoppings between the kagome lattice and triangular lattice, we construct the following TB model in the local coordinates basis $\psi_{\text{loc}}$, where we use $p_{xy}^t$ to denote $p_x,p_y$ of triangular Ge and $d_1, d_2$ to denote $d_{xy}, d_{x^2-y^2}$ in the local coordinates:
\begin{equation}
H_1(\bm{k})=
\left(
\begin{matrix}
H_{p_{xy}^t}(\bm{k}) & S_{p_{xy}^t, d_1}(\bm{k}) & S_{p_{xy}^t, d_2}(\bm{k}) \\
          & H_{d_1}(\bm{k})  & S_{d_1,d_2}(\bm{k}) \\
H.c.	& 				& H_{d_2}(\bm{k}) \\
\end{matrix}
\right),
\label{Eq:TB1_ham}
\end{equation}

\begin{equation}
\begin{aligned}
H_{p_{xy}^t}(\bm{k})&=\mu_{p_{xy}^t} \mathbf{1}_{2},\\
H_{d_1}(\bm{k})&=\mu_{d_1}\mathbf{1}_{3}
+ 2 t_{d_1}^{NN}
H_{\text{Kagome}}^{\text{inplane,NN}}(\bm{k})
+ 2 t_{d_1}^{NNN}
H_{\text{Kagome}}^{\text{inplane,NNN}}(\bm{k}) \\
&+ 2t_{d}^{4N1} \text{Diag}\left[\cos(k_1), \cos(k_1+k_2), \cos(k_2)\right] \\
&+ 2t_{d}^{4N2}
\text{Diag}\left[\cos(k_2)+\cos(k_1+k_2), \cos(k_1)+\cos(k_2), \cos(k_1)+\cos(k_1+k_2)\right]
\\
H_{d_2}(\bm{k})&=\mu_{d_2}\mathbf{1}_{3}
+ 2 t_{d_2}^{NN}
H_{\text{Kagome}}^{\text{inplane,NN}}(\bm{k})
+ 2 t_{d_2}^{NNN}
H_{\text{Kagome}}^{\text{inplane,NNN}}(\bm{k})  \\
&+ 2t_{d}^{4N4} \text{Diag}\left[\cos(k_1), \cos(k_1+k_2), \cos(k_2)\right] \\
&+ 2t_{d}^{4N3}
\text{Diag}\left[\cos(k_2)+\cos(k_1+k_2), \cos(k_1)+\cos(k_2), \cos(k_1)+\cos(k_1+k_2)\right]
\\
S_{d_1,d_2}(\bm{k})&=
2 t_{d_1,d_2}^{NN}
\left(
\begin{matrix}
0 &  \cos(\frac{k_2}{2})  & -\cos(\frac{k_1+k_2}{2}) \\
-\cos(\frac{k_2}{2}) & 0 & \cos(\frac{k_1}{2})  \\
\cos(\frac{k_1+k_2}{2}) & -\cos(\frac{k_1}{2}) & 0\\
\end{matrix}
\right)\\
&+ 2 t_{d_1,d_2}^{NNN}
\left(
\begin{matrix}
0 &  \cos(k_1+\frac{k_2}{2})  & -\cos(\frac{k_1-k_2}{2}) \\
-\cos(k_1+\frac{k_2}{2})  & 0 & \cos(\frac{k_1}{2}+k_2)  \\
\cos(\frac{k_1-k_2}{2})  & -\cos(\frac{k_1}{2}+k_2) & 0\\
\end{matrix}
\right) \\
&+ \sqrt{3} t_{d}^{4N5}
\text{Diag}\left[\cos(k_2)-\cos(k_1+k_2), \cos(k_1)-\cos(k_2), -\cos(k_1)+ \cos(k_1+k_2)\right]
\\
S_{p_{xy}^t, d_1}(\bm{k})&=
t_{p_{xy}^t,d_1}^{NN}
\left(
\begin{matrix}
0 &  i\sqrt{3}  \sin(\frac{k_1+k_2}{2}) & i\sqrt{3} \sin(\frac{k_2}{2}) \\
-2i \sin(\frac{k_1}{2}) &  	-i \sin(\frac{k_1+k_2}{2}) & 
i \sin(\frac{k_2}{2}) \\
\end{matrix}
\right)\\
&+ t_{p_{xy}^t,d_1}^{NNN}\left(
\begin{matrix}
	-2i \sin(\frac{k_1}{2}+k_2) & i \sin(\frac{k_1-k_2}{2}) & -i \sin(k_1+\frac{k_2}{2})\\
	0& i \sqrt{3} \sin(\frac{k_1-k_2}{2}) & i \sqrt{3} \sin(k_1+\frac{k_2}{2})\\
\end{matrix}
\right)\\
S_{p_{xy}^t, d_2}(\bm{k})&=
t_{p_{xy}^t,d_2}^{NN}
\left(
\begin{matrix}
-2i \sin(\frac{k_1}{2}) & -i\sin(\frac{k_1+k_2}{2}) & i \sin(\frac{k_2}{2}) \\
0 &  -i\sqrt{3} \sin(\frac{k_1+k_2}{2}) & -i\sqrt{3} \sin(\frac{k_2}{2}) ) \\
\end{matrix}
\right)\\
%&+ t_{p_{xy}^t,d_2}^{NNN}\left(
%\begin{matrix}
%	0 &  -i\sqrt{3} \sin(\frac{k_1-k_2}{2}) & -i \sqrt{3} \sin(k_1+\frac{k_2}{2}) \\
%	-2i \sin(\frac{k_1}{2}+k_2) &  i \sin(\frac{k_1-k_2}{2}) & -i \sin(k_1+\frac{k_2}{2})\\
%\end{matrix}
%\right)\\
H_{\text{Kagome}}^{\text{inplane,NN}}(\bm{k})
&=\left(
\begin{matrix}
    0 &  \cos(\frac{k_2}{2})  &\cos(\frac{k_1+k_2}{2}) \\
    & 0 & \cos(\frac{k_1}{2})  \\
    c.c. &  & 0\\
\end{matrix}
\right), \quad 
H_{\text{Kagome}}^{\text{inplane,NNN}}(\bm{k}) =
\left(
\begin{matrix}
    0 &  \cos(k_1+\frac{k_2}{2})  &\cos(\frac{k_1-k_2}{2}) \\
    & 0 & \cos(\frac{k_1}{2}+k_2)  \\
    c.c. &  & 0\\
\end{matrix}
\right) 
\label{Eq_ham1_matrix_blocks}
\end{aligned}
\end{equation}
where $c.c.$ denotes complex conjugation. In this model, 
$\mu_{p_{xy}^t}$ and $\mu_{d_i} (i=1,2)$ are the onsite energies of the $p_{x, y}^t$ and $d_{i=1,2}$ orbitals, respectively, while
$t_{p_{xy}^t, d_{i}}^{NN} (i=1,2)$ are the NN inter-sublattice hoppings from the $p_x, p_y$ to the $d_{i}$ orbitals.
The intra-orbital NN and NNN hoppings of the $d_{i}$ orbitals are given by $t_{d_i}^{NN}$ and $t_{d_i}^{NNN} (i=1,2)$, respectively, while
$t_{d_1,d_2}^{NN}, t_{d_1,d_2}^{NNN}$ denote, respectively, the inter-orbital NN and NNN hoppings from the $d_{1}$ to the $d_{2}$ orbitals. Finally,  
$t_d^{4Ni}$ ($i=1,2,3,4,5$) are longer-range hoppings for the $d_1$ and $d_2$ orbtials that stretch across the unit cell. 

We do not include $z$-directional hoppings so that the band structures on $k_3=0,\pi$ planes are identical. A few long-range hoppings are also considered here to give a more faithful fit to the DFT bands. The $k_3$-dependence is introduced later in Sec.\ref{Sec:Full_TB_model} by considering coupling with $p_z$ orbitals of honeycomb Ge.

We discuss several ideal cases of the Hamiltonian \ref{Eq_ham1_matrix_blocks} that host perfectly flat bands. Case \#1 and \#2 are unrealistic cases for FeGe but can help gain insight into the flat band limits and serve as applications of the $S$-matrix formalism\cite{cualuguaru2022general}.

\begin{figure}[htbp]
    \centering
    \includegraphics[width=1\textwidth]{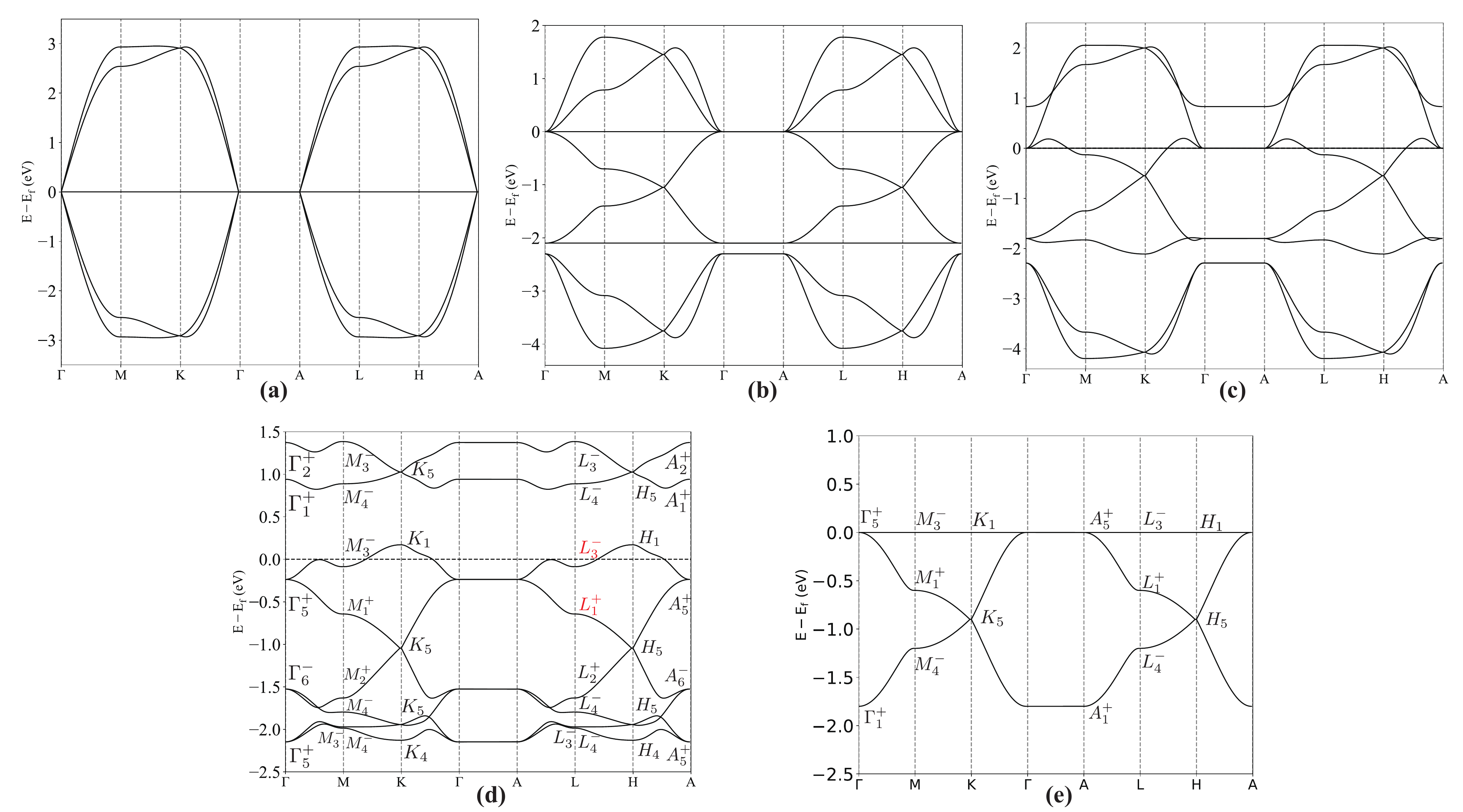}
	\caption{\label{Fig:TB_bandset1}TB band structures of $d_{xy}, d_{x^2-y^2}$ orbitals of Fe and $p_x,p_y$ orbitals of triangular Ge. (a) Case \#1, chiral symmetric limit with 4 perfectly flat bands. (b) Case \#3, two flat bands. (c) Case \#4, one flat band. (d) General case to fit DFT band structure, with one quasi-flat band near Fermi level.
    (e) The band structure of $s$ orbital on kagome sites together with IRREPs. 
    We label the IRREPs for bands in (d) and (e). In (d), the red IRREPs are mismatched with DFT bands (shown in \cref{Fig:FeGe_PM_struct_band}), which is either far from $E_f$ (at $\Gamma$) or will be fixed later (at $L$) in Sec.\ref{Sec:Full_TB_model}.
    The parameters in (d) are given in \cref{Table:TBpara_FeGe}. 
    The parameters in (c) are obtained from (d) by setting $H_{d_2}=\mu_{d_2}\bm{1}_3, S_{d_1,d_2}=\bm{0}$, $\mu_{p_{xy}^t}=-1.8, \mu_{d_1}=-1.25, \mu_{d_2}=0, t_{d_1}^{NN}=0.49, t_{d_1}^{NNN}=0.03, t_{p_{xy}^t,d_1}^{NN}=0.82, t_{p_{xy}^t,d_1}^{NN}=1.10$. The parameters in (b) are obtained from (c) by further setting $t_{d_1}^{NNN}=0, \mu_{p_{xy}^t}=-2.3$. The parameters in (a) are $t_{p_{xy}^t, d_1}^{NN}=0.82, t_{p_{xy}^t, d_2}^{NN}=1.23$ and all others zero. 
    }
\end{figure}

\emph{Case \#1.} 
When only considering inter-sublattice hoppings, the Hamiltonian is chiral symmetric, with the form
\begin{equation}
    H_1(\bm{k})=
    \left(
    \begin{matrix}
        \bm{0} & S_{p_{xy}^t, d_1}(\bm{k}) & S_{p_{xy}^t, d_2}(\bm{k}) \\
        S_{p_{xy}^t, d_1}(\bm{k})^\dagger  & \bm{0} & \bm{0}\\
        S_{p_{xy}^t, d_2}(\bm{k})^\dagger  & \bm{0}	& \bm{0}\\
    \end{matrix}
    \right),
\end{equation}
The number of orbitals in triangular and kagome lattices is $N_p=2$ and $N_d=6$, respectively. There are 4 perfectly flat bands at the Fermi level as shown in \cref{Fig:TB_bandset1}(a), which is the difference of orbital numbers in two sublattices\cite{cualuguaru2022general, regnault2022catalogue}, i.e., $N_d-N_p=4$ \citeSI{SI:S-matrix}. 

\emph{Case \#2.} 
When there is no hopping between three sets of orbitals and the kagome lattice has only NN hoppings, each kagome orbital has one perfectly flat band. In this case, the Hamiltonian is block-diagonal:
\begin{equation}
    H_1(\bm{k})=
    \left(
    \begin{matrix}
        H_{p_{xy}^t}(\bm{k})  & \bm{0} & \bm{0} \\
        \bm{0} & H_{d_1}(\bm{k})  & \bm{0}\\
        \bm{0}	& \bm{0}	& H_{d_2}(\bm{k}) \\
    \end{matrix}
    \right),
\end{equation}

\emph{Case \#3.} 
When $S_{p_{xy}^t,d_1}=0, H_{d_2}=\mu_{d_2}\mathbf{1}_3, S_{d_1,d_2}=0$, and $H_{d_1}$ has only NN hoppings, i.e.,
\begin{equation}
    H_1(\bm{k})=
    \left(
    \begin{matrix}
        H_{p_{xy}^t}(\bm{k})  & \bm{0} & S_{p_{xy}^t, d_2}(\bm{k}) \\
        \bm{0} & H_{d_1}(\bm{k})  & \bm{0}\\
        S_{p_{xy}^t, d_2}(\bm{k})^\dagger	& \bm{0}	& \mu_{d_2}\mathbf{1}_3 \\
    \end{matrix}
    \right),
\end{equation}
there will be 2 perfectly flat bands, one from $H_{d_1}(\bm{k})$ (decoupled) and the other one from $d_{x^2-y^2}$ (because $N_{d_2}-N_p=3-2=1$, where $N_{d_2}$ is the number of flat bands in $H_{d_2}$), as shown in \cref{Fig:TB_bandset1}(b).

\emph{Case \#4.} 
When $H_{d_2}=\mu_{d_2}\mathbf{1}_3, S_{d_1,d_2}=0$, i.e., 
\begin{equation}
    H_1(\bm{k})=
    \left(
    \begin{matrix}
        H_{p_{xy}^t}(\bm{k})  & S_{p_{xy}^t, d_1}(\bm{k})  & S_{p_{xy}^t, d_2}(\bm{k}) \\
        S_{p_{xy}^t, d_1}(\bm{k})^\dagger  & H_{d_1}(\bm{k})  & \bm{0}\\
        S_{p_{xy}^t, d_2}(\bm{k})^\dagger	& \bm{0}	& \mu_{d_2} \mathbf{1}_3 \\
    \end{matrix}
    \right),
\end{equation}
there will be one perfectly flat band from $d_{x^2-y^2}$, as shown in \cref{Fig:TB_bandset1}(c). This is because $H_{d_2}$ contributes 3 flat bands with the same energy $\mu_{d_2}$, which gives $3-2=1$ perfectly flat bands in $H_1(\bm{k})$. This agrees with the orbital projections in DFT in \cref{Fig:wannier_d_projections}, where the quasi-flat band near the Fermi level mainly comes from $d_{x^2-y^2}$ orbital. In the following, we will verify why the effective hoppings values in DFT are close to this limit.

Case \#4 is closest to the DFT band structure among the four cases and serves as an explanation for the origin of the quasi-flat band in DFT. 
The perfect flat band becomes weakly dispersive when hoppings in $H_{d_2}(\bm{k})$ and $S_{d_1,d_2}(\bm{k})$ are turned on. 
We fit the TB parameters to the DFT band structure based on dispersion and wavefunctions\cite{hu2023kagome}. The fitted bands have a quasi-flat band near the Fermi level, as shown in \cref{Fig:TB_bandset1}(d). The IRREPs are marked in the figure. The IRREPs of the TB bands agree well with DFT results near $E_f$, except for the IRREPs marked in red at $L$.  
This mismatch between $L_3^-$ and $L_1^+$ near $E_f$ will be fixed in Sec.\ref{Sec:Full_TB_model} by combining $H_1(\kk)$ with the honeycomb $p_z$ orbitals in $H_2(\kk)$, where $p_z$ couples with $d_{x^2-y^2}$ and leads to the band inversion at $L$. The honeycomb $p_z$ can be perturbed out to give effective $z$-directional hoppings for $d_{x^2-y^2}$. 
Remark that the two upper and two lower bands have strong hybridization with other bands in DFT, resulting in a wide distribution of orbital weights, and thus cannot be identified strictly (see \cref{Fig:FeGe_PM_struct_band} for DFT bands and IRREPs).

In the fitted parameters, hoppings in $H_{d_2}(\bm{k})$ ($d_1$, $d_2$ represent $d_{xy}, d_{x^2-y^2}$, respectively) and $S_{d_1,d_2}(\bm{k})$ are relatively small, which explains the quasi-flat band near the Fermi level. However, it is not straightforward to see why these hoppings take small values.
The hoppings from the 27-band Wannier TB in \cref{Table:wannier_hoppings} can serve as a reference, where the hopping between $d_{xy}$ and $d_{x^2-y^2}$ is indeed very small. 
However, the NN hoppings inside $d_{xy}$ and $d_{x^2-y^2}$ have comparable absolute values of about $0.5$ eV in the 27-band Wannier TB, which seems contradictory to these hoppings being small. This is because the hoppings in the 8-band model are effective hoppings and can be seen as the superposition of many other indirect hoppings in the Wannier TB model. 
In the Wanier TB model, the $d_{x^2-y^2}$ and $p_z$ of honeycomb Ge have close onsite energies and a relatively large NN hopping of about $0.5$ eV, while the NN hopping between $d_{xy}$ and honeycomb $p_z$ is forbidden (e.g., by $M_{100}, M_{010}$ and $M_{110}$ symmetries). We also construct a smaller Wannier TB model that excludes these two $p_z$ orbitals, which fits well with the bands close to the Fermi level. In this new Wannier TB model, the NN hopping between $d_{xy}$ orbitals is almost unchanged, but the NN hopping between $d_{x^2-y^2}$ is reduced to nearly zero. Theoretically, the NN coupling $S$-matrix between honeycomb $p_z^h$ and $d_{x^2-y^2}$ has the form
\begin{equation}
\begin{aligned}
S_{p_z^h, d_2}(\bm{k})&=
t_{p_z^h,d_2}^{NN}\cdot 2i \sin(\frac{k_3}{2})
    \left(
    \begin{matrix}
        e^{\frac{i}{6}(k_1+2k_2)}  & e^{\frac{i}{6}(k_1-k_2)}  & e^{-\frac{i}{6}(2k_1+k_2)} \\
        e^{-\frac{i}{6}(k_1+2k_2)} & e^{-\frac{i}{6}(k_1-k_2)}  & e^{\frac{i}{6}(2k_1+k_2)} \\
    \end{matrix}
    \right),
\end{aligned}
\label{Eq:S_pzh_d2}
\end{equation}
which leads to the following second-order perturbation term (see \cref{SI:S-matrix} for second-order perturbation theory):
\begin{equation}
\begin{aligned}
    H_{d_2, p_z^h}^{(2)}(\bm{k})&=
    \frac{1}{\mu_{d_2}-\mu_{p_z^h}}
    S^\dagger_{p_z^h, d_2}(\bm{k})S_{p_z^h, d_2}(\bm{k})\\
    &=\frac{(t_{p_z^h,d_2}^{NN})^2}{\mu_{d_2}-\mu_{p_z^h}}\cdot 4\left(
    (1-\cos(k_3))\mathbf{1}_{3}
     +(1-\cos(k_3))
		\left(
		\begin{matrix}
			0 &  \cos(\frac{k_2}{2})  & \cos(\frac{k_1+k_2}{2}) \\
			& 0 & \cos(\frac{k_1}{2})  \\
			c.c. &  & 0\\
		\end{matrix}
		\right)\right).
\end{aligned}
\label{Eq:pzhd2_perturb_term}
\end{equation}
This second-order perturbation Hamiltonian consists of four terms, i.e., the onsite term (the first term), NN (second) and NNN (fourth) terms crossing a unit cell along the $z$-direction with $\cos(k_3)$, and a NN (third) term in the $xy$ plane. The $xy$-plane NN term gives an effective hopping $t^\prime=\frac{2(t_{p_z^h,d_2}^{NN})^2}{\mu_{d_2}-\mu_{p_z^h}}> 0$ (positive because $\mu_{d_2}>\mu_{p_z^h}$ as seen from the 27-band Wannier hopping in \cref{Table:wannier_hoppings}). The original NN hopping of $d_{x^2-y^2}$ is -0.47, thus this perturbation term serves as an indirect hopping that reduces the effective hopping between $d_{x^2-y^2}$ to a small value. Note that in the 27-band Wannier TB, the onsite difference between $d_{x^-y^2}$ and $p_z^h$ is about 0.5 which is roughly the same as $t_{p_z^h,d_2}^{NN}$. Thus the second-order perturbation formula cannot be used directly to perturb out the $p_z^h$ orbital, as it will overestimate the hopping values (give $t^\prime\approx 1$). 
The Wannierization algorithm can be seen as a non-perturbative method to obtain effective hoppings when excluding certain orbitals, which indeed gives a vanishing NN hopping of $d_{x^2-y^2}$ when $p_z^h$ is excluded. The decoupling of $p_z^h$ can be achieved because the orbital weights of $p_z^h$ are mainly away from $E_f$ and excluding $p_z^h$ has little effect for bands near $E_f$, but at the price of larger spreads of Wannier functions in the corresponding Wannier TB which leads to longer-range hoppings.

Remark that in the previous paragraph, we mention using the same honeycomb $p_z$ to cause the band inversion at $L$. The fitted hopping between them is about $0.3$ eV in Sec.\ref{Sec:Full_TB_model}, which is smaller compared with Wannier hoppings in \cref{Table:wannier_hoppings} (i.e., reduce from 0.5 to 0.3). The argument in this paragraph still holds as only part of the $p_z$ effects is removed to reduce the hopping between $d_{x^2-y^2}$.

In the fitted bands shown in \cref{Fig:TB_bandset1}(d), the two lower bands mainly come from $p_x,p_y$ orbitals of triangular Ge, while the two upper bands and four middle bands mainly from $(d_{xy},d_{x^2-y^2})$. 
The two lower and two upper bands in the final fitted bands are fragile and cannot be decomposed into a positive integer combination of EBRs, with one possible EBR decomposition for two lower bands being $A_{g}@3f-A_{1g}@1a$ and one for two upper bands being $B_{1g}@3f+E_{1u}@1a+A_{1g}@1a-E^\prime @2c$, 
while the four middle bands from EBR $E^\prime @2c$, which also nonzero real space invariant (RSI)\cite{xu2021three} at $2c$ position.

\emph{Comparison with $s$-orbital model. }
We compare the band structure of the 8-band model with the band structure from the simplest 3-band model of $s$ orbital on kagome sites. The $s$-orbital Hamiltonian has the form
\begin{equation}
   H_{s}(\bm{k})=\mu_{s}\mathbf{1}_{3} + 2 t_{s}^{NN}
    \left(
    \begin{matrix}
        0 &  \cos(\frac{k_2}{2})  &\cos(\frac{k_1+k_2}{2}) \\
        & 0 & \cos(\frac{k_1}{2})  \\
	c.c. &  & 0\\
    \end{matrix}
    \right).
\end{equation}
By taking $\mu_s=-0.6,t_s^{NN}=-0.3$, the band structure is shown in \cref{Fig:TB_bandset1}(e), together with the IRREPs on HSPs. The wavefunction of the flat band takes the following form
\begin{equation}
\phi^{s}_{\text{FB}}(\bm{k}) =(\csc(\frac{k_2}{2}) \sin(\frac{k_1}{2}), -\csc(\frac{k_2}{2}) \sin(\frac{k_1+k_2}{2}), 1)^T
\label{Eq:psi_s_flatband}
\end{equation}
The $s$ orbital has the same IRREP $A_g$ as $d_{x^2-y^2}$ in site symmetry group $D_{2h}$ of kagome sites, and thus they form the same EBR $A_g@3f$. As the flat band and vHS near $E_f$ in the $H_1(\kk)$ sector are mainly given by $d_{x^2-y^2}$ (see \cref{Fig:wannier_d_projections}), the $s$-orbital model gives the correct IRREPs for the flat band and vHS. However, the single $s$-orbital model cannot describe the complicated connectivity of DFT band structures given by the hybridization between $d_{xy},d_{x^2-y^2}$, and $p_{xy}^t$. Moreover, the $d_{xy}$ orbital also forms a quasi-flat band along $\Gamma-M,\Gamma-A$ at about 1 eV that will move close to $E_f$ in the AFM phase. Thus the minimal TB model in the PM phase should give a faithful description of both $d_{xy}$ and $d_{x^2-y^2}$ orbitals in order to capture the physics in the AFM phase, which cannot be achieved by a single $s$-orbital model.

\emph{Overlap of flat band wavefunction. }
In order to verify the minimal model, we calculate the overlap of the flat band between fitted $H_1(\bm{k})$ and the 27-band Wanneir TB. Define the flat band projector as
\begin{equation}
P(\bm{k})=\sum_{|E_{n\bm{k}}|\le E_{c}} |\psi_{n\bm{k}}^{\text{DFT}}\rangle\langle \psi_{n\bm{k}}^{\text{DFT}}|
\end{equation}
where $\psi_{n\bm{k}}^{\text{DFT}}$ is the wavefunction from 27-band Wannier TB. $E_c$ is a cutoff energy which we take as $0.5$ eV in order to include the quasi-flat band in DFT. Then the overlap between flat band wavefunction $\phi_{\bm{k}}^{\text{FB}}$ in the minimal TB model and Wannier TB (DFT) is defined via the projection operator as
\begin{equation}
    O=\sqrt{\frac{1}{N_{\bm{k}}}\sum_{\bm{k}} \langle \phi_{\bm{k}}^{\text{FB}}|P(\bm{k})|\phi_{\bm{k}}^{\text{FB}}\rangle}
    \label{Eq:overlap_def}
\end{equation}
where $N_{\kk}$ is the number of points in the BZ.

We compute the overlap of flat band wavefunction as summarized in  \cref{Table:H1_flatband_overlap}, including the single $s$-orbital model (\cref{Eq:psi_s_flatband}), fitted $H_1(\bm{k})$, and also $H_1(\bm{k})$ with $p_z^h$ perturbation term that gives $k_3$-dependence (to be introduced in \cref{Sec:Full_TB_model}). $H_1(\bm{k})$ has higher overlap compared with the single $s$-orbital model. The flat band wavefunction from $s$-orbital (see \cref{Eq:psi_s_flatband}) also has a relatively high overlap with DFT because the flat band in $H_1(\bm{k})$ sector is mainly formed by $d_{x^2-y^2}$, and $s$ has the same EBR as $d_{x^2-y^2}$. 

\begin{table}[htbp]
\begin{tabular}{c|c|c|c}
\hline\hline
model                    & $s$-orbital & $H_1(\bm{k})$ & $H_1(\bm{k})$ with $p_z^h$ perturbation \\ \hline
flat band overlap & 91.3\% & 97.2\% & 96.8\% 
\\ \hline\hline
\end{tabular}
\caption{\label{Table:H1_flatband_overlap}
Comparison of the overlap of flat band wavefunction between single $s$-orbital, fitted $H_1(\bm{k})$ (defined in \cref{Eq_ham1_matrix_blocks}), and final $H_1(\bm{k})$ with $p_z^h$ perturbation term (defined in \cref{Eq:FeGe_TB_H123}). 
The square of the overlap is 82.8\%, 94.5\%, and 93.1\% for the three models, respectively. 
The overlap is computed based on \cref{Eq:overlap_def} with a $20\times20\times20$ mesh in the BZ.
}
\end{table}

Remark that we first fit $H_1(\bm{k})$ using eigenvalues and only IRREPs at high-symmetry points (HSPs), and obtain a model that shows good agreement with DFT band structure. This model, however, has a relatively low overlap of flat band wavefunction of about 75\%. The discrepancy mainly comes from the BZ region near the $M$ point. This is because the IRREP of the flat band at $M$ point, i.e., $M_3^-$, is the same as one upper band mainly formed by $d_{xy}$. The same IRREP results in certain mixtures between their wavefunctions, which have contributions from both $d_{x^2-y^2}$ and $d_{xy}$, as seen from the orbital projections in \cref{Fig:wannier_d_projections}. The $s$-orbital model also has a low overlap near the $M$ point because it fails to describe such mixtures. 
Thus, it is important to use both eigenvalues and wavefunctions when fitting TB models. Using only IRREPs at HSPs may be problematic, especially when neighboring bands have the same IRREP.

\subsubsection{$H_2(k)$: $d_{xz}, d_{yz}$ of Fe, and $p_z$ of Ge}\label{Sec:TB_H2}

In this section, we construct the TB Hamiltonian for $d_{xz}, d_{yz}$ of Fe in the local coordinates (denoted by $d_3$ and $d_4$, respectively) and $p_z$ of both honeycomb and triangular Ge (denoted by $p_z^h$ and $p_z^t$, respectively).
The BRs are the superposition of four EBRs, i.e., $A_{2}^{\prime\prime}@2d,  A_{2u}@1a, B_{2g}@3f, B_{3g}@3f$, where the IRREPs correspond to $p_z^h$, $p_z^t$, $d_{xz}$, $d_{yz}$ orbitals, respectively, as shown in \cref{SG191-PG-IRREPs}.

We start from the local coordinate system defined in \cref{Eq:local_coordinate}. The basis is chosen as
\begin{equation}
    \psi_{\text{loc}}=(p_z@2d, p_z^t@1a, d_{xz}@3f, d_{yz}@3f)
   \label{eq_TB2_local_basis}
\end{equation}
Notice that the transformation matrix $C_1$ from the global basis to the local basis for $(d_{xz}, d_{yz})$ is different from that for $(d_{xy}, d_{x^2-y^2})$, i.e., 
\begin{equation}
C_1=
\left(
\begin{matrix}
1& 0& & & &\\
0& 1& & & &\\
& & \frac{1}{2} & \frac{\sqrt{3}}{2}  & & \\
& & -\frac{\sqrt{3}}{2} & \frac{1}{2}  &   &\\		
& & & & -\frac{1}{2} & \frac{\sqrt{3}}{2}   \\
& & & & -\frac{\sqrt{3}}{2} & -\frac{1}{2}  \\
\end{matrix}
\right).
\label{eq_TB2_coord_transform}
\end{equation}

Using the basis $\psi^{\text{loc}}$, the representation matrices of the generators are
\begin{equation}
\begin{aligned}
D_{\text{loc}}(C_{6z})&=
\left(
\begin{matrix}
        0 & 1\\
        1 & 0\\
    \end{matrix}
    \right)\oplus
    \left(
\begin{matrix}
        1
    \end{matrix}
    \right)\oplus
\left(
\begin{matrix}
        0 & 0 & -1\\
        1 & 0 & 0\\
        0 & 1 & 0 \\
    \end{matrix}
    \right)\oplus 
\left(
\begin{matrix}
        0 & 0 & -1\\
        1 & 0 & 0\\
        0 & 1 & 0 \\
    \end{matrix}
    \right),\\
D_{\text{loc}}(C_{2,110})&=
\left(
\begin{matrix}
        0 & -1\\
        -1 & 0\\
    \end{matrix}
    \right)\oplus
    \left(
\begin{matrix}
        -1
    \end{matrix}
    \right)\oplus
\left(
\begin{matrix}
        0 & 0 & -1\\
        0 & -1 & 0\\
        -1 & 0 & 0 \\
    \end{matrix}
    \right)\oplus 
\left(
\begin{matrix}
        0 & 0 & 1\\
        0 & 1 & 0\\
        1 & 0 & 0 \\
    \end{matrix}
    \right),\\
    D_{\text{loc}}(P)&=	\left(
\begin{matrix}
        0 & -1\\
        -1 & 0\\
    \end{matrix}
    \right)\oplus
    \left(
\begin{matrix}
        -1
    \end{matrix}
    \right)\oplus \bm{1}_3 \oplus \bm{1}_3,\\
\end{aligned}
\end{equation}
and $D_{\text{loc}}(T)=\bm{1}_{9}$.

Considering the NN and possible NNN hoppings, we construct the following 9-band TB model:
\begin{equation}
    H_2(\bm{k})=
    \left(
    \begin{array}{cccc}
	H_{p_z^h}(\bm{k}) & \bm{0} & \color{red}{\bm{0}} & S_{p_z^h, d_4}(\bm{k})  \\
        & H_{p_z^t}(\bm{k}) & S_{p_{z}^t, d_3} & \color{red}{\bm{0}}  \\
	&  & H_{d_3}(\bm{k})  & S_{d_3,d_4}(\bm{k}) \\
	H.c.& &	& H_{d_4}(\bm{k}) \\
    \end{array}
    \right),
 \label{Eq:TB2_ham}
\end{equation}

\begin{equation}
\begin{aligned}
H_{p_z^h}(\bm{k})&=\mu_{p_z^h} \mathbf{1}_{2}
    +t_{p_z^h}^{NN} 		
    \left(
    \begin{matrix}
        0 &  e^{-\frac{i}{3}(2k_1+k_2)} (1+e^{ik_1}+e^{i(k_1+k_2)})\\
        c.c. & 0\\
    \end{matrix}
    \right),\\
    H_{d_i}(\bm{k})&=
        \left(\mu_{d_i} + 2 t_{d_i}^{zNN}\cos(k_3)\right)\mathbf{1}_{3}
    + 2t_{d_i}^{NN}
        \left(
    \begin{matrix}
        0 &  \cos(\frac{k_2}{2})  & -\cos(\frac{k_1+k_2}{2}) \\
    & 0 & \cos(\frac{k_1}{2})  \\
      H.c. &  & 0\\
    \end{matrix}
    \right),\\
    S_{d_3,d_4}(\bm{k})&=
    2 t_{d_3,d_4}^{NN}
    \left(
    \begin{matrix}
        0 &  \cos(\frac{k_2}{2})  & \cos(\frac{k_1+k_2}{2}) \\
        -\cos(\frac{k_2}{2}) & 0 & \cos(\frac{k_1}{2})  \\
        -\cos(\frac{k_1+k_2}{2}) & -\cos(\frac{k_1}{2}) & 0\\
    \end{matrix}
    \right), \\
S_{p_z^h, d_4}(\bm{k})&=
    2 t_{p_z^h, d_4}^{NN} \cos(\frac{k_3}{2})
    \left(
    \begin{matrix}
        -e^{\frac{i}{6}(k_1+2k_2)}  & e^{\frac{i}{6}(k_1-k_2)}  & -e^{-\frac{i}{6}(2k_1+k_2)} \\
        e^{-\frac{i}{6}(k_1+2k_2)} &  -e^{-\frac{i}{6}(k_1-k_2)}  & e^{\frac{i}{6}(2k_1+k_2)} \\
    \end{matrix}
    \right),\\
H_{p_{z}^t}(\mathbf{k}) &=
\mu_{p_z^t} + 2 t_{p_z^t}^{NN} \cos(k_3),\\
S_{p_z^t,d_3}(\bm{k})&=
2i\cdot t_{p_z^t, d_3}^{NN}
    \left(
    \begin{matrix}
          \sin(\frac{k_1}{2})  & \sin(\frac{k_1+k_2}{2})  & \sin(\frac{k_2}{2}) \\
    \end{matrix}
    \right),\\
\label{Eq_ham2_matrix_blocks}
\end{aligned}
\end{equation}
In the model, $\mu_{i}$ is the onsite energy, $t_{i}^{NN}, t_{i,j}^{NN}$ are intra-orbital and inter-orbital NN hoppings, and $t_{i,k}^{NNN}$ are inter-orbital NNN hoppings.
We use the black $\bm{0}$ to represent the $S$-matrix that is less relevant for bands near $E_f$ and is ignored, while the red $\color{red}{\bm{0}}$ to represent the symmetry-forbidden NN hoppings. 
For example, $S_{p_z^t, d_4}=\bm{0}$ is enforced by $M_{120}(=M_y)$, and 
$S_{p_z^h, d_3}=\bm{0}$ is enforced by $M_{100}$, $M_{010}$, and $M_{110}$, as $d_{xz},d_{yz}$ have opposite representation matrices under these rotations and a detailed examination shows that these symmetries enforce $t_{p_z^h,d_3}^{NN}=0$.

We first discuss several limits that can produce perfectly flat bands.

\begin{figure}[htbp]
    \centering
    \includegraphics[width=1\textwidth]{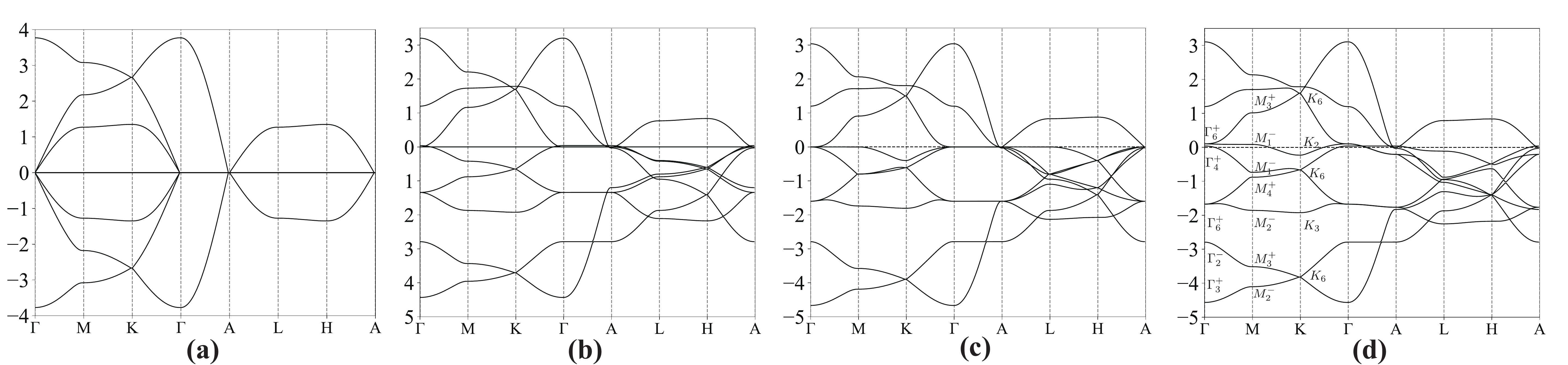}	
    \caption{\label{Fig:TB_bandset2}TB band structures of $H_2(\bm{k})$. (a) The chiral symmetric limit in perfect-flat band case \#1 discussed in this section, with 3 perfectly flat bands.
    (b) The flat band limit discussed in case \#3, with one perfectly flat band. 
    (c) The flat band limit discussed in case \#4, where a flat band exists along $\Gamma$-$M$, $\Gamma$-$A$, and $A$-$L$ lines. The parameters are taken as $\mu_{d_3}=\mu_{d_4}=-0.8, -t_{d_3}^{NN}=t_{d_4}^{NN}= -t_{d_3,d_4}^{NN} =0.2, t_{d_4}^{zNN}=0$ and others the same as in \cref{Table:TBpara_FeGe}
    (d) The fitted bands to DFT results, with one quasi-flat band near $E_f$. The IRREPs are marked for bands on the $k_3=0$ plane. 
    The TB parameters for (d) are summarized in \cref{Table:TBpara_FeGe}.
    }
\end{figure}

\emph{Case \#1.} 
When only considering hoppings between $d$ and $p_z$ orbitals, the Hamiltonian is chiral symmetric, with the form
\begin{equation}
H_2(\bm{k})=
\left(
\begin{matrix}
    \bm{0} & \bm{0} & \bm{0} & S_{p_z^h, d_4}(\bm{k}) \\
    & \bm{0} & S_{p_z^t, d_3}(\bm{k}) & \bm{0}\\
      &  & \bm{0} & \bm{0}\\
        H.c. &  & & \bm{0}\\
\end{matrix}
\right),
\end{equation}
In this case, we can decouple the orbitals into two decoupled sets, i.e., $p_z^h,d_4$ and $p_z^t,d_3$, which have $3-2=1$ and $3-1=2$ perfectly flat bands\cite{cualuguaru2022general,regnault2022catalogue}, i.e., 3 flat bands in total. On the $k_3=\pi$ plane, there are 7 perfectly flat bands, because $S_{p_z^h, d_4}(k_1,k_2,k_3=0)=0$ and thus the $p_z^h,d_4$ sector is a null matrix. Along $\Gamma-A$, there are also 7 flat bands, because $S_{p_z^h, d_4}(k_{1}=0,k_2=0,k_3)$ has rank=1 (gives $3+2-2=3$ flat bands) and $S_{p_z^t,d_3}(k_{1}=0,k_2=0, k_3)=0$ (gives $3+1=4$ flat bands). \citeSI{Sec:SI_Smatrix}. The band structure is shown in  \cref{Fig:TB_bandset2}(a), where the parameters are taken as $t_{p_z^h,d_4}^{NN}=0.77, t_{p_z^t,d_3}^{NN}=0.45$ and all others as zero.

\emph{Case \#2.} 
When $H_{d_4}(\bm{k})=\mu_{d_4}\mathbf{1}_3, S_{d_3,d_4}(\bm{k})=0$, i.e.,
\begin{equation}
        H_2(\bm{k})=
        \left(
        \begin{array}{cccc}
        H_{p_z^h}(\bm{k}) & \bm{0} & \bm{0} & S_{p_z^h, d_4}(\bm{k})  \\
            & H_{p_z^t}(\bm{k}) & S_{p_{z}^t, d_3} &  \bm{0}  \\
        &  & H_{d_3}(\bm{k})  & \bm{0} \\
        H.c.& &	& \mu_{d_4}\mathbf{1}_3 \\
        \end{array}
    \right),
\end{equation}
there will be 1 perfectly flat band from $d_4$, as $H_{d_4}(\bm{k})=\mu_{d_4}\mathbf{1}_3$ contains 3 perfectly flat bands and $H_{p_z^h}(\bm{k})$ has two orbitals, which results in $3-2=1$ perfectly flat band\cite{cualuguaru2022general}. \citeSI{Sec:SI_Smatrix}

\emph{Case \#3.} 
When $S_{d_3,d_4}(\bm{k})=\bm{0}$ and $d$ orbitals on kagome lattice have only NN hoppings, i.e., 
    \begin{equation}
        H_2(\bm{k})=
        \left(
        \begin{array}{cccc}
        H_{p_z^h}(\bm{k}) & \bm{0} & \bm{0} & S_{p_z^h, d_4}(\bm{k})  \\
            & H_{p_z^t}(\bm{k}) & S_{p_{z}^t, d_3} &  \bm{0}  \\
        &  & H_{d_3}(\bm{k})  & \bm{0} \\
        H.c.& &	& H_{d_4}(\bm{k}) \\
        \end{array}
    \right).
\end{equation}
There will still be one perfectly flat band. This results from the specific form of $S_{p_z^h,d_4}(\bm{k})$: the eigenvector of the flat band of $H_{d_4}(\bm{k})$ has the form 
\begin{equation}
    \phi^{\text{Kagome}}_{\text{FB}}(\bm{k})=(\csc(\frac{k_2}{2}) \sin(\frac{k_1}{2}), \csc(\frac{k_2}{2}) \sin(\frac{k_1+k_2}{2}), 1)^T,
    \label{Eq:H2_kagome_fb_wfc}
\end{equation}
which is identical to the null vector of $S_{p_z^h, d_4}(\bm{k})$, i.e., $S_{p_z^h, d_4}(\bm{k}) \phi^{\text{Kagome}}_{\text{FB}}(\bm{k})=\mathbf{0}$. 
Remark that 
\begin{equation}
S^\dagger_{p_z^h,d_4}(\bm{k})S_{p_z^h,d_4}(\bm{k})
\propto 4(1+\cos(k_3))\mathbf{1}_{3}
 -8\cos^2(\frac{k_3}{2})
    \left(
    \begin{matrix}
        0 &  \cos(\frac{k_2}{2})  &-\cos(\frac{k_1+k_2}{2}) \\
        & 0 & \cos(\frac{k_1}{2})  \\
        c.c. &  & 0\\
    \end{matrix}
    \right),
\end{equation}
which has a similar form with $H_{d_4}(\bm{k})$, and has $\phi(\bm{k})$ as a null-vector.
This results in one perfectly flat band, as shown in  \cref{Fig:TB_bandset2}(b), with parameters taken as $t_{d_3,d_4}^{NN}=0, \mu_{d_4}=-0.4, t_{d_4}^{zNN}=0$, and other parameters the same as in \cref{Table:TBpara_FeGe}.

\emph{Case \#4.} 
We also discuss a special case where a flat band exists on the $k_2=0$ plane (including $\Gamma$-$M$, $\Gamma$-$A$, and $A$-$L$ lines, and other symmetry-related planes), but dispersive in other regions of BZ, as shown in \cref{Fig:TB_bandset2}(c). To realize this flat band limit, the two kagome $d$ orbitals need to have the same onsite energy, i.e., $\mu_{d_3}=\mu_{d_4}=:\mu_d$, and only inplane NN hoppings which satisfy $-t_{d_3}^{NN}=t_{d_4}^{NN}= -t_{d_3,d_4}^{NN} =:t$. In this limit, a flat band arises on the $(k_1,0,k_3)$ plane, which has an energy $\mu_d+4t$. Its wavefunction is an equal-weight superposition of $d_3$ and $d_4$, i.e., $\phi_{\text{FB}}=\frac{1}{2}(-d_3@3f_1+d_3@3f_2+d_4@3f_1+d_4@3f_2)$, or written in the full basis of $H_2(\kk)$ as
\begin{equation}
    %\phi_{\text{FB}}=\frac{1}{2}(0,0,0,1,-1,0,1,1,0)^T
    \phi_{\text{FB}}=\frac{1}{2}(0,0,0,-1,1,0,1,1,0)^T
    \label{Eq:H2_fb_wfc}
\end{equation}
More rigorously, we can perform a basis transform for two $d$ orbitals. 
By taking $\mu_{d_3}=\mu_{d_4}=\mu_d, -t_{d_3}^{NN}=t_{d_4}^{NN}= -t_{d_3,d_4}^{NN} =t, t_{d_4}^{zNN}=0$, the matrix block for two $d$ orbitals in $H_2(\kk)$ becomes diagonal when $k_2=0$:
\begin{equation}
    S\begin{pmatrix}
        H_{d_3}(\kk) & S_{d_3,d_4}(\kk) \\
        S_{d_3,d_4}^\dagger(\kk) & H_{d_4}(\kk) \\
    \end{pmatrix}
    S^\dagger=\text{Diag}(\mu_d\pm 4t, \mu_d\pm 4t\cos(\frac{k_1}{2}), \mu_d\mp 4t\cos(\frac{k_1}{2})),
\end{equation}
where $S$ is given by the eigenvectors that diagonalize the Hamiltonian. 
It can be seen that there are two $\kk$-indepentent eigenvalues $\mu_d\pm 4t$. Only $\mu_d+4t$ remains an eigenvalue once $p_z^t$ and $p_z^h$ are included, which gives rise to the flat band on $(k_1,0,k_3)$ plane. The wavefunctions for $\mu_d+4t$ is nothing but $\phi_{\text{FB}}$ restricted in the $d_3,d_4$ subspace. 
Note that this flat limit relies on the specific form of $S_{p_z^h,d_4}(\kk)$ and $S_{p_z^t,d_3}(\kk)$ in $H_2(\kk)$, but does not hold in $H_1(\kk)$ as the $S$-matrices there, i.e., $S_{p_{xy}^t, d_{1,2}}(\kk)$, are different from the ones in $H_2(\kk)$. 
We also remark that when $-t_{d_3}^{NN}=t_{d_4}^{NN}=t_{d_3,d_4}^{NN}$, the flat limit also holds, and the flat band wavefunctions becomes $\phi_{\text{FB}}=\frac{1}{2}(0,0,0,1,-1,0,1,1,0)^T$.

This flat band limit serves as a good approximation for the 1:1 class, including FeGe, FeSn, and CoSn (to be discussed in Sec.\ref{Sec:FeSn_CoSn}). As seen from the 27-band Wannier TB parameters of FeGe in \cref{Table:wannier_hoppings} (formed by $d$ orbitals of Fe and $s, p$ orbitals of Ge), the five $d$ orbitals of Fe all share close onsite energies, and the hopping condition $-t_{d_3}^{NN}=t_{d_4}^{NN}= -t_{d_3,d_4}^{NN}$ is also roughly satisfied. These NN hoppings inside and between $d_3$ and $d_4$ have small absolute values of about $0.2$ eV. These relatively small NN hopping values come from the fact that the wavefunctions of $d_{xz}/d_{yz}$ orbitals distribute mainly along the $z$-direction. 
We remark that these are effective hoppings whose values may change depending on the choice of basis set for Wannier functions (see \cref{Tab:compare_H2_wannier_hopping_values}). 

\begin{table}[htpb]
\begin{tabular}{c|c|c|c|c|c|c}
\hline\hline
Wannier basis set & $\mu_{d_3}$ & $\mu_{d_4}$ & $t_{d_3}^{NN}$ & $t_{d_3, d_4}^{NN}$ & $t_{d_4}^{NN}$ & Spread of $d_3, d_4$ \\ \hline
Fe s,p,d, Ge s,p  & -1.04       & -1.21       & -0.26     & -0.17     & 0.45          & 0.66, 0.58           \\ \hline
Fe d, Ge s,p      & -0.89       & -0.84       & -0.20     & -0.16     & 0.29          & 0.68, 0.69           \\ \hline
Fe d, Ge p        & -0.91       & -0.23       & -0.21     & -0.14     & 0.04          & 0.69, 1.00           \\ 
\hline\hline
\end{tabular}
\caption{The values of onsite energies, NN hoppings, and Wannier function spreads of $d_3, d_4$ in three different Wannier TBs. The parameters in the second Wannier TB are closest to the flat band limit Case \#4. 
}
\label{Tab:compare_H2_wannier_hopping_values}
\end{table}

\emph{Final fitted model.} 
With several flat band limits discussed, we then fit TB parameters in $H_2(\kk)$ to DFT. The fitted bands are shown in \cref{Fig:TB_bandset2}(d) and parameters are summarized in \cref{Table:TBpara_FeGe}.
The fitted bands host a quasi-flat band close to $E_f$, which is very flat along $\Gamma$-$M$. This quasi-flat band can be explained by the flat band limits: 
\begin{itemize}
    \item Starting from the perfect-flat band limit case \#3, we turn on $S_{d_3,d_4}(\kk)$ with relatively small values which breaks the perfect-flat condition, and also shift the onsite energy and $z$-directional NN hopping of $d_{4}$ in order to fit with DFT, with the modified parameters being $t_{d_3,d_4}^{NN}=-0.2$ and $\mu_{d_4}=-0.83, t_{d_4}^{zNN}=0.1$. 

    In the fitted parameters shown in \cref{Table:TBpara_FeGe}, the hopping between $d_4$ and $p_z^h =0.77$ is the largest. This is because the spread of the $p$ orbitals is larger than $d$ orbitals, and the inplane distance between honeycomb and kagome sites is the shortest, giving a large overlap between $d_4$ and $p_z^h$. This large hopping gives $d_4$ a kagome bandwidth of about $3$ eV on the $k_3=0$ plane, which is much larger than the bandwidth given by the kagome hopping $6t_{d_4}^{NN}\approx 1$ eV. 
    Thus setting  $t_{d_3,d_4}^{NN}=-0.2$, which is comparable with $t_{d_4}^{NN}$, only introduces a bandwidth of about $0.4$ eV for the flat band of $d_4$. The band along $\Gamma$-$M$ remains flat.
    
    \item In the fitted parameters, the onsite energies of $d_3$ and $d_4$ are close. The NN hoppings inside $d_3,d_4$ and between $d_3,d_4$ have close absolute values of about $0.2$, which approximately satisfies the flat band limit case \#4. Thus an extremely flat band exists along $\Gamma$-$M$. 
    In \cref{TBband_FeGe}(h)(i) we also show the orbital projections of $d_3$ and $d_4$ which have comparable weights on the quasi-flat band along $\Gamma$-$M$. This is in agreement with the analysis of the flat band wavefunction in case \#4. 
\end{itemize}

In \cref{Fig:TB_bandset2}(d), we mark the IRREPs at HSPs on the $k_3=0$ plane which agrees with DFT results. However, the DFT bands of $d_{xz}$ and $d_{yz}$ on the $k_3=\pi$ plane have strong hybridization with other orbitals including the bonding and anti-bonding states of honeycomb Ge and cannot be fitted faithfully in the current TB model. The $d_3,d_4$ bands on the $k_3=0$ plane have no hybridization with (anti-)bonding states of honeycomb Ge because the $S$-matrices between them are proportional to $\sin(\frac{k_3}{2})$ and are vanishing when $k_3=0$. 
Including more orbitals and longer-range hoppings could lead to a better fitting on the $k_3=\pi$ plane. 
Nonetheless, the quasi-flat band near $E_f$ is maintained in the fitted model and the perfectly flat band condition is uncovered based on this minimal model. 
Remark that we use the hopping values in the 27-band Wannier TB as initial values for fitting, and only the DFT data on the $k_3=0$ plane is used for fitting. The $sp^2$ bonding and anti-bonding states are perturbed out to give effective $z$-directional hopping for $d_3,d_4$, which is fixed during the fitting.

\emph{Overlap of flat band wavefunction. }
To further validate the flat band limit, we compute the overlap (defined in \cref{Eq:overlap_def}) between flat band wavefunctions of $H_2(\kk)$ and DFT. 
For comparison, we also compute the overlap between DFT and the flat band given by the kagome $d_{yz}$ orbital NN model (with the form of \cref{Eq:H2_kagome_fb_wfc}). 
The results are summarized in \cref{Table:H2_flatband_overlap}. It can be seen that $H_2(\kk)$ gives higher overlap compared with the single kagome orbital model, and has an almost perfect match on the $k_3=0$ plane. Thus $H_2(\kk)$ gives a more faithful description of the system.

\begin{table}[htbp]
\begin{tabular}{c|c|c|c}
\hline\hline
model  & single kagome model & $H_2(\bm{k})$ in flat band limit \#4 & fitted $H_2(\bm{k})$ \\ \hline
flat band overlap on $k_3=0$ plane & 89.6\% & 98.7\% & 99.9\%   \\ \hline
flat band overlap in 3D BZ & 80.7\% & 85.4\% & 86.5\% 
\\ \hline\hline
\end{tabular}
\caption{\label{Table:H2_flatband_overlap}
Comparison of the overlap of flat band wavefunction between kagome $d_{yz}$ orbital NN model, $H_2(\bm{k})$ in flat band limit \#4, and final fitted $H_2(\bm{k})$. 
The square of the overlap in 3D BZ is 65.1\%, 72.9\%, and 74.8\% for the three models, respectively. 
$H_2(\kk)$ has a higher overlap compared with the single kagome model.}
\end{table}

\subsubsection{$H_3(k)$: $d_{z^2}$ of Fe and bonding states of honeycomb Ge}\label{Sec:TB_H3}

In this section, we consider the TB model of $d_{z^2}@3f$ and the $sp^2$ bonding state formed by $s@3g$. They form two EBRs, i.e., $A_g @3f$ and $A^\prime @3g$. 

For $d_{z^2}$ and $s$ orbitals on kagome sites, the local coordinate system defined in \cref{Eq:local_coordinate}, which only mixes the $\bm{x}$ and $\bm{y}$ axes, is the same as the global Cartesian coordinate system. Thus we start from the local coordinate system directly. Choose the TB basis as 
\begin{equation}
    \psi_{\text{loc}}=(b@3g, d_{z^2}@3f).
\end{equation}
Under this basis, the generators of SG 191 have the following matrix form (permutations):
\begin{equation}
D_{\text{loc}}(C_{6z})=
\left(
\begin{matrix}
    0 & 0 & 1 \\
    1 & 0 & 0 \\
    0 & 1 & 0 \\
\end{matrix}
\right)\oplus\left(
\begin{matrix}
    0 & 0 & 1 \\
    1 & 0 & 0 \\
    0 & 1 & 0 \\
\end{matrix}
\right),\quad
D_{\text{loc}}(C_{2,110})=
\left(
\begin{matrix}
    0 & 0 & 1 \\
    0 & 1 & 0 \\
    1 & 0 & 0 \\
\end{matrix}
\right)\oplus\left(
\begin{matrix}
    0 & 0 & 1 \\
    0 & 1 & 0 \\
    1 & 0 & 0 \\
\end{matrix}
\right),\quad
D_{\text{loc}}(P)=\bm{1}_6.
\end{equation}
The TRS in the NSOC setting is $D_{\text{loc}}(T)=\bm{1}_6$.

\begin{figure}[htbp]
	\centering
	\includegraphics[width=0.7\textwidth]{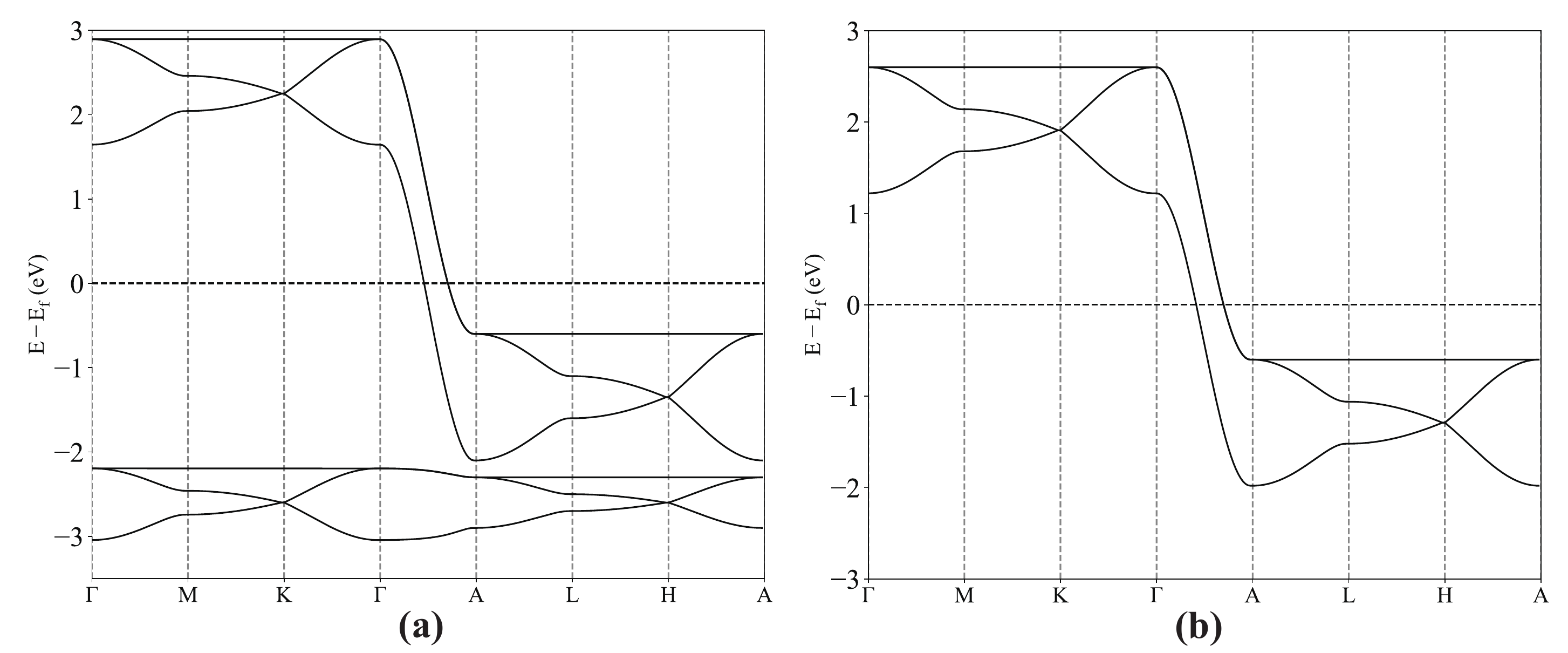}
	\caption{\label{TB_bandset3} (a) Fitted TB band structure of $H_3(\bm{k})$ defined in \cref{Eq:TB3_ham_withbonding}.
    (b) Fitted TB band structure of $d_{z^2}$ only, defined in \cref{Eq:TB3_ham}.}
\end{figure}

We construct the following 6-band TB model:
\begin{equation}
\begin{aligned}
    H_3(\bm{k})&=
    \left(
    \begin{matrix}
	H_{b}(\bm{k}) & S_{b, d_5}(\bm{k}) \\
	H.c.	&  H_{d_5}(\bm{k}) \\
    \end{matrix}
    \right),
    \\
    H_{d_5}(\bm{k})&= \mu_{d_5} \bm{1}_3 + t_{d_5}^{NN} H_{\text{kagome}}^{\text{inplane},z^2}(k_1,k_2),
    \\
    H_{b}(\bm{k}) &= (\mu_b + 2 t_{b}^{zNN}\cos(k_3)) \bm{1}_3 + t_b^{NN} H_{\text{kagome}}^{\text{inplane},z^2}(k_1,k_2),
    \\
    S_{b,d_5}(\bm{k}) &= 2 \cos(\frac{k_3}{2}) \left(t_{b,d_5}^{NN} \bm{1}_3 + t_{b,d_5}^{NNN}
    H_{\text{kagome}}^{\text{inplane},z^2}(k_1,k_2) \right),
    \\
    H_{\text{kagome}}^{\text{inplane},z^2}(k_1,k_2) &=
    2\left(
    \begin{matrix}
	0 & \cos(\frac{k_2}{2}) & \cos(\frac{k_1+k_2}{2})   \\
	& 0 & \cos(\frac{k_1}{2})   \\
	c.c. & & 0 \\
    \end{matrix}
    \right). \\
\end{aligned}
\label{Eq:TB3_ham_withbonding}
\end{equation}
In the model, $\mu_{i=d_5,b}$ are the onsite energies of the $d_5$ and $b$ orbitals, $t_{i=d_5, b}^{NN}$ are the intra-orbital NN intra-kagome hoppings of the $d_5$ and $b$ orbtials, while $t_{b}^{zNN}$ is the $z$-directional hopping of the bonding states. Finally, $t_{b, d_5}^{NN}, t_{b, d_5}^{NNN}$ are the inter-orbital NN and NNN hoppings between the $d_5$ and $b$ orbitals.  
As shown in \cref{TB_bandset3}(a), we fit these parameters to the dispersion in DFT. The values of the parameters are listed in \cref{Table:TBpara_FeGe}.

The main reason to combine the bonding state of honeycomb Ge with $d_{z^2}$ is that the $d_{z^2}$ orbital weights on $k_3=0$ plane have a wide distribution from -4 to 4 eV as seen from the orbital projections of DFT in \cref{Fig:wannier_d_projections}, and using $d_{z^2}$ orbital alone cannot faithfully reproduce this band structure. By combining with the bonding state, the band structure and orbital weights of $d_{z^2}$ can be fitted better, and the DOS of $d_{z^2}$ near $E_f$ is also reduced by the bonding state as more $d_{z^2}$ weights are introduced below $E_f$ (see \cref{tab_FeGe_TB_filling}).

However, if one is only interested in the low energy physics near $E_f$, it is more convenient to use $d_{z^2}$ only to build a simpler model, as the $d_{z^2}$ weights below $-2$ eV are far from $E_f$ and can be perturbed out. 
We then constructed a model for $d_{z^2}$ only: 
\begin{equation}
    H_{d_5}(\bm{k})=\left(\mu_{d_5} + 2 t_{d_5}^{zNN} \cos(k_3)\right)\bm{1}_3
    + t_{d_5}^{NN} H_{\text{kagome}}^{\text{inplane},z^2}(k_1,k_2).
    \label{Eq:TB3_ham}
\end{equation}
where $t_{d_5}^{zNN}$ is $z$-directional hopping.
Remark that in the fitted combined model with bonding states in \cref{Eq:TB3_ham_withbonding}, the onsite energies of $d_{z^2}$ and bonding state are close and the hopping between them is also strong, thus the second-order perturbation can not be applied directly.  We thus refit parameters for $H_{d_5}(\bm{k})$ directly. The fitting parameters are $\mu_{d_5}=0.54, t_{d_5}^{NN}=-0.23, t_{d_5}^{zNN}=0.8$, using which the dispersion of $d_{z^2}$ close to $E_f$ is recovered, as shown in \cref{TB_bandset3}(b). 

Compared with the 6-band model in \cref{Eq:TB3_ham_withbonding} (with bonding states), the simplified model of $d_{z^2}$ only in \cref{Eq:TB3_ham} has about twice the $d_{z^2}$ orbital weights near $E_f$ (see \cref{tab_FeGe_TB_filling} and \cref{table:FeGe_DFT_filling_compare}). The weight at $E_f$ in the 3-band model is also larger than the orbital weights in DFT. Thus the 3-band TB model cannot faithfully capture the wavefunctions of DFT near $E_f$, which will lead to larger Coulomb interaction in the band basis. To remedy this, one needs to renormalize the Coulomb interaction of $d_{z^2}$ to smaller values when using the 3-band model.

\subsubsection{Full TB model}\label{Sec:Full_TB_model}

In this section, we first combine the three decoupled Hamiltonian $H_{1,2,3}$ together with an extra $S$-matrix between $p_z^h$ and $d_2$, and then use second-order perturbation theory to decouple the full Hamiltonian into three groups. 

We combine the three models $H_{1,2,3}(\bm{k)}$ defined in \cref{Eq:TB1_ham}, \cref{Eq:TB2_ham}, \cref{Eq:TB3_ham} together, with the orbital basis
\begin{equation}
\begin{aligned}
    (&p_x@1a, p_y@1a, d_{xy}@3f, d_{x^2-y^2}@3f)\oplus 
    (p_z@2d,  p_z@1a, d_{xz}@3f, d_{yz}@3f)\oplus 
   %(b@3g, d_{z^2}@3f),
    (d_{z^2}@3f),
\end{aligned}
\end{equation}
which gives the full model:
\begin{equation}
H(\bm{k})=
\left(
\begin{array}{ccc|cccc|c}
    H_{p_{xy}^t}(\bm{k}) & S_{p_{xy}^t, d_1}(\bm{k}) & S_{p_{xy}^t, d_2}(\bm{k}) & \bm{0} & {\color{red}\mathbf{0}} & {\color{red}\mathbf{0}} & {\color{red}\mathbf{0}} & \bm{0}\\
     & H_{d_1}(\bm{k})  & S_{d_1,d_2}(\bm{k}) 
     & {\color{red}\mathbf{0}} & {\color{red}\mathbf{0}} & {\color{red}\mathbf{0}} & {\color{red}\mathbf{0}} & \bm{0} \\
     &  & H_{d_2}(\bm{k}) & S_{p_z^h, d_2}^\dagger(\bm{k}) & {\color{red}\mathbf{0}} & {\color{red}\mathbf{0}} & 
     {\color{red}\mathbf{0}} & \bm{0} \\\hline
     &  &  & H_{p_z^h}(\bm{k})& \bm{0}  & {\color{red}\mathbf{0}} & S_{p_z^h, d_4}(\bm{k}) & \bm{0} \\
     & & & & H_{p_z^t}(\mathbf{k}) & S_{p_{z}^t,d_3}(\bm{k}) & {\color{red}\mathbf{0}} & {\color{red}\mathbf{0}} \\
     & & & & & H_{d_3}(\bm{k})  & S_{d_3,d_4}(\bm{k}) 
      &  {\color{red}\mathbf{0}}\\
     & & & & & & H_{d_4}(\bm{k}) & {\color{red}\mathbf{0}}\\\hline
     H.c. & & & & & & & H_{d_5}(\bm{k})\\
\end{array}
\right),
\label{eq_FeGe_TB_full_model}
\end{equation}
where $p_{i}^t$ ($p_i^h$) denotes the $p_i$ orbitals of triangular (honeycomb) Ge, and $d_1$ to $d_5$ denote $d_{xy}, d_{x^2-y^2}, d_{xz}, d_{yz}, d_{z^2}$, on kagome sites, respectively. The ${\color{red}\mathbf{0}}$ in red means the NN coupling is forbidden by symmetries, and $\bm{0}$ in black means the coupling has negligible values or is unimportant for fitting bands near the Fermi level. 

The $S$-matrix of NN hopping between $p_z^h$ of honeycomb Ge and $d_{x^2-y^2}$ of kagome Fe is defined in \cref{Eq:S_pzh_d2}. 
Similar to the $S_{p_z^h, d_{4}}(\bm{k})$, $S_{p_z^h, d_2}(\bm{k})=\mathbf{0}$ when $k_3=0$, and has rank 1 along the $k_3$ axis. 
This $S$-matrix brings $k_3$ dependence to $H_1(\bm{k})$, and the perfect flat band in Case (4) (see Sec.\ref{Sec:TB_H1}) will not be perfectly flat on $k_3=\pi$ plane when $t_{p_z^h, d_2}^{NN}\ne 0$. 
The coupling between $d_2$ and $p_z^h$ causes a band inversion at $L$ near the Fermi level for $d_2$ bands, which agrees with the IRREPs of DFT bands.
The onsite energy difference of $p_z^h$ and $d_{x^2-y^2}$ is about $3$ eV, which justifies that excluding the $p_z^h$ orbital in $H_1(\bm{k})$ is reasonable as $p_z^h$ only slightly modifies the bands on $k_3=\pi$ plane. 
Remark that the orbitals in $H_1(\kk)$ mainly distribute on $xy$ plane, thus are quasi-2D and have weak $k_3$-dependence. However, the $d_{xy/yz}$ orbitals in $H_2(\kk)$ are mainly $z$-directional and naturally have strong $k_3$-dependence.

In the following, we use the second-order perturbation theory to decouple the combined model in order to obtain three simple decoupled Hamiltonians. 
As the onsite energy differences of $p_z^h$ orbitals and $d_2$ orbitals are large, we use the second-order perturbation method to decouple these orbitals. The perturbed Hamiltonians have the following form 
\begin{equation}
\begin{aligned}
    H_{d_2, p_z^h}^{(2)}(\bm{k})&=
    \frac{1}{\mu_{d_2}-\mu_{p_z^h}}
    S^\dagger_{p_z^h, d_2}(\bm{k})S_{p_z^h, d_2}(\bm{k})\\
    &=\frac{(t_{p_z^h,d_2}^{NN})^2}{\mu_{d_2}-\mu_{p_z^h}}\cdot 4\left(
    (1-\cos(k_3))\mathbf{1}_{3}
     +(1-\cos(k_3))
		\left(
		\begin{matrix}
			0 &  \cos(\frac{k_2}{2})  & \cos(\frac{k_1+k_2}{2}) \\
			& 0 & \cos(\frac{k_1}{2})  \\
			c.c. &  & 0\\
		\end{matrix}
		\right)\right),\\
    H_{p_z^h,d_2}^{(2)}(\bm{k})&=
    \frac{1}{\mu_{p_z^h}-\mu_{d_2}}
    S_{p_z^h, d_2}(\bm{k})S^\dagger_{p_z^h, d_2}(\bm{k})
\end{aligned}
\label{Eq:pzh_perturb_term}
\end{equation}

The onsite energy difference of $p_z^t$ and $d_3$ is also large, and they can be decoupled using similar second-order perturbation, with perturbed Hamiltonians:
\begin{equation}
\begin{aligned}
    H_{d_3, p_z^t}^{(2)}(\bm{k})&=
    \frac{1}{\mu_{d_3}-\mu_{p_z^t}}
    S^\dagger_{p_z^t, d_3}(\bm{k})S_{p_z^t, d_3}(\bm{k})\\
    &=\frac{2(t_{p_z^t,d_3}^{NN})^2}{\mu_{d_3}-\mu_{p_z^t}}
    \left(
    \begin{matrix}
	1-\cos(k_1) &  2\sin(\frac{k_1}{2})\sin(\frac{k_1+k_2}{2})  &2\sin(\frac{k_1}{2})\sin(\frac{k_2}{2}) \\
	& 1-\cos(k_1+k_2) & 2\sin(\frac{k_2}{2})\sin(\frac{k_1+k_2}{2})  \\
	c.c. &  & 1-\cos(k_2)\\
    \end{matrix}
    \right),\\
\end{aligned}
\label{Eq:pzt_perturb_term}
\end{equation}

We remark that the $p_{x,y}^t$ orbitals in $H_1(\mathbf{k})$ cannot be decoupled using second-order perturbation, as they have close onsite energies and strong coupling with $d_1,d_2$ thus fail to meet the condition for 2nd-order perturbation to work (i.e., $\mu_d-\mu_p\gg t_{pd}$), and the perturbed band structure will change significantly. The honeycomb $p_z^h$ in $H_2(\mathbf{k})$ is similar. 
Moreover, these second-order perturbation formulas only apply rigorously when $H_{d_i}(\bm{k})$ and $H_{p_z^h}(\bm{k})$ are diagonal (see \citeSI{SI:S-matrix}). 
When they are not diagonal, one can still do the numerical perturbation at each $k$-point but the perturbed Hamiltonians do not have a simple analytic form in the BZ.
However, when $H_{d_i}(\bm{k})$ and $H_{p_z^h}(\bm{k})$ deviate not too large from the diagonal form, the 2nd-order perturbation formulas can be used as an approximation.

\begin{figure}[htbp]
    \centering
    \includegraphics[width=1\textwidth]{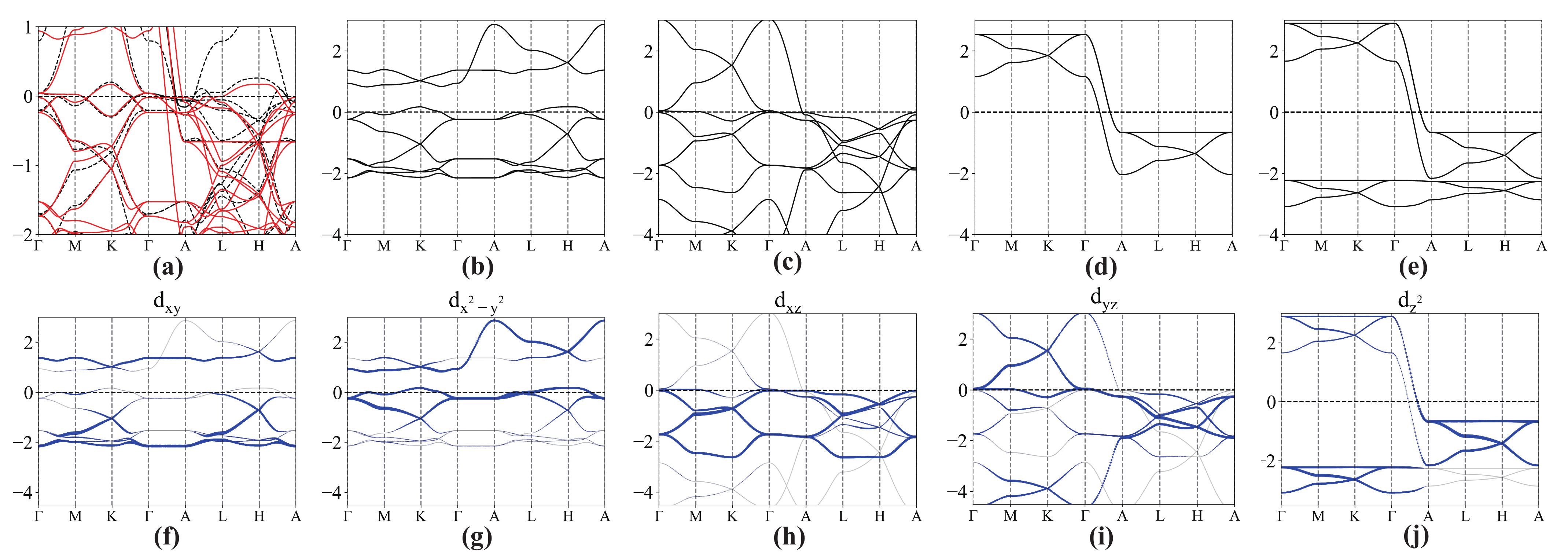}
    \caption{\label{TBband_FeGe} (a) Band structure of the direct-summed decoupled model in \cref{Eq:FeGe_TB_directsum} (red lines), with a comparison with DFT bands in the PM phase (black dashed lines).
    (b)-(d) Band structure of the decoupled $H_1(\bm{k})$, $H_2(\bm{k})$, $H_3(\bm{k})$ defined in \cref{Eq:FeGe_TB_H123}, respectively. (e) Band structure of $H_3(\bm{k})$ with bonding states defined in \cref{Eq:TB3_ham_withbonding}. 
    (e)-(i) The orbital projection of each $d$ orbital in three decoupled models, where (i) shows the $d_{z^2}$ projections in $H_3(\bm{k})$ with bonding states. The orbital projections agree well with DFT results in \ref{Fig:wannier_d_projections}.}
\end{figure}

Using the aforementioned perturbed Hamiltonians, the full model is decoupled into the three following Hamiltonians
\begin{equation}
\begin{aligned}
H_{1}(\bm{k})&=
\left(
\begin{matrix}
    H_{p_{xy}^t}(\bm{k}) & S_{p_{xy}^t, d_1}(\bm{k}) & S_{p_{xy}^t, d_2}(\bm{k}) \\
     & H_{d_1}(\bm{k})  & S_{d_1,d_2}(\bm{k}) \\
     H.c. &  & H_{d_2}(\bm{k})+H_{d_2, p_z^h}^{(2)}(\bm{k}) \\
\end{matrix}
\right),\\
H_2(\bm{k})&=
\left(
\begin{matrix}
    H_{p_z^h}(\bm{k})+H_{p_z^h,d_2}^{(2)}(\bm{k}) & \bm{0} & S_{p_z^h,d_4}(\bm{k})\\
    & H_{d_3}(\bm{k})+H_{d_3, p_z^t}^{(2)}(\bm{k}) & S_{d_3,d_4}(\bm{k})\\
    H.c. & & H_{d_4}(\bm{k})\\
\end{matrix}
\right),\\
H_3(\bm{k})&= H_{d_5}(\bm{k})
% H_3(\bm{k})&=
%     \left(
%     \begin{matrix}
% 	H_{b}(\bm{k}) & S_{b, d_5}(\bm{k}) \\
% 	H.c.	&  H_{d_5}(\bm{k}) \\
%     \end{matrix}
%     \right).
\end{aligned}
\label{Eq:FeGe_TB_H123}
\end{equation}
where $H_{i=1,2,3}(\kk)$ are defined in \cref{Eq:TB1_ham}, \cref{Eq:TB2_ham}, \cref{Eq:TB3_ham}, with second-order perturbed terms defined in \cref{Eq:pzh_perturb_term}, \cref{Eq:pzt_perturb_term}. 
By direct-summing these three Hamiltonians, we arrive at 
\begin{equation}
    H(\bm{k})=H_1(\bm{k})\oplus H_2(\bm{k}) \oplus H_3(\bm{k}),
    \label{Eq:FeGe_TB_directsum}
\end{equation}
which gives the final decoupled single-particle Hamiltonian
\begin{equation}
    \hat{H}_0 = \sum_{\bm{k}}\sum_{ij}
    H_{ij}(\bm{k}) c^\dagger_{\bm{k}i}c_{\bm{k}j},
    \label{Eq:FeGe_H_single_particle_H}
\end{equation}
where the Bloch electron operators $c_{\bm{k}i}$ are defined as the Fourier transform of the following orbital basis
\begin{equation}
\begin{aligned}
    (&p_x@1a, p_y@1a, d_{xy}@3f, d_{x^2-y^2}@3f)\oplus (p_z^h@2d, d_{xz}@3f, d_{yz}@3f)\oplus
    (d_{z^2}@3f),
    %(b@3g, d_{z^2}@3f),
    \label{eq_final_TB_basis}
\end{aligned}
\end{equation}
with orbitals defined in the local coordinate system in \cref{Eq:local_coordinate}, and TB parameters values summarized in \cref{Table:TBpara_FeGe}. 
The corresponding band structure is shown in \cref{TBband_FeGe}.
The filling number of each orbital and the DOS at $E_f$ in the decoupled model is given in \cref{tab_FeGe_TB_filling}, where most of the values agree well DFT (see \cref{table:FeGe_DFT_filling_compare}) 

The largest error comes from the filling of $d_{x^2-y^2}$, which is smaller than the DFT value. This is due to a larger weight of $d_{x^2-y^2}$ above $E_f$ compared with DFT. The $d_{z^2}$ filling in the simplified $H_{3}(\kk)$ is also small compared with the DFT value. The $d_{z^2}$ filling becomes 1.55 when the bonding state is included which agrees well with DFT.  
We also list the filling within $[-1.5, 1.5]$ eV, which shows better agreement with DFT with the root-mean-square error of $d$ orbitals fillings being $18\%$. Thus the model captures the low energy physics near $E_f$. 

\begin{table}[htbp]
\begin{tabular}{c|c|c|c|c|c|c|c|c|c|c|c|c|c|c|c|c}
\hline\hline
Parameter & $\mu_{p_{xy}^t}$ & $\mu_{d_1}$ & $\mu_{d_2}$ & $t_{d_1}^{NN}$  & $t_{d_2}^{NN}$ & $t_{d_2}^{NNN}$    & $t_{d_1,d_2}^{NN}$ & $t_{d_1,d_2}^{NNN}$ & $t_{p_{xy}^t, d_1}^{NN}$ & $t_{p_{xy}^t, d_1}^{NNN}$ &  $t_{p_{xy}^t, d_2}^{NN}$ & $t_d^{4N1}$ & $t_d^{4N2}$ & $t_d^{4N3}$  & $t_d^{4N4}$ & $t_d^{4N5}$  
\\ \hline
Value/eV & -1.53 & -0.96 & 0.06 & 0.58 & 0.11  & 0.09 & -0.09 & 0.19 & -0.33 & 0.14 & 0.35 & -0.16 & 0.09 & 0.10 & -0.16 & -0.01
\\ \hline
Parameter & $\mu_{p_z^h}$ & $\mu_{d_3}$ & $\mu_{d_4}$ & $t_{d_3,d_4}^{NN}$ & $t_{d_3}^{NN}$  & $t_{d_4}^{NN}$ & $t_{d_4}^{zNN}$ & $t_{p_z^h,d_4}^{NN}$ & $t_{p_z^h}^{NN}$ & $t_{p_z^h, d_2}^{NN}$ & $\mu_{p_z^t}$ & $t_{p_z^t}^{NN}$ & $t_{p_z^t, d_3}^{NN}$  &&
\\ \hline
Value/eV & -1.47 & -0.94 & -0.89 & -0.20 & -0.23 & 0.2 & 0.10 & 0.77  & -0.46 & 0.35 & 0.54 & 0.30 & 0.45  &&
\\ \hline
Parameter & $\mu_{d_5}$  & $t_{d_5}^{NN}$ & $\mu_b$ & $t_b^{NN}$ & $t_b^{zNN}$ & $t_{b, d_5}^{NN}$ & $t_{b,d_5}^{NNN}$ &&
\\ \hline
Value/eV & -1.16 & -0.25 & -0.66 & -0.1 & 0.9 & -1.1 & 0.04 &&
\\ \hline\hline
\end{tabular}
\caption{\label{Table:TBpara_FeGe} Onsite energies and hoppings used in the full TB model (with bonding state in $H_3(\kk)$ defined in \cref{Eq:TB3_ham_withbonding}), where $d_1$ to $d_5$ denote $d_{xy}, d_{x^2-y^2}, d_{xz}, d_{yz}, d_{z^2}$, respectively, $p_{i}^t$ ($p_i^h$) denotes the $p_i$ orbitals of the triangular (honeycomb) Ge, and $b$ is the $sp^2$ bonding state of honeycomb Ge. For the simplified $H_3(\kk)$ of $d_{z^2}$ only defined in \cref{Eq:TB3_ham}, the parameters are $\mu_{d_5}=0.60, t_{d_5}^{NN}=-0.23, t_{d_5}^{zNN}=0.8$. 
Parameters not listed in the table have zero values.}
\end{table}

\begin{table}[htbp]
\begin{tabular}{c|c|c|c|c|c|c|c|c}
\hline\hline
Orbital  & $p_{xy}^t$ & $p_z^h$ & $d_{z^2}$ & $d_{xz}$ & $d_{yz}$ & $d_{x^2-y^2}$ & $d_{xy}$ & Total \\ \hline
TB filling $\in[-\infty,2]$ &  1.82 & 1.47 & 0.77 & 1.96  & 1.51 & 0.82  & 1.52 & 26.34 \\\hline
TB filling $\in[-1.5,1.5]$ &  0.05 & 0.18 & 0.69 & 1.22  & 0.98 & 0.76 & 0.43 & 12.72 \\\hline
TB DOS@$E_f$ & 0.00 & 0.02 & 0.14 & 0.08 & 0.21 & 0.44 & 0.06 & 1.00
\\\hline\hline
\end{tabular}
\caption{\label{tab_FeGe_TB_filling}The filling and DOS at $E_f$ of each orbital in the decoupled TB model \cref{Eq:FeGe_TB_directsum}.
The filling numbers are computed for bands $\in[-\infty,2]$ and $\in[-2,2]$ eV, averaged over atoms on the same Wyckoff positions, and time 2 to account for two degenerate spins. The DOS is normalized to 1. The filling in $[-2,2]$ eV agrees well with the DFT values in \cref{Tab:filling_magmom_wannier}. 
The DOS at $E_f$ mainly comes from $d_{xz},d_{yz},d_{x^2-y^2}$ orbitals, which is also close to DFT results in \cref{Tab:filling_magmom_wannier}.
For the $H_3(\kk)$ with bonding state defined in \cref{Eq:TB3_ham_withbonding}, the $d_{z^2}$ filling in $[-\infty,2]$ is 1.55, in $[-2,2]$ is 0.46, and DOS$@E_f$ is 0.02 (not listed in the table for simplicity), which is more faithful as more $d_{z^2}$ weight below $E_f$ is introduced by the bonding states. 
}
\end{table}

\begin{table}[htbp]
\begin{tabular}{c|c|c|c|c|c|c|c|c}
\hline\hline
Orbital & $p_{xy}^t$ & $p_z^h$ & $d_{z^2}$ & $d_{xz}$ & $d_{yz}$ & $d_{x^2-y^2}$ & $d_{xy}$ & Total \\ \hline
DFT filling $\in[-\infty,2]$ &  0.75 & 0.74 & 1.54  & 1.72  & 1.56 & 1.42 & 1.57 & 26.41  \\\hline
%DFT filling $\in[-2,2]$ & 0.12 & 0.04 & 0.73 & 1.25 & 0.89 & 0.66 & 0.84 &  13.46 \\\hline
DFT filling $\in[-1.5, 1.5]$ & 0.02 & 0.03 & 0.48 & 1.02 & 0.78 & 0.56 & 0.39 &  9.79 \\\hline
DFT DOS@$E_f$ & 0.00 & 0.00 & 0.02 & 0.21 & 0.31 & 0.33 & 0.06 & 0.94
\\\hline\hline
\end{tabular}
\caption{\label{table:FeGe_DFT_filling_compare} The filling and DOS$@E_f$ of each orbital computed using the Wannier TB model (can be seen as DFT values) constructed using $s,p,d$ orbitals of Fe and $s,p$ orbitals of Ge.
}
\end{table}

\begin{figure}[tbp]
\centering
\includegraphics[width=0.9\textwidth]{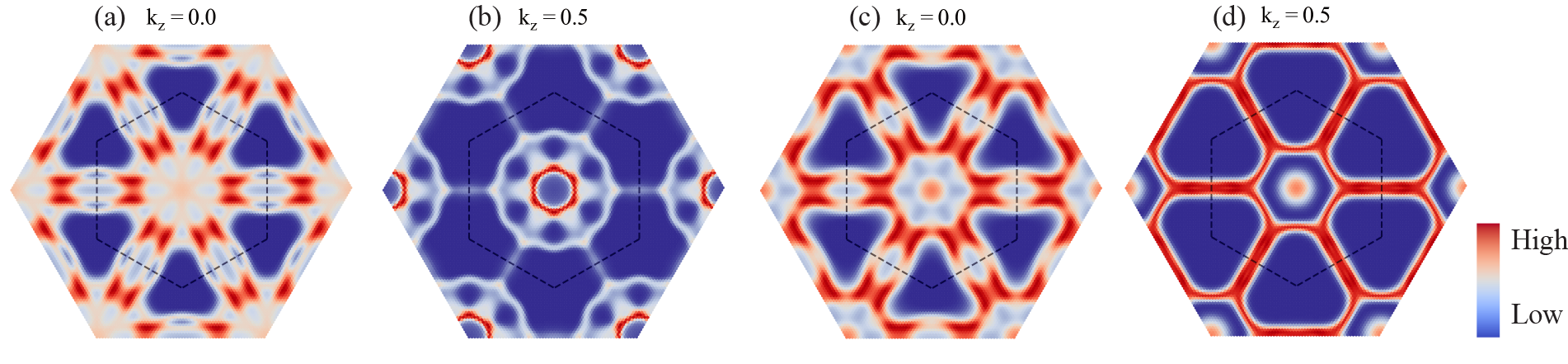}
	\caption{\label{app:fig:compare-FS} Comparison of Fermi surfaces (FSs) from DFT and TB model. 
    (a)(b) shows the DFT FS on the $k_z=0$ and $k_z=\pi$ planes, respectively. (c)(d) are the same but for the effective model. 
 }
\end{figure}

Finally, we compare the Fermi surfaces (FS) obtained from the minimal TB model and DFT, as shown in Fig.~\ref{app:fig:compare-FS}. The results show quantitative agreement, particularly on the $k_z=0$ plane. On the $k_z=\pi$ plane, while the FS shapes are consistent, there are intensity discrepancies because the quasi-flat band from the $d_{x^2-y^2}$ orbital is closer to $E_f$ in the TB model compared to DFT. Including longer-range hopping terms could improve the agreement further.

\newpage
\section{Interaction parameters of $\text{FeGe}$}\label{Sec:CRPA}

In this section, we compute the Coulomb interaction for FeGe using the \textit{ab initio} constraint random phase approximation (cRPA) method\cite{aryasetiawan2004frequency, solovyev2005screening, aryasetiawan2006calculations, miyake2009ab, di2023electronic}. We only compute the Coulomb interaction matrix for $d$ orbitals of Fe, as the Coulomb interactions of $s$ and $p$ orbitals of Fe and Ge have weaker interaction (for example, the intra-orbital onsite Hubbard $U$ for Ge $p$ is about two times smaller than that of Fe $d$, evaluated using the cRPA method to be introduced in the following), and the bands of them are also far away from $E_f$.

\subsection{Constraint random phase approximation (cRPA) method}

We first introduce the theoretical background of the cRPA method. 
In the cRPA method, one chooses a correlated subspace near the Fermi level that is usually generated by $3d$ (or $4f$) electrons. Because the $d$ (or $f$) orbitals in general have larger interactions than $p$ (or $s$) orbitals, one focuses on the interaction in the $d$-subspace. However, the $p$ orbitals always hybridize with $d$ orbitals in realistic materials which gives effective screening of the interaction of $d$ orbitals.

We start with the fully screened Coulomb interaction, which is given by
\begin{equation}
    W(\bm{r}, \bm{r}^\prime;\omega)
    =\epsilon^{-1}(\bm{r}, \bm{r}^\prime;\omega) v(\bm{r}, \bm{r}^\prime),
    \label{eq_fully_screened_W}
\end{equation}
where $v(\bm{r}, \bm{r}^\prime)$ is the bare Coulomb interaction, and $\epsilon(\bm{r}, \bm{r}^\prime;\omega)=1-v(\bm{r}, \bm{r}^\prime) P(\bm{r}, \bm{r}^\prime;\omega)$ is the dielectric function in RPA, with $P(\bm{r}, \bm{r}^\prime;\omega)$ being the non-interacting polarization\cite{miyake2009ab} given by:
\begin{equation}
    \begin{aligned}
	P\left(\mathbf{r}, \mathbf{r}^{\prime} ; \omega\right)
    =\sum_{\bm{k} n}^{\text {occ }} \sum_{\bm{k}^{\prime} n^{\prime}}^{\text {unocc }}
    \left\{\frac{\psi_{\bm{k} n}^*(\mathbf{r}) \psi_{\bm{k}^{\prime} n^{\prime}}(\mathbf{r}) 
    \psi_{\bm{k}^{\prime} n^{\prime}}^*\left(\mathbf{r}^{\prime}\right) 
    \psi_{\bm{k} n}\left(\mathbf{r}^{\prime}\right)}
    {\omega-\varepsilon_{\bm{k}^{\prime} n^{\prime}}+\varepsilon_{\bm{k} n}+i \delta}\right. 
	\left.-
    \frac{\psi_{\bm{k} n}(\mathbf{r}) 
    \psi_{\bm{k}^{\prime} n^{\prime}}^*(\mathbf{r}) 
    \psi_{\bm{k}^{\prime} n^{\prime}}\left(\mathbf{r}^{\prime}\right) 
    \psi_{\bm{k} n}^*\left(\mathbf{r}^{\prime}\right)}{\omega+\varepsilon_{\bm{k}^{\prime} n^{\prime}}-\varepsilon_{\bm{k} n}-i \delta}\right\},
    \end{aligned}
    \label{eq_polarization_P}
\end{equation}
where $\psi_{\bm{k}n}$ and $\epsilon_{\bm{k}n}$ are single-particle eigenstates and eigenvalues, and `occ' and `unocc' denotes occupied and unoccupied states. For systems with narrow bands of $3d$ (or $4f$) orbitals, one could divide the polarization into two parts, i.e., $P(\bm{r}, \bm{r}^\prime;\omega)=P_d(\bm{r}, \bm{r}^\prime;\omega) + P_r(\bm{r}, \bm{r}^\prime;\omega)$, where $P_d(\bm{r}, \bm{r}^\prime;\omega)$ includes only $3d$ to $3d$ transitions,
and $P_r(\bm{r}, \bm{r}^\prime;\omega)=P(\bm{r}, \bm{r}^\prime;\omega)-P_d(\bm{r}, \bm{r}^\prime;\omega)$ is the rest of the polarization.
When the bands of $d$ orbitals are separated from other bands, $P_d(\bm{r}, \bm{r}^\prime;\omega)$ is computed by replacing the sum over all occupied and unoccupied states in \cref{eq_polarization_P} by the sum over the occupied and unoccupied states of $d$ orbitals only. However, when the bands of $d$ orbitals are entangled with other bands, $P_d(\bm{r}, \bm{r}^\prime; \omega)$ is not well-defined. Many methods are proposed in the literature to solve this problem, including the \textit{disentanglement method}\cite{miyake2009ab} and the \textit{Projector method}\cite{kaltak2015dmft}. 
The \textit{Projector method} defines a projection operator that can extract the $d$ orbital contributions on each Bloch state and is the default method adopted in \textit{VASP}, which we use in this work.

It was shown\cite{aryasetiawan2004frequency, solovyev2005screening, aryasetiawan2006calculations, miyake2009ab} that the partially screened Coulomb interaction of the electrons in the selected $d$ subspace is:
\begin{equation}
    W_r(\bm{r}, \bm{r}^\prime; \omega)=
    \left[1-v(\bm{r}, \bm{r}^\prime) P_r(\bm{r}, \bm{r}^\prime;\omega)\right]^{-1}
    v(\bm{r}, \bm{r}^\prime).
    \label{eq_partially_screened_W}
\end{equation}
Compared with the fully screened Coulomb interaction defined in \cref{eq_fully_screened_W}, the partially screened interaction $W_r(\bm{r}, \bm{r}^\prime; \omega)$ is computed using the partial polarization $P_r(\bm{r}, \bm{r}^\prime; \omega)=P(\bm{r}, \bm{r}^\prime; \omega) - P_d(\bm{r}, \bm{r}^\prime; \omega)$, in which the contribution of $d-d$ screening has been removed from the total polarization. This subtraction could avoid the double counting of the $d-d$ screening since the $d-d$ screening is an inherent part of the $d$ subspace Hamiltonian and $W_r(\bm{r}, \bm{r}^\prime; \omega)$ can be interpreted as the effective interaction for electrons in the $d$-subspace. 
By combining \cref{eq_fully_screened_W} and \cref{eq_partially_screened_W}, the fully screened interaction $W(\bm{r}, \bm{r}^\prime; \omega)$ can also be expressed by $W_r(\bm{r}, \bm{r}^\prime; \omega)$ as the result of further screening by $P_d(\bm{r}, \bm{r}^\prime; \omega)$:
\begin{equation}
    W(\bm{r}, \bm{r}^\prime; \omega)
    =\left[ 1-W_r(\bm{r}, \bm{r}^\prime; \omega) P_d(\bm{r}, \bm{r}^\prime; \omega)\right]^{-1} W_r(\bm{r}, \bm{r}^\prime; \omega)
\end{equation}
Among the bare, partially screened, and fully screened Coulomb interactions, the bare Coulomb interaction $v(\bm{r}, \bm{r}^\prime)$ has the largest value while fully screened $W(\bm{r}, \bm{r}^\prime; \omega)$ has the smallest value, which means the more screening considered, the smaller the Coulomb interaction is for a given system.

With $W_r(\bm{r}, \bm{r}^\prime; \omega)$, the Coulomb interaction matrix can be evaluated using a localized basis set of the $d$ subspace, e.g., the maximally localized Wanneir functions $\{\varphi_i(\bm{r})\}$: 
\begin{equation}
    U_{ijkl}\left(\omega\right)= \iint d^3 r d^3 r^{\prime} \varphi_{i}^*(\mathbf{r}) \varphi_{j}(\mathbf{r}) W_r\left(\mathbf{r}, \mathbf{r}^{\prime} ; \omega\right) \varphi_{k}^*\left(\mathbf{r}^{\prime}\right) \varphi_{l}\left(\mathbf{r}^{\prime}\right).
    \label{eq_Uijkl_definition}
\end{equation}
Remark that the subscript convention follows the one used in \textit{VASP}, and is different from the more commonly used convention of the Coulomb matrix (see \cref{Eq:coulomb_def2}). 
The computed interaction parameter values depend on the choice of Wannier functions. For basis with more orbitals, the Wannier functions are more localized in general, which produce larger U values.

In the Coulomb matrix, the following matrix elements are useful:
\begin{equation}
\begin{aligned}
    U_{iijj}(\omega) &= \iint d^3 r d^3 r^{\prime}\left|\varphi_{i \mathbf{0}}(\mathbf{r})\right|^2 W_r\left(\mathbf{r}, \mathbf{r}^{\prime} ; \omega\right)\left|\varphi_{j \mathbf{0}}\left(\mathbf{r}^{\prime}\right)\right|^2, \\
    U_{ijji}(\omega) &= \iint d^3 r d^3 r^{\prime} 
	\varphi_{i \mathbf{0}}^*(\mathbf{r}) 
	\varphi_{j \mathbf{0}}(\mathbf{r}) W_r\left(\mathbf{r}, \mathbf{r}^{\prime} ; \omega\right) 
	\varphi_{j \mathbf{0}}^*\left(\mathbf{r}^{\prime}\right)
	\varphi_{i \mathbf{0}}\left(\mathbf{r}^{\prime}\right),\\
U_{ijij}(\omega) &= \iint d^3 r d^3 r^{\prime} 
	\varphi_{i \mathbf{0}}^*(\mathbf{r}) 
	\varphi_{j \mathbf{0}}(\mathbf{r}) W_r\left(\mathbf{r}, \mathbf{r}^{\prime} ; \omega\right) 
	\varphi_{i \mathbf{0}}^*\left(\mathbf{r}^{\prime}\right)
	\varphi_{j \mathbf{0}}\left(\mathbf{r}^{\prime}\right),\\
\end{aligned}
\label{definition_Uijkl}
\end{equation}
where $U_{iijj}=:U_{ij}$ is the onsite (inter-orbital) Hubbard $U$, 
and $U_{ijji}=:J_{ij}$ is the onsite exchange (assume $i,j$ is the index of orbitals on the same site). $U_{ijji}=U_{ijij}$ if the basis $\{\varphi_i\}$ are real, which holds in spinless systems with TRS symmetry. 
FeGe in the PM phase has TRS symmetry, enforcing its Wannier functions to be real in the spinless setting, and thus $U_{ijji}=U_{ijij}$.

Using the Coulomb matrix elements, the Hubbard-Kanamori parameters are defined as
\begin{equation}
    \begin{aligned}
	\mathcal{U} &=
		\frac{1}{N} \sum_{i \in \mathcal{T}} U_{i i i i}\\
	\mathcal{U}^\prime&=
		\frac{1}{N(N-1)} \sum_{i, j \in \mathcal{T}, i \neq j}^N U_{i i j j}\\
		\mathcal{J}&=
		\frac{1}{N(N-1)} \sum_{i, j \in \mathcal{T}, i \neq j}^N U_{i j j i}
    \end{aligned}
    \label{Eq:Hubbard_Kanmori_def}
\end{equation}
where $N$ is the number of orbitals in the target $d$ subspace.
For cubic systems, $\mathcal{U}^{\prime}= \mathcal{U} - 2\mathcal{J}$. 

The general Coulomb interaction in real space is 
\begin{equation}
    \hat{H}_{\text{int}}=\frac{1}{2}\sum_{i,\sigma\sigma^\prime}
    \left\langle i_1,i_2 |W_r| i_4,i_3 \right\rangle
    c_{i_1\sigma}^\dagger c_{i_2\sigma^\prime}^\dagger 
    c_{i_3\sigma^\prime} 
    c_{i_4\sigma},
\end{equation}
where $i$ is a composite index denoting both sites and orbitals, and 
\begin{equation}
\begin{aligned}
    \left\langle i_1,i_2 |W_r| i_4,i_3 \right\rangle &=
    \iint d^3 r d^3 r^{\prime} 
    \varphi_{i_1}^*(\mathbf{r}) 
    \varphi_{i_2}^*\left(\mathbf{r}^{\prime} \right)
    W_r\left(\mathbf{r}, \mathbf{r}^{\prime}\right) 
    \varphi_{i_4}\left(\mathbf{r}\right)
     \varphi_{i_3}(\mathbf{r}^{\prime}) \\
    &=
    \iint d^3 r d^3 r^{\prime} 
    \varphi_{i_1}^*(\mathbf{r}) 
    \varphi_{i_4}\left(\mathbf{r}\right)
    W_r\left(\mathbf{r}, \mathbf{r}^{\prime}\right) 
    \varphi_{i_2}^*\left(\mathbf{r}^{\prime} \right)
     \varphi_{i_3}(\mathbf{r}^{\prime}) \\     
    &= U_{i_1 i_4 i_2 i_3}
    %&= U_{i_1 i_3 i_2 i_4}
\end{aligned}
\label{Eq:coulomb_def2}
\end{equation}

The interacting Hamiltonian can be simplified into the generalized Kanamori form\cite{georges2013strong} by considering the following onsite interactions
\begin{equation}
	\begin{aligned}
	\hat{H}_{\text{int}} &=
    %\sum_{ijmm^\prime\sigma} t_{ij}^{mm^\prime} c_{im\sigma}^\dagger c_{jm^\prime \sigma} + 
    \mathcal{U}\sum_{im} n_{im\uparrow}n_{im\downarrow}+
    \mathcal{U}^\prime \sum_{i,m\ne m^\prime} n_{i m\uparrow}n_{i m^\prime\downarrow}
    +(\mathcal{U}^\prime -\mathcal{J})\sum_{i,m<m^\prime \sigma} n_{i m \sigma} n_{i m^\prime \sigma}	\\
	&-\mathcal{J}_X \sum_{i, m<m^\prime} 
	(c_{im\uparrow}^\dagger c_{i m \downarrow}  c_{i m^\prime \downarrow}^\dagger c_{i m^\prime \uparrow}+H.c.)
	+\mathcal{J}_P \sum_{i, m<m^\prime}  
	(c_{im\uparrow}^\dagger c_{im\downarrow}^\dagger c_{i m^\prime \downarrow} c_{i m^\prime \uparrow} + H.c.)
	\end{aligned}
\end{equation}
where $i,j$ denote the different lattice sites, and $m, m^\prime$ denote different orbitals on the same site, i.e., we only consider the interaction between different orbitals on the same site. The $\mathcal{J}_X$ term is the spin-flipping term and the $\mathcal{J}_P$ term is the pair-hopping term.
The Coulomb matrix element 
$U_{ijkl} = \left\langle i,k|W_r|l,j\right\rangle$,
which is the coefficient of the two-body operator 
$c_{i\sigma}^\dagger c_{k\sigma^\prime}^\dagger c_{l\sigma^\prime} c_{j \sigma}$. 
Thus 
\begin{itemize}
    \item $U_{iijj}$ corresponds to $n_{i\sigma} n_{j\sigma^\prime}$ ($\mathcal{U}, \mathcal{U}^\prime$).
    \item $U_{ijji}$ corresponds to 
	$-c_{i\sigma}^\dagger c_{i\sigma^\prime}  c_{j\sigma^\prime}^\dagger c_{j \sigma}$ ($\mathcal{J}_X$ if $\sigma=-\sigma^\prime$, $\mathcal{J}$ if $\sigma=\sigma^\prime$).
    \item $U_{ijij}$ corresponds to 
	$c_{i\sigma}^\dagger c_{i\bar{\sigma}}^\dagger c_{j\bar{\sigma}} c_{j \sigma}$ ($\mathcal{J}_P$).
\end{itemize}

In the case of real Wannier basis, $\mathcal{J}=\mathcal{J}_X=\mathcal{J}_P$, and the Kanamori Hamiltonian has the simplified form
\begin{equation}
\begin{aligned}
    \hat{H}_{\text{int}} &=
    \mathcal{U}\sum_{im} n_{im\uparrow}n_{im\downarrow}+
    \mathcal{U}^\prime \sum_{i,m\ne m^\prime} n_{i m\uparrow}n_{i m^\prime\downarrow}
    +(\mathcal{U}^\prime -\mathcal{J})\sum_{i,m<m^\prime \sigma} n_{i m \sigma} n_{i m^\prime \sigma}	\\
	&-\mathcal{J} \sum_{i, m\ne m^\prime} 
	c_{im\uparrow}^\dagger c_{i m \downarrow}  c_{i m^\prime \downarrow}^\dagger c_{i m^\prime \uparrow}
	+\mathcal{J} \sum_{i, m\ne m^\prime}  
    c_{im\uparrow}^\dagger c_{im\downarrow}^\dagger c_{i m^\prime \downarrow} c_{i m^\prime \uparrow}
\end{aligned}
\end{equation}

If one considers orbital-dependent onsite interactions $U_{mm^\prime}$ and $J_{mm^\prime}$, the interacting Hamiltonian is generalized to 
\begin{equation}
\begin{aligned}
    \hat{H}_{\text{int}} &=
    \sum_{im} U_{mm} n_{im\uparrow}n_{im\downarrow}
    + \sum_{i,m\ne m^\prime} 
    U_{mm^\prime} n_{i m\uparrow}n_{i m^\prime\downarrow}
    +\sum_{i,m<m^\prime \sigma} 
    (U_{mm^\prime} - J_{mm^\prime})
    n_{i m \sigma} n_{i m^\prime \sigma}	\\
    &-\sum_{i, m\ne m^\prime} 
    J_{mm^\prime} c_{im\uparrow}^\dagger c_{i m \downarrow}  c_{i m^\prime \downarrow}^\dagger c_{i m^\prime \uparrow}
    +\sum_{i, m\ne m^\prime} 
    J_{mm^\prime} c_{im\uparrow}^\dagger c_{im\downarrow}^\dagger c_{i m^\prime \downarrow} c_{i m^\prime \uparrow}
\end{aligned}
\end{equation}
In the following, we will use the cRPA method to compute $U_{ij}$ and $J_{ij}$ to obtain the interacting Hamiltonian for FeGe.

\subsection{cRPA interactions parameters for FeGe}
We compute the Coulomb interaction matrix using the cRPA method implemented in VASP\cite{kresse1996efficient, kaltak2015dmft}, where the partially screened Coulomb interaction kernel \cref{eq_partially_screened_W} is first computed, and Coulomb interaction is evaluated using a set of maximally localized Wannier functions (obtained from WANNIER90\cite{mostofi2008wannier90}) based on \cref{eq_Uijkl_definition}.

We first construct MLWFs using $d$ orbitals of Fe and $p$ orbitals of Ge, which have 24 orbitals in total. Two models can be constructed from them. One is called the $d-dp$ model in the literature\cite{vaugier2012hubbard}, which excludes only the polarization inside the $d$ orbitals when computing the partially screened Coulomb interaction. The other model is called the $dp-dp$ model\cite{vaugier2012hubbard}, which excludes the polarization of both $d$ and $p$ orbitals. The $dp-dp$ model produces a larger Coulomb interaction as more screening is removed.

We also construct another set of MLWFs using $s, p, d$ orbitals of Fe and $s, p$ orbitals of Ge, which has 39 orbitals in total. The MLWFs are more localized as more orbitals are included. A third model, which we call $d-$full, is constructed based on this large MLWF set which screens only $d$ orbitals.

In \cref{d-dp}, \ref{dp-dp}, and \ref{d-full}, we list the Coulomb interaction matrix $U_{ij}$ and $J_{ij}$ for the five $d$ orbitals on the same Fe, the NN and NNN Coulomb interaction $U_{ij}^{NN}$ and $U_{ij}^{NNN}$ of five $d$ orbitals, and the Hubbard-Kanamori parameters computed for five $d$ orbitals on the same site for the $d-dp$, $dp-dp$, and $d$-full model, respectively.

For the three models, $d-dp$ and $d$-full models have comparable interaction values and can be used in interacting Hamiltonians with compatible Wannier basis sets. The $dp-dp$ model, however, has significantly larger interaction values compared with $d-dp$ and $d$-full models, and can only be used in the interacting Hamiltonians where both $d$ and $p$ orbitals have interactions.

\begin{table}[htbp]
\begin{tabular}{c|ccccccc|ccccc}
\hline\hline
$U_{ij}$  & $z^2$ & $xz$ & $yz$ & $x^2-y^2$ & $xy$ &  & $J_{ij}$ & $z^2$ & $xz$ & $yz$ & $x^2-y^2$ & $xy$ \\\hline
$z^2$     & 3.81  & 2.87 & 2.93 & 2.41      & 2.41 &  &          &       & 0.47 & 0.50 & 0.75      & 0.73 \\
$xz$      &       & 3.81 & 2.57 & 2.59      & 2.58 &  &          &       &      & 0.68 & 0.67      & 0.65 \\
$yz$      &       &      & 4.06 & 2.64      & 2.64 &  &          &       &      &      & 0.70      & 0.67 \\
$x^2-y^2$ &       &      &      & 4.04      & 3.2  &  &          &       &      &      &           & 0.40 \\
$xy$      &       &      &      &           & 3.96 &  &          &       &      &      &           &     
\\\hline\hline
$U_{ij}^{NN}$ & $z^2$ & $xz$ & $yz$ & $x^2-y^2$ & $xy$ & &$U_{ij}^{NNN}$ & $z^2$ & $xz$ & $yz$ & $x^2-y^2$ & $xy$ \\\hline
$z^2$         & 1.48  & 1.47 & 1.48 & 1.50      & 1.51 &   &             & 1.27  & 1.27 & 1.26 & 1.27      & 1.27 \\
$xz$          &       & 1.48 & 1.48 & 1.50      & 1.51 &   &             &       & 1.27 & 1.26 & 1.28      & 1.28 \\
$yz$          &       &      & 1.50 & 1.52      & 1.52 &     &           &       &      & 1.26 & 1.27      & 1.27 \\
$x^2-y^2$     &       &      &      & 1.55      & 1.56 &     &           &       &      &      & 1.28      & 1.29 \\
$xy$          &       &      &      &           & 1.58 &    &            &       &      &      &           & 1.29\\
\hline\hline
\end{tabular}
\caption{\label{d-dp} The Coulomb interaction $U_{ij}$ and $J_{ij}$ of $d$ orbitals, and $U_{ij}^{NN}$ and $U_{ij}^{NNN}$ between NN and NNN $d$ orbitals in $d-dp$ model. where $U_{ij}=U_{iijj}$ and $J_{ij}=U_{ijji}$, with $U_{ijkl}$ defined in \cref{definition_Uijkl}.
The onsite Hubbard-Kanamori parameters are $\mathcal{U}=3.94, \mathcal{U}^\prime=2.69, \mathcal{J}=0.62$. All numbers are in eV.}
\end{table}

\begin{table}[htbp]
\begin{tabular}{c|ccccccc|ccccc}
\hline\hline
$U_{ij}$  & $z^2$ & $xz$ & $yz$ & $x^2-y^2$ & $xy$ &  & $J_{ij}$ & $z^2$ & $xz$ & $yz$ & $x^2-y^2$ & $xy$ \\\hline
$z^2$     & 8.00  & 7.15 & 7.21 & 6.55      & 6.44 &  &          &       & 0.49 & 0.52 & 0.83      & 0.79 \\
$xz$      &       & 8.25 & 6.89 & 6.88      & 6.74 &  &          &       &      & 0.74 & 0.72      & 0.71 \\
$yz$      &       &      & 8.51 & 6.91      & 6.81 &  &          &       &      &      & 0.77      & 0.73 \\
$x^2-y^2$ &       &      &      & 8.39      & 7.41 &  &          &       &      &      &           & 0.40 \\
$xy$      &       &      &      &           & 8.05 &  &          &       &      &      &           &     
\\\hline\hline
$U_{ij}^{NN}$ & $z^2$ & $xz$ & $yz$ & $x^2-y^2$ & $xy$ & &$U_{ij}^{NNN}$ & $z^2$ & $xz$ & $yz$ & $x^2-y^2$ & $xy$ \\\hline
$z^2$         & 2.68  & 2.67 & 2.69 & 2.72      & 2.73 &   &             & 1.87  & 1.88 & 1.86 & 1.88      & 1.88 \\
$xz$          &       & 2.67 & 2.68 & 2.71      & 2.72 &   &             &       & 1.89 & 1.87 & 1.89      & 1.90 \\
$yz$          &       &      & 2.71 & 2.74      & 2.75 &     &           &       &      & 1.86 & 1.88      & 1.88 \\
$x^2-y^2$     &       &      &      & 2.78      & 2.79 &     &           &       &      &      & 1.90      & 1.90 \\
$xy$          &       &      &      &           & 2.81 &    &            &       &      &      &           & 1.91\\
\hline\hline
\end{tabular}
\caption{\label{dp-dp} The Coulomb interaction $U_{ij}$ and $J_{ij}$ of $d$ orbitals,
and $U_{ij}^{NN}$ and $U_{ij}^{NNN}$ between NN and NNN $d$ in $dp-dp$ model.
The onsite Hubbard-Kanamori parameters are $\mathcal{U}=8.24, \mathcal{U}^\prime=6.90, \mathcal{J}=0.67$. All numbers are in eV.}
\end{table}

\begin{table}[htbp]
\begin{tabular}{c|ccccccc|ccccc}
\hline\hline
$U_{ij}$  & $z^2$ & $xz$ & $yz$ & $x^2-y^2$ & $xy$ &  & $J_{ij}$ & $z^2$ & $xz$ & $yz$ & $x^2-y^2$ & $xy$ \\\hline
$z^2$     & 4.16  & 2.87 & 3.04 & 2.42      & 2.45 &  &          &       & 0.54 & 0.57 & 0.90      & 0.92 \\
$xz$      &       & 3.80 & 2.51 & 2.53      & 2.57 &  &          &       &      & 0.74 & 0.74      & 0.76 \\
$yz$      &       &      & 4.21 & 2.70      & 2.74 &  &          &       &      &      & 0.77      & 0.80 \\
$x^2-y^2$ &       &      &      & 4.28      & 3.43 &  &          &       &      &      &           & 0.47 \\
$xy$      &       &      &      &           & 4.46 &  &          &       &      &      &           &   \\
\hline\hline
$U_{ij}^{NN}$ & $z^2$ & $xz$ & $yz$ & $x^2-y^2$ & $xy$ & &$U_{ij}^{NNN}$ & $z^2$ & $xz$ & $yz$ & $x^2-y^2$ & $xy$ \\\hline
$z^2$         & 1.38  & 1.38 & 1.40 & 1.40      & 1.41 &   &             & 1.22  & 1.22 & 1.22 & 1.22      & 1.22 \\
$xz$          &       & 1.39 & 1.39 & 1.40      & 1.40 &   &             &       & 1.22 & 1.22 & 1.23      & 1.23 \\
$yz$          &       &      & 1.42 & 1.42      & 1.42 &     &           &       &      & 1.21 & 1.22      & 1.22 \\
$x^2-y^2$     &       &      &      & 1.44      & 1.44 &     &           &       &      &      & 1.23      & 1.23 \\
$xy$          &       &      &      &           & 1.45 &    &            &       &      &      &           & 1.23\\
\hline\hline
\end{tabular}
\caption{\label{d-full} The Coulomb interaction $U_{ij}$ and $J_{ij}$ of $d$ orbitals,
and $U_{ij}^{NN}$ and $U_{ij}^{NNN}$ between NN and NNN $d$ in $d-$full model.
The onsite Hubbard-Kanamori parameters are $\mathcal{U}=4.18, \mathcal{U}^\prime=2.73, \mathcal{J}=0.72$.  These values satisfy the relation  $\mathcal{U}^\prime\approx \mathcal{U}-2\mathcal{J}$ which holds exactly in cubic systems. The averaged NN and NNN interactions are $\bar{U}^{NN}=1.41, \bar{U}^{NNN}=1.22$.
All numbers are in eV.}
\end{table}

\newpage
\subsection{Approximated interaction parameters}\label{Sec:approximated_interaction}
In this section, we simplify the Coulomb interaction matrix via approximate symmetries.

In the $d$-full model shown in \cref{d-full}, the averaged onsite Hubbard-Kanamori parameters are $\mathcal{U}=4.18, \mathcal{U}^\prime=2.73, \mathcal{J}=0.72$ eV for five $d$ orbitals on the same site, as shown in \cref{d-full}. These values satisfy the relation  $\mathcal{U}^\prime\approx \mathcal{U}-2\mathcal{J}$ which holds exactly in cubic systems.  
The Hubbard-Kanamori parameters are averaged values (defined in \cref{Eq:Hubbard_Kanmori_def}) and are thus symmetric for five $d$ orbitals, thus having $\text{SU}(2)$ spin-rotation symmetry and $\text{SO}(5)$ orbital symmetry.

However, the five $d$ orbitals are in fact symmetry-independent, as the site symmetry group of the kagome site is $D_{2h}$, which has no symmetry that could relate different $d$ orbitals. Thus the Coulomb interactions $U_{ij}$ and $J_{ij}$ of $d$ orbitals can take independent values and only have $\text{SU}(2)$ spin-rotation symmetry. 

Although five $d$ orbitals are symmetry-independent, their interactions still share similarities, e.g., the onsite intra-orbital Hubbard $U_{ii}$ values are close, which indicates approximated symmetries. 
To simplify the Coulomb matrix, we will assume cubic and spherical symmetries in the following. 

For the NN and NNN interactions $U_{ij}^{NN}$ and $U_{ij}^{NNN}$ in the $d$-full model, we adopt the spherical approximation and average them over orbitals: 
\begin{equation}
    \begin{aligned}
        \overline{U}^{NN} &= \frac{1}{25}\sum_{ij}U_{ij}^{NN}= 1.41 \text{ eV}\\
        \overline{U}^{NNN} &= \frac{1}{25}\sum_{ij}U_{ij}^{NNN}= 1.22 \text{ eV}\\
    \end{aligned}
\end{equation}
The root mean square error for $U_{ij}^{NN}$ and $U_{ij}^{NNN}$ is 0.019 and 0.004 eV, respectively.

For the onsite Coulomb matrix, we will first assume cubic symmetries and compute the symmetrized values, and then further assume spherical symmetries and adopt the Slater integrals to evaluate the approximated Coulomb matrix, as shown in the following sections.

\subsubsection{Symmetrized Coulomb matrix using $O_h$ symmetries}

\begin{figure}[htbp]
    \centering
    \includegraphics[width=0.4\textwidth]{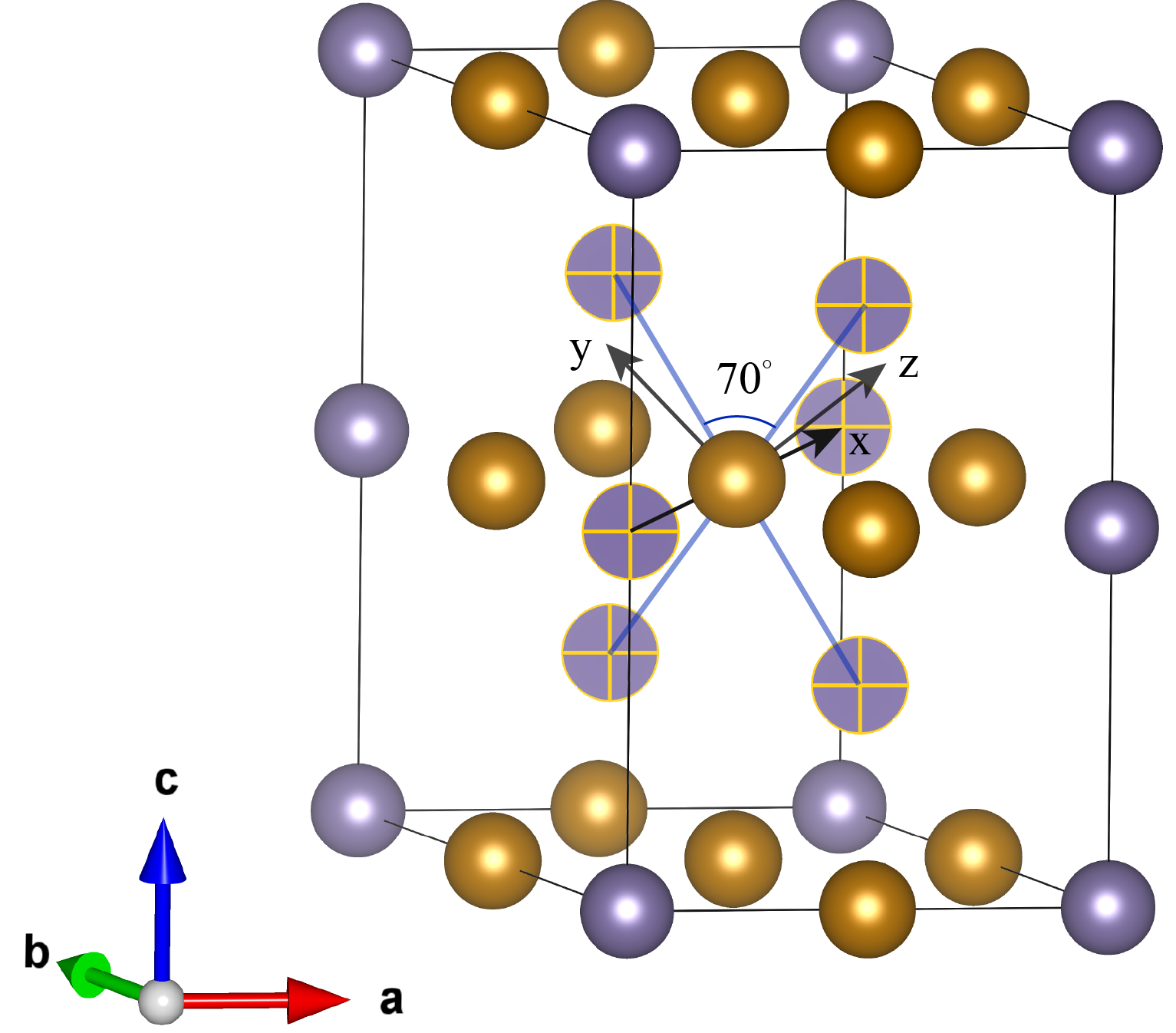}
    \caption{\label{Fig:quasi_cubic_coord} Schematic show of the quasi-cubic local coordinate system for Fe atom at $(\frac{1}{2},\frac{1}{2},\frac{1}{2})$, defined in \cref{Eq:quasi_cubic_lc}. Each Fe atom has six closest surrounding Ge atoms that form a distorted octahedron and thus give an approximated cubic crystal field.}
\end{figure}

The assumption of cubic symmetries can be justified using the following local coordinate, as shown in \cref{Fig:quasi_cubic_coord}.
For each Fe atom on the kagome site in FeGe, there are six nearest Ge atoms, i.e., two triangular Ge's and four honeycomb Ge's, which form a distorted octahedron and give an approximated cubic field. The local coordinate system is defined as
\begin{equation}
    \begin{aligned}
        3f_1&: \bm{x}_1=(1,0,0), \bm{y}_1=(0,\frac{1}{\sqrt{2}},\frac{1}{\sqrt{2}}), \bm{z}_1=(0,-\frac{1}{\sqrt{2}},\frac{1}{\sqrt{2}})\\
        3f_2&: \bm{x}_2=(\frac{1}{2}, \frac{\sqrt{3}}{2}, 0), 
        \bm{y}_2=(-\frac{\sqrt{3}}{2\sqrt{2}}, \frac{1}{2\sqrt{2}}, \frac{1}{\sqrt{2}}), 
        \bm{z}_2=(\frac{\sqrt{3}}{2\sqrt{2}}, -\frac{1}{2\sqrt{2}}, \frac{1}{\sqrt{2}}) \\
        3f_3&: \bm{x}_3=(-\frac{1}{2}, \frac{\sqrt{3}}{2}, 0), 
        \bm{y}_3=(-\frac{\sqrt{3}}{2\sqrt{2}}, -\frac{1}{2\sqrt{2}}, \frac{1}{\sqrt{2}}), 
        \bm{z}_3=(\frac{\sqrt{3}}{2\sqrt{2}}, \frac{1}{2\sqrt{2}}, \frac{1}{\sqrt{2}}) \\
    \end{aligned}
    \label{Eq:quasi_cubic_lc}
\end{equation}
We call this local coordinate system the ``quasi-cubic local coordinate system" as six neighboring Ge atoms form an approximated cubic field. We recompute the cRPA Coulomb matrix in the $d$-full model using the quasi-cubic local coordinate system, with $U_{ij}$ and $J_{ij}$ tabulated in \cref{d-full-cubic-lc-U}. This Coulomb matrix is more symmetric, e.g., the intra-orbital Hubbard $U_{ii}$ of each $d$ orbital is more close to the averaged $\mathcal{U}$.

We remark that this quasi-cubic local environment only holds well by considering six neighboring Ge atoms for each Fe atom, while the inclusion of NN Fe atoms will spoil the cubic environment. Since in the cRPA calculations, only the $p$ (and s) orbitals of Ge are projected out while the $d-d$ screening of Fe is kept, this quasi-cubic local environment of Fe given by the Ge atoms can still approximately hold. 
In the single-particle Wannier TB model (see \cref{Table:wannier_hoppings}), the hopping parameters do not show good agreements with this cubic environment, possibly because both Fe and Ge contribute to the local environment in the TB and the $d$ orbitals in the TB are also not aligned with this cubic local coordinates.

The original onsite Coulomb matrix $U_{ijkl}^o$ in the $d$-full model (computed under basis $\psi^{\text{loc}}$ under local coordinates \cref{Eq:local_coordinate}) is related to the onsite Coulomb matrix $U_{ijkl}^c$ computed using basis $\psi^{\text{quasi-cubic}}$ in the quasi-cubic local coordinate systems by the following transformation $S$, derived from the transformation 
$\psi^{\text{quasi-cubic}}_i=S_{ij}\psi_{j}^{\text{loc}}$
\begin{equation}
    \begin{aligned}
        S &=\left(
        \begin{matrix}
        \frac{1}{4} & 0 & -\frac{\sqrt{3}}{2} & -\frac{\sqrt{3}}{4} & 0 \\
        0 & \frac{1}{\sqrt{2}} & 0 & 0 & -\frac{1}{\sqrt{2}} \\
        \frac{\sqrt{3}}{2} & 0 & 0 & \frac{1}{2} & 0 \\
        -\frac{\sqrt{3}}{4} & 0 & -\frac{1}{2} & \frac{3}{4} & 0 \\
        0 & \frac{1}{\sqrt{2}} & 0 & 0 & \frac{1}{\sqrt{2}}\\
        \end{matrix}
    \right),\\
    \end{aligned}
\end{equation}
\begin{equation}
    U^c_{ijkl} = \sum_{i^\prime j^\prime k^\prime l^\prime} S^*_{ii^\prime} S^*_{jj^\prime} U^o_{i^\prime j^\prime k^\prime l^\prime} S_{kk^\prime} S_{ll^\prime}
\end{equation}

\begin{table}[htbp]
\begin{tabular}{c|cccccc|ccccc}
\hline\hline
$U_{ij}$  & $z^2$ & $xz$ & $yz$ & $x^2-y^2$ & $xy$ & $J_{ij}$ & $z^2$ & $xz$ & $yz$ & $x^2-y^2$ & $xy$ \\\hline
$z^2$     & 4.18 &  3.03 &  3.11 &  2.40  &  2.40 &    &    & 0.56& 0.57 & 0.88  &  0.87 \\
$xz$      &       &  4.11 &  2.66 &  2.59  &  2.59 &    &    &      & 0.78 & 0.77  &  0.78 \\
$yz$      &       &        &  4.33 &  2.65  &  2.66 &    &    &      &       & 0.79  &  0.78 \\
$x^2-y^2$ &       &        &        &  4.13  &  3.21 &    &    &      &       &        &  0.45 \\
$xy$      &       &        &        &         &  4.11 &    &    &      &       &        &        \\
\hline\hline
\end{tabular}
\caption{\label{d-full-cubic-lc-U}The Coulomb interaction $U_{ij}$ and $J_{ij}$ in the quasi-cubic local coordinate system that aligned with the bonds between Fe and Ge, which form an approximated cubic lattice field.}
\end{table}

We then further symmetrize the onsite Coulomb matrix $U_{ijkl}^o$ computed in the $d$-full model by assuming the cubic $O_h$ symmetry group. 
$O_h$ can be generated by the following three generators:
\begin{equation}
\begin{aligned}
    C_{3,111}&=\left(
    \begin{matrix}
	0 & 0 & 1 \\
	1 & 0 & 0\\
	0 & 1 & 0 \\
    \end{matrix}
    \right),\quad
    C_{4,z}=\left(
    \begin{matrix}
	0 & -1 & 0 \\
	1 & 0 & 0\\
	0 & 0 & 1 \\
    \end{matrix}
    \right),\quad
    P=\left(
    \begin{matrix}
	-1 & 0 & 0 \\
	0 & -1 & 0\\
	0 & 0 & -1 \\
    \end{matrix}
    \right),
\end{aligned}
\end{equation}
which have representation matrices under five $d$ orbitals $(d_{z^2}, d_{xz}, d_{yz}, d_{x^2-y^2}, d_{xy})$:
\begin{equation}
\begin{aligned}
    D(C_{3,111})&=\left(
    \begin{matrix}
	-\frac{1}{2} & 0 & 0 & -\frac{\sqrt{3}}{2} & 0 \\
	0 & 0 & 1 & 0 & 0\\
	0 & 0 & 0 & 0 & 1 \\
        \frac{\sqrt{3}}{2} & 0 & 0 & -\frac{1}{2} & 0\\
        0 & 1 & 0 & 0 & 0 \\
    \end{matrix}
    \right),\quad
    D(C_{4,z})=\left(
    \begin{matrix}
        1 & 0 & 0 & 0 & 0 \\
	0 & 0 & -1 & 0 & 0 \\
	0 & 1 & 0 & 0 & 0 \\
	0 & 0 & 0 & -1 & 0 \\
        0 & 0 & 0 & 0 & -1 \\
    \end{matrix}
    \right),\quad
    D(P)=\left(
    \begin{matrix}
	1 & 0 & 0 & 0 & 0\\
	0 & 1 & 0& 0 & 0\\
	0 & 0 & 1 & 0 & 0\\
        0 & 0 & 0 & 1 & 0 \\
        0 & 0 & 0 & 0 & 1 \\
    \end{matrix}
    \right).
\end{aligned}
\end{equation}
where the transformation is defined as $d_i(g^{-1}\bm{r})=d_{j}(\bm{r})D_{ji}(g)$, or equivalently, $d_i(g\bm{r})=D_{ij}(g)d_{j}(\bm{r})$, as $D^T(g)=D^{-1}(g)=D(g^{-1})$. 

Using these generators, we generate all 48 operations in the $O_h$ group and symmetrize the Coulomb matrix:
\begin{equation}
    U^{\text{sym}}_{ijkl} = \frac{1}{N_\mathcal{G}}\sum_{g\in\mathcal{G}}\sum_{i^\prime j^\prime k^\prime l^\prime} D(g)^*_{ii^\prime} D(g)^*_{jj^\prime} U^o_{i^\prime j^\prime k^\prime l^\prime} D(g)_{kk^\prime} D(g)_{ll^\prime},
\end{equation}
where $\mathcal{G}=O_h$, $N_\mathcal{G}=48$, and $U_{ijkl}^o$ denote the original Coulomb matrix computed using the cRPA method. The symmetrized Coulomb matrix elements $U_{ij}$ and $J_{ij}$ for the $d$-full model are tabulated in \cref{d-full-symmetrized-U}. The onsite Hubbard $U$ values have $T_{2g}$ ($d_{xz},d_{yz},d_{xy}$) and $E_g$ ($d_{z^2}, d_{x^2-y^2}$) splitting. 
We remark that in practice using $O$ group symmetries is equivalent for five $d$ orbitals, as $D(P)=\bm{1}_5$ which has no effects.

\begin{table}[htbp]
\begin{tabular}{c|cccccc|ccccc}
\hline\hline
$U_{ij}$  & $z^2$ & $xz$ & $yz$ & $x^2-y^2$ & $xy$ & $J_{ij}$ & $z^2$ & $xz$ & $yz$ & $x^2-y^2$ & $xy$ \\\hline
$z^2$     & 4.22 &  3.05 &  3.05 &  2.42  &  2.41 &    &    & 0.56& 0.56 & 0.90  &  0.88 \\
$xz$      &       &  4.16 &  2.61 &  2.62  &  2.61 &    &    &      & 0.77 & 0.78  &  0.77 \\
$yz$      &       &        &  4.16 &  2.62  &  2.61 &    &    &      &       & 0.78  &  0.77 \\
$x^2-y^2$ &       &        &        &  4.22  &  3.27 &    &    &      &       &        &  0.46 \\
$xy$      &       &        &        &         &  4.16 &    &    &      &       &        &        \\
\hline\hline
\end{tabular}
\caption{\label{d-full-symmetrized-U} The onsite Coulomb interaction $U_{ij}$ and $J_{ij}$ of $d$ orbitals symmetrized using $O_h$ symmetries.
The root mean square error of the symmetrized $U_{ij}$ and $J_{ij}$ to the original ones is 0.132 and 0.023, respectively.}
\end{table}

\subsubsection{Slater integrals}

In \cref{d-full-symmetrized-U}, the onsite Coulomb matrix symmetrized by $O_h$ symmetries still has some matrix elements that take very close values, e.g., the five intra-orbital $U_{ii}$, which suggests a higher symmetry group, i.e., spherical $\text{SO}(3)$ rotation symmetry. For the spherical Coulomb matrix, Slater integrals\cite{slater1960quantum, sugano2012multiplets} can be adopted to parameterize. We first give a brief introduction to the Slater integrals in this section and then fit the value of the Slater integrals in the next section.

The Coulomb matrix in a spherical lattice field can be parametrized by Slater integrals $F^k$ in the complex spherical harmonics basis with the form\cite{wang2015interaction, vaugier2012hubbard}
\begin{equation}
    U_{m_1, m_2, m_3, m_4}=\delta_{m_1+m_2, m_3+m_4}\sum_k c_k^{m_1,m_3}c_k^{m4,m2}F^k,
    \label{eq_slater_Uijkl}
\end{equation}
where $c_k^{m_i,m_j}$ is the Gaunt coefficient, whose value is tabulated in Table 1.2 in Ref.\cite{sugano2012multiplets}, and $F^k$ are Slater integrals that need to be fitted for specific materials. For $d$ orbitals, there are only three nonzero Slater integrals $F_0$, $F_2$, and $F_4$ as $l=2$ (for $f$ orbitals, $F_6$ is also needed), and it is a good approximation to take the average value $F_4=0.625F_2$ for $3d$ transition metals according to literature\cite{de19902p, anisimov1997first}.

For $3d$ shell, \cref{eq_slater_Uijkl} is written in complex spherical harmonics basis $\phi_{l=2,m}(\bm{r})=R_{3d}(r)Y_{l=2}^{m}(\bm{r})$, and needs to be transformed to cubic harmonic basis $\psi_{m}(\bm{r})=R_{3d}(r)X_{2c}(\bm{r})$ which are used in Wannier functions, where the radical parts $R_{3d}(r)$ are identical while the angular parts $X_{2c}(\bm{r})$ are related to $Y_2^m$ by the following equations
\begin{equation}
    \begin{aligned}
        d_{z^2} &= N_{2}^c \frac{3z^2-r^2}{2r^2\sqrt{3}}=Y_2^0\\
        d_{xz} &= N_2^c \frac{xz}{r^2} = \frac{1}{\sqrt{2}}(Y_2^{-1}-Y_2^{1})\\
        d_{yz} &= N_2^c \frac{yz}{r^2} = \frac{i}{\sqrt{2}}(Y_2^{-1} + Y_2^{1})\\
        d_{xy} &= N_2^c \frac{xy}{r^2} = \frac{i}{\sqrt{2}}(Y_2^{-2} -Y_2^{2})\\
        d_{x^2-y^2} &= N_2^c \frac{x^2-y^2}{2r^2} = \frac{1}{\sqrt{2}}(Y_2^{-2} + Y_2^{2})\\
    \end{aligned}
\end{equation}

The following relations can be derived for onsite Hubbard U and Hund's exchange $J_{ij}$ of five $d$ orbitals for spherical-symmetric systems using \cref{eq_slater_Uijkl} and the value of Gaunt coefficients\cite{wang2015interaction}:
\begin{equation}
    \begin{aligned}
        U &= F^0 + \frac{4}{49}(F^2 + F^4) \\
        J(d_{z^2},d_{xz}) &= J(d_{z^2},d_{yz})=\frac{1}{49}F^2+\frac{30}{441}F^4 \\
        J(d_{z^2},d_{x^2-y^2}) &= J(d_{z^2},d_{xy})=\frac{4}{49}F^2+\frac{15}{441}F^4 \\
        J(d_{xz},d_{yz}) &= J(d_{xz},d_{x^2-y^2}) =J(d_{xz},d_{xy}) \\ &=J(d_{yz},d_{x^2-y^2}) =J(d_{yz},d_{xy}) 
        =\frac{3}{49}F^2+\frac{20}{441}F^4 \\
        J(d_{x^2-y^2}, d_{xy}) &= \frac{35}{441}F^4
    \end{aligned}
\end{equation}

\subsubsection{Fitting Slater integrals}

We fit the Slater integrals $F^0, F^2, F^4$ to the Coulomb matrix symmetrized using $O_h$ symmetries as well as the one computed in the quasi-cubic local coordinate system, which yields identical values of Slater integrals.
The resultant Coulomb matrix computed from the Slater integrals is shown in \cref{d-full-slater-U-F0F2F4}. It has spherical symmetries and is invariant under any $\text{SO}(3)$ coordinate transformations. The ratio for the fitted parameters is $F^4/F^2=0.686$, which is close to $0.625$.

We also use the approximated relation $F^4=0.625F^2$ in literature\cite{de19902p} and fit only the Slater integrals $F^0$ and $F^2$, with results shown in \cref{d-full-slater-U-F0F2}. The resultant Coulomb matrix has a slightly larger fitting error compared with \cref{d-full-slater-U-F0F2F4}, but only has two parameters $F^0$ and $F^2$, which is simpler for further mean-field calculations.

\begin{table}[htbp]
\begin{tabular}{c|cccccc|ccccc}
\hline\hline
$U_{ij}$  & $z^2$ & $xz$ & $yz$ & $x^2-y^2$ & $xy$ & $J_{ij}$ & $z^2$ & $xz$ & $yz$ & $x^2-y^2$ & $xy$ \\\hline
$z^2$     & 4.18 &  3.05 &  3.05 &  2.41  &  2.41 &    &    & 0.56& 0.56 & 0.88  &  0.88 \\
$xz$      &       &  4.18 &  2.62 &  2.62  &  2.62 &    &    &      & 0.78 & 0.78  &  0.78 \\
$yz$      &       &        &  4.18 &  2.62  &  2.62 &    &    &      &       & 0.78  &  0.78 \\
$x^2-y^2$ &       &        &        &  4.18  &  3.26 &    &    &      &       &        &  0.46 \\
$xy$      &       &        &        &         &  4.18 &    &    &      &       &        &        \\
\hline\hline
\end{tabular}
\caption{\label{d-full-slater-U-F0F2F4} The Coulomb interaction $U_{ij}$ and $J_{ij}$ of $d$ orbitals fitted using \cref{eq_slater_Uijkl}, where the Slater integrals $F^0$, $F^2$, and $F^4$ are fitted to the full Coulomb matrix $U_{ijkl}$ computed using cRPA method in the quasi-cubic local coordinate system. The fitted parameters are $F^0=3.018, F^2=8.409, F^4=5.766$.
The root mean square error (the error between the fitted values and the DFT values) of the fitted $U_{ij}$ and $J_{ij}$ is 0.133 and 0.024, respectively.}
\end{table}

\begin{table}[htbp]
\begin{tabular}{c|cccccc|ccccc}
\hline\hline
$U_{ij}$  & $z^2$ & $xz$ & $yz$ & $x^2-y^2$ & $xy$ & $J_{ij}$ & $z^2$ & $xz$ & $yz$ & $x^2-y^2$ & $xy$ \\\hline
$z^2$     & 4.15 &  3.08 &  3.08 &  2.39  &  2.39 &    &    & 0.54& 0.54 & 0.88  &  0.88 \\
$xz$      &       &  4.15 &  2.62 &  2.62  &  2.62 &    &    &      & 0.77 & 0.77  &  0.77 \\
$yz$      &       &        &  4.15 &  2.62  &  2.62 &    &    &      &       & 0.77  &  0.77 \\
$x^2-y^2$ &       &        &        &  4.15  &  3.30 &    &    &      &       &        &  0.42 \\
$xy$      &       &        &        &         &  4.15 &    &    &      &       &        &        \\
\hline\hline
\end{tabular}
\caption{\label{d-full-slater-U-F0F2} The Coulomb interaction $U_{ij}$ and $J_{ij}$ of $d$ orbitals fitted using \cref{eq_slater_Uijkl}, where the Slater integrals $F^0$, $F^2$ are fitted to the full Coulomb matrix $U_{ijkl}$ computed using cRPA method in the quasi-cubic local coordinate system, while $F^4=0.625F^2$ are fixed. The fitted parameters are $F^0=3.018, F^2=8.544, F^4=5.346$.
The root mean square error of the fitted $U_{ij}$ and $J_{ij}$ is 0.134 and 0.028, respectively. }
\end{table}

\section{Summary of TB models and interaction parameters of $\text{FeGe}$}\label{app:sec:summary_model_FeGe}

In this section, we give a brief summary of the TB models and interaction parameters in FeGe discussed in \cref{Sec:SI_FeGe_TB} and \cref{Sec:CRPA}.

To construct TB models for the complicated band structure of FeGe, we first choose three sets of orbitals, i.e., (i) $d_{x^2-y^2},d_{xy}$ of Fe and $p_x, p_y$ of triangular Ge, (ii) $d_{xz}, d_{yz}$ of Fe and $p_z$ of honeycomb and triangular Ge, and (iii) $d_{z^2}$ of Fe and the $sp^2$ bonding state of honeycomb Ge, and build TB models with NN and a few longer-range hoppings for them. The fitted TB models match the two quasi-flat bands and the van Hove singularities below the Fermi level. We then couple the three models together to construct the full model in \cref{eq_FeGe_TB_full_model}. At last, we decouple them using second-order perturbations and obtain three independent models in \cref{Eq:FeGe_TB_H123}, with hopping parameters listed in \cref{Table:TBpara_FeGe}. The band structures of these models are shown in \cref{TBband_FeGe}.

To compute the interaction parameters, we use the cRPA method implemented in VASP to obtain the screened Coulomb matrix using Wannier functions. In \cref{d-dp}, \ref{dp-dp}, and \ref{d-full}, we list the onsite Coulomb interaction matrix elements $U_{ij}$, $J_{ij}$, as well as the NN and NNN Coulomb interactions $U_{ij}^{NN}$ and $U_{ij}^{NNN}$ for the so-called $d-dp$ model, $dp-dp$ model, and $d$-full models.

We take the Coulomb matrix in the $d$-full model and do further approximations. We take spherical approximations for Coulomb interactions that are not onsite, i.e., the NN and NNN Coulomb interactions $U_{ij}^{NN}$ and $U_{ij}^{NNN}$ are set to be equal, respectively. For the onsite Coulomb matrix, we first assume cubic $O_h$ point group symmetries and show the symmetrized results shown in \cref{d-full-FeSn-cubic}. Lastly, we assume spherical symmetry for the onsite Coulomb matrix and fit with Slater integrals $F^{0,2,4}$, with results shown in \cref{d-full-FeSn-slater-U-F0F2F4},\ref{d-full-FeSn-slater-U-F0F2}.

The final simplified Hamiltonian has the form 
\begin{equation}
    \hat{H} = \hat{H}_0 + \hat{H}_{\text{int}},
\end{equation}
where the single-particle Hamiltonian is
\begin{equation}
    \hat{H}_0 = \sum_{\bm{k}}\sum_{ij}
    H_{ij}(\bm{k}) c^\dagger_{\bm{k}i}c_{\bm{k}j},
\end{equation}
with $H(\bm{k})=H_1(\bm{k})\oplus H_2(\bm{k}) \oplus H_3(\bm{k})$ defined in \cref{Eq:FeGe_TB_H123} and the orbital basis given in \cref{eq_final_TB_basis}.

The interacting Hamiltonian with orbital-dependent onsite interactions $U_{m_1m_2}$ and $J_{m_1m_2}$ and averaged NN and NNN interaction $\overline{U}_1$ and $\overline{U}_2$ is
\begin{equation}
\begin{aligned}
    \hat{H}_{\text{int}} &=
    \sum_{im} U_{mm} n_{im\uparrow}n_{im\downarrow}
    + \sum_{i,m\ne m^\prime} 
    U_{mm^\prime} n_{i m\uparrow}n_{i m^\prime\downarrow}
    +\sum_{i,m<m^\prime \sigma} 
    (U_{mm^\prime} - J_{mm^\prime})
    n_{i m \sigma} n_{i m^\prime \sigma}	\\
    &-\sum_{i, m\ne m^\prime} 
    J_{mm^\prime} c_{im\uparrow}^\dagger c_{i m \downarrow}  c_{i m^\prime \downarrow}^\dagger c_{i m^\prime \uparrow}
    +\sum_{i, m\ne m^\prime} 
    J_{mm^\prime} c_{im\uparrow}^\dagger c_{im\downarrow}^\dagger c_{i m^\prime \downarrow} c_{i m^\prime \uparrow}\\
    & + \overline{U}_1 \sum_{<ij>}\sum_{mm^\prime, \sigma\sigma^\prime} n_{i m \sigma} n_{j m^\prime \sigma^\prime}
    + \overline{U}_2 \sum_{\ll ij \gg}\sum_{mm^\prime, \sigma\sigma^\prime} n_{i m \sigma} n_{j m^\prime \sigma^\prime}
\end{aligned}
\end{equation}
where $<ij>$ and $\ll ij \gg$ denotes NN and NNN sites (each $i,j$ pair count only once, thus no $\frac{1}{2}$ coefficient). The averaged NN and NNN Coulomb interaction $\overline{U}_1=1.41, \overline{U}_2=1.22$ eV, and the values of spherical-symmetric onsite interaction $U_{ij}$ and $J_{ij}$ parameterized by $F^0$ and $F^2$ with $F^4=0.625F^2$ are given in \cref{d-full-slater-U-F0F2}.

\section{Mean-field study of interacting Hamiltonian in FeGe}
In this section, we apply the Hartree-Fock mean-field method to investigate the interacting Hamiltonian presented in \cref{app:sec:summary_model_FeGe}. We demonstrate that the antiferromagnetic (AFM) phase observed in the experiment and DFT can be accurately reproduced using the simplified Hamiltonian.

\subsection{Interaction} 
The interaction term of the Hamiltonian can be decomposed into two parts: the local (onsite) interactions and the non-local interactions
\ba 
\hH_{int} = \hH_{int,loc} +\hH_{int,non-local}
\ea 
The local interactions take the form of 
\ba  
% &\hH_{loc}  = \hH_{int} 
% +\sum_{\RR,i,\sigma}  \mu_{d_i} c_{\RR,i,\sigma}^\dag c_{\RR,i,\sigma} \nonumber \\ 
&\hH_{int,loc}=\frac{1}{2}\sum_{\RR,i_1,i_2,i_3,i_4,\sigma,\sigma'} U_{i_1i_2i_3i_4} 
:c_{\RR,i_1,\sigma}^\dag
c_{\RR,i_3,\sigma'}^\dag c_{\RR,i_4,\sigma'}
c_{\RR,i_2,\sigma} :
\ea  
where $:\hat{O}:$ denotes the normal order of the operator with respect to the non-interacting system.  

For convenience, we can introduce $U_{ij},J_{ij},K_{ij}$
\ba  
U_{ij} = U_{iijj},\quad J_{ij} = U_{ijji} ,\quad K_{ij} = U_{ijij}
\ea  
where $K_{ij} =J_{ij}$ in the case without spin-orbit coupling. 
Then the interaction term can be written as
\ba 
\hH_{int,loc} 
=& \frac{1}{2}\sum_{\RR,i,j,\sigma,\sigma'}U_{ij} : c_{\RR,i,\sigma}^\dag c_{\RR,i,\sigma} c_{\RR,j,\sigma'}^\dag c_{\RR,j,\sigma'}: \nonumber \\ 
&
-\frac{1}{2}\sum_{\RR,i,j,\sigma,\sigma'}J_{ij}: c_{\RR,i,\sigma}^\dag c_{\RR,i,\sigma'} c_{\RR,j,\sigma'}^\dag c_{\RR,j,\sigma}:
+ \frac{1}{2}\sum_{\RR,i,j,\sigma,\sigma's}
J_{ij}:c_{\RR,i,\sigma}^\dag c_{\RR,i,\sigma'}^\dag c_{\RR,j,\sigma'}c_{\RR,j,\sigma}:
\ea

We can introduce the Fourier transformation
\ba 
c_{\RR,i,\sigma} = \frac{1}{\sqrt{N}}\sum_{\kk} c_{\kk,i,\sigma}e^{i\kk\cdot \RR }
\label{eq:cop_ft}
\ea 
Then the interaction in the momentum space can be written as
\ba 
\hH_{int,loc} = &\frac{1}{2}\sum_{\RR,i_1,i_2,i_3,i_4,\sigma,\sigma'}
U_{i_1i_2i_3i_4} \frac{1}{N^2} \sum_{\kk_1,\kk_2,\kk_3,\kk_4}
:c_{\kk_1,i_1,\sigma}^\dag
c_{\kk_3,i_3,\sigma'}^\dag c_{\kk_4,i_4,\sigma'}
c_{\kk_2,i_2,\sigma}: e^{i\RR\cdot (\kk_2 +\kk_4-\kk_3 -\kk_1) } \nonumber \\ 
=&\frac{1}{2}\sum_{i_1,i_2,i_3,i_4,\sigma,\sigma'}
U_{i_1i_2i_3i_4}  \frac{1}{N} \sum_{\kk_1,\kk_2,\kk_3,\kk_4}
:c_{\kk_1,i_1,\sigma}^\dag
c_{\kk_3,i_3,\sigma'}^\dag c_{\kk_4,i_4,\sigma'}
c_{\kk_2,i_2,\sigma} :\delta_{\kk_2+\kk_4-\kk_3-\kk_1,\bm{0}}\nonumber \\ 
=&\frac{1}{2N}\sum_{i_1,i_2,i_3,i_4,\sigma,\sigma'} U_{i_1i_2i_3i_4} \sum_{\kk,\kk',\qq} 
:c_{\kk+\qq,i_1,\sigma}^\dag
c_{\kk'-\qq,i_3,\sigma'}^\dag c_{\kk',i_4,\sigma'}
c_{\kk,i_2,\sigma}: 
\label{eq:h_int_loc}
\ea 

The non-local interaction takes the form of 
\ba 
\hH_{int,non-local} = \frac{1}{2}\sum_{\RR,\RR',i,j,\sigma,\sigma'} 
V_{\RR-\RR',ij}:c_{\RR,i,\sigma}^\dag c_{\RR,i,\sigma} c_{\RR',j,\sigma'}^\dag c_{\RR',j',\sigma'}:
\ea 
Using Eq.~\ref{eq:cop_ft} and 
\ba 
V_{\qq,ij} =\frac{1}{N} \sum_{\RR,\RR'}V_{\RR-\RR',ij}e^{-i\qq\cdot(\RR-\RR')}
\ea 
we have 
\ba 
\hH_{int,non-local} =& \frac{1}{2N^3}
\sum_{\RR,\RR',\qq,\kk_1,\kk_2,\kk_3,\kk_4}V_{\qq,ij}
:c_{\kk_1,i,\sigma}^\dag c_{\kk_2,i,\sigma}c_{\kk_3,j,\sigma'}^\dag c_{\kk_4,j,\sigma'}:
e^{i\RR'\cdot (\kk_4-\kk_3 -\qq) +i\RR\cdot(\kk_2-\kk_1+\qq) } \nonumber \\ 
=&\frac{1}{2N}
\sum_{\qq,\kk_1,\kk_2,\kk_3,\kk_4}V_{\qq,ij}
:c_{\kk_1,i,\sigma}^\dag c_{\kk_2,i,\sigma}c_{\kk_3,j,\sigma'}^\dag c_{\kk_4,j,\sigma'}
:
\delta_{\kk_4-\kk_3,\qq}\delta_{\kk_1-\kk_2,\qq} \nonumber \\ 
=& \frac{1}{2N}\sum_{\kk,\kk',\qq}V_{\qq,ij}:c_{\kk+\qq,i,\sigma}^\dag c_{\kk,i,\sigma}c_{\kk',j,\sigma'}^\dag c_{\kk'+\qq,j,\sigma'}:
\label{eq:h_int_nonloc}
\ea

\subsection{Hartree-Fock study of the AFM phase}
We discuss the Hartree-Fock solutions of the AFM phase. We consider the magnetic order with wavevector 
\ba 
\QQ = \frac{1}{2}\bm{b}_3
\ea 
The Hartree-Fock treatment of local interaction (Eq~\ref{eq:h_int_loc}) reads
\ba 
\hH^{MF}_{int,loc} =&\sum_{ijml,\sigma\sigma'} \frac{1}{2}U_{ijml}\bigg\{ -
N\bigg[O^{\QQ_0}_{i\sigma,j\sigma}O^{\QQ_0}_{m\sigma',l\sigma'}  + O^{\QQ}_{i\sigma,j\sigma}O^{\QQ}_{m\sigma',l\sigma'} \bigg] 
+N\bigg[ O^{\QQ_0}_{i\sigma,l\sigma'}O^{\QQ_0}_{m\sigma',j\sigma}  + O^{\QQ}_{il}O^{\QQ}_{mj} \bigg] \nonumber \\ 
&+ O^{\QQ_0}_{i\sigma,j\sigma}\sum_{\kk \in RBZ} [:c_{\kk,m,\sigma'}^\dag c_{\kk,l,\sigma'}: + :c_{\kk+\QQ,m,\sigma'}^\dag c_{\kk+\QQ,l,\sigma'}: ] + O^{\QQ}_{i\sigma,j\sigma}\sum_{\kk \in RBZ} [:c_{\kk,m,\sigma'}^\dag c_{\kk+\QQ,l,\sigma'} : + :c_{\kk,m,\sigma'}^\dag c_{\kk+\QQ,l,\sigma'} :] \nonumber \\
&+ O^{\QQ_0}_{m\sigma',l\sigma'}\sum_{\kk \in RBZ} [
:c_{\kk,i,\sigma}^\dag c_{\kk,j,\sigma}  :+ :c_{\kk+\QQ,i,\sigma}^\dag c_{\kk+\QQ,j,\sigma} :]
+ O^{\QQ}_{m\sigma',l\sigma'}\sum_{\kk \in RBZ} [:c_{\kk,i,\sigma}^\dag c_{\kk+\QQ,j,\sigma} : + :c_{\kk,i,\sigma}^\dag c_{\kk+\QQ,j,\sigma}: ] \nonumber \\
&- O^{\QQ_0}_{i\sigma,l\sigma'}\sum_{\kk \in RBZ} [:c_{\kk,m,\sigma'}^\dag c_{\kk,j,\sigma} : + :c_{\kk+\QQ,m,\sigma'}^\dag c_{\kk+\QQ,j,\sigma} :]- O^{\QQ}_{i\sigma,l\sigma'}\sum_{\kk \in RBZ} [:c_{\kk,m,\sigma'}^\dag c_{\kk+\QQ,j,\sigma} : + :c_{\kk,m,\sigma'}^\dag c_{\kk+\QQ,j,\sigma}: ] \nonumber \\
&- O^{\QQ_0}_{m\sigma',j\sigma}\sum_{\kk \in RBZ} [:c_{\kk,i,\sigma}^\dag c_{\kk,l,\sigma'}: + :c_{\kk+\QQ,i,\sigma}^\dag c_{\kk+\QQ,l,\sigma'}: ]-O^{\QQ}_{m\sigma',j\sigma}\sum_{\kk \in RBZ} [:c_{\kk,i,\sigma}^\dag c_{\kk+\QQ,l,\sigma'}:  + :c_{\kk,i,\sigma}^\dag c_{\kk+\QQ,l,\sigma'} :] \bigg\}
\ea 
where 
\ba 
&O_{i\sigma,j\sigma'}^{\QQ_0} = \frac{1}{N} \sum_{\kk \in RBZ} \langle :c_{\kk,i,\sigma}^\dag c_{\kk,j,\sigma'} + c_{\kk+\QQ,i,\sigma}^\dag c_{\kk+\QQ,j,\sigma'}:\rangle \nonumber \\
&O_{i\sigma,j\sigma'}^{\QQ} = \frac{1}{N} \sum_{\kk \in RBZ} \langle :c_{\kk+\QQ,i,\sigma}^\dag c_{\kk,j,\sigma'} + c_{\kk+\QQ,i,\sigma}^\dag c_{\kk,j,\sigma'}:\rangle 
\ea 
and RBZ denotes the folded (reduced) Brillouin zone. 

For the non-local interactions in Eq.~\ref{eq:h_int_nonloc}, we only include the Hartree contributions which take the form of 
\ba 
\hH^{MF}_{int,non-loc} = &\frac{1}{2}\sum_{\qq,i,j,\sigma,\sigma'}V_{ij}(\qq) 
\bigg\{ -
\delta_{\qq,\bm{0}} NO_{i\sigma,i\sigma}^{\QQ_0} O_{i\sigma,i\sigma}^{\QQ_0} 
-\delta_{\qq,\QQ}NO_{i\sigma,i\sigma}^{\QQ_0}O_{j\sigma',j\sigma'}^{\QQ_0} \nonumber \\
&+(O_{i\sigma,i\sigma}^{\QQ_0}\delta_{\qq,\bm{0}}+O^{\QQ}_{i\sigma,i\sigma}\delta_{\qq,\QQ})\sum_{\kk \in RBZ} [:c_{\kk,j,\sigma'}^\dag c_{\kk+\qq,j,\sigma'} :+ :c_{\kk+\QQ,j,\sigma'}^\dag c_{\kk+\QQ+\qq,j,\sigma'}:]\nonumber\\ 
&+(O_{j\sigma',j\sigma'}^{\QQ_0}\delta_{\qq,\bm{0}}+O^{\QQ}_{j\sigma',j\sigma'}\delta_{\qq,\QQ})\sum_{\kk \in RBZ} [:c_{\kk+\qq,i,\sigma}^\dag c_{\kk,i,\sigma}: + :c_{\kk+\qq+\QQ,i,\sigma}^\dag c_{\kk+\QQ,i,\sigma}:]\bigg\}
\ea

\subsection{Normalized interaction of $z^2$ orbitals} 

Even though the third sector of the current non-interacting Hamiltonian, $\hH_{3}$, includes only the $d_{z^2}$ orbitals of Fe, these orbitals are expected to hybridize with the $sp^2$ bonding states in practice. This hybridization reduces the $d_{z^2}$ orbital weights in the bands described by the $\hH_{3}$ sector, effectively weakening the interaction strength, as the interactions act only on the $d$ orbitals. To account for this effect, we normalize the interaction strength of the $d_{z^2}$ orbitals.

To achieve this, we first consider the model with both $d_{z^2} $ and $sp^2$ bonding state. We let $[H_{3,d-sp}(\kk)]_{\alpha\gamma}$ denote the corresponding single-particle Hamiltonian with $\alpha,\gamma$ denotes the orbitals. We introduce the eigenvalues and eigenstates 
\ba 
\sum_{\gamma}H_{\alpha\gamma}^{3,d-sp}(\kk)u_{\gamma n}(\kk) = E_n(\kk) u_{\gamma n}(\kk) 
\ea 
We also include the electron operators in the band basis 
\ba 
\gamma_{\kk,n,\sigma} = \sum_{\gamma} u^*_{\gamma n}(\kk) c_{\kk,\gamma,\sigma}
\ea 
where $c_{\kk,\gamma,\sigma}$ denote the electron operator with moment $\kk$, orbital $\gamma$ and spin $\sigma$. In practice $c_{\kk,\gamma,\sigma}$ describes electron in both $d_{z^2}$ orbital and $sp^2$ bonding state. 

We focus on the three bands whose energy is closest to the Fermi energy (six bands if spin indices are considered). We use $(\gamma_{\kk,n,\sigma},n=1,2,3)$ to denote such three bands. We note that the band structure of these bands has been correctly reproduced by $\hH_3$. However, in the model with $sp^2$ bonding state, these bands are formed by both $d_{z^2}$ and $sp^2$ bonding state. We further consider the interactions acting on the $\gamma_{\kk,n,\sigma}$ $(n=1,2,3)$ bands. We note that the $d_{z^2}$ orbitals can be written as
\ba 
c_{\kk,\gamma,\sigma} = \sum_{n} u_{\gamma n}(\kk) \gamma_{\kk,n,\sigma},\quad \gamma \in \{\text{$d_{z^2}$ orbitals}\}
\label{eq:dz2_to_bands}
\ea 
The interaction acting on the $d_{z^2}$ orbitals takes the form of
\ba 
\hH_{d_{z^2}} = \frac{1}{N}\sum_{\kk,\kk',\qq}\sum_{\sigma,\sigma'}\sum_{\alpha_1,\alpha_2,\alpha_3,\alpha_4\in \{\text{$d_{z^2}$ orbitals}\}} 
U_{\alpha_1\alpha_2\alpha_3\alpha_4}(\qq) 
:c_{\kk+\qq,\alpha_1,\sigma}^\dag c_{\kk,\alpha_2,\sigma} c_{\kk',\alpha_3,\sigma'}^\dag c_{\kk'+\qq,\alpha_4,\sigma} :
\ea 
Using Eq.~\ref{eq:dz2_to_bands}, we have
\ba 
\hH_{d_{z^2}} \approx &\frac{1}{N} \sum_{\kk,\kk',\qq,\sigma,\sigma'} \sum_{ i_1,i_2,i_3,i_4\in\{1,2,3\}}\sum_{\alpha_1\alpha_2\alpha_3\alpha_4\in \{\text{$d_{z^2}$ orbitals}\}}U_{\alpha_1\alpha_2\alpha_3\alpha_4}(\qq):\gamma_{\kk+\qq,i_1,\sigma}^\dag \gamma_{\kk,i_2,\sigma} \gamma_{\kk',i_3,\sigma'}^\dag \gamma_{\kk'+\qq,i_4,\sigma}:  \nonumber \\ 
&u_{\alpha_1i_1}^*(\kk+\qq)u_{\alpha_2i_2}(\kk+\qq)
u_{\alpha_3i_3}^*(\kk')u_{\alpha_4i_4}(\kk'+\qq)
\label{eq:int_dz2_proj}
\ea 
where we only consider the interaction acting on the lowest three bands. Due to the fact that the lowest three bands are only partially formed by $d$ orbitals, the interaction strength is reduced. 
We now estimate the strength of interactions acting on the three lowest bands. In the band basis, we notice that the interaction has strength (Eq.~\ref{eq:int_dz2_proj})
\ba
U_{\alpha_1\alpha_2\alpha_3\alpha_4}(\qq)u_{\alpha_1i_1}^*(\kk+\qq)u_{\alpha_2i_2}(\kk+\qq)
u_{\alpha_3i_3}^*(\kk')u_{\alpha_4i_4}(\kk'+\qq)
\ea 
We then calculate the average weight of the $d_{z^2}$ orbitals of the lowest three bands
\ba 
w_{d_{z^2}} = \frac{1}{3}\frac{1}{N}\sum_{\kk, \alpha\in \{\text{$d_{z^2}$ orbits}\}} \sum_{i=1,2,3} |u_{\alpha i}(\kk)|^2 
\ea 
where the $1/3$ prefactor comes from the average of the three bands. Numerically, we find $w_{d_z^2} = 0.499$. The corresponding weight of the $sp^2$ bonding state is $w_{sp} = 0.501$. 

Then approximately, we can introduce the normalized interaction as 
\ba 
U^{norm}_{\alpha_1\alpha_2\alpha_3\alpha_4}(\qq) = 
U_{\alpha_1\alpha_2\alpha_3\alpha_4}(\qq) w_{d_{z^2}}^2 ,\quad \alpha_1,\alpha_2,\alpha_3,\alpha_4  \in \{\text{ $d_{z^2}$ orbitals}\}
\ea 

Similarly, for the density-density and spin-spin interaction between $z^2$ orbitals and other orbitals, we also need to normalize them by a factor of $w_{d_z^2}$. In summary, the on-site interactions and the non-local Coulomb interactions become
\ba 
&U^{norm}_{ij} = U_{ij}w_{d_z^2}, J^{norm}_{ij}=J_{ij}w_{d_z^2}, V_{ij}^{norm}=V_{ij}w_{d_z^2}\quad \text{ if $i=d_{z^2},j\ne d_{z^2}$ or $i\ne d_{z^2},j= d_{z^2}$}\nonumber \\ 
&U^{norm}_{ij} = U_{ij}w^2_{d_z^2}, J^{norm}_{ij}=J_{ij}w^2_{d_z^2}, V_{ij}^{norm}=V_{ij}w^2_{d_z^2}\quad \text{ if $i=d_{z^2},j=d_{z^2}$}\nonumber \\ 
&U^{norm}_{ij} = U_{ij}, J^{norm}_{ij}=J_{ij}, V_{ij}^{norm}=V_{ij}\quad \quad\quad \quad\quad \quad \text{ if $i\ne d_{z^2},j\ne d_{z^2}$}
\ea

\subsection{Mean-field results}
 
We solve the interacting Hamiltonian of FeGe by incorporating the AFM order, using the Hamiltonian  $\hat{H}_3$ with only $d_{z^2}$ orbital, and normalizing the interaction as discussed in the previous section.

\begin{figure}[htbp]
	\centering
	\includegraphics[width=1\textwidth]{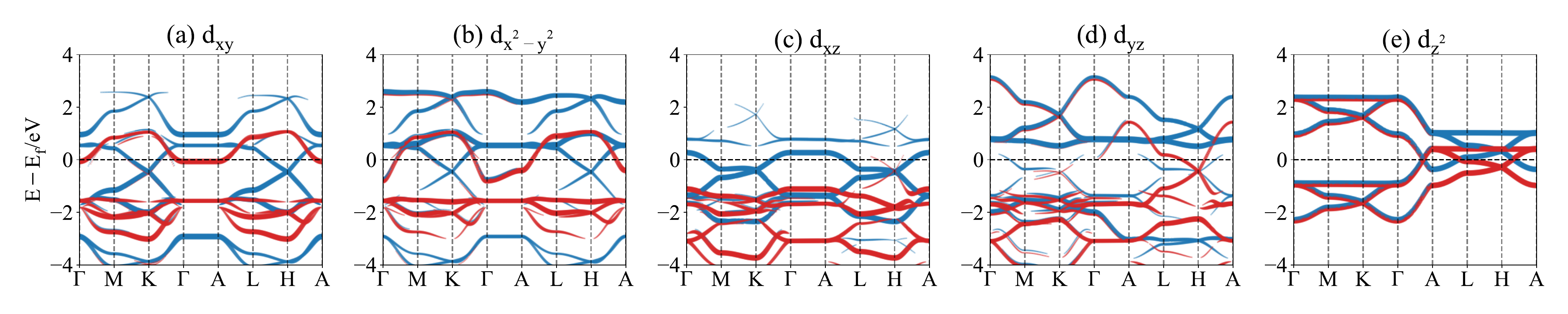}
	\caption{\label{app:fig:AFM-MF-orb-weight} The mean-field band structures in the AFM phase, with orbital weights from (a) $d_{xy}$, (b) $d_{x^2-y^2}$, (c) $d_{xz}$, (d) $d_{yz}$, and (e) $d_{z^2}$. In the figure, the blue and red colors denote $d$ orbitals of opposite spins of Fe on one kagome plane. The other kagome plane is related by $\mathcal{T}\cdot\{E|001\}$, which has the same dispersion but opposite spin distribution. In the plot, we slightly shift the spin-up bands by 0.1 eV for better visualization. 
 }
\end{figure}

In \cref{app:fig:AFM-MF-orb-weight}, we present the mean-field band structures in the AFM phase, including the orbital weights from the $d$-orbitals of Fe. These results show good agreement with the DFT band structure in the AFM phase, as illustrated in \cref{Fig:FeGe_AFM_struct_band}.

\begin{figure}[htbp]
	\centering
	\includegraphics[width=1\textwidth]{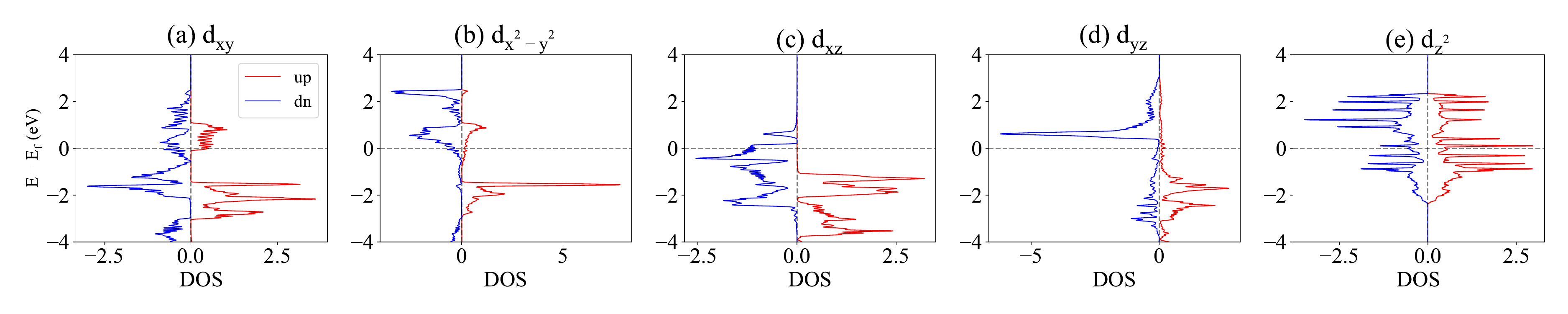}
	\caption{\label{app:fig: AFM-MF-DOS} The mean-field density of states in the AFM phase, from orbital of (a) $d_{xy}$, (b) $d_{x^2-y^2}$, (c) $d_{xz}$, (d) $d_{yz}$, and (e) $d_{z^2}$. In the figure, the blue and red colors denote $d$ orbitals of opposite spins of Fe on one kagome plane.}
\end{figure}

To further demonstrate the validity of the mean-field band structure, we compute the magnetic moments of each $d$ orbital of Fe in \cref{app:table:mean-field-filling-magmom}, and orbital-resolved DOS in \cref{app:fig: AFM-MF-DOS}. They also show good agreement with DFT fillings in \cref{Tab:filling_magmom_wannier} and DOS in \cref{Fig:wannier_d_projections}. 

Thus we conclude that our interacting Hamiltonian faithfully reproduces the AFM phase in FeGe with close agreement to DFT. This demonstrates the robustness of our formalism in capturing interaction-driven phases.

\begin{table}[htbp]
\centering
\begin{tabular}{c|c|c|c|c|c|c}
\hline\hline
Orbital & $d_{z^2}$ & $d_{xz}$ & $d_{yz}$ & $d_{x^2-y^2}$ & $d_{xy}$ & Total \\ \hline
Filling & 1.560 & 1.030 & 1.883 & 1.232 & 0.909 & 6.61 \\ \hline
Magnetic moment & 0.085 & 0.629 & 0.117  & 0.646 & 0.211  & 1.69 \\ \hline\hline
\end{tabular}%
\caption{\label{app:table:mean-field-filling-magmom} The filling and magnetic moments of each $d$ orbitals per Fe atom in the AFM phase, obtained from mean-field calculations. }
\end{table}

\section{Building models for 1:1 class from FeGe: application to FeSn and CoSn}\label{Sec:FeSn_CoSn} 

In \cref{app:sec:summary_model_FeGe}, we construct minimal TB models for FeGe by separating $d$ orbitals on kagome sites into three groups and combining them with specific orbitals of triangular and honeycomb Ge. In this section, we extend this strategy to its kagome siblings FeSn and CoSn. Their band structures are very close to FeGe: Co has one more occupied $d$ orbital compared with Fe which shifts up Fermi level about $0.5$ eV, and Sn has the same number of occupied $p$ orbitals as Ge and almost the same $E_f$. We adopt the same strategy for FeSn and CoSn to build minimal TB Hamilontians of three decoupled blocks and also compute the Coulomb interaction parameters. 
The minimal TB models for FeSn and CoSn take the same analytic form as FeGe, with onsite energies of orbitals shifted according to the $E_f$ and hopping parameters refitted. The interaction parameters of FeSn and CoSn also share similar values with FeGe.

\subsection{Crystal structures and band structures}
FeSn and CoSn have similar crystal structures with FeGe of SG 191, but different lattice constants, which are listed in \cref{latt_const_compare}. FeSn develops a planar AFM order below temperature $T_N \approx 365$ $K$\cite{haggstrom1975studies, sales2019electronic, multer2022imaging}, with magnetic moments in the kagome layer along $(1,0,0)$ or $(2, 1, 0)$ direction and alternating between layers along the $c$ axis. The magnetic moment is 1.85 $\mu_B$ at 100 $K$\cite{sales2019electronic}. No CDW in FeSn has been reported as far as of now. CoSn is paramagnetic and also has no CDW reported.

The MSG of FeSn is 63.466 $C_c mcm$ by assuming that the magnetic moments are along $(1,0,0)$ direction. The in-plane magnetic moments break the $C_6$ and $C_3$ symmetry in SG 191. However, the spin-space group (SSG) of FeSn is 191.2.1.1.L \cite{jiang2023enumeration}, which is the same as FeGe, as the axis in spin-space can be assigned arbitrarily in SSGs. It can be seen that the SSG of FeSn contains all spacial symmetry of SG 191 including $C_6$, which is a pure-lattice operation, while the MSG breaks $C_6$. This shows the advantage of using SSGs in describing the symmetry of magnetic materials.

\begin{figure}[htbp]
	\centering
	\includegraphics[width=0.9\textwidth]{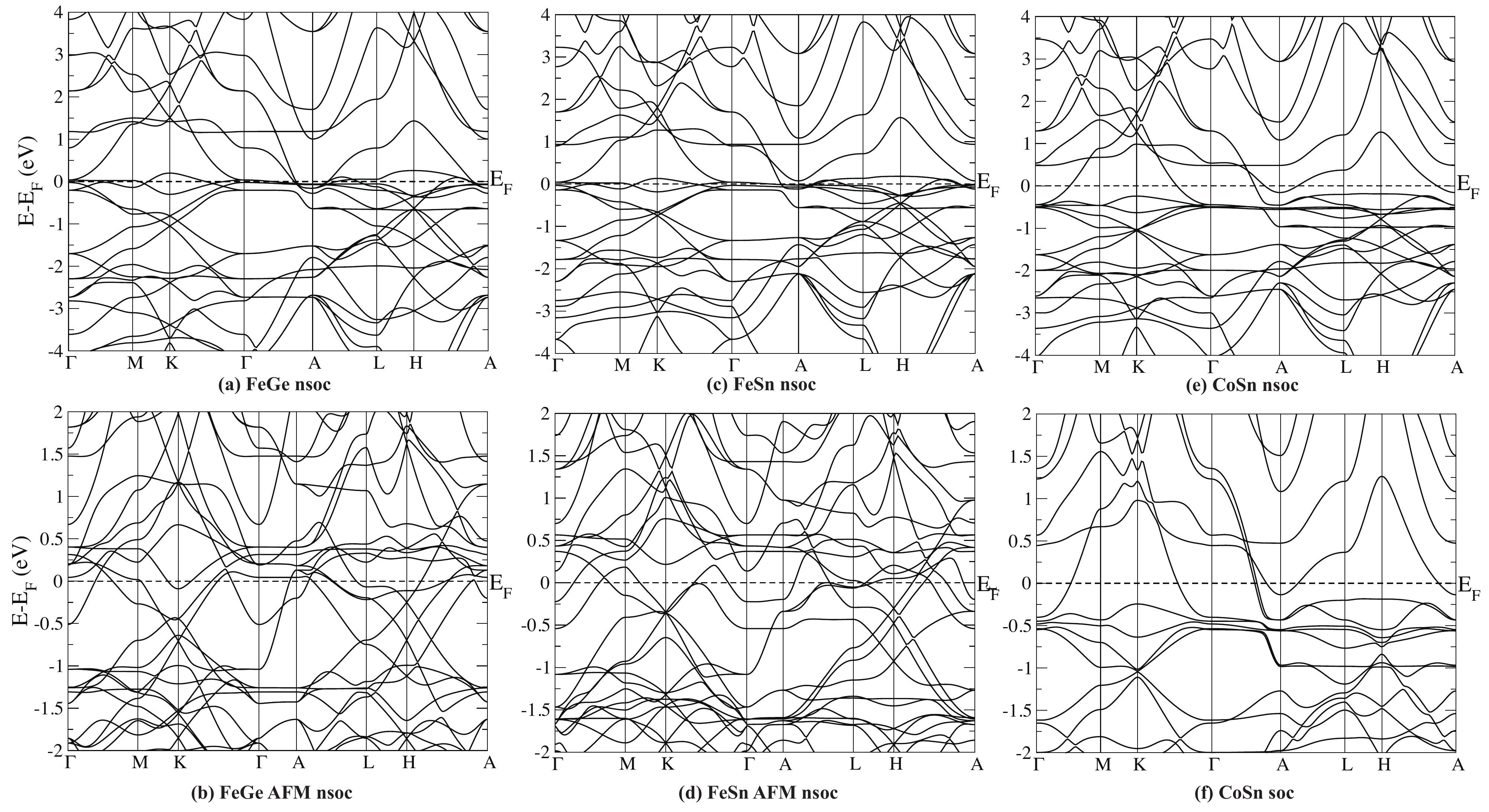}
	\caption{\label{Fig:CoSn_FeSn_band} The band structures of (a) FeGe NSOC, (b) FeGe AFM NSOC, (c) FeSn NSOC, (d) FeSn AFM NSOC, (e) CoSn NSOC, and (f) CoSn SOC cases. }
\end{figure}

The DFT band structures of FeSn and CoSn are given in \cref{Fig:CoSn_FeSn_band}. FeSn has a very close band structure with FeGe, while CoSn has $E_f$ shifted upwards about 0.5 eV, mainly due to the extra $d$ electron in Co compared with Fe.
Similar to FeGe, there are also two quasi-flat bands around $E_f$ in FeSn and CoSn. For FeGe and FeSn, the two flat bands cross the $E_f$ and are partially filled, while for CoSn the flat bands are fully filled. In FeGe, the quasi-flat band in $H_1(\kk)$ sector has $0.284$ filling (without spin) in the whole BZ, and $0.626$ on the $k_3=0$ plane, while the $H_2(\kk)$ sector flat band has $0.880$ filling in the whole BZ, and $0.716$ on the $k_3=0$ plane. The filling of the flat bands in FeSn is similar. The partially filled flat bands in FeGe and FeSn lead to anti-ferromagnetic instability.

\begin{table}[htbp]
\begin{tabular}{c|c|c}
\hline\hline
Material & $a/\AA$ & $c/\AA$ \\\hline
FeGe & 4.985 & 4.049 \\\hline
FeSn & 5.298 & 4.448 \\\hline
CoSn & 5.279 & 4.260 \\
\hline\hline
\end{tabular}
\caption{\label{latt_const_compare} The lattice constants of FeGe\cite{teng2022discovery}, FeSe\cite{giefers2006high}, and CoSn\cite{larsson1996single}.}
\end{table}

\subsection{Tight-binding models of FeSn and CoSn}

\begin{figure}[htbp]
	\centering
	\includegraphics[width=\textwidth]{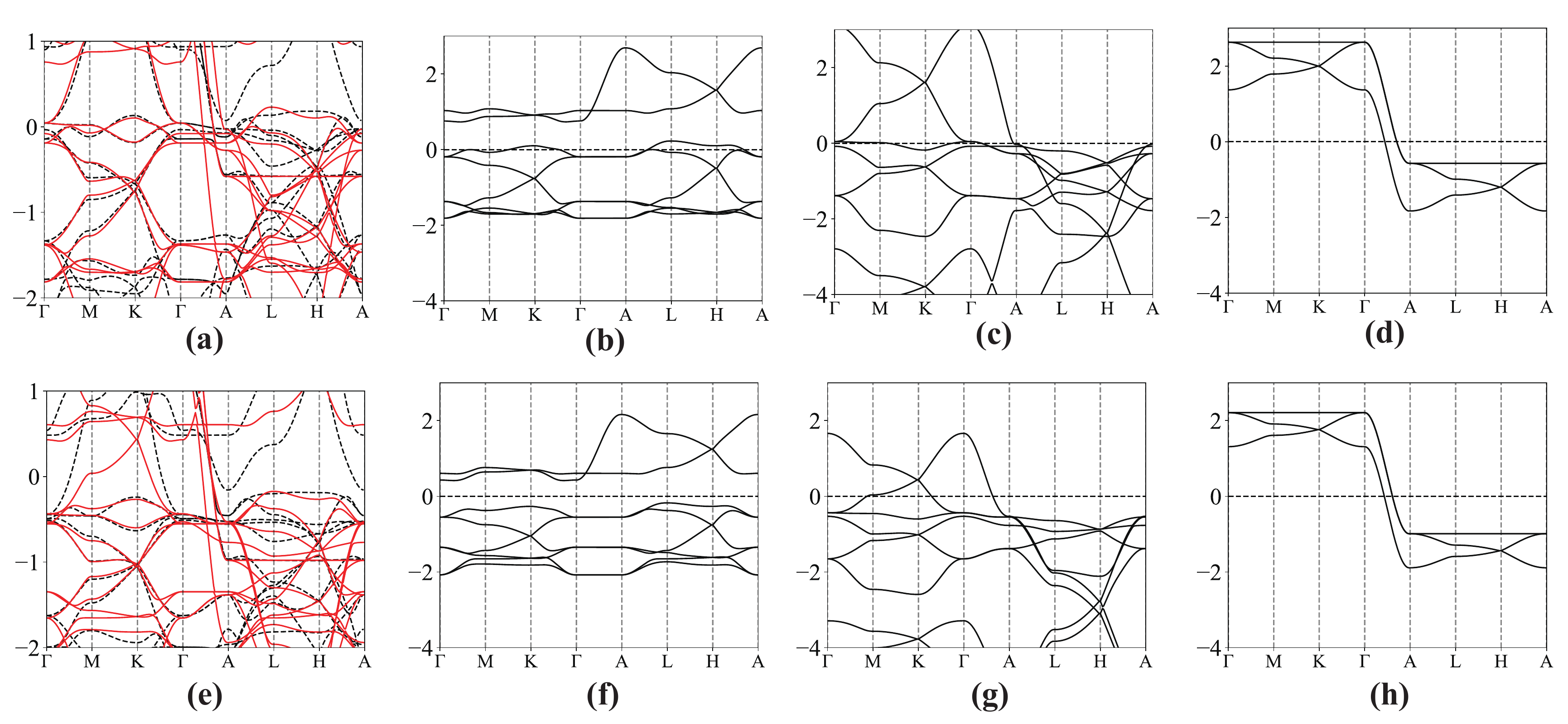}
	\caption{\label{Fig:TBband_FeSn_CoSn} The fitted TB band structures for FeSn (a)-(e) and CoSn (e)-(h).  (a). Fitted bands for FeSn of the decoupled model in \cref{Eq:FeGe_TB_directsum}, where red lines are fitted TB bands and black dashed lines are DFT bands. 
    (b)-(d): Band structure of the decoupled $H_1(\bm{k})$, $H_2(\bm{k})$, and $H_3(\bm{k})$ defined in \cref{Eq:FeGe_TB_H123}, respectively. (e)-(h) are similar figures for CoSn.}
\end{figure}

As FeSn and CoSn share similar band structures with FeGe, we build TB models for them using the same model of FeGe in \cref{eq_FeGe_TB_full_model} and \cref{Eq:FeGe_TB_H123}. The fitted TB parameters are slightly different from that of FeGe, as tabulated in \cref{Table_TBpara_FeSn}, \ref{Table_TBpara_CoSn}, and the corresponding band structures are shown in \cref{Fig:TBband_FeSn_CoSn}.
Compared with FeGe, FeSn has very close onsite and hopping parameters, while CoSn has lower onsite energies due to the shifted $E_f$, and hoppings being close to FeGe.

\begin{table}[htbp]
\begin{tabular}{c|c|c|c|c|c|c|c|c|c|c|c|c|c|c|c|c}
\hline\hline
Parameter & $\mu_{p_{xy}^t}$ & $\mu_{d_1}$ & $\mu_{d_2}$ & $t_{d_1}^{NN}$  & $t_{d_2}^{NN}$ & $t_{d_2}^{NNN}$    & $t_{d_1,d_2}^{NN}$ & $t_{d_1,d_2}^{NNN}$ & $t_{p_{xy}^t, d_1}^{NN}$ & $t_{p_{xy}^t, d_1}^{NNN}$ &  $t_{p_{xy}^t, d_2}^{NN}$ & $t_d^{4N1}$ & $t_d^{4N2}$ & $t_d^{4N3}$  & $t_d^{4N4}$ & $t_d^{4N5}$  \\ \hline
Value/eV & -1.37 & -0.79 & 0.11 & 0.47 & 0.11  & 0.05 & -0.08 & 0.15 & -0.26 & 0.10 & 0.31 & -0.13 & 0.05 & 0.07 & -0.13 &-0.02
\\ \hline
Parameter & $\mu_{p_z^h}$ & $\mu_{d_3}$ & $\mu_{d_4}$ & $t_{d_3,d_4}^{NN}$ & $t_{d_3}^{NN}$  & $t_{d_4}^{NN}$  & $t_{d_4}^{zNN}$ & $t_{p_z^h,d_4}^{NN}$ & $t_{p_z^h}^{NN}$ & $t_{p_z^h, d_2}^{NN}$ & $\mu_{p_z^t}$ & $t_{p_z^t}^{NN}$ & $t_{p_z^t, d_3}^{NN}$  &&
\\ \hline
Value/eV & -1.41 & -0.80 & -0.78 & -0.15 & -0.18 & 0.20 & 0.10 & 0.77  & -0.46 & 0.28 & 0.60 & 0.30 & 0.45 &&
\\ \hline
Parameter & $\mu_{d_5}$  & $t_{d_5}^{NN}$ & $\mu_b$ & $t_b^{NN}$ & $t_b^{zNN}$ & $t_{b, d_5}^{NN}$ & $t_{b,d_5}^{NNN}$ &&
\\ \hline
Value/eV & -0.98 & -0.20 & -0.6 & -0.1 & 0.9 & -1.1 & 0.04 &&
\\ \hline\hline
\end{tabular}
\caption{\label{Table_TBpara_FeSn} Onsite energies and hoppings used in the decoupled TB model of FeSn, where $d_1$ to $d_5$ denote $d_{xy}, d_{x^2-y^2}, d_{xz}, d_{yz}, d_{z^2}$, respectively, $b$ the bonding state, and $p_{i}^t$ ($p_i^h$) denotes the $p_i$ orbitals of the triangular (honeycomb) Ge.  For the simplified $H_3(\kk)$ of $d_{z^2}$ only, the parameters are $\mu_{d_5}=0.61, t_{d_5}^{NN}=-0.21, t_{d_5}^{zNN}=0.8$. Parameters not listed in the table have zero values.}
\end{table}

\begin{table}[htbp]
\begin{tabular}{c|c|c|c|c|c|c|c|c|c|c|c|c|c|c|c|c}
\hline\hline
Parameter & $\mu_{p_{xy}^t}$ & $\mu_{d_1}$ & $\mu_{d_2}$ & $t_{d_1}^{NN}$  & $t_{d_2}^{NN}$ & $t_{d_2}^{NNN}$    & $t_{d_1,d_2}^{NN}$ & $t_{d_1,d_2}^{NNN}$ & $t_{p_{xy}^t, d_1}^{NN}$ & $t_{p_{xy}^t, d_1}^{NNN}$ &  $t_{p_{xy}^t, d_2}^{NN}$ & $t_d^{4N1}$ & $t_d^{4N2}$ & $t_d^{4N3}$  & $t_d^{4N4}$ & $t_d^{4N5}$  \\ \hline
Value/eV & -1.35 & -1.01 & -0.21 & 0.44 & 0.13  & 0.04 & -0.10 & 0.16 & -0.25 & 0.07 & 0.26 & -0.14 & 0.03 & 0.06 & -0.14 & -0.03
\\ \hline
Parameter & $\mu_{p_z^h}$ & $\mu_{d_3}$ & $\mu_{d_4}$ & $t_{d_3,d_4}^{NN}$ & $t_{d_3}^{NN}$  & $t_{d_4}^{NN}$ & $t_{d_4}^{zNN}$ & $t_{p_z^h,d_4}^{NN}$ & $t_{p_z^h}^{NN}$ & $t_{p_z^h, d_2}^{NN}$ & $\mu_{p_z^t}$ & $t_{p_z^t}^{NN}$ & $t_{p_z^t, d_3}^{NN}$  &&
\\ \hline
Value/eV & -1.91 & -1.18 & -1.23 & -0.12 & -0.16 & 0.20 & 0.10 & 0.83  & -0.46 & 0.28 & 0.60 & 0.30 & 0.45 &&
\\ \hline
Parameter & $\mu_{d_5}$  & $t_{d_5}^{NN}$ & $\mu_b$ & $t_b^{NN}$ & $t_b^{zNN}$ & $t_{b, d_5}^{NN}$ & $t_{b,d_5}^{NNN}$ &&
\\ \hline
Value/eV & -1.30 & -0.16 & -1.0 & -0.1 & 0.9 & -1.1 & 0.04 &&
\\ \hline\hline
\end{tabular}
\caption{\label{Table_TBpara_CoSn} Onsite energies and hoppings used in the decoupled TB model of CoSn, where $d_1$ to $d_5$ denote $d_{xy}, d_{x^2-y^2}, d_{xz}, d_{yz}, d_{z^2}$, respectively, $b$ the bonding state, and $p_{i}^t$ ($p_i^h$) denotes the $p_i$ orbitals of the triangular (honeycomb) Ge.  For the simplified $H_3(\kk)$ of $d_{z^2}$ only, the parameters are $\mu_{d_5}=0.31, t_{d_5}^{NN}=-0.15, t_{d_5}^{zNN}=0.8$. Parameters not listed in the table have zero values.}
\end{table}

\clearpage
\subsection{Spin-orbital coupling effects}

\begin{figure}[htbp]
\centering
\includegraphics[width=1\textwidth]{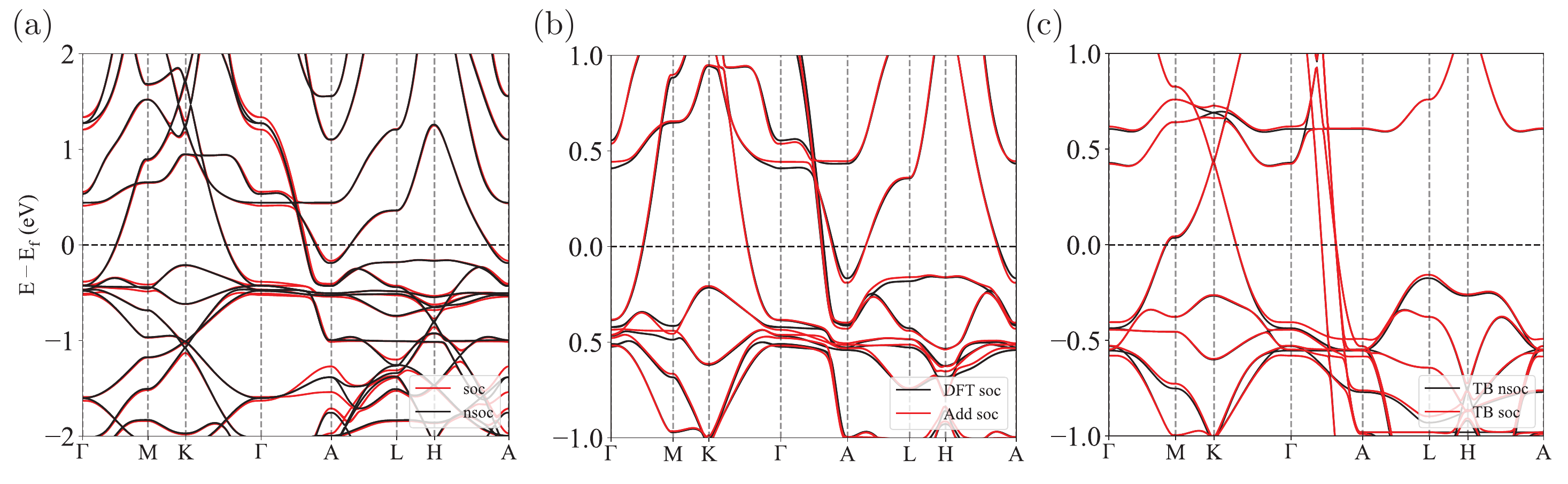}
\caption{\label{app:fig:CoSn_add_soc}
The SOC effects in CoSn. (a) shows the comparison of DFT band structures with and without SOC, where SOC opens gaps for some crossing points. (b) shows the comparison between the DFT SOC bands and the DFT NSOC bands with a manually added SOC term for Co $d$ orbitals (red). (c) shows the bands from the minimal TB model of CoSn (without SOC, black) defined in \cref{Table_TBpara_CoSn} and the TB bands with an added SOC term (red). 
}
\end{figure}

In the minimal tight-binding (TB) model for the 1:1 family described in \cref{eq_FeGe_TB_full_model}, the spin degree of freedom is not initially included. This is justified for materials like FeGe where the spin-orbit coupling (SOC) effects are negligible and can be safely disregarded. However, in FeSn and CoSn, the influence of SOC is more significant, as shown in the SOC band structure of CoSn in \cref{app:fig:CoSn_add_soc} (a). To incorporate the SOC effects, we introduce an onsite SOC term to the minimal TB model, ensuring a more accurate representation of the electronic structure in materials where SOC cannot be neglected.

For each kagome site, we take the following $d$ orbital basis
\begin{equation}
\begin{aligned}
&|d_{z^2},\uparrow \rangle,
|d_{xz},\uparrow \rangle, 
|d_{yz},\uparrow \rangle, 
|d_{x^2-y^2},\uparrow \rangle, 
|d_{xy},\uparrow \rangle, \\
&|d_{z^2},\downarrow \rangle,
|d_{xz},\downarrow \rangle, 
|d_{yz},\downarrow \rangle, 
|d_{x^2-y^2},\downarrow \rangle, 
|d_{xy},\downarrow \rangle.
\end{aligned}
\label{app:eq:spinful_d_basis}
\end{equation}
The onsite SOC Hamiltonian for $d$ orbital basis defined in \cref{app:eq:spinful_d_basis} has the form~\cite{konschuh2010tight}
\begin{equation}
H_{SOC}^d= \frac{\lambda_{soc}^d}{2}
\begin{bmatrix}
0 & 0 & 0 & 0 & 0 & 0 & -\sqrt{3} & i\sqrt{3} & 0 & 0\\
0 & 0 & -i & 0 & 0 & \sqrt{3} & 0 & 0 & -1 & i\\
0 & i & 0 & 0 & 0 & -i\sqrt{3} & 0 & 0 & -i & -1\\
0 & 0 & 0 & 0 & -2i & 0 & 1 & i & 0 & 0\\
0 & 0 & 0 & 2i & 0 & 0 & -i & 1 & 0 & 0\\
0 & \sqrt{3} & i\sqrt{3} & 0 & 0 & 0 & 0 & 0 & 0 & 0\\
-\sqrt{3} & 0 & 0 & 1 & i & 0 & 0 & i & 0 & 0\\
-i\sqrt{3} & 0 & 0 & -i & 1 & 0 & -i & 0 & 0 & 0\\
0 & -1 & i & 0 & 0 & 0 & 0 & 0 & 0 & 2i\\
0 & -i & -1 & 0 & 0 & 0 & 0 & 0 & -2i & 0 
\end{bmatrix}
\label{app:eq:H_soc_d}
\end{equation}
Notice that SOC couples d-orbitals with opposite spins and opposite (spinless) $M_z$ sectors. Specifically, the $M_z$-even sector, including $d_{xy}$, $d_{x^2-y^2}$, and $d_{z^2}$, couples to the $M_z$-odd sector of $d_{xz}$ and $d_{yz}$. This behavior is distinct from the spinless case, where inter-sector hoppings between different $M_z$ sectors are forbidden. 
The difference arises because the spinful $M_z$ eigenvalues differ from their spinless counterparts. For the basis orbitals under consideration, the spinful $M_z$ eigenvalues are:
\begin{equation}
\begin{aligned}
    [&-i,+i,+i,-i,-i,\\
    &+i,-i,-i,+i,+i]    
\end{aligned}
\end{equation}
We observe that SOC couples orbitals with matching spinful $M_z$ eigenvalues, thereby enabling interactions between spinless $M_z$-even and $M_z$-odd sectors. 

For $p$ orbitals with the basis
\begin{equation}
\begin{aligned}
&|p_z,\uparrow \rangle,
|p_x,\uparrow \rangle, 
|p_y,\uparrow \rangle, \\
&|p_z,\downarrow \rangle,
|p_x,\downarrow \rangle, 
|p_y,\downarrow \rangle
\end{aligned}
\label{app:eq:spinful_p_basis}
\end{equation}
The corresponding onsite SOC Hamiltonian has the form
\begin{equation}
H_{SOC}^p= \frac{\lambda_{soc}^p}{2}
\begin{bmatrix}
0 & 0 & 0 & 0 & -1 & i \\
0 & 0 & -i & 1 & 0 & 0 \\
0 & i & 0 & -i & 0 & 0 \\
0 & 1 & i & 0 & 0 & 0 \\
-1 & 0 & 0 & 0 & 0 & i \\
-i & 0 & 0 & 0 & -i & 0
\end{bmatrix}
\label{app:eq:H_soc_p}
\end{equation}

To determine the SOC strength parameters, $\lambda_{soc}^{p}$ and $\lambda_{soc}^{d}$, we start from a Wannier TB model derived from DFT calculations without SOC. This model includes Co $d$-orbitals and Sn $s, p$-orbitals. SOC terms, as defined in \cref{app:eq:H_soc_p} and \cref{app:eq:H_soc_d}, are then added, and the SOC parameters are fitted by comparing the resulting TB band structure with the DFT SOC band structure. The fitted SOC strengths are:
\begin{equation}
    \lambda_{soc}^{d}=0.123 \text{ eV},\quad \lambda_{soc}^{p}=0.269 \text{ eV}.
\end{equation}
We note that $\lambda_{soc}^{p}$ is stronger than $\lambda_{soc}^{d}$, as Sn is much heavier than Co and exhibits a stronger SOC effect. However, because the Sn $p$-orbitals are far from $E_f$, we focus primarily on the SOC effects of the Co $d$-orbitals. As shown in \cref{app:fig:CoSn_add_soc}~(b), the DFT SOC band structure near $E_f$ is well reproduced by including only the SOC term for Co $d$-orbitals. Adding the SOC term for Sn $p$-orbitals introduces minimal changes to the bands within $\pm 1$~eV of $E_f$.

Using the fitted SOC strength parameter $\lambda_{soc}^{d}$ from \textit{ab initio} calculations, we incorporate this SOC term into the minimal TB model for CoSn, defined in \cref{Table_TBpara_CoSn}. The resulting band structure, shown in \cref{app:fig:CoSn_add_soc}~(c), shows excellent agreement with the DFT SOC band structure. SOC-induced gaps are observed at key crossing points, such as $\Gamma$ and $K$.

For FeSn, the SOC gap openings at $\Gamma$ and $K$ are very similar to CoSn, thus a similar SOC strength is expected.

We remark that one could consider the general ($\kk$-dependent) symmetry-allowed spin-off-diagonal coupling terms as SOC contributions to achieve a better agreement with DFT results. 

\subsection{Interaction parameters of FeSn and CoSn}

The interaction parameters are computed for FeSn and CoSn using the $d$-full model, i.e., Wannier functions are constructed from $s, p, d$ orbitals of Fe (Co) and $s, p$ orbitals of Sn for FeSn (CoSn), and exclude only the polarization between $d$ orbitals, as tabulated in \cref{d-full-FeSn} and \ref{d-full-CoSn}.

In \cref{d-full-FeSn-cubic} and \ref{d-full-CoSn-cubic}, we list the $O_h$-symmetrized onsite Coulomb matrices elements $U_{ij}$ and $J_{ij}$ for FeSn and CoSn, respectively. 

In \cref{d-full-FeSn-slater-U-F0F2F4} and \ref{d-full-CoSn-slater-U-F0F2F4}, we list the fitted Coulomb elements $U_{ij}$ and $J_{ij}$ using Slater integrals $F_0, F_2, F_4$ for FeSn and CoSn, while in \cref{d-full-FeSn-slater-U-F0F2} and \ref{d-full-CoSn-slater-U-F0F2}, we list the fitted Coulomb elements using Slater integrals $F_0$ and $F_2$ only and fix $F_4=0.625F_2$ for FeSn and CoSn, respectively.

In \cref{Table:compare_HK}, we list the Hubbard-Kanamori parameters for FeGe, FeSn, and CoSn, which share close values. 
Compared with FeGe, FeSn has onsite intra-orbital Hubbard $\mathcal{U}\approx 4$ eV which is about 0.2 eV smaller than FeGe, while for CoSn, $\mathcal{U},\mathcal{U}^\prime$ are both about 0.2 eV smaller than FeGe.

The two quasi-flat bands at $E_f$ and large Coulomb interactions in FeGe and FeSn give rise to the AFM order. FeSn also has multiple vHS close to $E_f$ in the AFM phase, as shown in \cref{Fig:CoSn_FeSn_band}(d), but unlike FeGe there is no CDW transition reported, which requires future investigation. 
For CoSn, despite the similar Coulomb interaction as FeGe and FeSn, there is no AFM or CDW reported, which is possibly due to that the two flat bands in CoSn are fully filled and no vHSs close to $E_f$.

\begin{table}[htbp]
\begin{tabular}{c|c|c|c}
\hline\hline
Hubbard-Kanamori & $\mathcal{U}$ & $\mathcal{U}^\prime$ & $\mathcal{J}$ \\\hline
FeGe & 4.18 & 2.73 & 0.72 \\\hline
FeSn & 3.99 & 2.66 & 0.66 \\\hline
CoSn & 3.95 & 2.54 & 0.70 \\
\hline\hline
\end{tabular}
\caption{\label{Table:compare_HK}
The Hubbard-Kanamori parameters in the $d$-full model in FeGe, FeSn, and CoSn. These three materials share close interaction parameters, with FeGe being slightly larger. All numbers are in eV. }
\end{table}

\begin{table}[htbp]
\begin{tabular}{c|ccccccc|ccccc}
\hline\hline
$U_{ij}$  & $z^2$ & $xz$ & $yz$ & $x^2-y^2$ & $xy$ &  & $J_{ij}$ & $z^2$ & $xz$ & $yz$ & $x^2-y^2$ & $xy$ \\\hline
$z^2$     & 3.93  & 2.79 & 2.98 & 2.35      & 2.37 &  &          &       & 0.47 & 0.49 & 0.86      & 0.86 \\
$xz$      &       & 3.59 & 2.43 & 2.46      & 2.48 &  &          &       &      & 0.68 & 0.68      & 0.68 \\
$yz$      &       &      & 4.02 & 2.66      & 2.67 &  &          &       &      &      & 0.72      & 0.72 \\
$x^2-y^2$ &       &      &      & 4.17      & 3.40 &  &          &       &      &      &           & 0.39 \\
$xy$      &       &      &      &           & 4.22 &  &          &       &      &      &           &   \\
\hline\hline
$U_{ij}^{NN}$ & $z^2$ & $xz$ & $yz$ & $x^2-y^2$ & $xy$ & &$U_{ij}^{NNN}$ & $z^2$ & $xz$ & $yz$ & $x^2-y^2$ & $xy$ \\\hline
$z^2$   & 1.33  & 1.32 & 1.35 & 1.36      & 1.36 &   &           & 1.16  & 1.16 & 1.16 & 1.17      & 1.17 \\
$xz$    &       & 1.32 & 1.33 & 1.34      & 1.35 &   &           &       & 1.17 & 1.16 & 1.17      & 1.17 \\
$yz$    &       &      & 1.37 & 1.37      & 1.38 &   &           &       &      & 1.16 & 1.17      & 1.17 \\
$x^2-y^2$&      &      &      & 1.39      & 1.40 &   &           &       &      &      & 1.18      & 1.18 \\
$xy$    &       &      &      &           & 1.40 &    &          &       &      &      &           & 1.18\\
\hline\hline
\end{tabular}
\caption{\label{d-full-FeSn} The Coulomb interaction $U_{ij}$ and $J_{ij}$ of $d$ orbitals, and $U_{ij}^{NN}$ and $U_{ij}^{NNN}$ between NN and NNN $d$ in $d-$full model of FeSn.
The onsite Hubbard-Kanamori parameters for five $d$ orbitals are $\mathcal{U}=3.99, \mathcal{U}^\prime=2.66, \mathcal{J}=0.66$. 
The averaged NN and NNN interactions are $\overline{U}^{NN}=1.36, \overline{U}^{NNN}=1.17$, with root mean square error being 0.023 and 0.006, respectively. All numbers are in eV.}
\end{table}

\begin{table}[htbp]
\begin{tabular}{c|cccccc|ccccc}
\hline\hline
$U_{ij}$  & $z^2$ & $xz$ & $yz$ & $x^2-y^2$ & $xy$ & $J_{ij}$ & $z^2$ & $xz$ & $yz$ & $x^2-y^2$ & $xy$ \\\hline
$z^2$     & 4.06 &  3.00 &  3.00 &  2.35  &  2.33 &    &    & 0.49& 0.49 & 0.86  &  0.82 \\
$xz$      &       &  3.94 &  2.53 &  2.55  &  2.53 &    &    &      & 0.70 & 0.71  &  0.70 \\
$yz$      &       &        &  3.94 &  2.55  &  2.53 &    &    &      &       & 0.71  &  0.70 \\
$x^2-y^2$ &       &        &        &  4.06  &  3.23 &    &    &      &       &        &  0.38 \\
$xy$      &       &        &        &         &  3.94 &    &    &      &       &        &        \\
\hline\hline
\end{tabular}
\caption{\label{d-full-FeSn-cubic} The Coulomb interaction $U_{ij}$ and $J_{ij}$ of $d$ orbitals for the $d$-full model symmetrized using $O_h$ symmetries in FeSn.
The root mean square error of the averaged $U_{ij}$ and $J_{ij}$ is 0.177 and 0.069, respectively.}
\end{table}

\begin{table}[htbp]
\begin{tabular}{c|cccccc|ccccc}
\hline\hline
$U_{ij}$  & $z^2$ & $xz$ & $yz$ & $x^2-y^2$ & $xy$ & $J_{ij}$ & $z^2$ & $xz$ & $yz$ & $x^2-y^2$ & $xy$ \\\hline
$z^2$     & 3.98 &  2.99 &  2.99 &  2.33  &  2.33 &    &    & 0.49& 0.49 & 0.82  &  0.82 \\
$xz$      &       &  3.98 &  2.55 &  2.55  &  2.55 &    &    &      & 0.71 & 0.71  &  0.71 \\
$yz$      &       &        &  3.98 &  2.55  &  2.55 &    &    &      &       & 0.71  &  0.71 \\
$x^2-y^2$ &       &        &        &  3.98  &  3.21 &    &    &      &       &        &  0.38 \\
$xy$      &       &        &        &         &  3.98 &    &    &      &       &        &        \\
\hline\hline
\end{tabular}
\caption{\label{d-full-FeSn-slater-U-F0F2F4} The Coulomb interaction $U_{ij}$ and $J_{ij}$ of $d$ orbitals fitted using \cref{eq_slater_Uijkl} in FeSn, where the Slater integrals $F^0$, $F^2$, and $F_4$ are fitted to the full Coulomb matrix $U_{ijkl}$ symmetrized by $O_h$ symmetries. The fitted parameters are $F^0=2.926, F^2=8.092, F^4=4.819$.
The root mean square error of the fitted $U_{ij}$ and $J_{ij}$ is 0.143 and 0.026, respectively.}
\end{table}

\begin{table}[htbp]
\begin{tabular}{c|cccccc|ccccc}
\hline\hline
$U_{ij}$  & $z^2$ & $xz$ & $yz$ & $x^2-y^2$ & $xy$ & $J_{ij}$ & $z^2$ & $xz$ & $yz$ & $x^2-y^2$ & $xy$ \\\hline
$z^2$     & 3.99 &  2.98 &  2.98 &  2.34  &  2.34 &    &    & 0.50& 0.50 & 0.83  &  0.83 \\
$xz$      &       &  3.99 &  2.55 &  2.55  &  2.55 &    &    &      & 0.72 & 0.72  &  0.72 \\
$yz$      &       &        &  3.99 &  2.55  &  2.55 &    &    &      &       & 0.72  &  0.72 \\
$x^2-y^2$ &       &        &        &  3.99  &  3.19 &    &    &      &       &        &  0.40 \\
$xy$      &       &        &        &         &  3.99 &    &    &      &       &        &        \\
\hline\hline
\end{tabular}
\caption{\label{d-full-FeSn-slater-U-F0F2} The Coulomb interaction $U_{ij}$ and $J_{ij}$ of $d$ orbitals fitted using \cref{eq_slater_Uijkl} in FeSn, where the Slater integrals $F^0$, $F^2$ are fitted to the full Coulomb matrix $U_{ijkl}$ symmetrized by $O_h$ symmetries, while $F^4=0.625F^2$ are fixed. The fitted parameters are $F^0=2.926, F^2=8.024, F^4=5.015$.
The root mean square error of the fitted $U_{ij}$ and $J_{ij}$ is 0.144 and 0.029, respectively.}
\end{table}

\begin{table}[htbp]
\begin{tabular}{c|ccccccc|ccccc}
\hline\hline
$U_{ij}$  & $z^2$ & $xz$ & $yz$ & $x^2-y^2$ & $xy$ &  & $J_{ij}$ & $z^2$ & $xz$ & $yz$ & $x^2-y^2$ & $xy$ \\\hline
$z^2$     & 3.94  & 2.77 & 2.88 & 2.21      & 2.22 &  &          &       & 0.51 & 0.53 & 0.89      & 0.90 \\
$xz$      &       & 3.65 & 2.36 & 2.39      & 2.40 &  &          &       &      & 0.73 & 0.73      & 0.73 \\
$yz$      &       &      & 3.97 & 2.48      & 2.49 &  &          &       &      &      & 0.76      & 0.77 \\
$x^2-y^2$ &       &      &      & 4.07      & 3.25 &  &          &       &      &      &           & 0.42 \\
$xy$      &       &      &      &           & 4.11 &  &          &       &      &      &           &   \\
\hline\hline
$U_{ij}^{NN}$ & $z^2$ & $xz$ & $yz$ & $x^2-y^2$ & $xy$ & &$U_{ij}^{NNN}$ & $z^2$ & $xz$ & $yz$ & $x^2-y^2$ & $xy$ \\\hline
$z^2$   & 1.32  & 1.32 & 1.33 & 1.36      & 1.36 &   &           & 1.17  & 1.16 & 1.17 & 1.17      & 1.18 \\
$xz$    &       & 1.33 & 1.33 & 1.36      & 1.37 &   &           &       & 1.17 & 1.17 & 1.18      & 1.18 \\
$yz$    &       &      & 1.35 & 1.37      & 1.38 &   &           &       &      & 1.17 & 1.17      & 1.18 \\
$x^2-y^2$&      &      &      & 1.41      & 1.42 &   &           &       &      &      & 1.19      & 1.19 \\
$xy$    &       &      &      &           & 1.42 &    &          &       &      &      &           & 1.19\\
\hline\hline
\end{tabular}
\caption{\label{d-full-CoSn} The Coulomb interaction $U_{ij}$ and $J_{ij}$ of $d$ orbitals, and $U_{ij}^{NN}$ and $U_{ij}^{NNN}$ between NN and NNN $d$ in $d-$full model of CoSn.
The onsite Hubbard-Kanamori parameters for five $d$ orbitals are $\mathcal{U}=3.95, \mathcal{U}^\prime=2.54, \mathcal{J}=0.70$. 
The averaged NN and NNN interactions are $\overline{U}^{NN}=1.36, \overline{U}^{NNN}=1.17$, with root mean square error being 0.029 and 0.009, respectively. All numbers are in eV.}
\end{table}

\begin{table}[htbp]
\begin{tabular}{c|cccccc|ccccc}
\hline\hline
$U_{ij}$  & $z^2$ & $xz$ & $yz$ & $x^2-y^2$ & $xy$ & $J_{ij}$ & $z^2$ & $xz$ & $yz$ & $x^2-y^2$ & $xy$ \\\hline
$z^2$     & 4.06 &  3.00 &  3.00 &  2.35  &  2.33 &    &    & 0.49& 0.49 & 0.86  &  0.82 \\
$xz$      &       &  3.94 &  2.53 &  2.55  &  2.53 &    &    &      & 0.70 & 0.71  &  0.70 \\
$yz$      &       &        &  3.94 &  2.55  &  2.53 &    &    &      &       & 0.71  &  0.70 \\
$x^2-y^2$ &       &        &        &  4.06  &  3.23 &    &    &      &       &        &  0.38 \\
$xy$      &       &        &        &         &  3.94 &    &    &      &       &        &        \\
\hline\hline
\end{tabular}
\caption{\label{d-full-CoSn-cubic} The Coulomb interaction $U_{ij}$ and $J_{ij}$ of $d$ orbitals for the $d$-full model symmetrized using $O_h$ symmetries in CoSn.
The root mean square error of the averaged $U_{ij}$ and $J_{ij}$ is 0.216 and 0.035, respectively.}
\end{table}

\begin{table}[htbp]
\begin{tabular}{c|cccccc|ccccc}
\hline\hline
$U_{ij}$  & $z^2$ & $xz$ & $yz$ & $x^2-y^2$ & $xy$ & $J_{ij}$ & $z^2$ & $xz$ & $yz$ & $x^2-y^2$ & $xy$ \\\hline
$z^2$     & 3.94 &  2.89 &  2.89 &  2.20  &  2.20 &    &    & 0.53& 0.53 & 0.87  &  0.87 \\
$xz$      &       &  3.94 &  2.43 &  2.43  &  2.43 &    &    &      & 0.76 & 0.76  &  0.76 \\
$yz$      &       &        &  3.94 &  2.43  &  2.43 &    &    &      &       & 0.76  &  0.76 \\
$x^2-y^2$ &       &        &        &  3.94  &  3.12 &    &    &      &       &        &  0.41 \\
$xy$      &       &        &        &         &  3.94 &    &    &      &       &        &        \\
\hline\hline
\end{tabular}
\caption{\label{d-full-CoSn-slater-U-F0F2F4} The Coulomb interaction $U_{ij}$ and $J_{ij}$ of $d$ orbitals fitted using \cref{eq_slater_Uijkl} in CoSn, where the Slater integrals $F^0$, $F^2$, and $F_4$ are fitted to the full Coulomb matrix $U_{ijkl}$ symmetrized by $O_h$ symmetries. The fitted parameters are $F^0=2.825, F^2=8.522, F^4=5.163$.
The root mean square error of the fitted $U_{ij}$ and $J_{ij}$ is 0.094 and 0.022, respectively.
}
\end{table}

\begin{table}[htbp]
\begin{tabular}{c|cccccc|ccccc}
\hline\hline
$U_{ij}$  & $z^2$ & $xz$ & $yz$ & $x^2-y^2$ & $xy$ & $J_{ij}$ & $z^2$ & $xz$ & $yz$ & $x^2-y^2$ & $xy$ \\\hline
$z^2$     & 3.95 &  2.88 &  2.88 &  2.21  &  2.21 &    &    & 0.53& 0.53 & 0.87  &  0.87 \\
$xz$      &       &  3.95 &  2.43 &  2.43  &  2.43 &    &    &      & 0.76 & 0.76  &  0.76 \\
$yz$      &       &        &  3.95 &  2.43  &  2.43 &    &    &      &       & 0.76  &  0.76 \\
$x^2-y^2$ &       &        &        &  3.95  &  3.11 &    &    &      &       &        &  0.42 \\
$xy$      &       &        &        &         &  3.95 &    &    &      &       &        &        \\
\hline\hline
\end{tabular}
\caption{\label{d-full-CoSn-slater-U-F0F2} The Coulomb interaction $U_{ij}$ and $J_{ij}$ of $d$ orbitals fitted using \cref{eq_slater_Uijkl} in CoSn, where the Slater integrals $F^0$, $F^2$ are fitted to the full Coulomb matrix $U_{ijkl}$ symmetrized by $O_h$ symmetries, while $F^4=0.625F^2$ are fixed. The fitted parameters are $F^0=2.825, F^2=8.475, F^4=5.297$.
The root mean square error of the fitted $U_{ij}$ and $J_{ij}$ is 0.095 and 0.023, respectively.}
\end{table}

\clearpage
\section{Building models for 1:6:6 class from 1:1 class: Application to \ch{MgFe6Ge6}}\label{Sec:MgFe6Ge6}

In previous sections, we studied in detail the electronic properties of FeGe, FeSn, and CoSn in detail. These materials belong to the 1:1 class in SG 191, and form the fundamental building blocks for the 1:6:6 material family~\cite{venturini2006filling,fredrickson2008} with chemical formula MT$_6$Z$_6$, where M is a metallic element (including Mg, Sc, and rare-earth elements, etc), T a transition metal, and Z a main group element (including Si, Ga, Ge, and Sn).
The 1:6:6 family materials can be seen as the doubled 1:1 materials along the $c$-axis and inserting $M$ atoms in the middle. Usually, the electrons of M atoms are away from the Fermi level, which means it is possible to construct bands of 1:6:6 materials from 1:1 materials by doubling the unit cell and treating the orbitals of M as perturbations. We test this strategy by focusing on one representative 1:6:6 material MgFe$_6$Ge$_6$ and build its band structure from FeGe in this section. This strategy is expected to work for the other 1:6:6 materials, and the minimal TB model we build for FeSn and CoSn could also be used to construct their corresponding 1:6:6 materials.

\subsection{Crystal structure and band structure of \ch{MgFe6Ge6}}
\begin{figure}[htbp]
	\centering
	\includegraphics[width=1\textwidth]{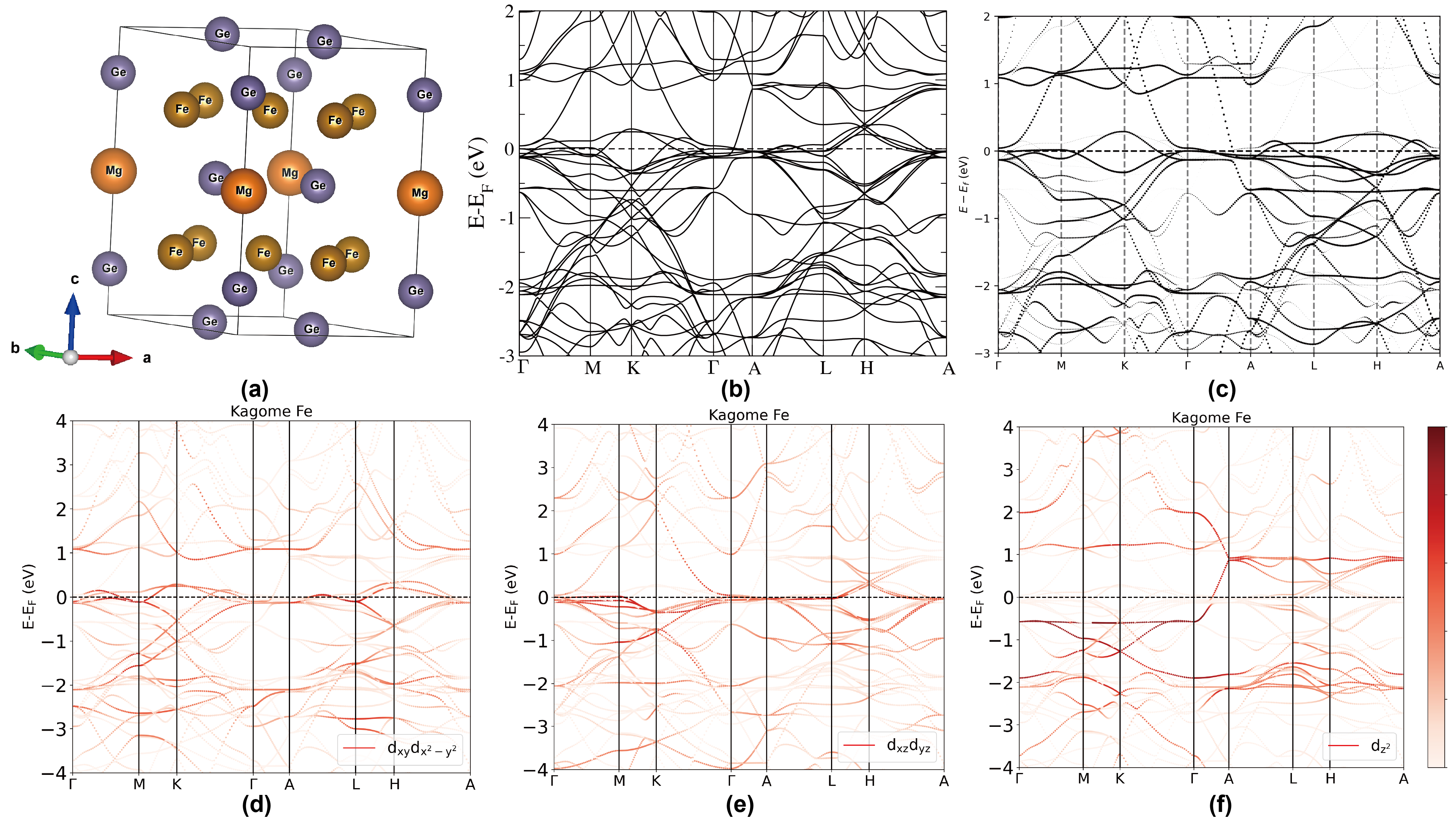}
	\caption{\label{Fig_MgFe6Ge6_PM_struct_band} (a) Crystal structure and (b) band structure without SOC of $\text{MgFe}_6\text{Ge}_6$ in the paramagnetic phase.
    (c) Unfolded band structure of $\text{MgFe}_6\text{Ge}_6$.
    (d)-(f) are orbital projections of the bands in the PM phase of $\text{MgFe}_6\text{Ge}_6$, where (d) is the projection of $d_{xy}, d_{x^2-y^2}$, (e) $d_{xz}, d_{yz}$, and (f) $d_{z^2}$ orbitals of Fe. The band structure and orbital weights of $\text{MgFe}_6\text{Ge}_6$ are very close to FeGe after folding the BZ in the $k_3$ direction. 
    }
\end{figure}

Similar to FeGe, $\text{MgFe}_6\text{Ge}_6$ has SG 191 $P6/mmm$ symmetry in the paramagnetic phase. The lattice constants are $a=5.071\AA$ and $c=8.046\AA$\cite{mazet2013magnetic}. The crystal structure is shown in \cref{Fig_MgFe6Ge6_PM_struct_band}(a), where
\begin{itemize}
    \item Mg atoms form a triangular lattice on $z=0.5$ plane in the unit cell, i.e., $1b$ Wyckoff position.
    \item Fe atoms form two kagome layers on $z=0.252$ and $z=0.748$ planes, i.e., $6i$ Wyckoff position.
    \item Two Ge atoms form two triangular lattices on $z=0.159$ and $z=0.841$ planes, i.e., $2e$ Wyckoff position, which differs from FeGe where the triangular Ge atoms lie on the same plane with Fe. 
    \item Four Ge atoms form two honeycomb lattices on $z=0$ and $z=0.5$ planes, i.e., $2c$ and $2d$ Wyckoff positions.
\end{itemize}
The coordinates of these Wyckoff positions can be found in \cref{SG191-wyckoff}. We emphasize that $\text{MgFe}_6\text{Ge}_6$ can be viewed as a twofold supercell of FeGe by adding Mg atoms in the middle and (slightly) shifting the Ge atoms on triangular lattice away (about 0.7 \AA) from the kagome layer formed by Fe atoms. Indeed we find the band structure of $\text{MgFe}_6\text{Ge}_6$ is similar and related to the band structure of FeGe. 

The band structure of $\text{MgFe}_6\text{Ge}_6$ in the PM phase without SOC is shown in \cref{Fig_MgFe6Ge6_PM_struct_band}(b). 
As $\text{MgFe}_6\text{Ge}_6$ can be seen as a perturbed twofold supercell of FeGe, we unfold the band structure of $\text{MgFe}_6\text{Ge}_6$ to the BZ of FeGe in \cref{Fig_MgFe6Ge6_PM_struct_band}(c) using \textit{VaspBandUnfolding} package\cite{zheng2018vaspbandunfolding, popescu2012extracting}. The unfolded band structure is very close to the band structure of FeGe.
In \cref{Fig_MgFe6Ge6_PM_struct_band}(d)-(f), we show the projections of $d$ orbitals of Fe in $\text{MgFe}_6\text{Ge}_6$. 
The orbital weights of Fe $d$ orbitals in  $\text{MgFe}_6\text{Ge}_6$ are also very close to FeGe. 
The $p$ orbitals of Ge atoms also have similar distributions as in FeGe and are not shown for simplicity. 
The orbitals of Mg atoms are far away from the Fermi level, with the $s$ orbital being about 3 eV above $E_f$. 

\begin{figure}[htbp]
	\centering
    \includegraphics[width=0.5\textwidth]{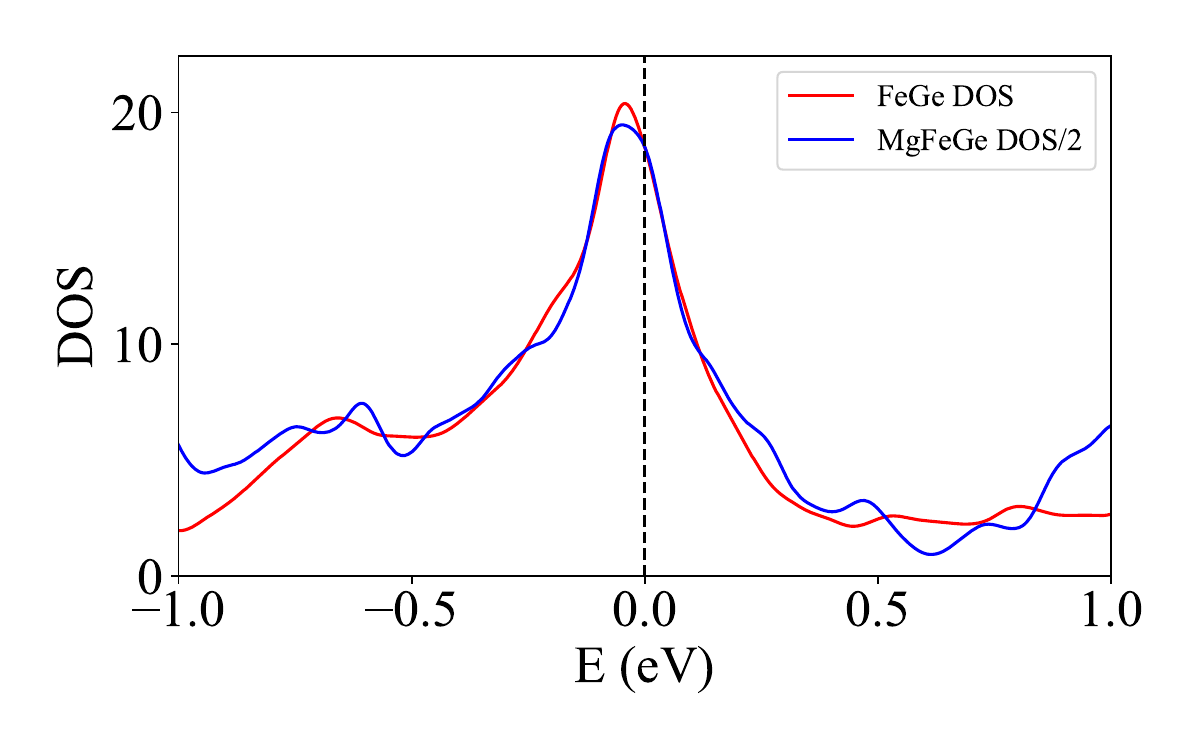}
	\caption{\label{Fig:compare_dos_FeGe_MgFeGe} Comparison of DOS in PM phase between FeGe and \ch{MgFe6Ge6}, where the DOS of \ch{MgFe6Ge6} has divided by 2 in order to compare with FeGe. The two DOS curves show good agreements at the Fermi level where the large peak is given by the flat band. }
\end{figure}

In \cref{Tab:MgFeGe_filling}, we show the filling of each orbital in \ch{MgFe6Ge6} in the PM phae. Compared with the filling of FeGe in \cref{Tab:filling_magmom_wannier}, \ch{MgFe6Ge6} has one more valence given by the $s$ orbital of Mg per FeGe layer (2 extra electrons in total). The extra valence is mainly contributed by Ge and Mg, while the filling of Fe is almost unchanged. This is also verified by the $d$ orbital projections of Fe which are very close for FeGe and \ch{MgFe6Ge6}. In \cref{Fig:compare_dos_FeGe_MgFeGe}, we compare the total DOS of FeGe and \ch{MgFe6Ge6}, which show a good match at the Fermi level. The extra valence electrons in \ch{MgFe6Ge6} mainly change the DOS below -0.5 eV.

The above observations inspire us to understand the band structure of $\text{MgFe}_6\text{Ge}_6$ via the band structure of FeGe. In the next section, we will show the procedure of constructing the TB models for $\text{MgFe}_6\text{Ge}_6$ based on the TB model of FeGe. In short, we start from the TB models of the FeGe but with an enlarged unit cell, such that two layers of FeGe have been included. Then we treat Mg atoms as a perturbation. 
We remark that the inserted 1 atom in the 1:6:6 family usually provides extra valence electrons and changes the Fermi level. This should be treated carefully when constructing the TB model for 1:6:6 using 1:1 building blocks. 
However, for \ch{MgFe6Ge6}, the fillings of $d$ orbitals have little change, and we use the same Fermi level in the minimal TB model for simplicity.

\begin{figure}[htbp]
	\centering
	\includegraphics[width=1\textwidth]{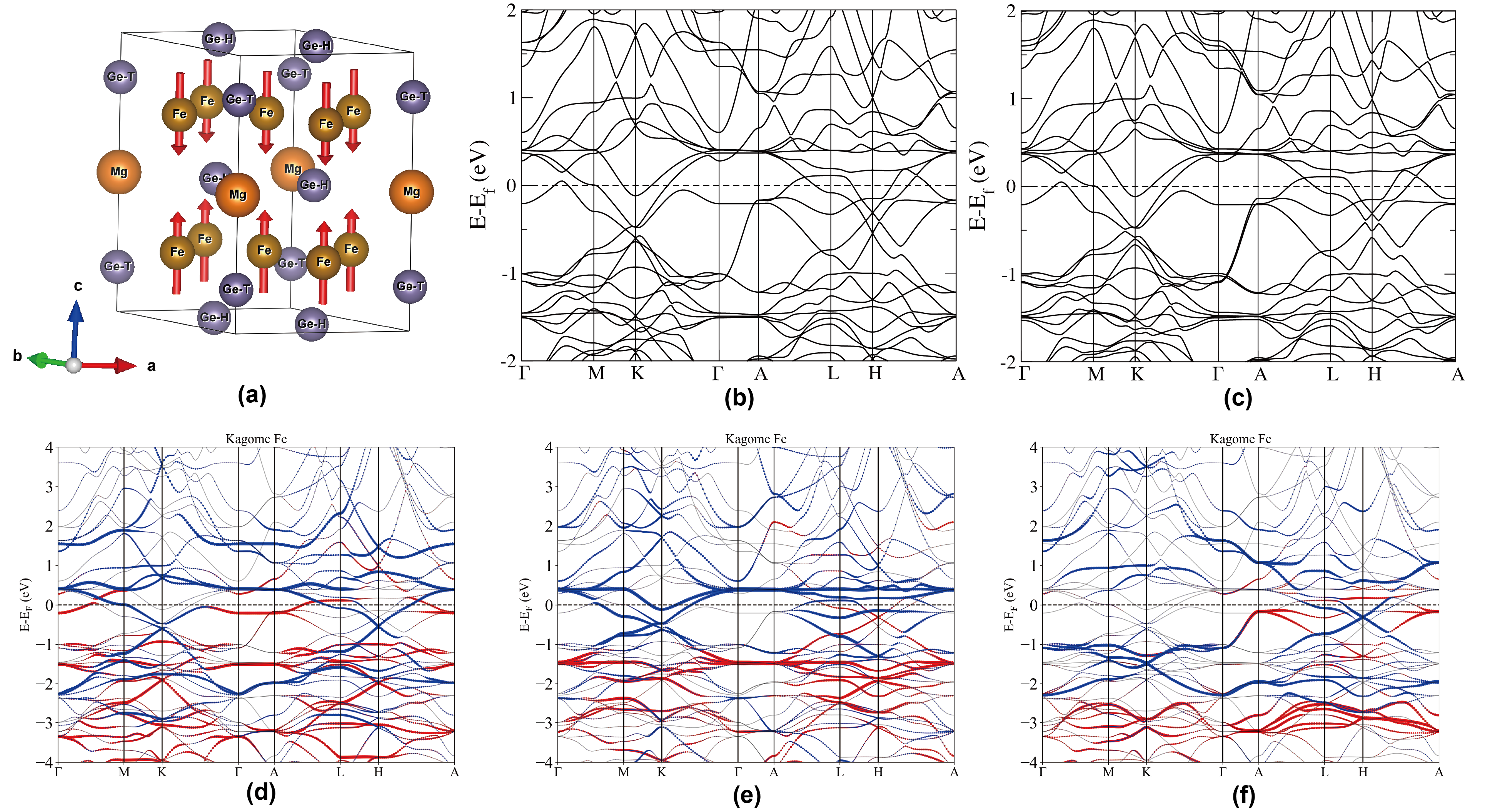}
	\caption{\label{Fig_MgFe6Ge6_AFM_struct_band} (a) Crystal structure, band structure (b) without SOC, and (c) with SOC of $\text{MgFe}_6\text{Ge}_6$ in the AFM phase.
    (d)-(f) are orbital projections of the bands in the AFM phase of $\text{MgFe}_6\text{Ge}_6$, where (d) is the projection of $d_{xy}, d_{x^2-y^2}$, (e) $d_{xz}, d_{yz}$, and (f) $d_{z^2}$ orbitals of Fe, where blue and red colors denote $d$ orbitals of opposite spins of Fe on the same kagome plane. 
    The band structure and orbital weights of $\text{MgFe}_6\text{Ge}_6$ are very close to FeGe in the AFM phase. The calculated magnetic moment in $\text{MgFe}_6\text{Ge}_6$ is 1.79 $\mu_B$ per Fe atom which is larger than that in FeGe which is 1.53 $\mu_B$.}
\end{figure}

Before constructing the TB models, we also mention that $\text{MgFe}_6\text{Ge}_6$ develops an AFM order at $T_N=501K$ with the magnetic moment on Fe atoms being $m=1.59$ $\mu_B$ at $2K$ along the $\bm{c}$-axis\cite{mazet2013magnetic}. The crystal structure, band structure, and $d$ orbital projections in the AFM phase are shown in \cref{Fig_MgFe6Ge6_AFM_struct_band}. The computed magnetic moment on Fe is $1.79$ $\mu_B$ in DFT. 
We point out that the magnetic structure of $\text{MgFe}_6\text{Ge}_6$ is almost identical to the magnetic structure of FeGe. Note that in the PM phase the unit cell of $\text{MgFe}_6\text{Ge}_6$ is a doubled FeGe with extra Mg atoms, while in the AFM phase of FeGe, the magnetic unit cell is the same for $\text{MgFe}_6\text{Ge}_6$ and FeGe. 
This further confirms the similarity and connection between $\text{MgFe}_6\text{Ge}_6$ and FeGe. 
We also remark that $\text{MgFe}_6\text{Ge}_6$ only appear as powder in literature\cite{mazet2013magnetic}.

\begin{table}[htbp]
\begin{tabular}{c|c|c|c|c|c|c|c|c|c|c|c}
\hline\hline
& Orbitals & $s$  & $p_z$ & $p_x$ & $p_y$ & $d_{z^2}$ & $d_{xz}$ & $d_{yz}$ & $d_{x^2-y^2}$ & $d_{xy}$ & Total \\ \hline
\multirow{4}{*}{PM filling}  & Fe       & 0.61 & 0.28  & 0.31  & 0.27  & 1.44      & 1.64     & 1.50     & 1.32          & 1.49     & 8.86  \\ \cline{2-12} 
& Tri-Ge   & 1.46 & 0.70  & 0.63  & 0.62  & -         & -        & -        & -             & -        & 3.40  \\ \cline{2-12} 
& Hon-Ge   & 1.49 & 0.72  & 0.62  & 0.64  & -         & -        & -        & -             & -        & 3.46  \\ \cline{2-12} 
& Mg & 0.4 & -        & -        & -   & -         & -        & -        & -             & -        & 0.4  \\ \hline
PM DOS@$E_f$ & Fe & - & - & - & - & 0.05 & 0.27 & 0.26 & 0.30 & 0.05 & 0.94
\\\hline\hline
\end{tabular}
\caption{\label{Tab:MgFeGe_filling}
The filling and density of states (DOS) of \ch{MgFe6Ge6} in the PM phase, calculated from a Wannier TB built from Mg $s$, Fe $s,p,d$, and Ge $s,p$ orbitals. The total filling is 74.2, which is close to the number of valence electrons, i.e., 74 (given by $4s^1 3d^7$ of 3 Fe,$4s^2 4p^2$ of 3 Ge, and $3s^2$ of Mg). 
}
\end{table}

\subsection{Tight-binding models}
TB models of $\text{MgFe}_6\text{Ge}_6$, denoted by $H^{[166]}(\bm{k})$, can be constructed using the TB models of FeGe defined in \cref{eq_FeGe_TB_full_model} and \cref{Eq:FeGe_TB_H123}, denoted as $H^{[11]}(\bm{k})$. 

We first consider the TB model of a perfect twofold supercell (SC) of FeGe with two identical unit cells in the $z$-direction, in which the basis $\bm{a}_3$ is doubled while $\bm{b}_3$ of BZ is halved. The TB model $H^{[\text{SC}]}(\bm{k})$ is
\begin{equation}
    H^{[\text{SC}]}(\bm{k})= H^{[11]}(k_1,k_2,\frac{k_3}{2})\oplus H^{[11]}(k_1,k_2,\frac{k_3}{2}+\pi),
    \label{eq_166TB_directsummed}
\end{equation}
where the bands on $k_3=0,\pi$ ($k_3=\pm\frac{\pi}{2}$) planes of $H^{[11]}(\bm{k})$ are folded onto the $k_3=0$ ($k_3=\pi$) plane of $H^{[\text{SC}]}$. 
In this case, the bands on the $k_3=\pi$ plane are always twofold degenerate, as a result of perfect SC.

More rigorously, we consider the effect of the inclusion of Mg atoms and the dislocation of triangular Ge atoms in the TB model for $\text{MgFe}_6\text{Ge}_6$, which breaks the perfect twofold SC. The inclusion of Mg atoms has larger effects on the band structure. 
We use $d_i^1$ and $d_i^2$ to denote the $d_i$ orbitals on the kagome layer of $z=0.252$ and $z=0.748$ plane, $p_i^{h_1}$ and $p_i^{h_2}$ to denote the $p_i$ orbitals on the honeycomb layer of $z=0$ and $z=\frac{1}{2}$ plane, and $p_i^{t_1}$ and $p_i^{t_2}$ to denote the $p_i$ orbitals on the triangular layer of $z=0.159$ and $z=0.841$ plane, respectively.
Following the method used in FeGe, we split the orbitals into three groups and construct TB models for them separately. As the onsite energy of the Mg $s$ orbital is much higher than other orbitals (onsite difference of about 4 eV) in the model, the coupling with the $s$ orbital of the Mg atom will be treated as a perturbation which will modify the hoppings between certain orbitals \citeSI{Sec:combine_Mg_s}.

\subsubsection{$H_1(k)$}
The first group of orbitals contains $p_x,p_y$ orbitals of Ge at $2e$ (denoted by $p_{xy}^{t}$), $d_{xy}, d_{x^2-y^2}$ (denoted by $d_1,d_2$) orbitals of Fe at $6i$. 
We also add $p_z$ orbitals of honeycomb Ge at $2c$ and $2d$ (denoted by $p_z^{h_i}$) here to introduce the $S$-matrix, where in FeGe they are considered later in the final model in \cref{eq_FeGe_TB_full_model}. The TB model has the form 
\begin{equation}
\begin{aligned}
H_1^{[166]}(\bm{k})=
&\left(
\begin{array}{ccc|ccc|cc}
    H_{p_{xy}^t}(k_1,k_2) & S_{p_{xy}^{t_1}, d_1^1}(\bm{k}) & S_{p_{xy}^{t_1}, d_2^1}(\bm{k}) & \bm{0} & \bm{0} & \bm{0} & \bm{0} & \bm{0} \\
    & H_{d_1}(k_1,k_2)  & S_{d_1,d_2}(k_1,k_2) & \bm{0} & \bm{0} & \bm{0} & \bm{0} & \bm{0} \\
    & & H_{d_2}(k_1,k_2) & \bm{0} & \bm{0} & \bm{0} & 
    S_{p_z^{h_1},d_2^1}^\dagger(\bm{k}) & S_{p_z^{h_2},d_2^1}^\dagger(\bm{k}) \\\hline
    & & & H_{p_{xy}^t}(k_1,k_2) & S_{p_{xy}^{t_2}, d_1^2}(\bm{k}) & S_{p_{xy}^{t_2}, d_2^2}(\bm{k}) & \bm{0} & \bm{0} \\
    & & & & H_{d_1}(k_1,k_2)  & S_{d_1,d_2}(k_1,k_2) & \bm{0} & \bm{0} \\
    & & & & & H_{d_2}(k_1,k_2) & 
    S_{p_z^{h_1},d_2^2}^\dagger(\bm{k}) & S_{p_z^{h_2},d_2^2}^\dagger(\bm{k}) \\\hline
    & & & & & & H_{p_z^{h_1}}(\bm{k}) & \\
    H.c. & & & & & & & H_{p_z^{h_2}}(\bm{k}) \\
\end{array}
\right)
\label{Eq:TB1_ham_166}
\end{aligned}
\end{equation}
The coupling terms between $p_{xy}^t$ and $d_{1,2}$ orbitals are
\begin{equation}
\begin{aligned}
    S_{p_{xy}^{t_1}, d_1^1}(\bm{k})&=e^{i\Delta_{KT} k_3} S_{p_{xy}^t, d_1}^{\text{inplane}}(k_1,k_2), \\
    S_{p_{xy}^{t_1}, d_2^1}(\bm{k})&=e^{i\Delta_{KT} k_3} S_{p_{xy}^t, d_2}^{\text{inplane}}(k_1,k_2), \\
    S_{p_{xy}^{t_2}, d_1^2}(\bm{k})&=e^{-i\Delta_{KT} k_3} S_{p_{xy}^t, d_1}^{\text{inplane}}(k_1,k_2), \\
    S_{p_{xy}^{t_2}, d_2^2}(\bm{k})&=e^{-i\Delta_{KT} k_3} S_{p_{xy}^t, d_2}^{\text{inplane}}(k_1,k_2), \\
\end{aligned}
\end{equation}
where $\Delta_{KT}=0.09$ is the dislocation of Ge atoms in the $z$-direction, and $S_{p_{xy}^t, d_i}^{\text{inplane}}(k_1,k_2)$ is the same as $S_{p_{xy}^t, d_i}(\bm{k})$ defined in \cref{Eq_ham1_matrix_blocks}. 
The coupling terms between $p_z^h$ and $d_{2}$ orbitals are:
\begin{equation}
    \begin{aligned}
        S_{p_z^{h_1}, d_2^1}(\bm{k})&=
	t_{p_z^{h^1}, d_2}^{NN} e^{i\frac{k_3}{4}}
            S_{p_z^h,d_2}^{\text{inplane}}(k_1,k_2),\\
        S_{p_z^{h_1}, d_2^2}(\bm{k})&=-t_{p_z^{h^1}, d_2}^{NN} e^{-i\frac{k_3}{4}}
            S_{p_z^h,d_2}^{\text{inplane}}(k_1,k_2),\\
        S_{p_z^{h_2}, d_2^1}(\bm{k})&=
	-t_{p_z^{h_2}, d_2}^{NN} e^{-i\frac{k_3}{4}}
            S_{p_z^h,d_2}^{\text{inplane}}(k_1,k_2),\\
        S_{p_z^{h_2}, d_2^2}(\bm{k})&=
        t_{p_z^{h^2}, d_2}^{NN} e^{i\frac{k_3}{4}}
            S_{p_z^h,d_2}^{\text{inplane}}(k_1,k_2),\\
        S_{p_z^h,d_2}^{\text{inplane}}(k_1,k_2)&=
        \left(
	\begin{matrix}
		e^{\frac{i}{6}(k_1+2k_2)}  & e^{\frac{i}{6}(k_1-k_2)}  & e^{-\frac{i}{6}(2k_1+k_2)} \\
		e^{-\frac{i}{6}(k_1+2k_2)} &  e^{-\frac{i}{6}(k_1-k_2)}  & e^{\frac{i}{6}(2k_1+k_2)}\\
	\end{matrix}
	\right),\\
    \end{aligned}
\end{equation}
where $t_{p_z^{h^1}, d_4}^{NN}$ and $t_{p_z^{h^2}, d_4}^{NN}$ can take independent values.
$H_{p_{z}^{h_1}}(k_1,k_2),H_{p_{z}^{h_2}}(k_1,k_2)$ have the same form as $H_{p_{z}^{h}}(k_1,k_2)$ in \cref{Eq_ham2_matrix_blocks}, but $\mu_{p_z^{h_i}},t_{p_z^{h_i}}^{NN}$ can be different. This is because the honeycomb Ge's in MgFe$_6$Ge$_6$ form two honeycomb lattices on $z=0$ and $z=0.5$ planes which are symmetry-independent. Notice that in FeGe the honeycomb layers are related by lattice translations, but in MgFe$_6$Ge$_6$ there is no symmetry that relates them. In fact, the Mg atoms lie on the honeycomb layer on $z=\frac{1}{2}$ plane which indeed makes them inequivalent. Other matrix blocks have the same form as in \cref{Eq_ham1_matrix_blocks}.

\subsubsection{$H_2(k)$}
The second group has $p_z$ orbitals of Ge at $2c$ and $2d$ (denoted by $p_z^{h_1}$ and $p_z^{h_2}$), $p_z$ orbitals of Ge at $2e$ (denoted by $p_z^t$), and $d_{xz},d_{yz}$ (denoted by $d_3,d_4$) orbitals of Fe at $6i$. 
The TB Hamiltonian has the form
\begin{equation}
\begin{aligned}
H_2^{[166]}(\bm{k})=
&\left(
\begin{array}{cc|cc|cc|cc}
    H_{p_{z}^{h_1}}(k_{1,2}) & \bm{0} & 0 & 0 & \bm{0} & S_{p_z^{h_1}, d_4^1}(\bm{k}) &  \bm{0} & S_{p_z^{h_1}, d_4^2}(\bm{k}) \\
    & H_{p_{z}^{h_2}}(k_{1,2}) & 0 & 0 & \bm{0}  & S_{p_z^{h_2}, d_4^1}(\bm{k}) & \bm{0} & S_{p_z^{h_2}, d_4^2}(\bm{k})\\\hline
    & & \epsilon_{p_z^t} & S_{p_z^t}^{z}(k_3) & S_{p_z^{t_1}, d_3^1}(\bm{k}) & \bm{0} & \bm{0} & \bm{0} \\
    & & & \epsilon_{p_z^t} & \bm{0} & \bm{0} & S_{p_z^{t_2}, d_3^2}(\bm{k}) & \bm{0} \\\hline
    &  &  &  & H_{d_3}(k_{1,2}) & S_{d_3,d_4}(k_{1,2}) & S_{d_3^1,d_3^2}(\bm{k}) & \bm{0} \\
    &  &  &  & & H_{d_4}(k_{1,2}) & \bm{0} & S_{d_4^1,d_4^2}(\bm{k}) \\\hline
    &  &  &  & & & H_{d_3}(k_{1,2}) & S_{d_3,d_4}(k_{1,2}) \\ 
    H.c. & & & & & & & H_{d_4}(k_{1,2}) \\
\end{array}
\right),\\
\label{Eq:TB2_ham_166}
\end{aligned}
\end{equation}
where we use $k_{1,2}=(k_1,k_2)$, and $H_{p_{z}^{h_i}}(k_{1,2})$ have the same form as $H_{p_{z}^{h}}(k_{1},k_2)$ in \cref{Eq_ham2_matrix_blocks}, but hopping parameters can be different.
The $d_{i=3,4}$ Hamiltonians are
\begin{equation}
    \begin{aligned}
        H_{d_i}(k_1,k_2) &= \mu_{d_i} \bm{1}_3 + t_{d_i}^{NN} H_{\text{kagome}}^{\text{inplane},xz}(k_1,k_2),\\
        S_{d_i^1,d_i^2}(\bm{k}) &=
        \left(t_{d_i}^{zNN,+} e^{i\frac{k_3}{2}} + t_{d_i}^{zNN,-} e^{-i\frac{k_3}{2}}\right)\bm{1}_3 \\
        %&+\left(t_{d_i}^{zNNN,+} e^{i\frac{k_3}{2}} + t_{d_i}^{zNNN,-} e^{-i\frac{k_3}{2}}\right) H_{\text{kagome}}^{\text{inplane},xz}(k_1,k_2),\\
        H_{\text{kagome}}^{\text{inplane},xz}(k_1,k_2)
        &=\left(
	\begin{matrix}
		0 &  \cos(\frac{k_2}{2})  & -\cos(\frac{k_1+k_2}{2}) \\
		& 0 & \cos(\frac{k_1}{2})  \\
		  H.c. &  & 0\\
	\end{matrix}
	\right),\\
    \end{aligned}
    \label{Eq:166_H2_blocks}
\end{equation}
Notice that we set the $z$-directional distance between two $d$ orbitals to be $\Delta_{KK}=\frac{1}{2}$ for simplicity (i.e., $e^{\pm i\frac{k_3}{2}}$ term in \cref{Eq:166_H2_blocks}), whose accurate value is 0.496. 

The coupling terms between $p_z^h$ and $d_4$ orbitals are:
\begin{equation}
    \begin{aligned}
        S_{p_z^{h_1}, d_4^1}(\bm{k})&=
	t_{p_z^{h^1}, d_4}^{NN} e^{i\frac{k_3}{4}}
            S_{p_z^h,d_4}^{\text{inplane}}(k_1,k_2),\\
        S_{p_z^{h_1}, d_4^2}(\bm{k})&=
        t_{p_z^{h^1}, d_4}^{NN} e^{-i\frac{k_3}{4}}
            S_{p_z^h,d_4}^{\text{inplane}}(k_1,k_2),\\
        S_{p_z^{h_2}, d_4^1}(\bm{k})&=
	t_{p_z^{h_2}, d_4}^{NN} e^{-i\frac{k_3}{4}}
            S_{p_z^h,d_4}^{\text{inplane}}(k_1,k_2),\\
        S_{p_z^{h_2}, d_4^2}(\bm{k})&=
        t_{p_z^{h_2}, d_4}^{NN} e^{i\frac{k_3}{4}}
            S_{p_z^h,d_4}^{\text{inplane}}(k_1,k_2),\\
        S_{p_z^h,d_4}^{\text{inplane}}(k_1,k_2)&=
        \left(
	\begin{matrix}
		-e^{\frac{i}{6}(k_1+2k_2)}  & e^{\frac{i}{6}(k_1-k_2)}  & -e^{-\frac{i}{6}(2k_1+k_2)} \\
		e^{-\frac{i}{6}(k_1+2k_2)} &  -e^{-\frac{i}{6}(k_1-k_2)}  & e^{\frac{i}{6}(2k_1+k_2)},\\
	\end{matrix}
	\right),\\
    \end{aligned}
\end{equation}
where $t_{p_z^{h^1}, d_4}^{NN}$ and $t_{p_z^{h^2}, d_4}^{NN}$ can take independent values, as honeycomb Ge's on $z=0$ and $z=0.5$ planes are symmetry-independent. 
The coupling terms between $p_z^t$ and $p_z^t$, as well as $p_z^t$ and $d$ orbitals, are
\begin{equation}
    \begin{aligned}
        S_{p_z^{t}}^z(k_3)&=t_{p_z^t}^{NN} e^{-i\Delta_{TT}k_3}, \\
        S_{p_z^{t_1},d_3^1}(\bm{k})&= e^{i\Delta_{KT} k_3}
	  \cdot t_{p_z^t, d_3}^{NN}\cdot 2i
	\left(
	\begin{matrix}
		\sin(\frac{k_1}{2})  & \sin(\frac{k_1+k_2}{2})  & \sin(\frac{k_2}{2}) \\
	\end{matrix}
	\right),\\
        S_{p_z^{t_2},d_3^2}(\bm{k})&= e^{-i\Delta_{KT} k_3}
	  \cdot t_{p_z^t, d_3}^{NN}\cdot 2i
	\left(
	\begin{matrix}
		\sin(\frac{k_1}{2})  & \sin(\frac{k_1+k_2}{2})  & \sin(\frac{k_2}{2}) \\
	\end{matrix}
	\right),\\ 
    \end{aligned}
\end{equation}
where $\Delta_{TT}=0.32$ is the distance in the $z$-direction between two closest triangular Ge atoms, and $\Delta_{KT}=0.09$ is the dislocation of triangular Ge atoms in the $z$-direction from the kagome layer. 
Other matrix blocks are the same as in \cref{Eq_ham2_matrix_blocks}. 
Notice that as in FeGe, the hopping between $d_4$ and $p_z^t$ is forbidden by the $M_{120}(=M_y)$ symmetry.

\subsubsection{$H_3(k)$}
For the third group of orbitals, i.e., $d_{z^2}$ of Fe at $6i$ and the bonding states of honeycomb Ge at $b^1@3f$ and $b^2@3g$ ($3f$ and $3g$ are on $z=0,\frac{1}{2}$ planes, respectively, as shown in \cref{SG191-wyckoff}), the TB Hamiltonian has the form
\begin{equation}
    \begin{aligned}
        H_3^{[166]}(\bm{k})&=
        \left(
        \begin{array}{cc|cc}
            H_{b^1}(k_1,k_2) & S_{b^1,b^2}(k_3) & S_{b^1,d_5^1}(\bm{k}) & S_{b^1,d_5^2}(\bm{k}) \\
            & H_{b^2}(k_1,k_2) & S_{b^2,d_5^1}(\bm{k}) & S_{b^2,d_5^2}(\bm{k}) \\\hline
            & & H_{d_5}(k_1,k_2) & S_{d_5^1, d_5^2}(k_3) \\
            H.c.& & & H_{d_5}(k_1,k_2) \\
        \end{array}
        \right),\\
        S_{d_5^1, d_5^2}(k_3) &= 
        (t_{d_5}^{zNN,+} e^{i\frac{k_3}{2}} + 
        t_{d_5}^{zNN,-} e^{-i\frac{k_3}{2}})\mathbf{1}_3\\ 
        S_{b^1, b^2}(k_3) &= 
        (t_{b}^{zNN,+} e^{i\frac{k_3}{2}} + 
        t_{b}^{zNN,-} e^{-i\frac{k_3}{2}})\mathbf{1}_3\\ 
        S_{b^1,d_5^1}(\bm{k}) &=e^{i\frac{k_3}{4}} (t_{b^1,d_5}^{NN} + t_{b^1,d_5}^{NNN} H_{\text{kagome}}^{\text{inplane},z^2}(k_1,k_2))\\
        S_{b^1,d_5^2}(\bm{k}) &= S_{b^1,d_5^1}(k_1,k_2,-k_3) \\
        S_{b^2,d_5^1}(\bm{k}) &=e^{-i\frac{k_3}{4}} (t_{b^2,d_5}^{NN} + t_{b^2,d_5}^{NNN} H_{\text{kagome}}^{\text{inplane},z^2}(k_1,k_2))\\
        S_{b^2,d_5^2}(\bm{k}) &= S_{b^2,d_5^1}(k_1,k_2,-k_3) \\        
    \end{aligned}
    \label{Eq:TB3_ham_166_withbonding}
\end{equation}
where $H_{b^{i=1,2}}(\bm{k}), H_{d_5}(\bm{k}), H_{\text{kagome}}^{\text{inplane},z^2}(\bm{k})$ are defined as in \cref{Eq:TB3_ham}. Note that the two sets of bonding states are symmetry-independent as they are on two different Wyckoff positions $3f$ and $3g$ on $z=0,\frac{1}{2}$ planes, respectively, and thus their onsite energies and hopping parameters can take independent values. This is similar to the honeycomb Ge $p_z^{h_1}@2c, p_z^{h_1}@2d$, which are symmetry-independent due to the extra Mg atom on $z=\frac{1}{2}$ layer.

Remember that in FeGe, the bonding states are used to introduce $d_{z^2}$ weights below $-2$ eV, which are far from $E_f$ and less relevant for the low energy physics near $E_f$. We build a simpler model of $d_{z^2}$ only that captures the band structure close to $E_f$. 
Similarly, for MgFe$_6$Ge$_6$, we build $H_3(\bm{k})$ using $d_{z^2}$ only and refit parameters to the bands near $E_f$:
\begin{equation}
    \begin{aligned}
        H_3^{[166]}(\bm{k})&=
        \left(
        \begin{array}{cc}
            H_{d_5}(k_1,k_2) & S_{d_5^1, d_5^2}(k_3) \\
            H.c.& H_{d_5}(k_1,k_2) \\
        \end{array}
        \right),\\      
    \end{aligned}
    \label{Eq:TB3_ham_166}
\end{equation}
The hopping parameters in this simplified model are the same as the ones used in FeGe in Sec.\ref{Sec:TB_H3}. 
Remark that this simplified model can be obtained by perturbing out the bonding states in \cref{Eq:TB3_ham_166_withbonding}. However, a straightforward 2nd-order perturbation cannot be applied as $d_{z^2}$ and the bonding states have close onsite energy and strong coupling. Thus the parameters in \cref{Eq:TB3_ham_166} are refitted for simplicity as in FeGe.

\subsubsection{Combining $s$ orbital of Mg}\label{Sec:combine_Mg_s}
In this section, we include the coupling of the $s$ orbital of Mg. The NN hopping parameters from Mg $s$ orbital to other orbitals are listed in \cref{table:hopping_Mg}. 
The hopping from $s$ of Mg to $p_z^{h_2}$ is forbidden by $M_z$ symmetry. 
It can be seen that the hoppings between $s$ and $d_{z^2}$ and $p_{z}^{h_1}$ are very small, and we ignore them. The distance between Mg (with $z=0.5$) and $p_z^{h_1}$ (with $z=0$) is large which explains their small hopping value. The $d_{z^2}$ orbital, however, is more localized in the $xy$ plane and thus has a small overlap with Mg $s$ as they have different $xy$ coordinates.
For $p_z^{t_1}$, the hopping has large values because $p_z^{t_1}$ and Mg $s$ both lie on the $(0,0,z)$-axis and have short distance of $0.34c$, which effectively push $p_z^{t_1}$ away from $E_f$.

\begin{table}[htbp]
\begin{tabular}{c|c|c|c|c|c|c|c}
\hline\hline
Hopping & $d_{z^2}$ & $d_{xz}$ & $d_{yz}$ & $d_{x^2-y^2}$ & $d_{xy}$ & $p_z^{t}$ & $p_z^{h_1}$\\ \hline
Mg $s$  & -0.01     & 0.31    & -        & -0.16         & -        & 1.81         & 0.08    \\ \hline\hline
\end{tabular}
\caption{\label{table:hopping_Mg}The NN hopping parameters of Mg $s$ orbitals to other orbitals, obtained from MLWFs. The onsite energy of $s$ orbital is $3.07$ eV. The five $d$ orbitals in the table denote the $d$ orbitals of Fe at $(0.5,0,0.748)$.}
\end{table}

We consider the S-matrices between Mg $s$ orbitals and $d_{x^2-y^2} (d_2), d_{xz} (d_3), p_z^t$, which have the form
\begin{equation}
\begin{aligned}
S_{s, d_2^1}(\bm{k})&= 2 t_{s,d_2} \cdot e^{-i\frac{k_3}{4}}
\left(\cos\left(\frac{k_1}{2}\right), \cos\left(\frac{k_1+k_2}{2}\right), \cos\left(\frac{k_2}{2}\right)\right) \\
S_{s, d_2^2}(\bm{k})&= 2 t_{s,d_2} \cdot e^{+i\frac{k_3}{4}}
\left(\cos\left(\frac{k_1}{2}\right), \cos\left(\frac{k_1+k_2}{2}\right), \cos\left(\frac{k_2}{2}\right)\right) \\
S_{s, d_3^1}(\bm{k})&= 2 t_{s,d_3} \cdot e^{-i\frac{k_3}{4}}
\left(\sin\left(\frac{k_1}{2}\right), \sin\left(\frac{k_1+k_2}{2}\right), \sin\left(\frac{k_2}{2}\right)\right) \\
S_{s, d_3^2}(\bm{k})&= 2 t_{s,d_3} \cdot e^{+i\frac{k_3}{4}}
\left(-\sin\left(\frac{k_1}{2}\right), -\sin\left(\frac{k_1+k_2}{2}\right), -\sin\left(\frac{k_2}{2}\right)\right) \\
%S_{s, d_5^1}(\bm{k})&= 2 t_{s,d_5} \cdot e^{-i\frac{k_3}{4}}\left(\cos\left(\frac{k_1}{2}\right), \cos\left(\frac{k_1+k_2}{2}\right), \cos\left(\frac{k_2}{2}\right)\right) \\
%S_{s, d_5^2}(\bm{k})&= 2 t_{s,d_5} \cdot e^{+i\frac{k_3}{4}}\left(\cos\left(\frac{k_1}{2}\right), \cos\left(\frac{k_1+k_2}{2}\right), \cos\left(\frac{k_2}{2}\right)\right) \\
S_{s,p_z^t}(\bm{k}) &= t_{s,p_z^t} \left(e^{-i \Delta_{MgT}k_3}, e^{+i \Delta_{MgT}k_3} \right) \\
%S_{s,p_z^{h_1}}(\bm{k}) &= t_{s,p_z^{h_1}}\cdot 2i\sin\left(\frac{k_3}{2}\right) \left(e^{-\frac{i}{3}(2k_1+k_2)}(1+e^{ik_1}+e^{i(k_1+k_2)}),e^{-\frac{i}{3}(k_1+2k_2)}(1+e^{ik_2}+e^{i(k_1+k_2)})\right)
\end{aligned}
\end{equation}
where $\Delta_{MgT}=0.34$. 
These coupling terms are added to previous models and will be perturbed out in the next section, as the onsite energy differences are large (about 4 eV) between the $s$ orbital of Mg and these orbitals.

\subsubsection{Full TB model}
We use second-order perturbation theory \citeSI{SI:S-matrix} to decouple the three aforementioned models, i.e., perturb out $p_z^h$ in $H_1(\kk)$ (give 2nd-order terms to $d_2$), $p_z^t$ from $H_2(\kk)$ (give 2nd-order terms to $d_3$), and $s$ of Mg that couples with $d_2$ and $d_3$. The three decoupled Hamiltonians have the form
\begin{equation}
\begin{aligned}
H_1^{[166]}(\bm{k})=
&\left(
\begin{array}{ccc|ccc}
    H_{p_{xy}^t}(k_{1,2}) & S_{p_{xy}^{t_1}, d_1^1}(\bm{k}) & S_{p_{xy}^{t_1}, d_2^1}(\bm{k}) 
    & \bm{0} & \bm{0} & \bm{0} \\
    & H_{d_1}(k_{1,2})  & S_{d_1,d_2}(k_{1,2})
    & \bm{0} & \bm{0} & \bm{0} \\
    & & {\color{blue} H_{d_2^1}^\prime(k_{1,2})} 
    & \bm{0} & \bm{0} & {\color{blue} S_{d_2^1,d_2^2}^\prime(\bm{k})} \\\hline
    & & & H_{p_{xy}^t}(k_{1,2}) & S_{p_{xy}^{t_2}, d_1^2}(\bm{k}) & S_{p_{xy}^{t_2}, d_2^2}(\bm{k})\\
    & & &  & H_{d_1}(k_{1,2})  & S_{d_1,d_2}(k_{1,2})\\
    H.c. & & & & & {\color{blue} H_{d_2^2}^\prime(k_{1,2})} \\
    \end{array}
    \right),\\
H_2^{[166]}(\bm{k})=
&\left(
\begin{array}{cc|cc|cc}
    H_{p_{z}^{h_1}}^{\prime}(k_{1,2}) & \bm{0} & \bm{0} & S_{p_z^{h_1}, d_4^1}(\bm{k}) &  \bm{0} & S_{p_z^{h_1}, d_4^2}(\bm{k}) \\
    & H_{p_{z}^{h_2}}^{\prime}(k_{1,2}) & \bm{0}  & S_{p_z^{h_2}, d_4^1}(\bm{k}) & \bm{0} & S_{p_z^{h_2}, d_4^2}(\bm{k})\\\hline
    & & {\color{blue} H_{d_3^1}^\prime (k_{1,2})} & S_{d_3,d_4}(k_{1,2}) & {\color{blue} S_{d_3^1,d_3^2}^\prime(\bm{k})} & \bm{0} \\
    & & & H_{d_4^1}(k_{1,2}) & \bm{0} & S_{d_4^1,d_4^2}(\bm{k}) \\\hline
    & & & & {\color{blue} H_{d_3^2}^\prime (k_{1,2})} & S_{d_3,d_4}(k_{1,2})\\
    H.c.& &  & & & H_{d_4^2}(k_{1,2}) \\
\end{array}
\right),\\
H_3^{[166]}(\bm{k})&=
\left(
\begin{array}{cc}
    H_{d_5}(k_1,k_2) & S_{d_5^1, d_5^2}(k_3) \\
    H.c.& H_{d_5}(k_1,k_2) \\
\end{array}
\right),\\    
\label{eq_TB_166_decoupled}
\end{aligned}
\end{equation}
The matrix blocks with prime contain perturbed terms, with blue-colored terms having contributions from Mg $s$ orbital, and are defined as 
\begin{equation}
\begin{aligned}
    H_{d_2^1}^\prime(k_{1,2}) &= H_{d_2}(k_{1,2}) + H_{d_2^1,p_z^{h_1}}^{(2)}(k_{1,2}) + H_{d_2^1,p_z^{h_2}}^{(2)}(k_{1,2}) + H_{d_2^1, s}^{(2)}(k_{1,2}) \\
    H_{d_2^2}^\prime(k_{1,2}) &= H_{d_2}(k_{1,2}) + H_{d_2^2,p_z^{h_1}}^{(2)}(k_{1,2}) + H_{d_2^2,p_z^{h_2}}^{(2)}(k_{1,2}) + H_{d_2^2, s}^{(2)}(k_{1,2}) \\
    S_{d_2^1,d_2^2}^\prime(\bm{k}) &= S_{d_2^1,d_2^2,p_z^{h_1}}^{(2)}(\bm{k}) + S_{d_2^1,d_2^2,p_z^{h_2}}^{(2)}(\bm{k}) + S_{d_2^1, d_2^2, s}^{(2)}(\bm{k})\\
    H_{p_{z}^{h_1}}^{\prime}(k_{1,2}) &= H_{p_{z}^{h_1}}(k_{1,2})+H_{p_z^{h_1},d_2^1}^{(2)}(k_{1,2})+H_{p_z^{h_1},d_2^2}^{(2)}(k_{1,2})\\
    H_{p_{z}^{h_2}}^{\prime}(k_{1,2}) &= H_{p_{z}^{h_2}}(k_{1,2})+H_{p_z^{h_2},d_2^1}^{(2)}(k_{1,2})+H_{p_z^{h_2},d_2^2}^{(2)}(k_{1,2})\\    
    H_{d_3^1}^\prime (k_{1,2}) &= H_{d_3}(k_{1,2}) + H_{d_3^1, p_z^{t_1}}^{(2)}(k_{1,2}) + H_{d_3^1, s}^{(2)}(k_{1,2}) \\
    H_{d_3^2}^\prime (k_{1,2}) &= H_{d_3}(k_{1,2}) + H_{d_3^2, p_z^{t_1}}^{(2)}(k_{1,2}) + H_{d_3^2, s}^{(2)}(k_{1,2}) \\
    S_{d_3^1,d_3^2}^\prime(\bm{k}) &= S_{d_3^1,d_3^2}(\bm{k}) + S_{d_3^1, d_3^2, s}^{(2)}(\bm{k}) \\
    %H_{d_4^1}^\prime(k_{1,2}) &= H_{d_4}(k_{1,2}) + H_{d_4^1,p_z^{h_1}}^{(2)}(k_{1,2}) + H_{d_4^1,p_z^{h_2}}^{(2)}(k_{1,2}) \\
    %H_{d_4^2}^\prime(k_{1,2}) &= H_{d_4}(k_{1,2}) + H_{d_4^2,p_z^{h_1}}^{(2)}(k_{1,2}) + H_{d_4^2, p_z^{h_2}}^{(2)}(k_{1,2})\\
    %S_{d_4^1,d_4^2}^\prime(\bm{k}) &= S_{d_4^1,d_4^2,p_z^{h_1}}^{(2)}(\bm{k}) + S_{d_4^1,d_4^2,p_z^{h_2}}^{(2)}(\bm{k}), \\
\end{aligned}
\end{equation}
where $k_{1,2}=(k_1,k_2)$, and the second-order perturbed terms are defined as
\begin{equation}
\begin{aligned}
    H_{d,p}^{(2)}(\bm{k}) &= \frac{1}{\mu_d-\mu_p}S_{p,d}^\dagger(\bm{k}) S_{p,d}(\bm{k})\\
    S_{d_i,d_j,p}^{(2)}(\bm{k}) &= \frac{1}{2}(\frac{1}{\mu_{d_i}-\mu_{p}} + \frac{1}{\mu_{d_j}-\mu_{p}}) S^\dagger_{p, d_i}(\bm{k}) S_{p,d_j}\\
\end{aligned}
\end{equation}
We ignore the $S$-matrix between $d_2^{1,2}$ and $d_3^{1,2}$ generated from perturbation of Mg $s$ orbital because it has negligible effect for bands near $E_f$ and will spoil the separation of three orbital groups.
Notice that $S_{d_2^1,d_2^2}^\prime(\bm{k})$ cannot be further perturbed out as $d_2^1$ and $d_2^2$ have the same onsite energy.

\begin{figure}[htbp]
	\centering
	\includegraphics[width=1\textwidth]{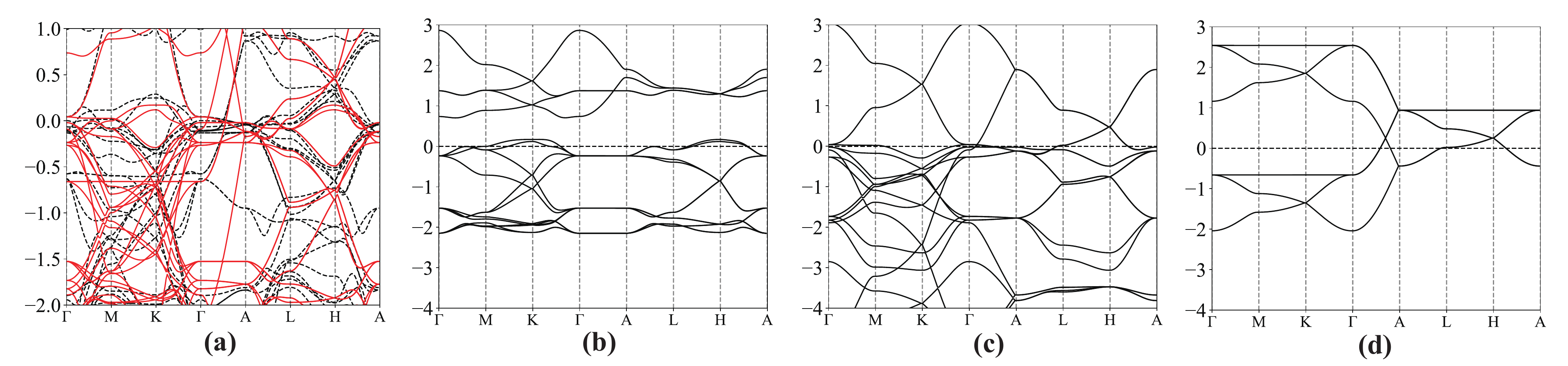}
	\caption{\label{fig_TBband_MgFe6Ge6} (a) The fitted TB band structures of $\text{MgFe}_6\text{Ge}_6$ using the decoupled TB model, i.e., combining three Hamiltonians $H_{i=1,2,3}^{[166]}$ defined in \cref{eq_TB_166_decoupled}(red lines), with a comparison with DFT bands in the PM phase (black dashed lines). 
    (b)-(d): Band structure of the decoupled $H_1^{[166]}$, $H_2^{[166]}$, and $H_3^{[166]}$ defined in \cref{eq_TB_166_decoupled}, respectively.
    }
\end{figure}

By direct-summing these three Hamiltonians, we arrive at 
\begin{equation}
    H^{[166]}(\bm{k})=H_1^{[166]}(\bm{k})\oplus H_2^{[166]}(\bm{k}) \oplus H_3^{[166]}(\bm{k}),
    \label{eq_decoupled_ham_166}
\end{equation}
which gives the final single-particle Hamiltonian
\begin{equation}
    \hat{H}_0 = \sum_{\bm{k}}\sum_{ij}
    H_{ij}(\bm{k}) c^\dagger_{\bm{k}i}c_{\bm{k}j},
    \label{Eq:FeGe_H_single_particle_H_166}
\end{equation}
where the Bloch electron operators $c_{\bm{k}i}$ are defined as the Fourier transform of the following orbital basis
\begin{equation}
\begin{aligned}
    (&p_x@2e_1, p_y@2e_1, 
    d_{xy}@6i_1, d_{xy}@6i_2, d_{xy}@6i_3, 
    d_{x^2-y^2}@6i_1, d_{x^2-y^2}6i_2, d_{x^2-y^2}@6i_3,\\
    &p_x@2e_2, p_y@2e_2, 
    d_{xy}@6i_4, d_{xy}@6i_5, d_{xy}@6i_6, 
    d_{x^2-y^2}@6i_4, d_{x^2-y^2}6i_5, d_{x^2-y^2}@6i_6)\\
    &\oplus(p_z^{h}@2c_1, p_z^h@2c_2, p_z^{h}@2d_1, p_z^h@2d_2, d_{xz}@6i_1, d_{xz}@6i_2, d_{xz}@6i_3, 
    d_{yz}@6i_1, d_{yz}@6i_2, d_{yz}@6i_3, \\
    &d_{xz}@6i_4, d_{xz}@6i_5, d_{xz}@6i_6, 
    d_{yz}@6i_4, d_{yz}@6i_5, d_{yz}@6i_6)\\
    %&\oplus(b@3f, b@3g, 
    &\oplus(d_{z^2}@6i_1, d_{z^2}@6i_2, d_{z^2}@6i_3,
    d_{z^2}@6i_4, d_{z^2}@6i_5, d_{z^2}@6i_6),\\
    \label{eq_final_TB_basis_166}
\end{aligned}
\end{equation}
where $2e_j$ and $6i_j$ denote the corresponding sites in $2e$ ($z=0.159$) and $6i$ ($z=0.252$) Wyckoff positions defined in \cref{SG191-wyckoff}, and $d$ orbitals defined in the local coordinate system in \cref{Eq:local_coordinate}. 
The band structure is shown in \cref{fig_TBband_MgFe6Ge6}, with parameters given in \cref{table_TBpara_MgFe6Ge6}.

\begin{table}[htbp]
\begin{tabular}{c|c|c|c|c|c|c|c|c|c|c|c|c|c}
\hline\hline
Parameter & $\mu_{p_{xy}^t}$ & $\mu_{d_1}$ & $\mu_{d_2}$ & $t_{d_1}^{NN}$     & $t_{d_1}^{NNN}$ & $t_{d_2}^{NN}$ & $t_{d_2}^{NNN}$    & $t_{d_1,d_2}^{NN}$ & $t_{d_1,d_2}^{NNN}$ & $t_{p_{xy}^t, d_1}^{NN}$ &  $t_{p_{xy}^t, d_2}^{NN}$  \\ \hline
Value/eV & -1.8 & -1.25 & -0.53 & 0.49 & 0.03  & 0.02 & 0.20 & -0.25 & -0.16 & 0.82 & 1.10
\\ \hline
Parameter & {\color{blue}$\mu_{p_z^{h_i}}$} & $\mu_{d_3}$ & $\mu_{d_4}$ & $t_{d_3,d_4}^{NN}$ & $t_{d_3}^{NN}$  & $t_{d_4}^{NN}$ & 
{\color{blue}$t_{d_4}^{zNN\pm}$}  & 
{\color{blue}$t_{p_z^{h_i},d_4}^{NN}$} & {\color{blue}$t_{p_z^{h_i}}^{NN}$} & 
{\color{blue}$t_{p_z^{h_i}, d_2}^{NN}$} & $\mu_{p_z^t}$ & $t_{p_z^t}^{NN}$ & $t_{p_z^t, d_3}^{NN}$  
\\ \hline
Value/eV & -1.41 & -0.88 & -0.83 & -0.22 & -0.23 & 0.20 & 0.10 & 0.77  & -0.46 & 0.31 & 0.60 & 0.30 & 0.45 
\\ \hline
Parameter & $\mu_{d_5}$  & $t_{d_5}^{NN}$ & 
{\color{blue}$t_{d_5}^{zNN\pm}$} & $\mu_b$ & $t_b^{NN}$ & {\color{blue}$t_b^{zNN\pm}$} & 
{\color{blue}$t_{b^i, d_5}^{NN}$} & 
{\color{blue}$t_{b^i,d_5}^{NNN}$} &
$\mu_s$ & $t_{s,d_2}$ & $t_{s, d_3}$ & $t_{s,p_z^t}$
\\ \hline
Value/eV & -1.10 & -0.25 & 0 & -0.6 & -0.1 & 0.9 & -1.1 & 0.04 &
3.07 & -0.16 & 0.31 & 1.81
\\ \hline\hline
\end{tabular}
\caption{\label{table_TBpara_MgFe6Ge6} Onsite energies and hoppings used in the TB model of $\text{MgFe}_6\text{Ge}_6$ defined in \cref{Eq:TB1_ham_166}, \cref{Eq:TB2_ham_166}, \cref{Eq:TB3_ham_166_withbonding}. 
$d_1$ to $d_5$ denote $d_{xy}, d_{x^2-y^2}, d_{xz}, d_{yz}, d_{z^2}$, respectively, $p_{xy}^t, p_z^t$ the $p_x, p_y$ and $p_z$ orbitals of the triangular Ge, $p_z^h$ and $b$ the $p_z$ and $sp^2$ bonding state of the honeycomb Ge, and $s$ is from Mg. 
For the simplified $H_3^{[166]}(\kk)$ of $d_{z^2}$ only defined in \cref{Eq:TB3_ham_166}, the parameters are $\mu_{d_5}=0.54, t_{d_5}^{NN}=-0.23, t_{d_5}^{zNN}=0.8$. 
Hoppings colored in blue denote that they can take independent values and can be further fine-tuned, but here we set them to be equal to the values from FeGe for simplicity, modified by the perturbations from Mg.}
\end{table}

\subsection{Interaction parameters of $\text{MgFe}_6\text{Ge}_6$}

The interaction parameters are computed for $\text{MgFe}_6\text{Ge}_6$ using the $d$-full model, i.e., Wannier functions are constructed from $s, p, d$ orbitals of Fe and $s, p$ orbitals of Ge, and exclude only the polarization between $d$ orbitals, as tabulated in \cref{d-full-MgFeGe}. 
In \cref{d-full-MgFeGe-cubic}, we list the $O_h$-symmetrized onsite Coulomb matrices elements $U_{ij}$ and $J_{ij}$. 
In \cref{d-full-MgFeGe-slater-U-F0F2F4} (\cref{d-full-MgFeGe-slater-U-F0F2}), we list the fitted Coulomb elements $U_{ij}$ and $J_{ij}$ using Slater integrals $F_0, F_2, F_4$ ($F_0,F_2$ only while fix $F_4=0.625F_2$). The values of interaction are similar to those of FeGe.

\begin{table}[htbp]
\begin{tabular}{c|cccccc|ccccc}
\hline\hline
$U_{ij}$  & $z^2$ & $xz$ & $yz$ & $x^2-y^2$ & $xy$ & $J_{ij}$ & $z^2$ & $xz$ & $yz$ & $x^2-y^2$ & $xy$ \\\hline
$z^2$     & 4.12 &  2.91 &  2.97 &  2.30  &  2.33 &    &    & 0.57& 0.57 & 0.93  &  0.94 \\
$xz$      &       &  3.97 &  2.46 &  2.48  &  2.52 &    &    &      & 0.79 & 0.78  &  0.80 \\
$yz$      &       &        &  4.12 &  2.56  &  2.60 &    &    &      &       & 0.79  &  0.80 \\
$x^2-y^2$ &       &        &        &  4.16  &  3.32 &    &    &      &       &        &  0.46 \\
$xy$      &       &        &        &         &  4.30 &    &    &      &       &        &        \\
\hline\hline
$U_{ij}^{NN}$  & $z^2$ & $xz$ & $yz$ & $x^2-y^2$ & $xy$ & $U_{ij}^{zNN}$ & $z^2$ & $xz$ & $yz$ & $x^2-y^2$ & $xy$ \\\hline
$z^2$     & 1.05 &  1.05 &  1.07 &  1.09  &  1.09 &    &    & 0.91& 0.91 & 0.91  &  0.91 \\
$xz$      &       &  1.05 &  1.06 &  1.08  &  1.09 &    &    &      & 0.91 & 0.91  &  0.91 \\
$yz$      &       &        &  1.10 &  1.11  &  1.11 &    &    &      &       & 0.90  &  0.91 \\
$x^2-y^2$ &       &        &        &  1.14  &  1.14 &    &    &      &       &        &  0.90 \\
$xy$      &       &        &        &         &  1.15 &    &    &      &       &        &        \\
\hline\hline
\end{tabular}
\caption{\label{d-full-MgFeGe} The Coulomb interaction $U_{ij}$ and $J_{ij}$ of $d$ orbitals, and $U_{ij}^{NN}$ and $U_{ij}^{NNN}$ between NN and NNN $d$ in $d-$full model of $\text{MgFe}_6\text{Ge}_6$.
The onsite Hubbard-Kanamori parameters for five $d$ orbitals are $\mathcal{U}=4.13, \mathcal{U}^\prime=2.64, \mathcal{J}=0.74$. 
The averaged NN and NNN interactions are $\overline{U}^{NN}=1.09, \overline{U}^{zNN}=0.91$, with root mean square error being 0.030 and 0.004, respectively. All numbers are in eV.}
\end{table}

\begin{table}[htbp]
\begin{tabular}{c|cccccc|ccccc}
\hline\hline
$U_{ij}$  & $z^2$ & $xz$ & $yz$ & $x^2-y^2$ & $xy$ & $J_{ij}$ & $z^2$ & $xz$ & $yz$ & $x^2-y^2$ & $xy$ \\\hline
$z^2$     & 4.15 &  2.99 &  2.99 &  2.30  &  2.30 &    &    & 0.57& 0.57 & 0.92  &  0.91 \\
$xz$      &       &  4.13 &  2.52 &  2.53  &  2.52 &    &    &      & 0.80 & 0.80  &  0.80 \\
$yz$      &       &        &  4.13 &  2.53  &  2.52 &    &    &      &       & 0.80  &  0.80 \\
$x^2-y^2$ &       &        &        &  4.15  &  3.23 &    &    &      &       &        &  0.45 \\
$xy$      &       &        &        &         &  4.13 &    &    &      &       &        &        \\
\hline\hline
\end{tabular}
\caption{\label{d-full-MgFeGe-cubic} The Coulomb interaction $U_{ij}$ and $J_{ij}$ of $d$ orbitals for the $d$-full model symmetrized using $O_h$ symmetries in $\text{MgFe}_6\text{Ge}_6$.
The root mean square error of the averaged $U_{ij}$ and $J_{ij}$ (error between the fitted values and the DFT values) is 0.153 and 0.050, respectively.}
\end{table}

\begin{table}[htbp]
\begin{tabular}{c|cccccc|ccccc}
\hline\hline
$U_{ij}$  & $z^2$ & $xz$ & $yz$ & $x^2-y^2$ & $xy$ & $J_{ij}$ & $z^2$ & $xz$ & $yz$ & $x^2-y^2$ & $xy$ \\\hline
$z^2$     & 4.13 &  2.99 &  2.99 &  2.30  &  2.30 &    &    & 0.57& 0.57 & 0.92  &  0.92 \\
$xz$      &       &  4.13 &  2.53 &  2.53  &  2.53 &    &    &      & 0.80 & 0.80  &  0.80 \\
$yz$      &       &        &  4.13 &  2.53  &  2.53 &    &    &      &       & 0.80  &  0.80 \\
$x^2-y^2$ &       &        &        &  4.13  &  3.22 &    &    &      &       &        &  0.45 \\
$xy$      &       &        &        &         &  4.13 &    &    &      &       &        &        \\
\hline\hline
\end{tabular}
\caption{\label{d-full-MgFeGe-slater-U-F0F2F4} The Coulomb interaction $U_{ij}$ and $J_{ij}$ of $d$ orbitals fitted using \cref{eq_slater_Uijkl} in $\text{MgFe}_6\text{Ge}_6$, where the Slater integrals $F^0$, $F^2$, and $F_4$ are fitted to the full Coulomb matrix $U_{ijkl}$ symmetrized by $O_h$ symmetries. The fitted parameters are $F^0=2.942, F^2=8.842, F^4=5.724$.
The root mean square error of the fitted $U_{ij}$ and $J_{ij}$ is 0.068 and 0.012, respectively.}
\end{table}

\begin{table}[htbp]
\begin{tabular}{c|cccccc|ccccc}
\hline\hline
$U_{ij}$  & $z^2$ & $xz$ & $yz$ & $x^2-y^2$ & $xy$ & $J_{ij}$ & $z^2$ & $xz$ & $yz$ & $x^2-y^2$ & $xy$ \\\hline
$z^2$     & 4.12 &  3.00 &  3.00 &  2.29  &  2.29 &    &    & 0.56& 0.56 & 0.92  &  0.92 \\
$xz$      &       &  4.12 &  2.53 &  2.53  &  2.53 &    &    &      & 0.80 & 0.80  &  0.80 \\
$yz$      &       &        &  4.12 &  2.53  &  2.53 &    &    &      &       & 0.80  &  0.80 \\
$x^2-y^2$ &       &        &        &  4.12  &  3.24 &    &    &      &       &        &  0.44 \\
$xy$      &       &        &        &         &  4.12 &    &    &      &       &        &        \\
\hline\hline
\end{tabular}
\caption{\label{d-full-MgFeGe-slater-U-F0F2} The Coulomb interaction $U_{ij}$ and $J_{ij}$ of $d$ orbitals fitted using \cref{eq_slater_Uijkl} in $\text{MgFe}_6\text{Ge}_6$, where the Slater integrals $F^0$, $F^2$ are fitted to the full Coulomb matrix $U_{ijkl}$ symmetrized by $O_h$ symmetries, while $F^4=0.625F^2$ are fixed. The fitted parameters are $F^0=2.942, F^2=8.899, F^4=5.562$.
The root mean square error of the fitted $U_{ij}$ and $J_{ij}$ is 0.069 and 0.013, respectively.}
\end{table}

\clearpage
\section{LEGO-like building blocks for the kagome 1:3:5 family}\label{app:sec:135-model}

In this section, we generalize the LEGO-like building blocks in the 1:1 and 1:6:6 families to the kagome 1:3:5 family and build minimal TB models. We consider three different material classes in the 1:3:5 family, i.e., \ch{CsCr3Sb5}\cite{liu2024superconductivity, guo2024ubiquitous, li2024correlated, xu2023frustrated, liu2023superconductivity}, with multiple quasi-flat bands near the Fermi level $E_f$, together with \ch{CsV3Sb5}\cite{ORT19,CHO21b,KAN21a,ORT21,ORT21a, kautzsch2023structural}, and \ch{CsTi3Bi5}\cite{yang2023observation, yang2024superconductivity, liu2023tunable, zhou2023physical, yi2023superconducting} which host multiple vHSs near $E_f$. All three compounds have been reported as superconductors with intriguing properties, exhibiting various types of charge density waves, spin density waves, or nematic transitions.

In the following, we first study \textit{ab initio} band structures in the kagome 1:3:5 family, emphasizing the hidden $d$ orbital decoupling. We then adapt the LEGO-like building blocks in FeGe to the 1:3:5 family and build minimal TB models to reproduce the DFT band structures faithfully. CRPA interaction is also computed to build the interacting Hamiltonian.

\subsection{Crystal and Band Structures in the 1:3:5 Family \ch{CsCr3Sb5}, \ch{CsV3Sb5}, and \ch{CsTi3Bi5}}
\begin{figure}[hbp]
\centering
\includegraphics[width=1\textwidth]{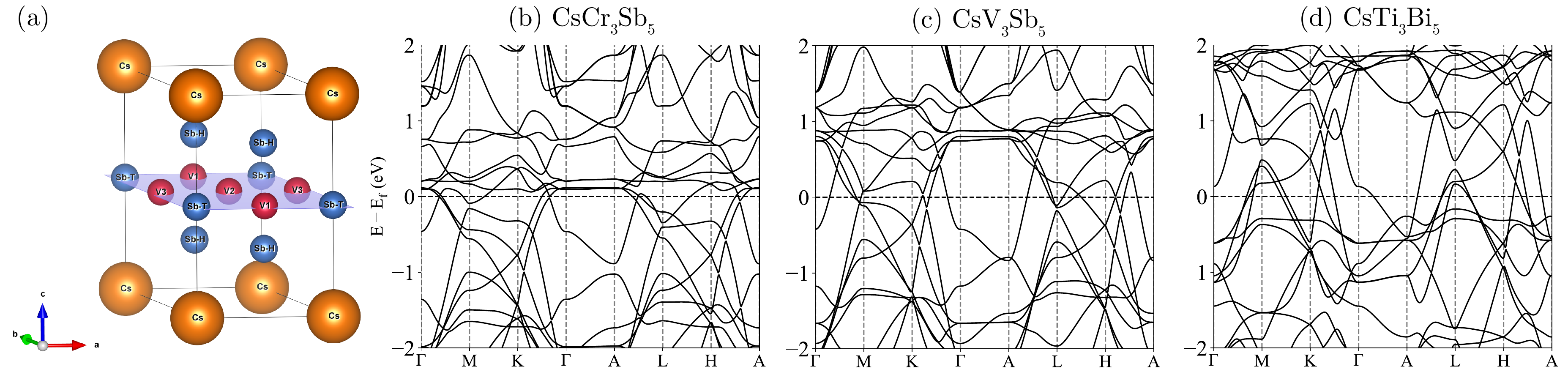}
\caption{\label{app:fig:135crystal_band} Crystal and band structures of kagome 1:3:5 family \ch{CsT3Z5} (T = V, Cr, Ti, Z = Sb, Bi).  (a) shows the crystal structure of \ch{CsV3Sb5}. (b)-(d) show the band structure \ch{CsCr3Sb5}, \ch{CsV3Sb5}, and \ch{CsTi3Bi5}, respectively. SOC is not included.
 }
\end{figure}

We begin with the crystal structure of the kagome 1:3:5 family. The structure has the same SG 191 symmetry, but unlike the 1:1 and 1:6:6 families, the 1:3:5 family features two honeycomb layers related by the $M_z$ symmetry, as shown in \cref{app:fig:135crystal_band} (a). The atomic positions are 
\begin{itemize}
\item Cs atoms at triangular site $1a=(0,0,0)$. 
\item Cr/V/Ti atoms at kagome site $3g=(\frac{1}{2},0,\frac{1}{2}), (\frac{1}{2}, \frac{1}{2}, \frac{1}{2}), (0, \frac{1}{2}, \frac{1}{2})$.
\item Sb/Bi atoms at triangular site $1b=(0, 0, \frac{1}{2})$, and honeycomb sites $4h=(\frac{1}{3}, \frac{2}{3}, \frac{1}{2}\pm z), (\frac{2}{3}, \frac{1}{3}, \frac{1}{2}\pm z)$. 
\end{itemize}

This structural difference leads to distinct band structures for the 1:3:5 family compared to the 1:1 and 1:6:6 families. In \cref{app:fig:135crystal_band} (b)-(d), we show the bands of \ch{CsCr3Sb5}, \ch{CsV3Sb5}, and \ch{CsTi3Bi5}, respectively. While their band structures are generally similar, they differ in Fermi level positions due to their different valence electron counts. In \ch{CsCr3Sb5}, multiple quasi-flat bands appear near $E_f$, whereas in \ch{CsV3Sb5} and \ch{CsTi3Bi5}, there are several van Hove singularities (vHSs) at $M$ and $L$, with quasi-flat bands located above $E_f$.

In contrast to the 1:1 and 1:6:6 families—where the out-of-plane $d$ orbitals ($d_{xz}$, $d_{yz}$, and $d_{z^2}$) are dispersive along $k_z$—the band structures of the 1:3:5 family are quasi-2D with weak $k_z$ dispersion. This reduced $k_z$-dispersion arises from the additional honeycomb Sb/Bi layer in the 1:3:5 materials, which increases the distance between neighboring kagome layers, thereby weakening interlayer coupling.

We then introduce one key simplification in the 1:3:5 family, focusing on \ch{CsCr3Sb5}. Because two honeycomb layers in the unit cell are related by $M_z$ symmetry, we transform them into effective $M_z$-even and -odd bases. The effective orbitals have Wannier centers on the kagome plane, allowing for the construction of simple TB models. 
To see this, we first build a faithful TB model from MLWFs in \ch{CsCr3Sb5}, based on the $d$ orbitals of Cr and $s,p$ orbitals of Sb. Denote the four honeycomb Sb atoms as
$Sb_{H_1}^{\pm}: (\frac{1}{3}, \frac{2}{3}, \frac{1}{2}\pm z), Sb_{H_2}^{\pm}: (\frac{2}{3}, \frac{1}{3}, \frac{1}{2}\pm z)$, and the electron operator of their $p_j$ orbitals as $c^\dagger_{H_i,\pm,p_j}$ $(i=1,2, j=x,y,z)$. 
The $M_z$-even/odd bases are formed by
\ba
&M_z~\text{even}:\quad 
c^\dagger_{H_i,\text{even}, p_{x/y}}=\frac{1}{\sqrt{2}}\left( c^\dagger_{H_i,+,p_{x/y}} + c^\dagger_{H_i,-,p_{x/y}}\right),\quad
c^\dagger_{H_i,\text{even}, p_{z}}=\frac{1}{\sqrt{2}}\left( c^\dagger_{H_i,+,p_{z}} - c^\dagger_{H_i,-,p_{z}}\right),\\
&M_z~\text{odd}:\quad 
c^\dagger_{H_i,\text{odd}, p_{x/y}}=\frac{1}{\sqrt{2}}\left( c^\dagger_{H_i,+,p_{x/y}} - c^\dagger_{H_i,-,p_{x/y}}\right),\quad
c^\dagger_{H_i,\text{odd}, p_{z}}=\frac{1}{\sqrt{2}}\left( c^\dagger_{H_i,+,p_{z}} + c^\dagger_{H_i,-,p_{z}}\right).\\
\label{app:eq:135-Mz-basis-Sb}
\ea
Note that $p_{x/y}$ and $p_z$ orbitals have opposite combination signs due to their opposite $M_z$ eigenvalues.

\begin{figure}[htbp]
\centering
\includegraphics[width=1\textwidth]{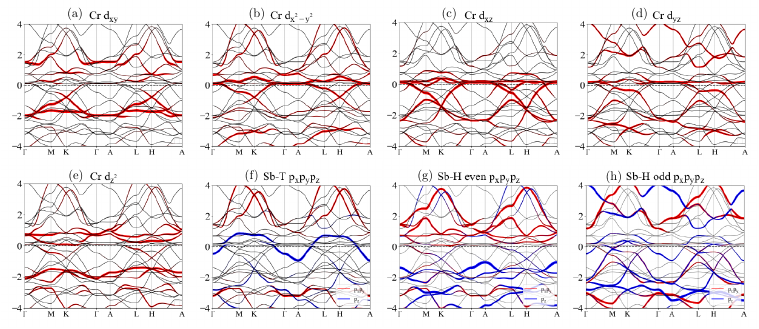}
\caption{\label{app:fig:Cr135_orb_weights} The orbital weights of \ch{CsCr3Sb5}. (a)-(e) are for five $d$ orbitals of Cr. (f) shows for the $p$ orbitals of triangular Sb. (g) and (h) show the $M_z$-odd and even combinations of honeycomb Sb $p$ orbitals, respectively, as defined in \cref{app:eq:135-Mz-basis-Sb}. 
 }
\end{figure}

We analyze the orbital weights in \ch{CsCr3Sb5} within the $M_z$-even/odd basis, as presented in \cref{app:fig:Cr135_orb_weights}. Near the Fermi level, the following quasi-flat bands are observed: one from Cr $d_{x^2-y^2}$, two from Cr $d_{xz}$ and $d_{yz}$, and one from $d_{z^2}$. The $M_z$-even and $M_z$-odd $p$ orbitals from the honeycomb Sb layers are decoupled, as illustrated in \cref{app:fig:Cr135_orb_weights} (g) and (h).

Comparing the band structures of the kagome 1:1 and 1:6:6 families, we observe similarities in the dispersions of the in-plane $d_{xy}$ and $d_{x^2-y^2}$ orbitals. However, the out-of-plane $d_{xz}$, $d_{yz}$, and $d_{z^2}$ orbitals exhibit distinct dispersions due to variations in $z$-directional hoppings influenced by the additional honeycomb Sb layers.

\subsection{\ch{CsCr3Sb5}}\label{app:sec:135-model-Cr}

We adapt the LEGO-like building block framework from FeGe to construct minimal tight-binding (TB) models for \ch{CsCr3Sb5}, incorporating adjustments due to the additional honeycomb layer of Sb atoms. The $M_z$-even and $M_z$-odd symmetric basis of the Sb $p$ orbitals play crucial roles in these models.

The orbital basis, serving as the LEGO-like building blocks in the minimal models of \ch{CsCr3Sb5}, are defined as follows:
\begin{itemize}
\item $H_1(\kk)$: Composed of $(p_x, p_y)$ orbitals from the triangular Sb atoms and $(d_{xy}, d_{x^2-y^2})$ orbitals from kagome Cr atoms, for a total of 8 orbitals. This block is identical to that of FeGe, involving in-plane $d$ orbitals of kagome Cr and $p$ orbitals of triangular Sb.

\item $H_2(\kk)$: Composed of $(d_{xz}, d_{yz})$ orbitals from Cr atoms and $M_z$-odd $(p_x, p_y, p_z)$ orbitals from honeycomb Sb atoms, for a total of 12 orbitals. 

\item $H_3(\kk)$: Composed of $d_{z^2}$ orbitals from Cr atoms and $M_z$-even $(p_x, p_y)$ orbitals from honeycomb Sb atoms, for a total of 7 orbitals. These orbitals form a bipartite crystalline lattice that yields a perfectly flat band if only inter-sublattice hoppings are present. 
\end{itemize}

The separation of orbitals into three decoupled groups is guided by symmetry analysis. $H_1(\kk)$ includes in-plane $d_{xy}, d_{x^2-y^2}$ orbitals of Cr, which couple strongly to triangular Sb $p_x, p_y$ orbitals due to the formation of $\sigma$-like bonds. 
$H_2(\kk)$ contains out-of-plane $d_{xz}, d_{yz}$ orbitals from Cr, which are $M_z$-odd and therefore couple exclusively to the $M_z$-odd $p_x, p_y, p_z$ orbitals of honeycomb Sb. These couplings dominate, as $z$-directional hopping is typically long-range and weaker.
$H_3(\kk)$ focuses on the $d_{z^2}$ orbitals of Cr, which are $M_z$-even and couple to $M_z$-even $p_x, p_y$ orbitals from honeycomb Sb. The $M_z$-even $p_z$ orbitals are far from Fermi level (see blue bands in \cref{app:fig:Cr135_orb_weights} (g)) and are omitted. 

In the following, we discuss in detail the three minimal TB models. In the model, $k_z$-dependence is ignored as the 1:3:5 family exhibits quasi-2D band structures. 

\subsubsection{$H_1(\kk)$}
The $H_1(\kk)$ sector has the same orbitals as in FeGe, i.e., the $(p_x,p_y)$ orbitals from triangular Sb and $(d_{xy}, d_{x^2-y^2})$ orbitals from Cr. 
Based on the Hamiltonian defined in\cref{Eq_ham1_matrix_blocks}, we find the following extra long-range hoppings are needed to obtain a faithful fitting to the DFT band structure in \ch{CsCr3Sb5}:
\begin{equation}
\tiny
\begin{aligned}
H_1^{\text{lr}}(\kk)=
&=
\begin{bmatrix}
    H_{p_{xy}^t}^{lr}(\kk) & S_{p_{xy}^t, d_{1}}^{lr}(\kk) & S_{p_{xy}^t, d_{2}}^{lr}(\kk) \\
    & H_{d_1}^{lr}(\kk) & S_{d_1, d_2}^{lr}(\kk)  \\
    H.c. & & H_{d_2}^{lr}(\kk)\\
\end{bmatrix}\\
H_{p_{xy}^t}(\kk) 
&= 
\begin{bmatrix}
 \frac{2}{\sqrt{3}} s_{p}^{1} (\cos (k_1)-2 (\cos (k_1+k_2)+\cos (k_2)))+2 s_{p}^{2} (\cos (k_1+k_2)+\cos (k_1)+\cos (k_2)) & 2 s_{p}^{1} (\cos (k_1+k_2)-\cos (k_2)) \\
 2 s_{p}^{1} (\cos (k_1+k_2)-\cos (k_2)) & 2 s_{p}^{2} (\cos (k_1+k_2)+\cos (k_1)+\cos (k_2))-2 \sqrt{3} s_{p}^{1} \cos (k_1) \\
\end{bmatrix}\\
&+ 2s_{p}^3 (\cos (k_1-k_2)+\cos (2 k_1+k_2)+\cos (k_1+2 k_2)) \mathbf{1}_2, 
\\
S_{p_{xy}^t, d_{1}}^{lr}(\kk)
&= 
\left[
\begin{array}{cc}
 4 i s_{p_{xy}^t, d_1}^1 \cos (k_1) \sin (\frac{k_1}{2}+k_2) & i ((s_{p_{xy}^t, d_1}^1-\sqrt{3} s_{p_{xy}^t, d_1}^2) \sin (\frac{1}{2} (k_1+3 k_2))-(s_{p_{xy}^t, d_1}^1+\sqrt{3} s_{p_{xy}^t, d_1}^2) \sin (\frac{1}{2} (3 k_1+k_2))) \\
 4 i s_{p_{xy}^t, d_1}^2 \sin (\frac{k_1}{2}) (\cos (k_1+k_2)+\cos (k_2)) & i (\sqrt{3} s_{p_{xy}^t, d_1}^1+s_{p_{xy}^t, d_1}^2) \sin (\frac{1}{2} (k_1+3 k_2))-i (\sqrt{3} s_{p_{xy}^t, d_1}^1-s_{p_{xy}^t, d_1}^2) \sin (\frac{1}{2} (3 k_1+k_2))  \\
\end{array}
\right.\\
&\left.
\begin{array}{cc}
\qquad\qquad\qquad & i ((s_{p_{xy}^t, d_1}^1-\sqrt{3} s_{p_{xy}^t, d_1}^2) \sin (k_1+\frac{3 k_2}{2})+(s_{p_{xy}^t, d_1}^1+\sqrt{3} s_{p_{xy}^t, d_1}^2) \sin (k_1-\frac{k_2}{2}))  \\
\qquad\qquad\qquad & -2 i (\sqrt{3} s_{p_{xy}^t, d_1}^1 \cos (k_2) \sin (k_1+\frac{k_2}{2})+s_{p_{xy}^t, d_1}^2 \sin (k_2) \cos (k_1+\frac{k_2}{2}))
\end{array}
\right]\\
S_{p_{xy}^t, d_{2}}^{lr}(\kk)
&= 
\left[
\begin{array}{cc}
 4 i s_{p_{xy}^t, d_2}^1 \sin (\frac{k_1}{2}) (\cos (k_1+k_2)+\cos (k_2)) &  
 i ((s_{p_{xy}^t, d_2}^1+\sqrt{3} s_{p_{xy}^t, d_2}^2) \sin (\frac{1}{2} (3 k_1+k_2))+(s_{p_{xy}^t, d_2}^1-\sqrt{3} s_{p_{xy}^t, d_2}^2) \sin (\frac{1}{2} (k_1+3 k_2))) \\
 
 4 i s_{p_{xy}^t, d_2}^2 \cos (k_1) \sin (\frac{k_1}{2}+k_2) 
 & i ((\sqrt{3} s_{p_{xy}^t, d_2}^1-s_{p_{xy}^t, d_2}^2) \sin (\frac{1}{2} (3 k_1+k_2))+(\sqrt{3} s_{p_{xy}^t, d_2}^1+s_{p_{xy}^t, d_2}^2) \sin (\frac{1}{2} (k_1+3 k_2))) \\
\end{array}
\right.\\
&\left.
\begin{array}{cc}
\qquad\qquad\qquad &  i ((\sqrt{3} s_{p_{xy}^t, d_2}^2-s_{p_{xy}^t, d_2}^1) \sin (k_1+\frac{3 k_2}{2})+(s_{p_{xy}^t, d_2}^1+\sqrt{3} s_{p_{xy}^t, d_2}^2) \sin (k_1-\frac{k_2}{2})) \\
 & 2 i (\sqrt{3} s_{p_{xy}^t, d_2}^1 \sin (k_2) \cos (k_1+\frac{k_2}{2})+s_{p_{xy}^t, d_2}^2 \cos (k_2) \sin (k_1+\frac{k_2}{2})) 
\end{array}
\right]\\
H_{d_1}^{lr}(\kk)
&=
\begin{bmatrix}
 0 & 2 s_{d12}^{1} \cos (k_2) \cos (k_1+\frac{k_2}{2}) & s_{d12}^{1} (\cos (\frac{1}{2} (3 k_1+k_2))+\cos (\frac{1}{2} (k_1+3 k_2))) \\
 2 s_{d12}^{1} \cos (k_2) \cos (k_1+\frac{k_2}{2}) & 0 & 2 s_{d12}^{1} \cos (k_1) \cos (\frac{k_1}{2}+k_2) \\
 s_{d12}^{1} (\cos (\frac{1}{2} (3 k_1+k_2))+\cos (\frac{1}{2} (k_1+3 k_2))) & 2 s_{d12}^{1} \cos (k_1) \cos (\frac{k_1}{2}+k_2) & 0 \\
\end{bmatrix} \\
H_{d_2}^{lr}(\kk)
&=
\begin{bmatrix}
 0 & 2 s_{d12}^{2} \cos (k_2) \cos (k_1+\frac{k_2}{2}) & s_{d12}^{2} (\cos (\frac{1}{2} (3 k_1+k_2))+\cos (\frac{1}{2} (k_1+3 k_2))) \\
 2 s_{d12}^{2} \cos (k_2) \cos (k_1+\frac{k_2}{2}) & 0 & 2 s_{d12}^{2} \cos (k_1) \cos (\frac{k_1}{2}+k_2) \\
 s_{d12}^{2} (\cos (\frac{1}{2} (3 k_1+k_2))+\cos (\frac{1}{2} (k_1+3 k_2))) & 2 s_{d12}^{2} \cos (k_1) \cos (\frac{k_1}{2}+k_2) & 0 \\
\end{bmatrix}\\
S_{d_1, d_2}^{lr}(\kk)
&=\frac{(s_{d12}^{1}+s_{d12}^{2})}{\sqrt{3}}
\begin{bmatrix}
 0 & -\cos (k_1+\frac{3 k_2}{2}) & \cos (\frac{1}{2} (k_1+3 k_2))\\
 \cos (k_1-\frac{k_2}{2}) & 0 & -\cos (\frac{1}{2} (k_1-2 k_2)) \\
 -\cos (\frac{1}{2} (3 k_1+k_2))& \cos (\frac{3 k_1}{2}+k_2) & 0 \\
\end{bmatrix}
\end{aligned}
\label{app:eq:135_H1}
\end{equation}

By fitting to DFT band structures, we obtain the parameters in $H_1(\kk)$ as tabulated in \cref{app:table:Cr135_H1_param}. The corresponding band structure is shown in \cref{app:fig:Cr135_fitted_TB} (a). There exists one unfilled quasi-flat band above $E_f$, which has the same origin as discussed in \cref{Sec:TB_H1} for FeGe.

\begin{table}[htbp]
\begin{tabular}{c|c|c|c|c|c|c|c|c|c|c|c|c|c|c}
\hline\hline
Parameter & $\mu_{p_{xy}^t}$ & $\mu_{d_1}$ & $\mu_{d_2}$ & $t_{d_1}^{NN}$ & $t_{d_1}^{NNN}$  & $t_{d_2}^{NN}$ & $t_{d_2}^{NNN}$   & $t_{d_1,d_2}^{NN}$ & $t_{d_1,d_2}^{NNN}$ & $t_{p_{xy}^t, d_1}^{NN}$ & $t_{p_{xy}^t, d_1}^{NNN}$ &  $t_{p_{xy}^t, d_2}^{NN}$ & $t_{p_{xy}^t, d_2}^{NNN}$
\\ \hline
Value/eV& -0.04 & -0.83& -0.33&  0.56&  0.06& -0.27&  0.17&  0.10&  0.04& -0.52&  0.19&  1.08&  0.19  \\ \hline
Parameter & $t_d^{4N1}$ & $t_d^{4N2}$ & $t_d^{4N3}$  & $t_d^{4N4}$ & $t_d^{4N5}$   & $s_{p}^1$ & $s_{p}^2$ & $s_{p}^3$ & $s_{p_{xy}^t,d_1}^1$ & $s_{p_{xy}^t,d_1}^2$ & $s_{p_{xy}^t,d_2}^1$ & $s_{p_{xy}^t,d_2}^2$ & $s_{d12}^1$ & $s_{d12}^2$ \\ \hline
Value/eV & -0.07&  0.05&  0.04&  0.15&  0.07 & -0.41& -0.77& 0.08&  0.0&  0.0& -0.08& -0.06& -0.04&  0.15
\\ \hline\hline
\end{tabular}
\caption{\label{app:table:Cr135_H1_param} The fitted hopping parameters in the $H_1(\kk)$ for \ch{CsCr3Sb5}. The parameters are defined in \cref{Eq_ham1_matrix_blocks} and \cref{app:eq:135_H1}.
}
\end{table}

\subsubsection{$H_2(\kk)$}
The $H_2(\kk)$ sector, however, takes different LEGO-like building blocks compared with FeGe. 
It contains $(d_{xz}, d_{yz})$ orbitals from Cr and $M_z$-odd $(p_x, p_y, p_z)$ orbitals of honeycomb Sb, i.e., 12 orbitals in total. 
The main modification is the inclusion of $M_z$-odd combinations of honeycomb Sb $p$ orbitals. 

The orbital bases in $H_2(\kk)$ are
\ba
(&d_{xz}@3g_1, d_{xz}@3g_2, d_{xz}@3g_3, d_{yz}@3g_1, d_{yz}@3g_2, d_{yz}@3g_3, \\
&p_z^o@2d_1, p_z^o@2d_2, 
p_x^o@2d_1, p_y^o@2d_1, p_x^o@2d_2, p_y^o@2d_2),
\label{app:eq:135_H2_basis}
\ea
where the superscript $o$ denotes the $M_z$-odd effective orbitals. 

The minimal TB model has the form
\begin{equation}
\footnotesize
\begin{aligned}
H_2(\kk) 
&=
\begin{bmatrix}
    H_{d_3}(\kk) & S_{d_3, d_4}(\kk) & \bm{0} & S_{d_3, p_{xy}^{o}}(\kk) \\
    & H_{d_4}(\kk) & S_{d_4, p_z^o}(\kk) & S_{d_4, p_{xy}^{o}}(\kk) \\
    & & H_{p_{z}^{o}}(\kk) & S_{p_{z}^{o}, p_{xy}^o}(\kk)\\
    H.c. & & & H_{p_{xy}^{o}}(\kk)
\end{bmatrix}\\
H_{p_{z}^{o}}(\kk) 
&=
\begin{bmatrix}
\epsilon_{p_z^o} & t_{p_z^o}^{NN} e^{-\frac{1}{3} i (k_1+2 k_2)} \left(e^{i (k_1+k_2)}+e^{i k_2}+1\right) \\
 c.c. & \epsilon_{p_z^{odd}}  \\
\end{bmatrix}\\
H_{p_{xy}^o}(\kk)
&= 
\epsilon_{p_{xy}^{odd}} \mathbf{1}_{4} \\
&+ \frac{1}{4} 
\begin{bmatrix}
0 & 0 & e^{-\frac{1}{3} i (k_1+2 k_2)} \left(4 t_{p_{xy}^o}^1+\left(1+e^{i k_1}\right) e^{i k_2} (t_{p_{xy}^o}^1+3 t_{p_{xy}^o}^2)\right) & -\sqrt{3} \left(-1+e^{i
   k_1}\right) e^{-\frac{1}{3} i (k_1-k_2)} (t_{p_{xy}^o}^1-t_{p_{xy}^o}^2) \\
0 & 0 & -\sqrt{3} \left(-1+e^{i k_1}\right) e^{-\frac{1}{3} i (k_1-k_2)} (t_{p_{xy}^o}^1-t_{p_{xy}^o}^2) & e^{-\frac{1}{3} i (k_1+2 k_2)} \left(4
   t_{p_{xy}^o}^2+\left(1+e^{i k_1}\right) e^{i k_2} (3 t_{p_{xy}^o}^1+t_{p_{xy}^o}^2)\right) \\
& & 0 & 0 \\
H.c. & & 0 & 0
\end{bmatrix}\\
S_{d_4,p_z^o}(\kk) 
&= t_{d_4,p_z^o}
\begin{bmatrix}
-e^{\frac{1}{6}i(k_1+2k_2)} & e^{-\frac{1}{6}i(k_1+2k_2)} \\
e^{\frac{1}{6}i(k_1-k_2)} & -e^{-\frac{1}{6}i(k_1-k_2)} \\
-e^{-\frac{1}{6}i(2k_1+k_2)} & e^{\frac{1}{6}i(2k_1+k_2)} \\
\end{bmatrix} \\
S_{p_{z}^{o}, p_{xy}^o}(\kk) 
&= t_{p_z^o, p_x^o}
\begin{bmatrix}
0 & 0 &  \left(-1+e^{i k_1}\right) e^{-\frac{1}{3} i (k_1-k_2)} & \frac{1}{\sqrt{3}} e^{-\frac{1}{3} i (k_1+2 k_2)} \left(e^{i (k_1+k_2)}+e^{i
   k_2}-2\right) \\
H.c. & & 0 & 0
\end{bmatrix}\\
S_{d_3, p_{xy}^{o}}(\kk)
&= t_{d_3,p_{xy}^o}
\begin{bmatrix}
  e^{\frac{1}{6} i (k_1+2 k_2)} & 0 &  e^{-\frac{1}{6} i (k_1+2 k_2)} & 0 \\
 \frac{1}{2}  e^{\frac{1}{6} i (k_1-k_2)} & \frac{1}{2} \sqrt{3}  e^{\frac{1}{6} i (k_1-k_2)} & \frac{1}{2}  e^{-\frac{1}{6} i
   (k_1-k_2)} & \frac{1}{2} \sqrt{3}  e^{-\frac{1}{6} i (k_1-k_2)} \\
 -\frac{1}{2}  e^{-\frac{1}{6} i (2 k_1+k_2)} & \frac{1}{2} \sqrt{3}  e^{-\frac{1}{6} i (2 k_1+k_2)} & -\frac{1}{2}  e^{\frac{1}{6} i (2
   k_1+k_2)} & \frac{1}{2} \sqrt{3}  e^{\frac{1}{6} i (2 k_1+k_2)}
\end{bmatrix}\\
S_{d_4, p_{xy}^{o}}(\kk)
&= t_{d_4, p_{xy}^o}
\begin{bmatrix}
 0 &  e^{\frac{1}{6} i (k_1+2 k_2)} & 0 &  e^{-\frac{1}{6} i (k_1+2 k_2)} \\
 -\frac{1}{2} \sqrt{3}  e^{\frac{1}{6} i (k_1-k_2)} & \frac{1}{2}  e^{\frac{1}{6} i (k_1-k_2)} & -\frac{1}{2} \sqrt{3}  e^{-\frac{1}{6} i
   (k_1-k_2)} & \frac{1}{2}  e^{-\frac{1}{6} i (k_1-k_2)} \\
 -\frac{1}{2} \sqrt{3}  e^{-\frac{1}{6} i (2 k_1+k_2)} & -\frac{1}{2}  e^{-\frac{1}{6} i (2 k_1+k_2)} & -\frac{1}{2} \sqrt{3}  e^{\frac{1}{6} i
   (2 k_1+k_2)} & -\frac{1}{2}  e^{\frac{1}{6} i (2 k_1+k_2)} 
\end{bmatrix}
\label{app:eq:135_H2}
\end{aligned}
\end{equation}
where $H_{d_3}(\kk), H_{d_4}(\kk), S_{d_3, d_4}(\kk)$ are defined in \cref{Eq_ham2_matrix_blocks}.

The fitted parameters in $H_2(\kk)$ are listed in \cref{app:table:Cr135_H2_param}, with the band structure shown in \cref{app:fig:Cr135_fitted_TB} (b). From the fitted parameters, we observe that the onsite energies of these orbitals are close, and coupling between $d_{xz/yz}$ and $M_z$-odd $p_{x/y/z}$ orbitals are at the order of 1 eV, which are very strong. Thus the resultant band structure shows strong hybridizations between orbitals and does not have clear kagome bands (of the NN kagome model). There exist two quasi-flat bands near $E_f$ with a Dirac crossing at $K$.

We also note that the $p_z$ orbital of triangular Sb (see blue bands in \cref{app:fig:Cr135_orb_weights} (f)) also couples to orbitals in $H_2(\kk)$, but we omit it for simplicity in $H_2(\kk)$. A simple one-orbital model for the triangular $p_z$ orbital has the form
\begin{equation}
    H_{p_z^t}(\kk) = \epsilon_{p_z^t} + 2 t_{p_z^t}^{NN} \left(\cos(k_1) + \cos(k_2) + \cos(k_1 + k_2)\right)
\end{equation}
By fitting to DFT bands, we obtain $\epsilon_{p_z^t}=0.51, t_{p_z^t}^{NN}=-0.15$ eV.

\begin{table}[htbp]
\centering
\begin{tabular}{c|c|c|c|c|c|c|c|c|c|c|c|c|c|c}
\hline\hline
Parameter & $\mu_{d_3}$ & $\mu_{d_4}$ & $\mu_{p_{z}^{o}}$ & $\mu_{p_{xy}^{odd}}$ & $t_{d_3}^{NN}$ & $t_{d_4}^{NN}$ & $t_{d_3,d_4}^{NN}$ & 
$s_{d_4, p_{z}^{o}}$ & $s_{d_3, p_{x}^{o}}$ & $s_{d_4, p_{xy}^{o}}$ & $t_{p_{z}^{o}}^{NN}$ 
& $t_{p_z^o, p_x^o}$ & $t_{p_{xy}^o}^1$ & $t_{p_{xy}^o}^2$
\\ \hline
Value/eV &  -0.699 & -0.251 & -0.507 & -0.216 & -0.209 & 0.199 & -0.153 & -0.822 & 0.700 & -0.727 & -0.523 & 0.057 & -0.406 & 1.754
\\ \hline\hline
\end{tabular}%
\caption{\label{app:table:Cr135_H2_param}The fitted parameters for the minimal TB model $H_2(\kk)$ defined in \cref{app:eq:135_H2} for \ch{CsCr3Sb5}.}
\end{table}

\subsubsection{$H_3(\kk)$}
The $H_3(\kk)$ sector is also different from FeGe. It contains the $d_{z^2}$ orbitals of Cr and $M_z$-even $(p_x, p_y)$ orbitals of honeycomb Sb, i.e., 7 orbitals in total. The orbital bases are
\begin{equation}
    (d_{z^2}@3g_1, d_{z^2}@3g_2, d_{z^2}@3g_3, 
    p_x^e@2d_1, p_y^e@2d_1, p_x^e@2d_2, p_y^e@2d_2),
\end{equation}
where the superscript $e$ denotes the $M_z$-even effective orbitals. 

The minimal TB model has the form
\begin{equation}
\footnotesize
\begin{aligned}
H_3(\kk) &=
\begin{bmatrix}
    H_{d_5}(\kk) & S_{d_5, p_{xy}^{e1}}(\kk) & S_{d_5, p_{xy}^{e1}}(\kk) \\
    & H_{p_{xy}^{e}}(\kk) & S_{p_{xy}^{e1,e2}}(\kk)\\
    H.c. & & H_{p_{xy}^{e}}(\kk)
\end{bmatrix}\\
H_{d_5}(\kk) &= \mu_{d_5}\mathbf{1}_{3}
+ 2 t_{d_5}^{NN}
H_{\text{Kagome}}^{\text{inplane,NN}}(\bm{k})\\
H_{p_{xy}^{e}}(\kk) 
&= \mu_{p_{xy}^{even}} \mathbf{1}_{2} \\
%\left[\mu_{p_{xy}^{even}}+2 t_{p_{xy}^{even}}^{1} (\cos (k_1+k_2)+\cos (k_1)+\cos (k_2))\right] \mathbf{1}_{2} \\
&+ t_{p_{xy}^{even}}^{1} 
\begin{bmatrix}
\frac{2 (\cos (k_1+k_2)+\cos (k_2))-\cos (k_1)}{\sqrt{3}} & i (-\sin (k_1+k_2)+\sin (k_1)+\sin (k_2))-\cos (k_1+k_2)+\cos (k_2) \\
-i (-\sin (k_1+k_2)+\sin (k_1)+\sin (k_2))-\cos (k_1+k_2)+\cos (k_2) & \sqrt{3} \cos (k_1) \\
\end{bmatrix}\\
S_{p_{xy}^{e1,e2}}
&= 
\left[
\begin{matrix}
 \frac{1}{12} e^{-\frac{1}{3} i (k_1+2 k_2)} \left(-4 \sqrt{3} s_{p_{xy}^{even}}^{3}  e^{i (k_1+2 k_2)}+3 \left(1+e^{i k_1}\right) e^{i k_2} (s_{p_{xy}^{even}}^{1} +3 s_{p_{xy}^{even}}^{2} )+16 \sqrt{3}
   s_{p_{xy}^{even}}^{3}  \cos (k_1)+12 s_{p_{xy}^{even}}^{1} \right)  \\
 -\frac{1}{4} \left(-1+e^{i k_1}\right) e^{-\frac{2}{3} i (2 k_1+k_2)} \left(\sqrt{3} e^{i (k_1+k_2)} (s_{p_{xy}^{even}}^{1} -s_{p_{xy}^{even}}^{2} )+4 \left(1+e^{i k_1}\right)
   s_{p_{xy}^{even}}^{3} \right) 
\end{matrix}   
\right. \\
&\left.
\begin{matrix}
-\frac{1}{4} \left(-1+e^{i k_1}\right) e^{-\frac{2}{3} i (2 k_1+k_2)} \left(\sqrt{3} e^{i (k_1+k_2)}
(s_{p_{xy}^{even}}^{1} -s_{p_{xy}^{even}}^{2} )+4 \left(1+e^{i k_1}\right) s_{p_{xy}^{even}}^{3} \right) \\
\frac{1}{4} e^{-\frac{1}{3} i (k_1+2 k_2)} \left(4 \sqrt{3} s_{p_{xy}^{even}}^{3}  e^{i (k_1+2 k_2)}+\left(1+e^{i k_1}\right) e^{i k_2} (3
s_{p_{xy}^{even}}^{1} +s_{p_{xy}^{even}}^{2} )+4 s_{p_{xy}^{even}}^{2} \right) 
\end{matrix}
\right] \\
S_{d_5, p_{xy}^{e1}}(\kk) &= t_{d_5,p_{xy}^e}
\begin{bmatrix}
 0 & -2  e^{\frac{1}{6} i (k_1+2 k_2)} \\
 -\sqrt{3}  e^{\frac{1}{6} i (k_1-k_2)} &  e^{\frac{1}{6} i (k_1-k_2)} \\
 \sqrt{3}  e^{-\frac{1}{6} i (2 k_1+k_2)} &  e^{-\frac{1}{6} i (2 k_1+k_2)} \\   
\end{bmatrix},\quad
S_{d_5, p_{xy}^{e2}}(\kk) = t_{d_5,p_{xy}^e}
\begin{bmatrix}
0 & 2  e^{-\frac{1}{6} i (k_1+2 k_2)} \\
\sqrt{3}  e^{-\frac{1}{6} i (k_1-k_2)} & 
-e^{-\frac{1}{6} i (k_1-k_2)} \\
-\sqrt{3}  e^{\frac{1}{6} i (2 k_1+k_2)} & 
-e^{\frac{1}{6} i (2 k_1+k_2)} \\   
\end{bmatrix}
\label{app:eq:135_H3}
\end{aligned}
\end{equation}

\begin{figure}[htbp]
\centering
\includegraphics[width=1\textwidth]{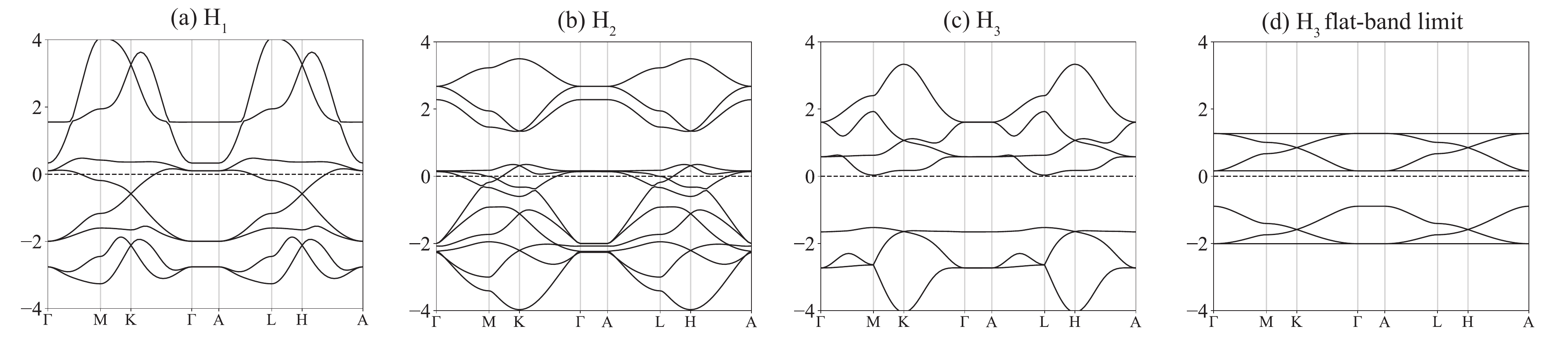}
\caption{\label{app:fig:Cr135_fitted_TB} The fitted minimal TB models for \ch{CsCr3Sb5}, where (a)-(c) are the bands of $H_1(\kk)$, $H_2(\kk)$, and $H_3(\kk)$, respectively. (d) is the flat-band limit of $H_3(\kk)$. 
 }
\end{figure}

The fitted TB parameters for $H_3(\kk)$ are listed in \cref{app:table:Cr135_H3_param}, and the corresponding bands are shown in \cref{app:fig:Cr135_fitted_TB} (c). One quasi-flat appears above $E_f$, which arises because the $d_{z^2}$ orbitals of Cr and the $M_z$-even $p_x, p_y$ orbitals of Sb form a bipartite crystalline lattice. To identify the perfect flat band limit, we consider only the inter-sublattice hopping terms and onsite terms. The resulting band structure, shown in \cref{app:fig:Cr135_fitted_TB}(d), reveals a perfect flat band near $E_f$, which is part of the four connected honeycomb bands from the $M_z$-even $p_x, p_y$ orbitals. This explains the quasi-flat band observed in the $H_3(\kk)$ sector of the DFT spectrum.

\begin{table}[htbp]
\centering
\begin{tabular}{c|c|c|c|c|c|c|c|c}
\hline\hline
Parameter & $\mu_{d_5}$ & $\mu_{p_{xy}^{even}}$ & $t_{d_5,p_{xy}^e}$ & $t_{d_5}^{NN}$ & $s_{p_{xy}^{even}}^{1}$ & $s_{p_{xy}^{even}}^{2}$ & $t_{p_{xy}^{even}}^{1}$ & $s_{p_{xy}^{even}}^{3}$  \\ \hline
Value/eV &  -0.893 & 0.158&  -0.449&  -0.191&  -0.472&  1.350&  -0.100&  0.178  \\ \hline\hline
\end{tabular}%
\caption{\label{app:table:Cr135_H3_param}The fitted parameters for the minimal TB model $H_3(\kk)$ defined in \cref{app:eq:135_H3} for \ch{CsCr3Sb5}.}
\end{table}

\subsubsection{CRPA interaction parameters}

The CRPA interaction parameters for the $d$ orbitals in \ch{CsCr3Sb5} are summarized in \cref{app:table:Cr135-CRPA}. Compared to FeGe, the interactions in \ch{CsCr3Sb5} are weaker, with the onsite Hubbard-Kanamori parameter $\mathcal{U}$ being approximately 60\% of that in FeGe. This reduction in interaction strength may result from the quasi-flat bands in \ch{CsCr3Sb5} being further from the Fermi level than those in FeGe.

\begin{table}[htbp]
\begin{tabular}{c|ccccccc|ccccc}
\hline\hline
$U_{ij}$  & $z^2$ & $xz$ & $yz$ & $x^2-y^2$ & $xy$ &  & $J_{ij}$ & $z^2$ & $xz$ & $yz$ & $x^2-y^2$ & $xy$ \\\hline
$z^2$     & 2.53  & 1.81 & 1.88 & 1.47      & 1.46 &  &          &       & 0.35 & 0.38 & 0.58      & 0.56 \\
$xz$      &       & 2.52 & 1.58 & 1.64      & 1.61 &  &          &       &      & 0.52 & 0.52      & 0.50 \\
$yz$      &       &      & 2.76 & 1.66      & 1.66 &  &          &       &      &      & 0.54      & 0.51 \\
$x^2-y^2$ &       &      &      & 2.79      & 2.13  &  &          &       &      &      &          & 0.29 \\
$xy$      &       &      &      &           & 2.67 &  &          &       &      &      &           &     
\\\hline\hline
\end{tabular}
\caption{\label{app:table:Cr135-CRPA} The Coulomb interaction $U_{ij}$ and $J_{ij}$ of $d$ orbitals in \ch{CsCr3Sb5}, where $U_{ij}=U_{iijj}$ and $J_{ij}=U_{ijji}$, with $U_{ijkl}$ defined in \cref{definition_Uijkl}.
The onsite Hubbard-Kanamori parameters are $\mathcal{U}=2.654, \mathcal{U}^\prime=1.691, \mathcal{J}=0.475$. 
The averaged NN and NNN interacting between  $d$ orbitals are $\bar{U}_{ij}^{NN}=0.934$ and $\bar{U}_{ij}^{NNN}=0.826$. The CRPA interaction is computed using a Wannier TB model of Cr $d$ and Sb $s,p$ orbitals, without SOC. All numbers are in eV.}
\end{table}

\subsection{\ch{CsV3Sb5}}

\begin{figure}[htbp]
\centering
\includegraphics[width=1\textwidth]{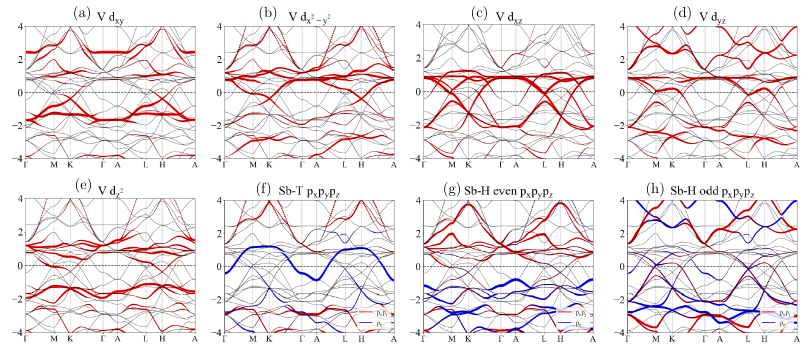}
\caption{\label{app:fig:V135_orb_weights} The orbital weights of \ch{CsV3Sb5}. (a)-(e) are for five $d$ orbitals of V. (f) shows for the $p$ orbitals of triangular Sb. (g) and (h) show the $M_z$-odd and even combinations of honeycomb Sb $p$ orbitals, respectively, as defined in \cref{app:eq:135-Mz-basis-Sb}. 
 }
\end{figure}

In this section, we study the kagome 1:3:5 family \ch{AV3Sb5} (A=K, Cs, Rb), by focusing on \ch{CsV3Sb5}. In \ch{CsV3Sb5}, the quasi-flat bands from V $d$ orbitals are at about 1 eV above the Fermi level. Instead, it holds multiple vHSs at $M$ near $E_f$ from the $H_2(\kk)$ sector. The orbital projected band structures of \ch{CsV3Sb5} are shown in \cref{app:fig:V135_orb_weights}. 
There are three vHSs at $M$ near the Fermi level, two originating from $d_{xz}, d_{yz}$, and one from the hybridization of $d_{x^2-y^2}$ with $d_{z^2}$ orbitals.

\subsubsection{Minimal TB models}
Following the minimal TB Hamiltonains defined in \ch{CsCr3Sb5}, i.e., \cref{app:eq:135_H1}, \cref{app:eq:135_H2}, and \cref{app:eq:135_H3}, we fit the parameters for \ch{CsV3Sb5}, 

In \ch{CsV3Sb5}, the $H_{1}(\kk)$ and $H_3(\kk)$ sectors are the same as \ch{CsCr3Sb5}, with parameters tabulated in \cref{app:table:V135_TB_param}.
The $H_2(\kk)$ sector, however, needs to add extra longer-range hopping terms to reproduce the vHSs near $E_f$ faithfully. These extra terms have the expression:
\begin{equation}
\tiny
\begin{aligned}
S_{d_3, p_{xy}^{o_1}}^{long}(\kk)
&= \frac{1}{8} 
\begin{bmatrix}
 2\left(1+e^{i k_1}\right) s_{d_3,p_{xy}^{o}}^1  e^{-\frac{1}{6} i (5 k_1+4 k_2)} & 2 e^{-\frac{1}{6} i (5 k_1+4 k_2)} \left(s_{d_3,p_{xy}^{o}}^2 -e^{i k_1}
   s_{d_3,p_{xy}^{o}}^2 \right) \\
e^{-\frac{1}{6} i (5 k_1+k_2)} \left(\left(s_{d_3,p_{xy}^{o}}^1 -\sqrt{3} s_{d_3,p_{xy}^{o}}^2 \right)  e^{i(k_1+k_2)} +s_{d_3,p_{xy}^{o}}^1 +\sqrt{3}
   s_{d_3,p_{xy}^{o}}^2 \right) & e^{-\frac{1}{6} i (5 k_1+k_2)} \left(e^{i (k_1+k_2)} \left(\sqrt{3} s_{d_3,p_{xy}^{o}}^1 +s_{d_3,p_{xy}^{o}}^2 \right)+\sqrt{3}
   s_{d_3,p_{xy}^{o}}^1 -s_{d_3,p_{xy}^{o}}^2 \right) \\
 - e^{\frac{1}{6} i (4 k_1-k_2)} \left(e^{i k_2} \left(s_{d_3,p_{xy}^{o}}^1 -\sqrt{3} s_{d_3,p_{xy}^{o}}^2 \right)+s_{d_3,p_{xy}^{o}}^1 +\sqrt{3} s_{d_3,p_{xy}^{o}}^2 \right) &  e^{\frac{1}{6} i(4 k_1-k_2)} \left(\sqrt{3} \left(1+e^{i k_2}\right) s_{d_3,p_{xy}^{o}}^1 +\left(-1+e^{i k_2}\right) s_{d_3,p_{xy}^{o}}^2 \right) \\
\end{bmatrix}\\
S_{d_3, p_{xy}^{o_2}}^{long}(\kk)
&= \frac{1}{8} 
\begin{bmatrix}
 2\left(1+e^{i k_1}\right) s_{d_3,p_{xy}^{o}}^1  e^{-\frac{1}{6} i (k_1-4 k_2)} & 2 \left(-1+e^{i k_1}\right) s_{d_3,p_{xy}^{o}}^2  e^{-\frac{1}{6} i (k_1-4 k_2)}
   \\
  e^{-\frac{1}{6} i (k_1+5 k_2)} \left(e^{i (k_1+k_2)} \left(s_{d_3,p_{xy}^{o}}^1 +\sqrt{3} s_{d_3,p_{xy}^{o}}^2 \right)+s_{d_3,p_{xy}^{o}}^1 -\sqrt{3} s_{d_3,p_{xy}^{o}}^2 \right) & 
   e^{-\frac{1}{6} i (k_1+5 k_2)} \left(e^{i (k_1+k_2)} \left(\sqrt{3} s_{d_3,p_{xy}^{o}}^1 -s_{d_3,p_{xy}^{o}}^2 \right)+\sqrt{3} s_{d_3,p_{xy}^{o}}^1 +s_{d_3,p_{xy}^{o}}^2 \right) \\
  e^{-\frac{1}{6} i (4 k_1+5 k_2)} \left(-e^{i k_2} \left(s_{d_3,p_{xy}^{o}}^1 +\sqrt{3} s_{d_3,p_{xy}^{o}}^2 \right)-s_{d_3,p_{xy}^{o}}^1 +\sqrt{3} s_{d_3,p_{xy}^{o}}^2 \right) & e^{-\frac{1}{6}
   i (4 k_1+5 k_2)} \left(\sqrt{3} \left(1+e^{i k_2}\right) s_{d_3,p_{xy}^{o}}^1 -e^{i k_2} s_{d_3,p_{xy}^{o}}^2 +s_{d_3,p_{xy}^{o}}^2 \right) \\
\end{bmatrix}\\
S_{d_4, p_{xy}^{o_1}}^{long}(\kk)
&= \frac{1}{8} 
\begin{bmatrix}
 2e^{-\frac{1}{6} i (5 k_1+4 k_2)} \left(s_{d_4,p_{xy}^{o}}^1 -e^{i k_1} s_{d_4,p_{xy}^{o}}^1 \right) & 2 \left(1+e^{i k_1}\right) s_{d_4,p_{xy}^{o}}^2  e^{-\frac{1}{6} i (5
   k_1+4 k_2)} \\
 e^{-\frac{1}{6} i (5 k_1+k_2)} \left(s_{d_4,p_{xy}^{o}}^1  \left(-1+e^{i (k_1+k_2)}\right)-\sqrt{3} s_{d_4,p_{xy}^{o}}^2  \left(1+e^{i (k_1+k_2)}\right)\right) &
 e^{-\frac{1}{6} i (5 k_1+k_2)} \left(e^{i (k_1+k_2)} \left(\sqrt{3} s_{d_4,p_{xy}^{o}}^1 +s_{d_4,p_{xy}^{o}}^2 \right)-\sqrt{3} s_{d_4,p_{xy}^{o}}^1 +s_{d_4,p_{xy}^{o}}^2 \right) \\
 -2 e^{\frac{1}{3} i (2 k_1+k_2)} \left(\sqrt{3} s_{d_4,p_{xy}^{o}}^2  \cos \left(\frac{k_2}{2}\right)-i s_{d_4,p_{xy}^{o}}^1  \sin \left(\frac{k_2}{2}\right)\right) & -e^{\frac{1}{6} i (4 k_1-k_2)} \left(e^{i k_2} \left(\sqrt{3} s_{d_4,p_{xy}^{o}}^1 +s_{d_4,p_{xy}^{o}}^2 \right)-\sqrt{3} s_{d_4,p_{xy}^{o}}^1 +s_{d_4,p_{xy}^{o}}^2 \right) \\
\end{bmatrix}\\
S_{d_4, p_{xy}^{o_2}}^{long}(\kk)
&=  \frac{1}{8} 
\begin{bmatrix}
2\left(-1+e^{i k_1}\right) s_{d_4,p_{xy}^{o}}^1  e^{-\frac{1}{6} i (k_1-4 k_2)} & 2\left(1+e^{i k_1}\right) s_{d_4,p_{xy}^{o}}^2  e^{-\frac{1}{6} i (k_1-4 k_2)}
   \\
 e^{-\frac{1}{6} i (k_1+5 k_2)} \left(-e^{i (k_1+k_2)} (s_{d_4,p_{xy}^{o}}^1 +\sqrt{3} s_{d_4,p_{xy}^{o}}^2 )+s_{d_4,p_{xy}^{o}}^1 -\sqrt{3} s_{d_4,p_{xy}^{o}}^2 \right) & 
   e^{-\frac{1}{6} i (k_1+5 k_2)} \left(e^{i (k_1+k_2)} (s_{d_4,p_{xy}^{o}}^2 -\sqrt{3} s_{d_4,p_{xy}^{o}}^1 )+\sqrt{3} s_{d_4,p_{xy}^{o}}^1 +s_{d_4,p_{xy}^{o}}^2 \right) \\
 e^{-\frac{1}{6} i (4 k_1+5 k_2)} \left(-e^{i k_2} \left(s_{d_4,p_{xy}^{o}}^1 +\sqrt{3} s_{d_4,p_{xy}^{o}}^2 \right)+s_{d_4,p_{xy}^{o}}^1 -\sqrt{3} s_{d_4,p_{xy}^{o}}^2 \right) & e^{-\frac{1}{6}
   i (4 k_1+5 k_2)} \left(\sqrt{3} \left(-1+e^{i k_2}\right) s_{d_4,p_{xy}^{o}}^1 -\left(1+e^{i k_2}\right) s_{d_4,p_{xy}^{o}}^2 \right) \\
\end{bmatrix}
\end{aligned}
\label{app:eq:135_H2_longrange}
\end{equation}

\begin{figure}[htbp]
\centering
\includegraphics[width=0.9\textwidth]{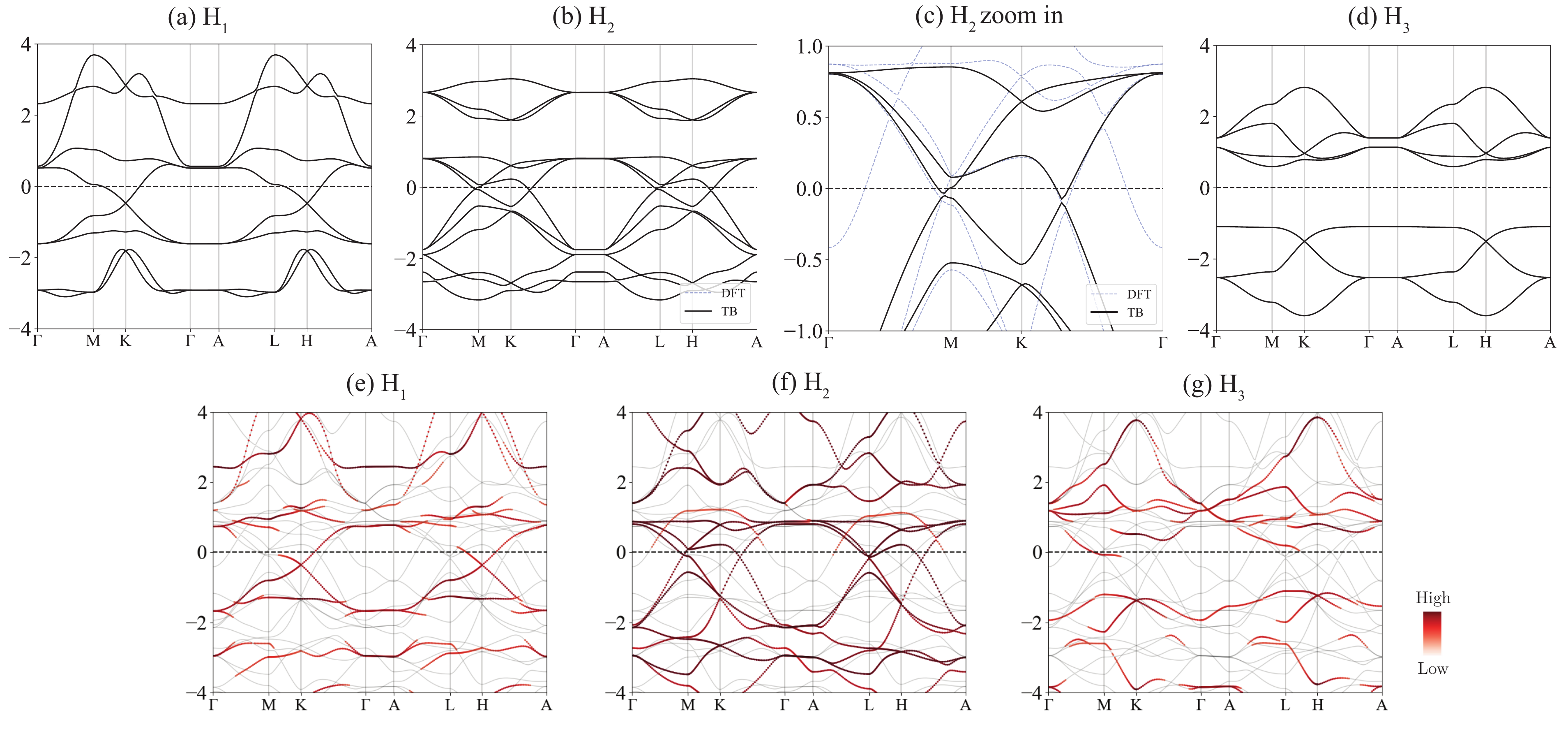}
\caption{\label{app:fig:V135_fitted_TB} The fitted minimal TB model for \ch{CsV3Sb5}, where (a), (b), (d) shows the bands of $H_1(\kk)$, $H_1(\kk)$, and  $H_1(\kk)$, respectively. (c) is the zoom-in comparison with DFT bands (blue dashed lines), where we only show the $M_z$-odd bands on the $k_z=0$ plane. A good agreement of the vHSs near $E_f$ is observed. 
Note the DFT band at about -0.5 eV at $\Gamma$ is mainly from the $p_z$ orbital of triangular Sb and is not considered in the current $H_2(\kk)$ model. 
The second raw (e)-(g) shows the DFT bands with orbital weights from orbitals in $H_{i=1,2,3}(\kk)$, respectively. A good agreement with DFT bands is observed. 
}
\end{figure}

By fitting the TB parameters to the DFT Hamiltonian, we obtain the minimal TB model with dispersion shown in \cref{app:fig:V135_fitted_TB}. The fitted parameters are tabulated in \cref{app:table:V135_TB_param}. A good agreement of the vHSs near $E_f$ is observed.

\begin{table}[htbp]
\footnotesize
\centering
\begin{tabular}{c|c|c|c|c|c|c|c|c|c|c|c|c|c|c|c|c|c|c}
\hline\hline
Parameter & $\mu_{p_{xy}^t}$ & $\mu_{d_1}$ & $\mu_{d_2}$ & $t_{d_1}^{NN}$ & $t_{d_1}^{NNN}$  & $t_{d_2}^{NN}$ & $t_{d_2}^{NNN}$   & $t_{d_1,d_2}^{NN}$ & $t_{d_1,d_2}^{NNN}$ & $t_{p_{xy}^t, d_1}^{NN}$ & $t_{p_{xy}^t, d_1}^{NNN}$ &  $t_{p_{xy}^t, d_2}^{NN}$ & $t_{p_{xy}^t, d_2}^{NNN}$
\\ \hline
Value/eV & -0.156 & -0.513 & 0.063 & 0.603 & 0.108 & -0.379 & 0.151 & 0.100 & 0.039 & -0.726 & 0.298 & 0.973 & 0.125 &   \\ \hline
Parameter & $t_d^{4N1}$ & $t_d^{4N2}$ & $t_d^{4N3}$  & $t_d^{4N4}$ & $t_d^{4N5}$   & $s_{p}^1$ & $s_{p}^2$ & $s_{p}^3$ & $s_{p_{xy}^t,d_1}^1$ & $s_{p_{xy}^t,d_1}^2$ & $s_{p_{xy}^t,d_2}^1$ & $s_{p_{xy}^t,d_2}^2$ & $s_{d12}^1$ & $s_{d12}^2$ \\ \hline
Value/eV & -0.018 & 0.069 & 0.056 & 0.106 & 0.089 & -0.419 & -0.727 & 0.025 & -0.009 & 0.032 & -0.112 & -0.089 & -0.063 & 0.244
\\\hline
Parameter & $\mu_{d_3}$ & $\mu_{d_4}$ & $\mu_{p_{z}^{o}}$ & $\mu_{p_{xy}^{odd}}$ & $t_{d_3}^{NN}$ & $t_{d_4}^{NN}$ & $t_{d_3,d_4}^{NN}$ & 
$s_{d_4, p_{z}^{o}}$ & $s_{d_3, p_{x}^{o}}$ & $s_{d_4, p_{xy}^{o}}$ & $t_{p_{z}^{o}}^{NN}$ 
& $t_{p_z^o, p_x^o}$ & $t_{p_{xy}^o}^1$ & $t_{p_{xy}^o}^2$ 
& $s_{d_3,p_{xy}^{o}}^1$ & $s_{d_3,p_{xy}^{o}}^2$ & $s_{d_4,p_{xy}^{o}}^1$  & $s_{d_4,p_{xy}^{o}}^2$ 
\\ \hline
Value/eV & -0.439&  0.337 & -0.787& 0.002&  -0.336&  0.171&  -0.226&  -1.041&  0.709&  -0.870&  -0.729& -0.052&  -0.362&  1.713& 0.105&  0.026&  0.684&  0.406 
\\\hline
Parameter & $\mu_{d_5}$ & $\mu_{p_{xy}^{even}}$ & $t_{d_5,p_{xy}^e}$ & $t_{d_5}^{NN}$ & $s_{p_{xy}^{even}}^{1}$ & $s_{p_{xy}^{even}}^{2}$ &  $t_{p_{xy}^{even}}^{1}$ & $s_{p_{xy}^{even}}^{3}$  \\ \hline
Value/eV &  -0.385 & 0.057 & -0.491 & -0.179 & -0.486 & 1.367 & -0.139 & 0.153
\\ \hline\hline
\end{tabular}%
\caption{\label{app:table:V135_TB_param}The fitted parameters for the minimal TB models $H_{1,2,3}(\kk)$ for \ch{CsV3Sb5}, defined in \cref{app:eq:135_H1}, \cref{app:eq:135_H2}, \cref{app:eq:135_H3}, and \cref{app:eq:135_H2_longrange}.}
\end{table}

\subsubsection{CRPA interaction parameters}

The CRPA interaction parameters for the $d$ orbital in \ch{CsV3Sb5} are summarized in \cref{app:table:V135-CRPA}. The values of interactions are similar to those in \ch{CsCr3Sb5}, but smaller than FeGe.

\begin{table}[htbp]
\begin{tabular}{c|ccccccc|ccccc}
\hline\hline
$U_{ij}$  & $z^2$ & $xz$ & $yz$ & $x^2-y^2$ & $xy$ &  & $J_{ij}$ & $z^2$ & $xz$ & $yz$ & $x^2-y^2$ & $xy$ \\\hline
$z^2$     & 2.36  & 1.79 & 1.84 & 1.51      & 1.48 &  &          &       & 0.31 & 0.35 & 0.54      & 0.51 \\
$xz$      &       & 2.50 & 1.64 & 1.69      & 1.65 &  &          &       &      & 0.50 & 0.48      & 0.47 \\
$yz$      &       &      & 2.78 & 1.74      & 1.71 &  &          &       &      &      & 0.53      & 0.49 \\
$x^2-y^2$ &       &      &      & 2.82      & 2.16 &  &         &       &      &      &            & 0.28 \\
$xy$      &       &      &      &           & 2.63 &  &          &       &      &      &           &     
\\\hline\hline
\end{tabular}
\caption{\label{app:table:V135-CRPA} The Coulomb interaction $U_{ij}$ and $J_{ij}$ of $d$ orbitals in \ch{CsV3Sb5}, where $U_{ij}=U_{iijj}$ and $J_{ij}=U_{ijji}$, with $U_{ijkl}$ defined in \cref{definition_Uijkl}.
The onsite Hubbard-Kanamori parameters are $\mathcal{U}=2.62, \mathcal{U}^\prime=1.72, \mathcal{J}=0.45$. 
The averaged NN and NNN interacting between $d$ orbitals are $\bar{U}_{ij}^{NN}=0.88$ and $\bar{U}_{ij}^{NNN}=0.78$. The CRPA interaction is computed using a Wannier TB model of V $d$, Sb $s,p$, without SOC. All numbers are in eV.}
\end{table}

\FloatBarrier
\subsection{\ch{CsTi3Bi5}}

\begin{figure}[htbp]
\centering
\includegraphics[width=1\textwidth]{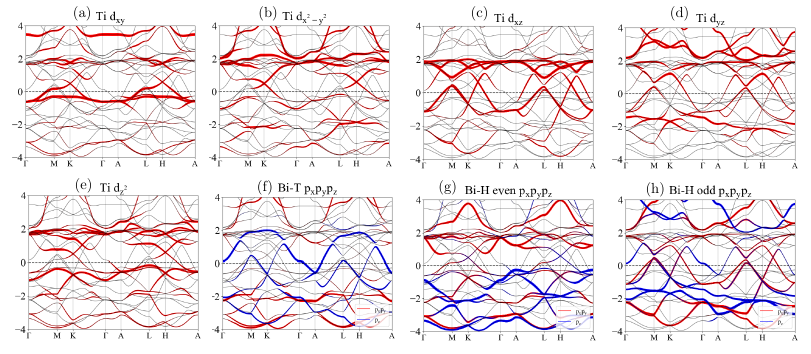}
\caption{\label{app:fig:Ti135_orb_weights} The orbital weights of \ch{CsTi3Bi5}. (a)-(e) are for five $d$ orbitals of V. (f) shows for the $p$ orbitals of triangular Sb. (g) and (h) show the $M_z$-odd and even combinations of honeycomb Sb $p$ orbitals, respectively, as defined in \cref{app:eq:135-Mz-basis-Sb}. SOC is not considered. 
 }
\end{figure}

In this section, we study the \ch{CsTi3Bi5}. Compared with \ch{CsV3Sb5}, Ti has one less valence electron compared with V, thus the Fermi level is about 1 eV lower in \ch{CsTi3Bi5} than in \ch{CsV3Sb5}, with the band structures being similar. 
The orbital projected bands of \ch{CsTi3Bi5} are shown in \cref{app:fig:Ti135_orb_weights}. 

\begin{figure}[htbp]
\centering
\includegraphics[width=0.35\textwidth]{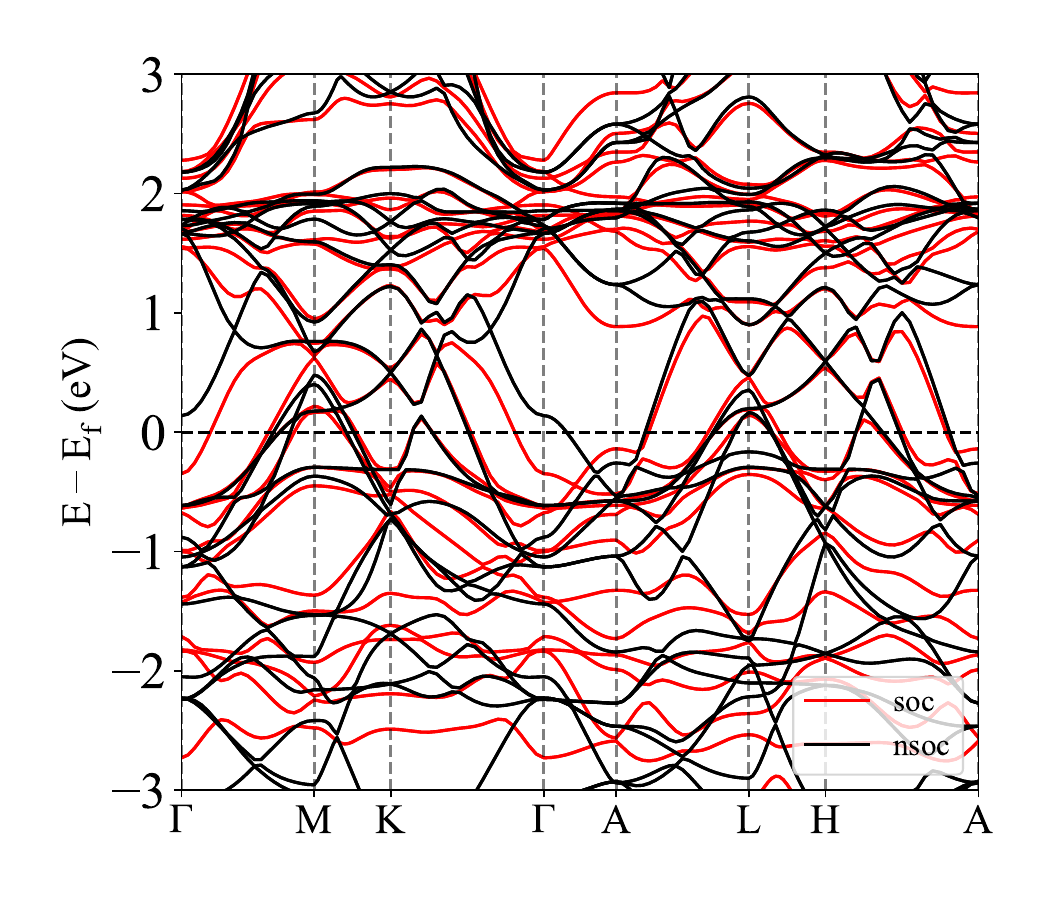}
\caption{\label{app:fig:Ti135_compare_soc} SOC effect in \ch{CsTi3Bi5}. The bands in black (red) are computed without (with) SOC effects in DFT. 
 }
\end{figure}

In \ch{CsTi3Bi5}, the SOC effects are significantly stronger compared to \ch{CsV3Sb5} and \ch{CsCr3Sb5}, primarily due to the presence of Bi atoms, which exhibit strong SOC. As shown in \cref{app:fig:Ti135_compare_soc}, the band structure of \ch{CsTi3Bi5} changes substantially in bands with large Bi orbital weights, while the bands dominated by Ti orbitals are only slightly affected, as indicated by the orbital weights in \cref{app:fig:Ti135_orb_weights}. 
The pronounced SOC effects on Bi bands cannot be accurately captured by the simple onsite SOC term introduced in \cref{app:eq:H_soc_p} for CoSn. To achieve better agreement with the DFT band structures that include SOC, one could consider incorporating general symmetry-allowed spin-off-diagonal coupling terms as SOC contributions. 
In the next subsection, we construct minimal TB models for \ch{CsTi3Bi5} without incorporating SOC effects. This simplification is justified as the bands near the Fermi level are predominantly derived from Ti $d$ orbitals. A more comprehensive treatment of SOC effects in the model of \ch{CsTi3Bi5} will be addressed in future work.

\subsubsection{Minimal TB model}

\begin{figure}[htbp]
\centering
\includegraphics[width=0.8\textwidth]{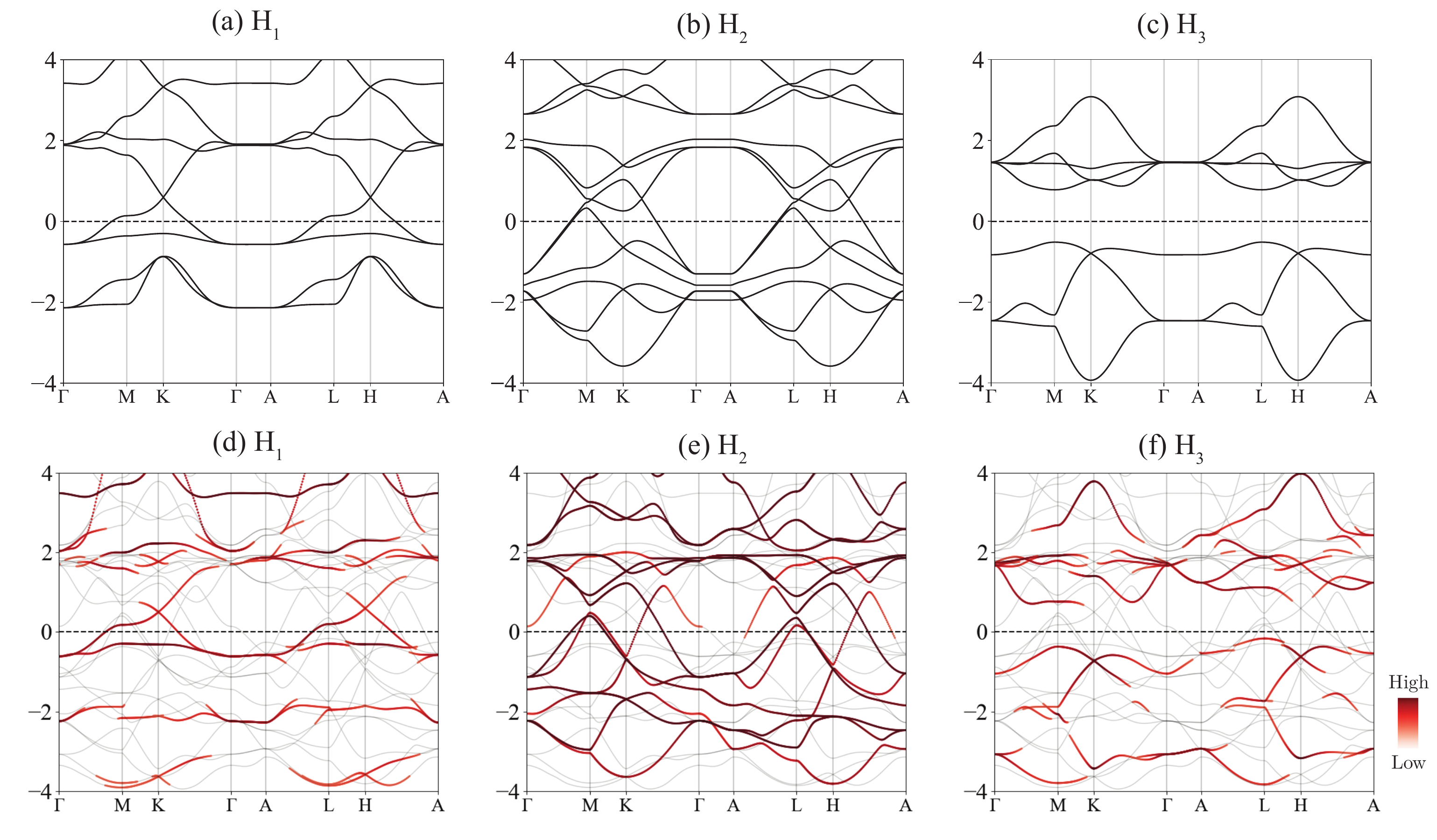}
\caption{\label{app:fig:Ti135_fitted_TB} The fitted minimal TB model for \ch{CsTi3Bi5}, where (a) is the bands of $H_1(\kk)$, (b) is $H_2(\kk)$, (c) is $H_3(\kk)$. 
The second raw (d)-(f) shows the DFT bands with orbital weights from orbitals in $H_{i=1,2,3}(\kk)$, respectively. A good agreement with DFT bands is observed. 
 }
\end{figure}

The three minimal TB models for \ch{CsTi3Bi5} take the same form as in \ch{CsV3Sb5}. By fitting the model with DFT data, we obtain the parameters tabulated in \cref{app:table:Ti135_TB_param} and the band structure in \cref{app:fig:Ti135_fitted_TB}. A good agreement with DFT bands is observed.

\begin{table}[htbp]
\footnotesize
\centering
\begin{tabular}{c|c|c|c|c|c|c|c|c|c|c|c|c|c|c|c|c|c|c}
\hline\hline
Parameter & $\mu_{p_{xy}^t}$ & $\mu_{d_1}$ & $\mu_{d_2}$ & $t_{d_1}^{NN}$ & $t_{d_1}^{NNN}$  & $t_{d_2}^{NN}$ & $t_{d_2}^{NNN}$   & $t_{d_1,d_2}^{NN}$ & $t_{d_1,d_2}^{NNN}$ & $t_{p_{xy}^t, d_1}^{NN}$ & $t_{p_{xy}^t, d_1}^{NNN}$ &  $t_{p_{xy}^t, d_2}^{NN}$ & $t_{p_{xy}^t, d_2}^{NNN}$
\\ \hline
Value/eV & -0.01 &  0.62 &  1.38 &  0.63 &  0.08 & -0.35 &  0.15 &  0.05 &  0.07 & -0.65 &  0.29 &  0.86 & -0.04
\\ \hline
Parameter & $t_d^{4N1}$ & $t_d^{4N2}$ & $t_d^{4N3}$  & $t_d^{4N4}$ & $t_d^{4N5}$   & $s_{p}^1$ & $s_{p}^2$ & $s_{p}^3$ & $s_{p_{xy}^t,d_1}^1$ & $s_{p_{xy}^t,d_1}^2$ & $s_{p_{xy}^t,d_2}^1$ & $s_{p_{xy}^t,d_2}^2$ & $s_{d12}^1$ & $s_{d12}^2$ \\ \hline
Value/eV &  -0.08 &  0.08 &  0.03 &  0.19 &  0.06 & -0.20 & -0.51 &  0.04 &  0.03 & -0.02 & -0.08 & -0.12 & -0.05 &  0.20
\\\hline
Parameter & $\mu_{d_3}$ & $\mu_{d_4}$ & $\mu_{p_{z}^{o}}$ & $\mu_{p_{xy}^{odd}}$ & $t_{d_3}^{NN}$ & $t_{d_4}^{NN}$ & $t_{d_3,d_4}^{NN}$ & 
$s_{d_4, p_{z}^{o}}$ & $s_{d_3, p_{x}^{o}}$ & $s_{d_4, p_{xy}^{o}}$ & $t_{p_{z}^{o}}^{NN}$ 
& $t_{p_z^o, p_x^o}$ & $t_{p_{xy}^o}^1$ & $t_{p_{xy}^o}^2$ 
& $s_{d_3,p_{xy}^{o}}^1$ & $s_{d_3,p_{xy}^{o}}^2$ & $s_{d_4,p_{xy}^{o}}^1$  & $s_{d_4,p_{xy}^{o}}^2$ 
\\ \hline
Value/eV & 0.51 &  1.29 &  0.24 & -0.03 & -0.38 &  0.24 & -0.21 & -1.13 &  0.76 & -0.72 & -0.73 &  0.16 & -0.45 &  1.58 &  0.04 & -0.00 &  0.32 &  0.47
\\\hline
Parameter & $\mu_{d_5}$ & $\mu_{p_{xy}^{even}}$ & $t_{d_5,p_{xy}^e}$ & $t_{d_5}^{NN}$ & $s_{p_{xy}^{even}}^{1}$ & $s_{p_{xy}^{even}}^{2}$ &  $t_{p_{xy}^{even}}^{1}$ & $s_{p_{xy}^{even}}^{3}$  \\ \hline
Value/eV & 0.37 & -0.12 & -0.37 & -0.30 & -0.47 &  1.48 & -0.08 &  0.12
\\ \hline\hline
\end{tabular}%
\caption{\label{app:table:Ti135_TB_param}The fitted parameters for the minimal TB models $H_{1,2,3}(\kk)$ for \ch{CsTi3Bi5}, defined in \cref{app:eq:135_H1}, \cref{app:eq:135_H2}, \cref{app:eq:135_H3}, and \cref{app:eq:135_H2_longrange}.}
\end{table}

\subsubsection{CRPA interaction}

The CRPA interaction parameters for the $d$ orbital in \ch{CsTi3Bi5} are summarized in \cref{app:table:Ti135-CRPA}, where SOC effects are not considered. The values of interactions are similar to those in \ch{CsCr3Sb5}, but smaller than FeGe.

\begin{table}[htbp]
\begin{tabular}{c|ccccccc|ccccc}
\hline\hline
$U_{ij}$  & $z^2$ & $xz$ & $yz$ & $x^2-y^2$ & $xy$ &  & $J_{ij}$ & $z^2$ & $xz$ & $yz$ & $x^2-y^2$ & $xy$ \\\hline
$z^2$     & 2.20  & 1.76 & 1.85 & 1.56      & 1.48 &  &          &       & 0.26 & 0.32 & 0.49      & 0.44 \\
$xz$      &       & 2.43 & 1.71 & 1.75      & 1.63 &  &          &       &      & 0.45 & 0.44      & 0.40 \\
$yz$      &       &      & 2.88 & 1.83      & 1.74 &  &          &       &      &      & 0.50      & 0.44 \\
$x^2-y^2$ &       &      &      & 2.81      & 2.10 &  &          &       &      &      &           & 0.25 \\
$xy$      &       &      &      &           & 2.44 &  &          &       &      &      &           &     
\\\hline\hline
\end{tabular}
\caption{\label{app:table:Ti135-CRPA} The Coulomb interaction $U_{ij}$ and $J_{ij}$ of $d$ orbitals in \ch{CsTi3Bi5}, where $U_{ij}=U_{iijj}$ and $J_{ij}=U_{ijji}$, with $U_{ijkl}$ defined in \cref{definition_Uijkl}.
The onsite Hubbard-Kanamori parameters are $\mathcal{U}=2.55, \mathcal{U}^\prime=1.74, \mathcal{J}=0.40$. 
The averaged NN and NNN interacting between $d$ orbitals are $\bar{U}_{ij}^{NN}=0.87$ and $\bar{U}_{ij}^{NNN}=0.82$. The CRPA interaction is computed using a Wannier TB model of Ti $d$, Bi $s,p$, without SOC. All numbers are in eV.}
\end{table}

\clearpage
\section{A brief review of quasi-degenerate second-order perturbation theory}\label{SI:S-matrix}

In this section, we give a brief summary of the quasi-degenerate second-order perturbation theory\cite{winkler2003spin}. Using this method, one can split a Hamiltonian into decoupled subspaces.

Assume we have a solvable Hamiltonian $H_0$ with known eigenvalues $\{E_n\}$ and eigenvectors $\{\psi_n\}$, and $H_0$ can be divided into two decoupled (or weakly coupled) subspaces $A$ and $B$. The whole Hamiltonian has the form
\begin{equation}
    H=H_0+H^\prime
\end{equation}
where $H^\prime$ is assumed to be a perturbation. 
Divide $H^\prime$ into diagonal and off-diagonal parts with respect to $A$ and $B$, i.e., 
$H^\prime=H_d^\prime+H_{od}^\prime$. One can find a similarity transformation $S$, s.t. $\bar{H}=e^{-S} H e^{S}$ is diagonal with respect to the $A, B$ subspaces. This means the off-diagonal term $H_{od}^\prime$ is perturbed out and $\bar{H}$ becomes block-diagonal.

The most useful formula is the second-order perturbation term, i.e.,
\begin{equation}
    \begin{aligned}
        S &= S^{(1)}, \\
        \bar{H} &= H_0 + H_{d}^\prime + \bar{H}^{(2)}, \\
    \end{aligned}
\end{equation}
where
\begin{equation}
    \begin{aligned}
        S^{(1)}_{\alpha\beta} &= -\frac{H_{od,\alpha\beta}^\prime}{E_\alpha-E_\beta}, \\
        \bar{H}^{(2)}_{\alpha\alpha^\prime} &=
        \frac{1}{2}\sum_\beta H_{od,\alpha\beta}^\prime H_{od,\beta\alpha^\prime}^\prime
        \left(\frac{1}{E_\alpha-E_\beta} + \frac{1}{E_\alpha^\prime - E_\beta} \right),
    \end{aligned}
\end{equation}
in which
\begin{equation}
    H_{od,\alpha\beta}^\prime = \langle \psi_\alpha | H_{od}^\prime | \psi_\beta \rangle
\end{equation}

We give two special cases where the second-order terms have simple analytical forms.
\begin{itemize}
    \item If 
    \begin{equation}
        \begin{aligned}
            H &= \left(
            \begin{matrix}
	        \mu_1 \bm{1}_n & S \\
	           S^\dagger & \mu_2 \bm{1}_m \\
            \end{matrix}
            \right),
        \end{aligned}
    \end{equation}
where $S$ is the S-matrix with shape $(n,m)$. This Hamiltonian can be decoupled into
    \begin{equation}
        \begin{aligned}
            H_0 &= \left(
            \begin{matrix}
	        \mu_1 \bm{1}_n & \bm{0} \\
	           \bm{0} & \mu_2 \bm{1}_m\\
            \end{matrix}
            \right),\quad
            H_d^\prime = \bm{0},\quad
            H_{od}^\prime &= \left(
            \begin{matrix}
	        \bm{0}  & S \\
	        S^\dagger & \bm{0}  \\
            \end{matrix}
            \right).
        \end{aligned}
    \end{equation}
Then the second-order perturbed terms have the form if $|\mu_1-\mu_2|\gg |S|$:
    \begin{equation}
        \begin{aligned}
           \bar{H}^{(2)}_1 &= \frac{1}{\mu_1-\mu_2}SS^\dagger,\\
           \bar{H}^{(2)}_2 &= \frac{1}{\mu_2-\mu_1}S^\dagger S,\\     
            \Rightarrow \bar{H} &= \left(
            \begin{matrix}
	        \mu_1 \bm{1}_n + \bar{H}^{(2)}_1 & \bm{0} \\
	        \bm{0} & \mu_2 \bm{1}_m + \bar{H}^{(2)}_2\\
            \end{matrix}
            \right).
        \end{aligned}
    \end{equation}

\item If 
    \begin{equation}
        \begin{aligned}
            H &= \left(
            \begin{matrix}
                \mu_0 \bm{1}_{n_0} & S_1 & S_2 \\
	        S_1^\dagger & \mu_1 \bm{1}_{n_1} & \bm{0} \\
	           S_2^\dagger & \bm{0} & \mu_2 \bm{1}_{n_2} \\
            \end{matrix}
            \right),
        \end{aligned}
    \end{equation}
where $S_i$ is the S-matrix with shape $(n_0,n_i)$. This Hamiltonian can be decoupled into
    \begin{equation}
        \begin{aligned}
            H_0 &= \left(
            \begin{matrix}
                \mu_0 \bm{1}_{n_0} &\bm{0}  &\bm{0}   \\
	          \bm{0} & \mu_1 \bm{1}_{n_1} & \bm{0} \\
	          \bm{0}   & \bm{0} & \mu_2 \bm{1}_{n_2} \\
            \end{matrix}
            \right),\quad
            H_d^\prime = \bm{0},\quad
            H_{od}^\prime &= \left(
            \begin{matrix}
	        \bm{0}  & S_1 & S_2 \\
	        S_1^\dagger & \bm{0}  & \bm{0}  \\
            S_2^\dagger & \bm{0}  & \bm{0} \\
            \end{matrix}
            \right).
        \end{aligned}
    \end{equation}
Then the second-order perturbed terms have the following form if $|\mu_i-\mu_0|\gg |S_i|$:
\begin{equation}
    \begin{aligned}
        \bar{H}^{(2)}_0 &= \frac{1}{\mu_0-\mu_1}S_1 S_1^\dagger + \frac{1}{\mu_0-\mu_2}S_2 S_2^\dagger,\\
       \bar{H}^{(2)}_1 &= \frac{1}{\mu_1-\mu_0}S_1^\dagger S_1,\\
       \bar{H}^{(2)}_2 &= \frac{1}{\mu_2-\mu_0}S_2^\dagger S_2,\\     
       \bar{H}^{(2)}_{12} &= \frac{1}{2}\left(\frac{1}{\mu_1-\mu_0} +\frac{1}{\mu_2-\mu_0}\right)S_1^\dagger S_2,\\  
        \Rightarrow \bar{H} &= \left(
        \begin{matrix}
            \mu_0 \bm{1}_{n_0}+ \bar{H}^{(2)}_0 &\bm{0}  &\bm{0}   \\
          \bm{0} & \mu_1 \bm{1}_{n_1} +\bar{H}_1^{(2)} & \bar{H}_{12}^{(2)} \\
          \bm{0}  & \bar{H}_{12}^{(2)\dagger} & \mu_2 \bm{1}_{n_2} +\bar{H}_{2}^{(2)} \\
        \end{matrix}
        \right).
    \end{aligned}
\end{equation}
\end{itemize}

\section{A brief review of the $S$-matrix formalism and flat band theory}\label{Sec:SI_Smatrix}

In this section, we give a brief review of the $S$-matrix formalism and the flat-band theory of Ref.\cite{cualuguaru2022general, ma2020spin,regnault2022catalogue}. We only summarize the main results here. Rigorous proofs can be found in Ref.\cite{cualuguaru2022general}.

A bipartite crystalline lattice (BCL) is a periodic lattice with two different sublattices $L$ and $\tilde{L}$. Assume that $N_L$ and $N_{\tilde{L}}$ orbitals per unit cell are placed in the $L$ and $\tilde{L}$ sublattices, respectively. With no loss of generality, we take $N_L\ge N_{\tilde{L}}$. 
The $S$-matrix is the inter-sublattice hopping matrix of the model having dimension $N_{L}\times N_{\tilde{L}}$, denoted by $S(\bm{k})$. The tight-binding (TB) Hamiltonian with only inter-sublattice hoppings has the form
\begin{equation}
    H(\bm{k})=
    \left(
    \begin{matrix}
	\bm{0}_{N_L} &  S(\bm{k}) \\
	S^\dagger(\bm{k}) &  \bm{0}_{N_{\tilde{L}}}  \\
    \end{matrix}
    \right),
\end{equation}
where each entry denotes a matrix block, with $\bm{0}_N$ being the zero matrix of dimension $N\times N$.
This Hamiltonian has chiral symmetry $C$ with the representation matrix
\begin{equation}
    D(C)=\text{Diag}(\bm{1}_{N_L}, -\bm{1}_{N_{\tilde{L}}}),
\end{equation}
i.e., $D(C)H(\bm{k})D^{-1}(C)=-H(\bm{k})$.
The chiral symmetry enforces the dispersion to be chiral-symmetric, which results in at least $N_L-N_{\tilde{L}}$ perfectly flat bands pinned at zero energy. If the rank of the inter-sublattice hopping matrix obeys $\text{rank}(S_{\bm{k}})=r_s\leq N_{\tilde{L}}$, then there will be $2r_s$ chirally-symmetric dispersive bands, leading to $N_L+N_{\tilde{L}}-2r_s$ perfectly flat bands at zero energy.

When intra-sublattice hoppings are added, i.e.,
\begin{equation}
    H(\bm{k})=
    \left(
    \begin{matrix}
	A(\bm{k}) &  S(\bm{k}) \\
	S^\dagger(\bm{k}) &  B(\bm{k})  \\
    \end{matrix}
    \right),
\end{equation}
the Hamiltonian is no longer chiral-symmetric. If $A(\bm{k})$ has a $\bm{k}$-independent eigenvalue of multiplicity $n_a$ ($n_a>N_{\tilde{L}}$), then $H(\bm{k})$ will have at least $n_a-N_{\tilde{L}}$ perfectly flat band, whose energy is not necessarily zero.